\documentclass[11pt,a4paper]{article}
\usepackage{amsmath, amsfonts,amssymb,mathtools,nicefrac,nccmath,cases}
\usepackage{microtype,booktabs,multirow}
\usepackage[usenames,dvipsnames,table]{xcolor}
\usepackage{colortbl}
\usepackage{tikz}
\usetikzlibrary{shapes,arrows}
\usetikzlibrary{calc,shapes.callouts,shapes.arrows,bending,intersections}
\usetikzlibrary{shapes.geometric}
\usetikzlibrary{arrows.meta,arrows}
\usetikzlibrary{decorations.pathmorphing}
\usetikzlibrary{decorations.markings}
\usetikzlibrary{shapes.multipart}
\usetikzlibrary{tikzmark}
\usepackage{caption}
\usepackage{subcaption}
\usepackage{bm}
\usepackage[nosort]{cite}
\usepackage{colortbl}
\usepackage{amsthm}
\usepackage{soul}

\usepackage{xcolor}
\usepackage{lipsum}

\usepackage[linktocpage=true]{hyperref}
\hypersetup{colorlinks=true,linkcolor=nicecolor,citecolor=nicecolor,urlcolor=nicecolor}
\definecolor{nicecolor}{rgb}{0.1, 0.3, 0.4}

\definecolor{blue}{rgb}{0.06, 0.3, 0.57}
\definecolor{Gray}{gray}{0.4}

\definecolor{nicecolor}{rgb}{0.1, 0.3, 0.4}
\definecolor{blue}{rgb}{0.06, 0.3, 0.57}
\definecolor{Gray}{gray}{0.4}
\colorlet{tableheadcolor}{gray!15} % Table header colour = 25% gray
\colorlet{tablerowcolor}{gray!7} % Table row separator colour = 10% gray

\def\hybrid{\topmargin -20pt    \oddsidemargin 0pt
	\headheight 0pt \headsep 0pt
	\textwidth 6.5in        % US paper
	\textheight 9in         % US paper
	\textwidth 6.25in       % A4 paper
	\textheight 9 in       % A4 paper
	\marginparwidth .875in
	\parskip 5pt plus 1pt 
	\jot = 1.5ex
}
\hybrid
\numberwithin{equation}{section}
\numberwithin{table}{section}
\usepackage{float}
\usepackage{tabularx}
\usepackage{multirow}
\usepackage{hhline}
\usepackage{stackengine}
\newcommand\xrowht[2][0]{\addstackgap[.5\dimexpr#2\relax]{\vphantom{#1}}}

\newcolumntype{D}{>{\centering\arraybackslash}X}
\newcolumntype{L}{>{$}l<{$}}
\newcolumntype{R}{>{$}r<{$}}
\newcolumntype{C}{>{$}c<{$}}
\usepackage{IEEEtrantools}
\newcommand{\beq}{\begin{equation}}  \newcommand{\eeq}{\end{equation}}
\newcommand{\bal}{\begin{aligned}}   \newcommand{\eal}{\end{aligned}}
\newcommand{\bea}{\begin{eqnarray}}  \newcommand{\eea}{\end{eqnarray}}
\def\beqa{\begin{eqnarray}}
\def\eeqa{\end{eqnarray}}

\newcommand{\bmat}{\left(\begin{array}}
\newcommand{\emat}{\end{array}\right)}

\newcommand{\bbC}{\mathbb{C}}
\newcommand{\bbR}{\mathbb{R}}

\newcommand{\cO}{\mathcal{O}}
\newcommand{\cT}{\mathcal{T}}
\newcommand{\cE}{\mathcal{E}}
\newcommand{\cP}{\mathcal{P}}
\newcommand{\cC}{\mathcal{C}}
\newcommand{\cD}{\mathcal{D}}

\newcommand{\cS}{\mathcal{S}}
\newcommand{\cK}{\mathcal{K}}
\newcommand{\cN}{\mathcal{N}}

\newcommand{\cG}{\mathcal{G}}
\newcommand{\cA}{\mathcal{A}}
\newcommand{\cH}{\mathcal{H}}
\newcommand{\cB}{\mathcal{B}}
\newcommand{\cF}{\mathcal{F}}
\newcommand{\cI}{\mathcal{I}}
\newcommand{\cJ}{\mathcal{J}}

\newcommand{\MM}{\mathcal{M}}
\newcommand{\cM}{\mathcal M}
\newcommand{\cQ}{\mathcal Q}

\usepackage{cancel}

\newcommand{\be}{\begin{equation}}
\newcommand{\ee}{\end{equation}}

\newcommand{\half}{\frac{1}{2}}

\newcommand{\bbZ}{\mathbb{Z}}

\definecolor{Gray}{gray}{0.95}
\definecolor{darkspringgreen}{rgb}{0.09, 0.45, 0.27}
\definecolor{darkseagreen}{rgb}{0.56, 0.74, 0.56}
\definecolor{darkmouthgreen}{rgb}{0.05, 0.5, 0.06}
\definecolor{darkcyan}{rgb}{0.0, 0.55, 0.55}

\newtheorem{conjecture}{Conjecture}

\def\d {{\rm d}}
\def\del          {\partial}
\def\delbar       {\bar\partial}
\def\ii           {{\rm i}}

\def\Re           {{\rm Re\hskip0.1em}}
\def\Im           {{\rm Im\hskip0.1em}}

\def\calh         {{\cal H}}

\def\calk         {{\cal K}}

\def\calo         {{\cal O}}

\def\calt         {{\cal T}}

\usepackage{xcolor}
\definecolor{colorloc1}{RGB}{0,0,102}  
\definecolor{colorloc2}{RGB}{0,125,253} 
\usepackage[framemethod=default]{mdframed}
\newmdenv[skipabove=10pt,
skipbelow=7pt,
rightline=false,
leftline=true,
topline=false,
bottomline=false,
linecolor=colorloc1,
backgroundcolor=colorloc2!5,
innerleftmargin=4pt,
innerrightmargin=0pt,
innertopmargin=0pt,
leftmargin=2pt,
rightmargin=0pt,
linewidth=2pt,
innerbottommargin=0pt]{lbBox}

\begin{document}

\baselineskip=14pt
\parskip 5pt plus 1pt

\vspace*{-1.5cm}
\begin{flushright}    % Publication numbers
  {\small 
  IFT-UAM/CSIC-21-36
  }
\end{flushright}

\vspace{2cm}
\begin{center}        % Main title

  {\huge The EFT stringy viewpoint on large distances \\
   [.3cm]  }
  %  {\huge Strings and Membranes in the Landscape}
\end{center}

\vspace{0.5cm}
\begin{center}        % Authors
{\large  Stefano Lanza,$^{1}$ Fernando Marchesano,$^2$ Luca Martucci,$^3$ and Irene Valenzuela$^4$}
\end{center}

\vspace{0.15cm}
\begin{center}  
\emph{$^1$Institute for Theoretical Physics,
Utrecht University}\\
\emph{Princetonplein 5, 3584 CE Utrecht, The Netherlands}\\[2mm] 
$^2$ \emph{Instituto de F\'{\i}sica Te\'orica UAM-CSIC, Cantoblanco, 28049 Madrid, Spain} 
\\[2mm] 
${}^3$ \emph{Dipartimento di Fisica e Astronomia ``Galileo Galilei",  Universit\`a degli Studi di Padova} \\ 
\emph{\& I.N.F.N. Sezione di Padova, Via F. Marzolo 8, 35131 Padova, Italy} 
\\[2mm]
${}^4$\emph{Jefferson Physical Laboratory, Harvard University, 
	Cambridge, MA 02138, USA}\\[.3cm]

\end{center}

\vspace{2cm}

%%%%%%%%%%%%%%%%%%%%%%%%%%%%%%%%%%%%%%%%%%%%%%%
%%%%%%%%%%%%%%%%%%%%%%%%%%%%%%%%%%%%%%%%%%%%%%%
%%%%%%%%%%%%%%%%%%%%%%%%%%%%%%%%%%%%%%%%%%%%%%%
%%%%%%%%%%%%%%%%%%%%%%%%%%%%%%%%%%%%%%%%%%%%%%%
%%%%%%%%%%%%%%%%%%%%%%%%%%%%%%%%%%%%%%%%%%%%%%%
%%%%%%%%%%%%%%%%%%%%%%%%%%%%%%%%%%%%%%%%%%%%%%%
%%%%%%%%%%%%%%%%%%%%%%%%%%%%%%%%%%%%%%%%%%%%%%%
%%%%%%%%%%%%%%%%%%%%%%%%%%%%%%%%%%%%%%%%%%%%%%%

\begin{abstract}

\noindent We observe a direct relation between the existence of fundamental axionic strings, dubbed EFT strings, and infinite distance limits in 4d $\mathcal{N}=1$ EFTs coupled to gravity. The  backreaction of EFT strings  can be interpreted as  RG flow of their couplings, and allows one to probe different regimes within the field space of the theory. We propose that any 4d EFT infinite distance limit can be realised as an EFT string flow. We show that along such limits the EFT string becomes asymptotically tensionless, and so the EFT eventually breaks down. This provides an upper bound for the maximal field range of an EFT with a finite cut-off, and reproduces the Swampland Distance Conjecture from a bottom-up perspective. Even if there are typically other towers of particles becoming light, we propose that the mass of the leading tower scales as $m^2\sim \mathcal{T}^w$ in Planck units, with $\mathcal{T}$ the EFT string tension and $w$ a positive integer.  Our results hold even in the presence of a non-trivial potential, as long as its energy scale remains well below the cut-off.  We check both proposals for large classes of 4d $\mathcal{N}=1$ string compactifications, finding that only the values $w=1,2,3$ are realised.  

\end{abstract}

\thispagestyle{empty}
\clearpage

\setcounter{page}{1}

%%%%%%%%%%%%%%%%%%%%%%%%%%%%%%%%%%%%%%%%%%%%%%%
%%%%%%%%%%%%%%%%%%%%%%%%%%%%%%%%%%%%%%%%%%%%%%%
%%%%%%%%%%%                 %%%%%%%%%%%%%%%%%%%
%%%%%%%%%%%  DOCUMENT BODY  %%%%%%%%%%%%%%%%%%%
%%%%%%%%%%%                 %%%%%%%%%%%%%%%%%%%
%%%%%%%%%%%%%%%%%%%%%%%%%%%%%%%%%%%%%%%%%%%%%%%
%%%%%%%%%%%%%%%%%%%%%%%%%%%%%%%%%%%%%%%%%%%%%%%
%%%%%%%%%%%%%%%%%%%%%%%%%%%%%%%%%%%%%%%%%%%%%%%

\newpage

  \tableofcontents

%\newpage

%%%%%%%%%%%%%%%%%%%%%%%%%%%%%%%%%%%%%%%%%%%%%%%
\section{Introduction}
\label{sec:intro}

One pressing problem for Effective Field Theories (EFTs) of quantum gravity is to determine which kind of physics one can obtain at regions of weak coupling. Since these regimes appear along large field distance limits, one may be led to assume that the size of the EFT field space can be arbitrarily large, which is very suggestive for, e.g., embedding macroscopic models of large field inflation in a UV complete theory of quantum gravity. Concrete string theory realisations however show that asymptotic limits of infinite distance come with a mechanism that prevents a 4d EFT description beyond a certain point, as predicted by the Swampland Distance Conjecture (SDC) \cite{Ooguri:2006in}. As a consequence of this conjecture, quantum gravity would impose a maximal size to any EFT field space. Determining what this size is and what is the precise mechanism that triggers the EFT breakdown is a central part of the Swampland Program \cite{Vafa:2005ui,Brennan:2017rbf,Palti:2019pca,vanBeest:2021lhn}, and substantial progress has been achieved through different studies and refinements of the SDC \cite{Grimm:2018ohb,Lee:2018urn,Grimm:2018cpv,Corvilain:2018lgw,Joshi:2019nzi,Marchesano:2019ifh,Font:2019cxq,Grimm:2019wtx,Lee:2019xtm,Lee:2019wij,Baume:2019sry,Gendler:2020dfp,Xu:2020nlh,Grimm:2020cda,Klaewer:2020lfg,Baume:2020dqd,Bastian:2020egp,Perlmutter:2020buo,Calderon-Infante:2020dhm}.

The standard strategy to test and improve our insight on the SDC is to consider a particular string theory compactification, and then attempt to classify and understand the physics of each of its infinite distance limits. While this has been a very successful approach, a more intrinsic EFT viewpoint on these asymptotic large distance limits and their associated phenomena is still lacking. Such a description would undoubtedly put the SDC and all the swampland conjectures connected to it on a firmer ground, even beyond the string theory realm. A promising avenue to achieve this goal is to consider large field variations induced by the presence of localised objects in our EFT, like black holes or bubbles of nothing, as done in \cite{Klaewer:2016kiy,Buratti:2018xjt,Draper:2019zbb,Geng:2019bnn,Draper:2019utz,Geng:2019zsx,Bonnefoy:2019nzv,Gendler:2020dfp}. In this spirit, it was  pointed out in \cite{Lanza:2020qmt} that for 4d EFTs most of the swampland conjectures are connected to the physics of low-codimension objects, namely strings and membranes. In particular, certain $\half$BPS strings are directly related to the presence of infinite distance limits in 4d $\cN=1$ EFTs.

A connection between strings and large field distances is already hinted by the 4d cosmic string solutions of \cite{cstring}. Indeed, in this work the authors construct codimension-two profiles of an axio-dilaton $\tau$, whose simplest solution asymptotes towards the infinite field distance limit $\tau \to \ii \infty$ at spatial infinity. It was later on realised in \cite{Vafa:1996xn} that the point at which $\tau = \ii \infty$ can be interpreted as the location of a D7-brane. From this F-theory perspective one may construct 4d string solutions by compactifying type IIB on a six-manifold $X$, wrapping several $(p,q)$ 7-branes on $X$ and placing them parallel to each other in 4d. When approaching a D7-brane location the axio-dilaton profile will draw an infinite distance path $\tau \to \ii \infty$ in field space. From the 4d EFT viewpoint, such profile is sourced by the presence of a localised operator, which is how D7-branes are treated in type IIB supergravity. Moreover, the solution around the D7-brane  describes a monodromy $\Re \tau \to \Re \tau + 1$, which is promoted to an axionic shift symmetry as we approach the D7-brane core. In other words, the source of the profile $\tau \to \ii \infty$ is a 4d fundamental string magnetically coupled to the axion $\Re \tau$.

The purpose of this work is to characterise 4d EFT large field distance limits in terms of the key EFT ingredients of the above example. More precisely, we consider fundamental axionic strings, treated as codimension-two operators of the EFT, and study their backreaction profile in the vicinity of the string core. In the context of $\cN=1$ theories this backreaction describes a trajectory in a field space region with a perturbative axionic symmetry, with a logarithmic profile for the saxionic partner. Such a profile allows to map the physics of the backreacted solution to the physics of vacua up to the distance $1/\Lambda$ from the string core, with $\Lambda$ the EFT cut-off scale. As we approach the core %\irene{[Careful with 'approaching the string core', it is about the resolution scale, not about approaching the core]} 
we asymptotically reach a regime in which those non-perturbative effects that break the axionic symmetry become negligible, pointing to the emergence of a continuous shift symmetry. By the no global symmetry conjecture \cite{Banks:1988yz,Banks:2010zn}, one expects that for EFTs coupled to quantum gravity such exact shift symmetries are located at infinite distance points in field space, with some mechanism preventing the EFT to ever reach them, similarly to the setup in \cite{Reece:2018zvv}. 
%\irene{[Maybe jumping from picture of string backreaction to moving on moduli space too quickly? ]} 
Therefore, these solutions are particularly suitable to study the physics of the SDC from an EFT perspective. In fact, the properties of these strings guarantee that the associated solutions can always be treated in an EFT weakly-coupled regime, which is why we refer to them as {\it EFT strings} throughout the paper.

Indeed, one difference with  \cite{cstring} is that in our case  the fields that vary along a $\half$BPS string flow will generically be subject to a potential. However, we argue that if the Hubble and mass scales of the potential are low compared with the EFT cut-off scale $\Lambda$, so that near this scale it makes sense to talk about a field space, our approach to extract the physics of large field distances still applies. One way to see this is to apply the philosophy of \cite{Michel:2014lva,Polchinski:2015bea} to our context, and interpret the string backreaction as a renormalisation group (RG) flow of the string couplings. In this scheme, the backreaction details closer to the string core determine the string couplings at higher energies, and also the asymptotic large field behaviour of our theory. Therefore, in a sensible setup the interesting part of the EFT string solution will be at wavelengths much shorter than the effect induced by a potential. The RG flow picture also helps to understand the universal behaviour of the EFT string tension $\cT$, which asymptotically vanishes either as we increase $\Lambda$ or we approach an infinite distance point. This monotonic behaviour provides a rationale for the EFT breakdown that the SDC predicts. Indeed, if an EFT string is a genuinely fundamental object of our 4d EFT, then it cannot be resolved by it for any scale $\Lambda$, and so by consistency the EFT semiclassical description must break down  before $\cT < \Lambda^2$.

Another generalisation with respect to \cite{cstring} is the arbitrary number of different perturbative axionic shift symmetries that may coexist in given region of a 4d $\cN=1$ EFT. We find that this richness is encoded in a lattice convex cone of EFT string charges, whose interplay with the saxionic field space of this region allows us to classify different kinds of string flows, which ultimately reflect the asymptotic properties of the K\"ahler potential. 
We have indeed checked that this structure is realised in a plethora of 4d $\cN=1$ examples built from string compactifications, some of which allow us to connect 4d EFT strings with the definition of {\it supergravity strings} recently formulated in \cite{Katz:2020ewz} for 5d theories.

This set of universal properties that stem from the notion of EFT string prompt us to propose the {\em Distant Axionic String Conjecture}, which essentially claims that all 4d EFT infinite distance limits are realised by the RG flow associated to an EFT string. This proposal was already advanced in \cite{Lanza:2020qmt}, and in the present paper is captured by Conjecture \ref{conj:DASC}. Thereafter, we will discuss at some length several implications of this conjecture, like for instance how it constrains the behaviour of the field space curvature at infinity. 

Obviously, the most direct consequences involve the SDC. As said before, EFT strings are asymptotically tensionless, and thus any consistent 4d EFT must break down along the infinite distance trajectories that they describe. This observation does not specify the precise breakdown mechanism: it could be via the tower of oscillation modes of the string itself, or via some other tower of states -- e.g. Kaluza-Klein (KK) modes -- that appear at a lower scale, as observed in the context of 4d $\cN=2$  theories in \cite{Lee:2019wij}. We analyse a large set of asymptotic limits in different classes of 4d $\cN=1$ string compactifications, finding that in each case it is  either a tower of KK-like modes or the oscillation modes of an EFT string that trigger the EFT breakdown, strengthening the Emergent String Conjecture  \cite{Lee:2019wij}. Furthermore, whenever the leading tower is not given by the oscillation modes of the EFT string, the asymptotic behaviour of the scale $m_*$ of this tower is tied up to the asymptotics of the EFT string tension that describes this limit. We believe this to be a general feature of 4d EFTs compatible with quantum gravity, with the precise relation between $m_*$ and the EFT string tension as captured by Conjecture \ref{conj:cutoff}, which states that in Planck units $m_*^2\sim \cT^w$ for some scaling weight $w \in \mathbb{Z}_{>0}$, asymptotically along each EFT string flow. Finally, as already pointed out in \cite{Lanza:2020qmt}, these two conjectures together with the Weak Gravity Conjecture (WGC) for EFT strings imply the SDC in 4d theories, with the exponential descent rate of $m_*$ specified by the EFT string extremality factor and the scaling weight.  Therefore, if this set of proposals is true, one could essentially estimate the maximal cut-off $m_*$ of a 4d EFT theory simply using EFT data! 

The paper is organised as follows. In section~\ref{s:fundamental} we describe fundamental axionic strings and their properties from the viewpoint of 4d EFTs, which lead to the definition of EFT string. In section~\ref{sec:strinfdis} we argue why EFT string solutions are directly related to trajectories of infinite distance in the EFT field space, giving them an RG flow interpretation. In section~\ref{s:instantons} we discuss the interplay between $\half$BPS strings and non-perturbative corrections, and capture them in terms of continuous and discrete conical structures. Section~\ref{s:conjecture} contains two of the main results of the paper, namely conjectures \ref{conj:DASC} and \ref{conj:cutoff}. They characterise the asymptotic limits of infinite distance of a 4d EFT in terms of EFT string flows, and have interesting implications for several swampland criteria. Section~\ref{s:examples} shows how these two conjectures and all the EFT structure developed to arrive to them are realised in several  classes of string compactifications, providing abundant  examples that support our claims. We finally draw our conclusions in section~\ref{s:conclusions}.

Several details have been relegated to the appendices.
Appendix \ref{app:glossary} contains a glossary of terms for string flows and for cones in algebraic  geometry  used throughout the paper.
Appendix \ref{app:graveff} shows that the gravitational contribution to the EFT string  energy density does not affect the results of section \ref{s:fundamental}. Appendix \ref{ap:geodesic} argues that EFT string flows correspond to asymptotically geodesic paths in field space. Appendix \ref{ap:BD} elaborates on the correspondence between EFT string solutions and their RG flows, by computing the latter from a field theory perspective.  Appendix \ref{app:inst} discusses the backreaction and flows associated to 4d $\half$BPS instantons.  Appendix \ref{app:toroidal} examines in detail string flows in different toroidal orbifold compactifications and their resolutions.

\section{Fundamental axionic strings in 4d \texorpdfstring{\hbox{$\mathcal{N}=1$}}{N=1} EFTs}
\label{s:fundamental}

In this section we describe, from a 4d ${\cal N}=1$ EFT perspective, solutions that correspond to $\half$BPS axionic fundamental strings. These strings are characterised by the following properties:
\begin{itemize}
    \item[-] They are fundamental localised objects, in the sense that they have a singular core that cannot be resolved within a quantum field theory approach. This is to distinguish them from other solitonic strings that can be described  within the EFT.
    \item[-] They are magnetically coupled to axions and enjoy an approximate continuous shift symmetry near their core. With \emph{approximate} we mean that it should be preserved at perturbative level by the K\"ahler potential and only broken by exponentially suppressed corrections. Hence, the axion behaves as a 0-form gauge field.
   
\end{itemize}

This allows one to describe their solution by means of a dual formulation, where the axion is replaced by the 2-form gauge field $\cB_2$ under which the strings are electrically charged. Preserving the $\mathcal{N}=1$ supersymmetric requires that the chiral multiplets are replaced by linear multiplets. One of the vantage points of working in this dual picture is that we may easily compute the string tension $\cT$, which should satisfy
\be\label{EFTregime}
 \Lambda^{2} <  \cT < M^{2} _{\rm P}\, ,
\ee 
in order not to imply the EFT breakdown, where $\Lambda$ is the EFT cut-off scale. Therefore, such strings must be included as localised operators in the theory, as they cannot be resolved within the 4d EFT regime of validity.

\subsection{Axionic strings in \texorpdfstring{\hbox{${\cal N}=1$}}{N=1} EFTs}
\label{ss:axionic}

Let us consider a 4d ${\cal N}=1$ EFT formulated in terms of chiral multiplets. Axionic string configurations naturally appear when the set of chiral fields can be split as $\{\phi^\alpha\} = \{ t^i, \chi^\kappa\}$, with the superpotential $W \equiv W(\chi)$ only depending on a subset of them, and a K\"ahler potential $K$ that for now we keep general. Under these circumstances, one may find a moduli space of supersymmetric vacua with a fixed value $\chi_*$ for the scalars $\{\chi^\kappa\}$ and arbitrary value for the fields $t^i$. Indeed, since the F-flatness conditions imply
\be
D_{t^i} W|_{\rm vac} = (\partial_{t^i} K) W|_{\rm vac} = 0\, ,
\label{Whip}
\ee
then the supersymmetry conditions amount to impose $W(\chi_*) = \partial_{\chi^\kappa} W(\chi_*)  =0$, leaving the $\{t^i\}$ unconstrained. Moreover, the Cremmer et al.\ \cite{Cremmer:1982en} F-term potential $V( t, \chi)$  evaluated at $\chi_*$ vanishes at least quadratically on the chiral fields, and so it can be ignored when considering string-like configurations with a varying profile on the $\{t^i\}$. A similar statement holds for no-scale models with a semidefinite positive potential and a moduli space of Minkowski vacua.

Ignoring the presence of $V$, the relevant terms of  the effective action  are
\be
S=M^2_{\rm P}\int \left(\frac{1}{2}R*1-K_{\alpha\bar \beta}\,\d \phi^\alpha\wedge*\d\bar\phi^{\bar \beta} \right)\, ,
\label{effaction}
\ee 
whose string-like solutions can be analysed following the discussion in \cite{cstring}. For this one splits the 4d coordinates into $(t,x,z,\bar z)$, and imposes 2d Poincar\'e invariance on $(t,x)$. That is,  we allow the varying fields $t^i$ to depend only on $z,\bar z$ and choose a 4d metric of the form 
\be\label{metric}
\d s^2= -\d t^2+\d x^2+ e^{2D}\d z\d\bar z\, ,
\ee
where $D$ depends only on the transverse coordinates $(z,\bar z)$. 
The equations of motion read
\be
K_{\bar \alpha \beta}\del\delbar \phi^\beta+K_{\bar \alpha \beta \gamma}\del \phi^\beta\wedge {\bar\del}\phi^{ \gamma}=0\, ,
\label{eomstring}
\ee
where  $\del\equiv \d z\wedge\del_z$. The simplest class of solutions corresponds to holomorphic
\be\label{hol1}
\delbar\phi^\alpha=0\, ,
\ee
or anti-holomorphic profiles: $\del\phi^{\alpha}=0$. In our setup, the choice \eqref{hol1} amounts to set  $\chi \equiv \chi_*$ and 
\be\label{hol2}
\delbar t^i=0\, .
\ee
We can then regard $t^i (z)$ as defining a holomorphic map from $\mathbb{C}$ to the moduli space $\MM$.  
From the Einstein equations it also follows that 
\be
e^{2D}=|f(z)|^2 e^{-K}\, ,
\label{stringwarp0}
\ee
with $f(z)$ a holomorphic non-vanishing function \cite{cstring}. 
One may also derive condition \eqref{hol1} by writing the energy per unit length of a solution  in a BPS form:
\be\label{energysol}
\cE = M_{\rm P}^{2} \int K_{\alpha \bar \beta}\left(\del\phi^\alpha \wedge *_2\delbar\bar\phi^{\bar \beta}+\delbar\phi^\alpha\wedge *_2\del\bar\phi^{\bar \beta}\right)=M_{\rm P}^{2}\int J_\MM + 2M_{\rm P}^{2}\int K_{\alpha\bar \beta}\,\delbar\phi^\alpha\wedge *_2\del\bar\phi^{\bar \beta}\, ,
\ee
with the integral performed over the $(z, \bar{z})$-plane, where  the Hodge star operator $*_2$ acts, and $ J_\MM \equiv \ii K_{\alpha\bar \beta}\,\d\phi^\alpha\wedge \d\bar\phi^{\bar \beta}$. 
For fixed boundary conditions the first term is topological, while the second is positive definite. We can then obtain the BPS bound
\be\label{EBPSbound}
\cE\geq  M_{\rm P}^{2}\int J_\MM \, ,
\ee
which is saturated for holomorphic  profiles.

Local solutions to \eqref{hol2} corresponding to strings located at $z=0$ and associated to a monodromy of the form 
\be\label{tmon}
t^i\rightarrow t^i +e^i\, , \qquad \qquad e^i \in \bbZ\, ,
\ee
are given by
\be\label{tsol}
t^i=t_0^i+\frac{1}{2\pi\ii}e^i\log \left(\frac{z}{z_0}\right)\, ,
\ee
where $t_0^i, z_0$ are constant.
Writing $t^i = a^i+ \ii  s^i$ and  $z=re^{\ii\theta}$ we can split this solution as
\begin{subequations}
\label{solsplit}
\bea\label{imt}
s^i & = & s_{0}^i-\frac1{2\pi}e^i\log \left(\frac{r}{r_0}\right)\, ,\\
\label{axionmon}
a^i &= &\frac{\theta}{2\pi}\,e^i+\text{const}\, ,
\eea
\end{subequations}
where, later on, $a^i$ is to be interpreted as an axion and $s^i$ as its saxionic partner. We see that the latter develop a harmonic profile with a logarithmic singularity at the location of the string. 
More precisely,
\be\label{infs}
s^i\rightarrow  e^i \cdot \infty \quad~~~~~\text{for}\quad r\rightarrow 0\,.
\ee
This implies that when approaching the strings location certain scalars within $\{t^i\}$ are driven towards the boundary of the moduli space patch that they describe.  

Let us for simplicity assume  that  $e^i\geq 0$, $\forall i$ -- the more general case works analogously and will be discussed in detail in the following sections. By considering $s_{0}^i \gg 1$ and $M_{\rm P} r_0\gg 1$, the string solution will describe a large saxion profile $s^i$ along a range of $r$ much larger than the four-dimensional Planck length, namely $r\leq r_0$. 
 Within this region, by appropriately redefining $z$ one may set the non-vanishing holomorphic function in \eqref{stringwarp0} to a constant: 
\be\label{stringwarp}
e^{2D}=e^{K_0-K(t,\chi_*)}\,  ,
\ee
which we choose as $K_0\equiv K(t_0,\chi_*)$. In this way the normalisation of the warp factor is fixed as $e^{2D}|_{z_0}=1$, and only $t_{0}^i$ and $z_0$ remain as free adjustable parameters of the solution. Moreover, let us assume that the trajectory \eqref{infs} takes the EFT to a region of field space in which $K$ exhibits an approximate continuous shift symmetries of the form $\Re t^i\rightarrow \Re t^i+ $const, so that the $a^i = \Re t^i$ involved in \eqref{axionmon} can be considered to be axions in this limit. As a consequence, the solution for the metric will display a radial symmetry, and the solution \eqref{tsol} then corresponds to a 4d $\half$BPS axionic string. 
 
Finally, the single string solution can be easily generalised to a multi-string solution:
\be
t^i=t_0^i+\frac1{2\pi\ii} \sum_a e_a^i \log\left(\frac{z-z_a}{z_0-z_a}\right)\, ,
\ee
where $e_a^i$ are the charges of the string located at $z_a$ in the $z$-plane, and $t_0^i$ is the value of $t^i$ at $z_0$.  If $s^i_0\equiv \Im t_0^i$ is large and charges $e^i\geq 0$, then $s^i$ will remain large in the domain $\bigcap_{a}D_a$, where $D_a=\{|z-z_a|\leq |z_0-z_a|\}$, which is non-vanishing if the strings are sufficiently close to each other. This shows that they are in equilibrium and do not exert any force among them, as expected for mutually $\half$BPS objects.

\subsection{Dual formulation}
\label{sec:dualform}

The above description of $\half$BPS axionic strings involves a  K\"ahler potential which, near their core, is invariant under axionic shifts $a^i\rightarrow a^i+c^i$, and then depends on the chiral field $t^i$ only through its saxionic component $s^i=\Im t^i$.   In this case,  
the local string solution admits a dual description in terms of {\em dual saxions} $\ell_i$ and two-form potentials $\cB_{2\, i}$ defined as
\be\label{dualfields}
\ell_i=-\frac12 \frac{\del K}{\del s^i}\, , \qquad \cH_{3\, i} = \d  \cB_{2\, i} =-M^2_{\rm P}\, \cG_{i j}*\d a^j\, ,
\ee
where 
\be\label{GGmetric}
 \cG_{i j}\equiv \frac12 \frac{\del^2 K}{\del s^i\del s^j}\ .
\ee 
More precisely, by using a metric of the form \eqref{metric}, the BPS condition $\delbar t^i=0$ implies that
\be\label{H3ell}
\cH_{3\, i}=-   M^2_{\rm P}\,\d t\wedge \d x\wedge\left(\cG_{ij}\,\d s^j\right)=M^2_{\rm P}\,\d t\wedge \d x\wedge \d \ell_i\, ,
\ee
so that we can identify
\be\label{B2id}
\cB_{2\, i}=M^2_{\rm P}\,  \ell_i\d t\wedge \d x\, ,
\ee
up to a closed contribution. Therefore, as expected, strings transverse to the $(z, \bar z)$-plane are electrically charged under the  two-form potentials $\cB_{2\, i}$. The dual action describing a single string with charges $e^i$ is given by \cite{Lanza:2019xxg,Lanza:2019nfa}
\be\label{dualaction}
-\frac12\int \cG^{ij}\left( M^2_{\rm P}\,\d\ell_i\wedge *\d\ell_j+\frac{1}{M^2_{\rm P}}\cH_{3\, i}\wedge * \cH_{3\, j} \right)+S_{\rm string}\, ,
\ee
with
\be\label{stringS}
S_{\rm string}=-\int_\cS \d^2\xi\, \cT_{\bf e}\sqrt{-h}+e^i \int_\cS \cB_{2\, i}\, ,
\ee
where $\xi^a$ are  coordinates along the world-sheet $\cS$, $h_{ab}$ is the pulled-back world-sheet metric  and
\be\label{thetension}
\cT_{\bf e}\equiv M^2_{\rm P}\,e^i\ell_i
\ee
is the field-dependent string tension.

By using \eqref{metric}, the corresponding equations of motions are 
\be\label{dualfloweq}
\begin{aligned}
\d\left(\cG^{ij}*\d\ell_j\right)&=e^i \d t\wedge \d x\wedge \delta_2(\cS)\ ,\\
\d\left(\cG^{ij}*\cH_{3\, j}\right)&=- M^2_{\rm P}\,e^i \delta_2(\cS)\, .
\end{aligned}
\ee
From \eqref{H3ell} we get $*\cH_{3\, j}=- M^2_{\rm P} *_2\d \ell_j$ and then both equations reduce to  
\be
\d\left(\cG^{ij}*_2\d\ell_j\right)=-\d*_2\d s^i=-2  \ii\del\delbar s^i=e^i \delta_2(\cS)\, ,
\ee
which is indeed satisfied by \eqref{imt}. 

In this formulation it is clear that the physical consistency condition $e^i\ell_i> 0$ must be satisfied at each point in field space.  In fact, for each $\frac12$BPS string of charges $e^i$ satisfying  $e^i\ell_i> 0$, there is an anti-string of charges $\bar e^i=-e^i$, which preserves the opposite fraction of  supersymmetry. Its contribution to the EFT is obtained by replacing $e^i\rightarrow \bar e^i$ and $e^i\ell_i\rightarrow -\bar e^i\ell_i=e^i\ell_i> 0$ in \eqref{stringS}. The corresponding flow solution is given by the anti-holomorphic counterpart of \eqref{tsol}: $t^i=t^i_0-\frac{1}{2\pi\ii}\bar e^i\log\left(\frac{\bar z}{\bar z_0}\right)$.

It is important to realise  that a correct physical interpretation of the localised  string terms appearing in \eqref{dualaction} and \eqref{dualfloweq}  strongly depends on the metric structure of the moduli space  around the point $s^i=e^i\times\infty$. Take for instance the flat K\"ahler potential $K=C\phi\bar\phi$ for $\phi=e^{2\pi\ii t}$. In the $t$ chiral coordinate the metric looks degenerate at $s=\Im t=\infty$, but this is just a coordinate singularity. In this case the localised terms in \eqref{dualaction} and \eqref{dualfloweq} are just the dual manifestation of this degenerate parametrisation of field space. Indeed, by using the more natural $\phi$ coordinate, the local string flow solution \eqref{tsol} corresponding to an elementary charge $e=1$ is just given by the perfectly smooth local solution $\phi=\phi_0\,\frac{z}{z_0}$, in which no localised source appears. This local solution can then be considered  `solitonic'.\footnote{This solution may be easily completed into a finite energy configuration by considering $\phi$ as a local coordinate for $\mathbb{P}^1$ and  $K=C|\phi|^2$ as an approximation for  $|\phi|\ll 1$ of the Fubini-Study K\"ahler potential $K=C\log (1+|\phi|^2)$ on $\mathbb{P}^1$. By using \eqref{EBPSbound} one can easily see that the  tension of this BPS solitonic string is $2\pi C M^2_{\rm P}$.} A similar conclusion holds for any solution around a smooth finite distance point in field space. Instead in this paper we  will focus on fundamental localised  strings, which induce a flow of the saxions $s^i$ to infinite field space  distance.

\subsection{Energy-momentum tensor and tension}
\label{sec:linearen}

We now compute the components of the energy-momentum tensor for a single local string flow solution in both dual pictures. We start with the chiral picture where, by imposing the holomorphic condition \eqref{hol2}, one obtains the following energy-momentum tensor associated with the scalar sector
\be
\begin{aligned}
T_{tt}&=-T_{xx}=2M^2_{\rm P}\, K_{i\bar\jmath}\,e^{-2D}\del_z t^i\del_{\bar z}\bar t^{\bar\jmath}\, ,\\
T_{z\bar z}&=T_{z  z}=T_{\bar z \bar z}=0\, .
\end{aligned}
\ee
Then, by integrating it on a certain domain we get the associated linear energy density: 
\be
\label{Eback0}
\begin{aligned}
\cE_{\rm back}(r)&=\frac\ii2
\int_{D(r)} \d z\wedge \d \bar z\, e^{2D}T_{tt}=\ii M^2_{\rm P}\int_{D(r)} K_{i\bar\jmath}\,\del t^i\wedge \delbar\bar t^{\bar\jmath} =M_{\rm P}^2\int_{D(r)} J_{\MM}\, ,
\end{aligned}
\ee
where in the last equality we have omitted terms containing $\d\chi^\kappa$, $\d\bar\chi^{\bar\kappa}$, as they vanish on the string solution. As in \eqref{energysol}, $\cE_{\rm back}$ is related to the kinetic energy density of the string solution. Note that here we have only performed the integral on a disc $D(r)$ of radius $r$ in the $(z, \bar z)$-plane, where non-perturbative corrections are suppressed. In this region one may also assume an axionic symmetry, which further simplifies this result because then $K_{i\bar\jmath}=\frac12\cG_{ij}$ and so
\be
J_\MM|_{\rm sol}=\ii K_{i\bar\jmath}\,\d t^i\wedge \d \bar t^{\bar\jmath}=\cG_{ij}\,\d a^i\wedge \d s^{j}=\d \ell_i\wedge \d a^i = \frac1{2\pi}\,e^i\d \ell_i\wedge \d\theta\, .
\ee
The linear energy density associated with a disk of radius $r$ can then be written as
\be
\label{Eback}
\begin{aligned}
\cE_{\rm back}(r)&=M_{\rm P}^2\int_{D(r)} J_{\MM}=M_{\rm P}^2\,e^i\int^{r}_{0} \d\ell_i=M_{\rm P}^2\,e^i[\ell_i(r)-\ell_i(0)]\, .
\end{aligned}
\ee

There is, however, a second contribution to the energy-momentum tensor coming from the localised fundamental string. This contribution is more easily computed using the dual linear multiplet Lagrangian \eqref{dualaction}. In general, the contribution of the localised fundamental string is 
\be
T^{mn}_{\rm string}=- M^2_{\rm P}\, e^i\ell_i\, h^{ab}\del_aX^m\del_b X^n\frac{\sqrt{-h}}{\sqrt{-g}}\delta^2(\cS)\, .
\label{sttringcoup}
\ee
In our solution this amounts to
\be
T^{\rm string}_{tt}=-T^{\rm string}_{xx}=M^2_{\rm P}e^{-2D}e^i\ell_i\,\delta^2(r)\, ,
\ee
which contributes to the linear energy density as
\be
\cE_{\rm string}=\frac\ii2
\int \d z\wedge \d \bar z\, e^{2D}T^{\rm string}_{tt}=M^2_{\rm P}e^i\ell_i(0)\, .
\ee
Adding up both contributions we obtain
\be
\cE(r)\equiv \cE_{\rm back} +\cE_{\rm string}=M_{\rm P}^2\,e^i\ell_i(r)\, .
\label{linearE}
\ee
In obtaining this result we have implicitly neglected the gravitational contribution to $\cE(r)$. In appendix \ref{app:graveff} we argue that such a contribution leaves \eqref{linearE} unchanged.

Note that the expression for $\cE$ is identical  to the tension $\cT_{\bf e}$ that one typically assigns to a BPS string in an unperturbed vacuum, cf. \eqref{thetension}, for some choice of saxion vevs. In the string solution \eqref{solsplit}, the value of the saxions are determined by the choice of radius as $\ell_i=\ell_i(r)$. Therefore changing the radius corresponds to a flow in the value of the moduli $\ell_i$, and so to a flow in the string tension $\cT_{\bf e}$. The result \eqref{linearE} then implies that this tension flow is captured by the radial dependence of the linear energy density of the solution. As discussed in \cite{Lanza:2020qmt} and in the next section, this fits nicely with the 4d EFT description of string-like objects, if one interprets $1/r$ as the mass scale $\Lambda$ at which the effective string tension is computed.

\subsection{EFT strings and validity of the solution}
\label{sec:validity}

Given the local string solution \eqref{tsol} and \eqref{stringwarp}, one may reconsider its validity when some of the assumptions taken to derive it no longer apply. First of all, it is important to stress that the solution \eqref{tsol} is not unique. In principle one may add an arbitrary linear combination of terms of the form $e^{2\pi \ii m_i t^i}$, $m_i \in \mathbb{Z}$ without spoiling the monodromy \eqref{tmon}. Hence  the solution \eqref{tsol} is physically relevant if such terms can be neglected within a sufficiently large distance from the core of the string. In general this property depends on the specific form of the EFT K\"ahler potential and on the choice of string charges $e^i$.

Besides these intrinsic limitations of the above  string solution, one may consider the effect of a superpotential that does depend on all the fields $\phi^\alpha$, and in particular on the fields $t^i$. This induces a scalar potential $V$ that cannot be neglected in \eqref{effaction}, and which modifies the equations of motion and their solution. The question is under which circumstances the modification is small enough such that the above solution can still be considered to be a good approximation. In general, any potential $V$ can be easily associated with two kinds of field dependent energy scales: the mass $m\sim \sqrt{|\del^2 V|}$ and Hubble $H\sim \sqrt{V}/M_{\rm P}$ scales. Roughly speaking, we can then trust the  string solution \eqref{tsol} at length scales $L$ well below $1/m$ and $1/H$, within an error of order $mL$ or $HL$. Hence, as long as $m$ and $H$ remain below the EFT cut-off $\Lambda$, there is still a finite energy regime where we can trust the solution.
Quantifying the effect of a non-trivial potential in general is quite involved, although one can be more specific for the string solutions that we will consider in this paper.

In this work we will restrict to fundamental axionic strings, which imply the following two assumptions from an EFT perspective. First, the string singular core cannot be resolved with a 4d quantum field theory approach, so the string corresponds to a fundamental localised object in the theory. Second, the K\"ahler potential should preserve the continuous axionic shift symmetries at perturbative level in an expansion in $1/s$, where $s$ is the linear combination of saxions that diverges at the string core. In other words, the axionic shift symmetries should be broken only by those non-perturbative corrections of the form $e^{2\pi \ii m_i t^i}$ that are   exponentially  suppressed in a sufficiently large disk around the string core. In terms of the above string solutions this implies that we will focus on those of the form

\begin{equation}
    \cC^{\text{\tiny EFT}}_{\rm S} = \Bigg\{ \quad \parbox{23em}{\small{Flows \eqref{tsol} along which $K$ displays a perturbative axionic shift symmetry  and non-perturbative effects are suppressed}} \quad \Bigg\}\, .
    \label{CSEFT0}
\end{equation}
We will dub the corresponding strings as EFT strings, for reasons to become clear shortly. Note that, by definition, along an EFT  string flow solution we can neglect the  non-perturbative corrections which break the axionic symmetries and then we are legitimated to pass to the  dual formulation of subsection \ref{sec:dualform}. This means that  the EFT flows can be interpreted as sourced by fundamental localised strings charged  under the $\cB_{2\, i}$ gauge fields.

Both assumptions behind the definition of an EFT string  are actually correlated. The first one implies that the axion can be interpreted as a \emph{fundamental} axion (a 0-form gauge field) rather than an ordinary pseudo-Goldstone boson,  and it is then the gauge invariance of the 0-form gauge field which protects the theory from perturbative corrections spoiling the shift symmetry. 
If the axion is not fundamental, the dual description in terms of the $\cB_2$-field will stop being valid at some energy scale $\Lambda_{{\cal B}_2}$ associated with a finite number of new degrees of freedom.  Above this energy scale $\Lambda_{{\cal B}_2}$ the EFT gets UV completed into a different 4d EFT. In this case, the strings are solitonic in the sense that their core can be fully resolved within the 4d EFT above $\Lambda_{{\cal B}_2}$. Furthermore, the tension of strings charged under pseudo-Goldstone bosons is typically determined by the symmetry breaking scale. Contrary, the $\cB_2$-field description of a fundamental axion is always valid at least up to a scale $m_*$ at which the 4d EFT description breaks down because of an {\em infinite} number of new degrees of freedom.
So the string cannot be resolved within a 4d EFT approach, in which case we denote them as fundamental strings. Their tension is a priori arbitrary, although if we want to keep a semiclassical description of these objects we need to impose \eqref{EFTregime}. 

For fundamental strings, the gauge invariance of the axion should be preserved at any energy scale below $m_*$, implying that the  0-form global symmetry can only be broken in some specific ways that preserve the $(-1)$-form gauge invariance associated to a 0-form gauge field. This negative degree might sound strange, but notice that large  transformations of this gauge symmetry are simply the familiar discrete shifts of the axion, while the continuous shift correspond to the 0-form global symmetry. The 0-form global symmetry can be either broken by instantons (electrically charged states) or by coupling it to a \emph{$(-1)$-form gauge field} \cite{Hebecker:2017wsu,Heidenreich:2020pkc} with a flux-induced potential \`a la axion monodromy. This $(-1)$-form gauge field strength is simply a dynamical parameter (typically an internal flux) labelling different vacua of a multi-branched potential (see \cite{Dvali:2005an,Kaloper:2008fb,Marchesano:2014mla}). This is the analogous axionic version of a St\"uckelberg-like coupling providing a mass for a gauge field: The axionic symmetries get spontaneously broken and the axion gets massive by `eating up' the fluxes. These two possible mechanisms lead to a superpotential of the following form,
\beq
W=W_{\rm flux}(\phi)+\cO(e^{2\pi \ii m_ it^i})\, ,
\label{supoEFT}
\eeq
where $W_{\rm flux} (\phi)=M_{\rm P}^3\, f_A  \Pi^A(\phi)$ with $ \Pi^A(\phi)$ some functions of the chiral fields usually dubbed periods and $f_A \in \bbZ$ the so-called flux quanta. As above, the exponentially contributions are usually associated to non-perturbative effects due to the presence of BPS instantons charged under the axions, and are exponentially suppressed near the axionic string core by assumption.

Indeed, let us first consider the case where $W_{\rm flux}=0$ and the superpotential is a sum of terms of the form $\cO(e^{2\pi \ii m_it^i})$, $m_i \in \mathbb{Z}$. Such non-perturbative corrections are always expected to be present for at least some values of $m_i$, modifying both the superpotential and the K\"ahler potential, and breaking any continuous shift symmetry that the latter may develop in some field space regions. In this sense, assuming that \eqref{infs} takes the EFT to an axionic shift symmetry region amounts to require that $e^i m_i \geq 0$ for all the instanton charges $m_i$ that are relevant in the EFT region, and it essentially identifies the solution near the string core with a  perturbative EFT limit, in which all corrections of the form $\cO(e^{2\pi \ii m_it^i})$ can be neglected. In particular, the second derivative of the scalar potential will be suppressed like $\cO(e^{-4\pi m_i s^i})$ which is negligible in the region \eqref{infs}. String charges satisfying these positivity constraints correspond to the EFT strings in \eqref{CSEFT0}. As it will be discussed in section \ref{s:instantons}, using these observations one can characterise \eqref{CSEFT0} as a discrete cone of string charges, directly related with an EFT regime of perturbative axionic symmetries.

Let us now consider the case where the superpotential is of the form $W=W_{\rm flux} (\phi)$. The axionic string may become anomalous as the 0-form global symmetry may be spontaneously broken to a discrete remnant requiring the simultaneous shift of the axions and the fluxes. This anomaly is cured by attaching the string to membranes, whose charge $m_A$ modifies the flux quanta like $f_A \to f_A + m_A$ as we cross them. Hence, the membranes mediate non-perturbative transitions between the different branches of flux vacua. As pointed out in \cite{Lanza:2019xxg} and elaborated in \cite{Lanza:2020qmt}, not all of these transitions can be described dynamically at the level of the EFT, but only a sublattice $\Gamma_{\rm EFT}$ of them with a tension $\cT_{\rm mem}$ satisfying $ \frac{\cT_{\rm mem} }{M^2_{\rm P}} <  \Lambda  $. Those flux quanta not connected by $\Gamma_{\rm EFT}$ should be considered as separate sectors from the EFT viewpoint, each corresponding to a different choice of non-dynamical flux quanta. 
The question is then for which of these choices the string solution \eqref{tsol} is still a good approximation. By our discussion above, we may evaluate the effect of $W_{\rm flux}$ near the string core, where non-perturbative corrections are  assumed to die off. Intuitively, a first necessary condition for the string solution not to be spoiled  is that the Hubble scale vanishes asymptotically along \eqref{tsol}, which is the kind of potentials analysed in \cite{Grimm:2019ixq}. The associated membranes are selected by requiring $ \cT_{\rm mem} < M_P^3$ in the perturbative regime, and the induced potential exhibits a runaway behaviour in the limit \eqref{infs}. Additionally, one should require that  $(K_{i\bar{\jmath}}^{-1} \partial_{t^i} \partial_{\bar{t}^{\bar\jmath}} V)^{1/2}$ is small compared to the EFT cut-off $\Lambda$. Following the discussion in \cite[section 4.3]{Lanza:2020qmt},\footnote{This analysis assumes a particular class of  field metrics which we will encounter in the next section, cf. \eqref{Klog}.} one may see that this will be the case if $W_{\rm flux}$ involves periods $\Pi^A(\phi)$ that either do not depend on the direction \eqref{infs} or, if they do, they correspond to flux quanta within $\Gamma_{\rm EFT}$. All remaining fluxes will generate a potential which, asymptotically along \eqref{infs}, generates a mass for this field direction above $\Lambda$, so \eqref{infs} cannot actually be considered as an EFT field direction. Hence, we get a self-consistent picture in which EFT strings get attached to EFT membranes associated to fluxes inducing a potential that dies fast enough in the perturbative regime. For those choices of fluxes in which the limit \eqref{infs} is not obstructed and makes sense in the EFT, the string solution \eqref{tsol} should be a good approximation in the said limit.

To sum up, for  EFT strings all non-perturbative effects will die along the trajectory \eqref{infs}, and a sensible choice of flux quanta guarantees that the BPS string solution \eqref{tsol} can be trusted in a large region near the string core, even in the presence of a potential. For this reason, these EFT string solutions will always make sense from the viewpoint of an $\cN=1$ EFT. In fact, as will be discussed in section \ref{sec:strinfdis}, their profiles admit a compelling RG flow interpretation, as expected for localised operators of an EFT. This interpretation will support that, for our analysis to make sense, the fields entering the string solution need not be strictly massless, but only light compared to the EFT cut-off scale $\Lambda$.

 Finally, let us consider the case of axions  coupled \`a la St\"uckelberg to one or more $U(1)$ gauge fields.  That is, the axion may shift as  $a^i\rightarrow a^i+\lambda p^i$ under a $U(1)$ gauge transformation $A\rightarrow A+\d\lambda$, with $p^i\in\mathbb{Z}$. 
One can still dualise the axions $a^i$ into  $\cB_{2\,i}$, which couple to the gauge field $A$ via a term $p^i\int F\wedge \cB_{2\,i}$.  This coupling makes massive the $\cB_2$-field, so that a string of charges $p^i$ can break by nucleation of a pair of a magnetic monopoles. 
 The mass-squared is of order  $m^2\sim g^2 M^2_{\rm P}\cG_{ij}p^ip^j$,  with $g$ being the $U$(1) gauge coupling which is typically of order $g\sim 1/\sqrt{s}$, where $s$ is the linear combination of saxions that couples linearly to $F\wedge * F$. Following the reasoning above, as long as this mass is below the cut-off, there will exist some energy regime in which the EFT string solution will be a good approximation, and the string instability can be neglected. This seems a reasonable assumption in the perturbative regime; as it will be shown in the next section, the 2-form gauge coupling $\cG_{ij}$ vanishes asymptotically. Therefore the mass will be small as long as the 1-form gauge coupling $g$ does  not blow up. 
From the perspective of the axion,  the 0-form global axionic symmetry gets gauged as the gauge field $A$ gets massive by eating up the axion. Hence, instanton-breaking terms can only appear if the theory contains electrically charged particles. Here we can also make the distinction between fundamental axions or those coming from a sort of Abelian Higgs model. Recall that only the first case is of interest for us, as the associated strings are fundamental rather than solitonic. Finally, St\"uckelberg couplings generate D-term potentials with saxion-dependent Fayet-Iliopoulos terms in $\cN =1$ EFTs, which could in principle distort the corresponding string flow. In those cases we will however assume the presence of matter $\phi^\alpha$ charged under the $U$(1)-fields, such that there is always a flat D-term direction along $(t^i, \phi^\alpha)$ which defines the actual string flow.

\section{Strings and infinite field distance limits}
\label{sec:strinfdis}

As we have seen above, 4d EFT string solutions feature a logarithmic backreaction that describes a flow in the field space $\MM$ of our EFT, and more precisely on the saxionic fields $s^i$. Therefore, following the same philosophy as in  \cite{Klaewer:2016kiy,Draper:2019utz} (see also \cite{Dolan:2017vmn,Hebecker:2017wsu,Buratti:2018xjt,Draper:2019zbb,Geng:2019bnn,Geng:2019zsx,Bonnefoy:2019nzv,Gendler:2020dfp}), one may use these string solutions to study large variations in the field space of the EFT. In particular, as we will argue, EFT string flows are naturally associated to field space trajectories of infinite distance, and they provide a unique tool to test the physics of constant-field configurations. In addition, the string solution has a natural RG flow interpretation, which allows one to link variations along $\MM$ with changes of the EFT cut-off $\Lambda$. This dictionary between string backreacted solutions, RG flows and field space variations yields several important lessons. In particular, just like $\Lambda$ has a finite range and above some scale one must drop the 4d EFT  description, the same occurs after some distance along flows in $\MM$ generated by EFT string solutions.

\subsection{Physics along the string solution}
\label{sec:stringsol}

If we consider a certain point in field space $\bm{t}_0=\{t^i_0\} \in \MM$ and an EFT string of charge ${\bf e}$, then the solution \eqref{solsplit} will realise a flow on the saxions ${\bm s}=\{s^i\}$  from an initial value $\bm{s}_0$ at $r_0$ to the boundary of field space \eqref{infs}, as we approach the string location at $r=0$. Going in the opposite direction, some $s^i$ will become of order one and expected corrections of order $\cO(e^{2\pi \ii t^i})$ will become significant, a point at which the local solution \eqref{tsol} can no longer be trusted. Therefore, even if a general 4d string solution can be seen as a map from $\bbC$ to the EFT field space $\MM$, an EFT string solution selects a disc $D(r_0) \subset \bbC$, which is mapped to a perturbative region of $\MM$ in which all non-perturbartive effects can be neglected. In particular, non-perturbative corrections of the form $\cO(e^{2\pi \ii m_i t^i})$ with $m_i e^i  >0$  will be more and more suppressed as we proceed along the flow \eqref{imt}  towards $r\rightarrow 0$, until we reach an exact axionic shift symmetry in the limit \eqref{infs}. By standard quantum gravity arguments \cite{Banks:1988yz,Banks:2010zn}, we only expect to realise such global continuous symmetries at  infinite distance in field space. As a result, consistency with quantum gravity relates EFT string locations to infinite distance points on the boundary of $\MM$, and the backreacted saxionic trajectory \eqref{imt} to infinite distance paths in $\MM$.

This relation can be quantified by parametrising the radial saxionic flow \eqref{imt} as
\be\label{sssfl}
 s^i(\sigma)=s^i_0+\sigma e^i\, ,
 \ee
where we have defined $2\pi \sigma = \log (r_0/r)$.  For a given $s^i_0$,  let us denote by $\sigma_*$ the value of $\sigma$ at which the flow reaches the boundary of the  saxionic field space. The field space distance travelled by the radial flow is given by
 \be\label{d*}
 \begin{aligned}
 {\rm d}_*&=\int_{\rm flow}\sqrt{\cG_{i j}\d s^i\d s^ j}=\int^{\sigma_*}_0\d\sigma \sqrt{e^i e^ j\,\cG_{i j}(\sigma)} =\frac{1}{M_{\rm P}}\int^{\sigma_*}_0 \cQ_{\bf e}(\sigma)\,\d\sigma\, ,
 \end{aligned}
 \ee
 where in the last equality we have introduced the physical string charge $\cQ_{\bf e}$, defined as
 \be\label{calgdef}
\cQ^2_{\bf e}=M_{\rm P}^2\, \cG_{i j}e^i e^ j\, .
\ee
 This definition of $\cQ_{\bf e}$ is well-motivated in the field space region probed by the EFT flow along $D(r_0)$, since it lies within a weakly-coupled region where axionic shift symmetries are a good approximation. One can then resort to the dual formulation described in section \ref{sec:dualform} in terms of two-form potentials $\cB_{i\, 2}$ to which the strings couple electrically. Since the kinetic matrix for the $\cB_{2\,i}$ is ${M_{\rm P}^{-2}}\cG^{i j}$, see \eqref{dualaction}, the physical charge is then defined as in \eqref{calgdef}. Notice that along the flow \eqref{sssfl} $\cQ_{\bf e}(\sigma)$ is simply given by
\be\label{flowQ2}
\cQ_{\bf e}(\sigma)=M_{\rm P}\sqrt{\frac{\d^2 K(s(\sigma))}{2\d\sigma^2}}\, .
\ee

Quantum gravity arguments  constrain the possible asymptotic behaviour  of $\cQ_{\bf e}$ along the EFT flow. First, one can argue that necessarily $\sigma_* = \infty$, which is what we indeed expect for an EFT string flow of the form \eqref{imt}, since there the axionic shift symmetry gets restored at the string core $r=0$. For this, notice that $\frac{1}{2\pi}M_{\rm P}\cQ_{\bf e}$ can also be interpreted as axion decay constant $f_\vartheta$ of the  axionic direction $\vartheta$ that shifts by $2\pi$ around the EFT string: $a^i= \frac{1}{2\pi}e^i\,\vartheta$, $\vartheta\simeq \vartheta+2\pi$. The axion $\vartheta$ has the kinetic term  
\be
-\frac12\int f^2_\vartheta\,\d\vartheta\wedge * \d\vartheta=-\frac1{8\pi^2}M_{\rm P}^2\int \cQ^2_{\bf e}\,\d\vartheta\wedge * \d\vartheta\, .
\ee
An instanton of charges $m_i$ has action $S_{\bf m}=2\pi\, m_i s^i$, and the axionic version of the WGC \cite{ArkaniHamed:2006dz} applied to these instantons reads
\be\label{instWGC}
f_\vartheta\, S_{\bf m}\leq \gamma_{\bf m} M_{\rm P}\quad~~~\Leftrightarrow\quad~~~ \cQ_{\bf e} (m_i s^i)\leq \gamma_{\bf m}\,,
\ee
where $\gamma_{\bf m}$ is some constant depending only on the instanton charges $m_i$. 
The positive quantity $m_i s^i $ does not decrease along the EFT flow for allowed instanton charges $m_i$, and so \eqref{instWGC} implies that $\cQ_{\bf e}$ must remain finite along the corresponding EFT flow. As said, the absence of global symmetries requires that ${\rm d}_* = \infty$, which can then only happen if  $\sigma_* = \infty$. 

Second, since $m_i s^i\sim (m_i e^i)\sigma$ for $\sigma\rightarrow\infty$ with $m_i e^i\geq 0$, then \eqref{instWGC} implies that $\cQ_{\bf e}$ vanishes asymptotically  along the flow:  $\cQ_{\bf e}\rightarrow 0$. Notice however that \eqref{d*} and $ {\rm d}_* = \infty$ imply that $\cQ_{\bf e}$ cannot go to zero too quickly, although all conditions are satisfied if $\cQ_{\bf e} \sim \sigma^{-\eta}$, with $\eta \in (0,1]$.
This range can be reduced if we take into account that the asymptotic behaviour of the string tension $\cT_{\bf e}(\sigma)=M^2_{\rm P}\,e^{i}\ell_i (\sigma)$ is governed by
\be\label{monoT}
\frac{\d\cT_{\bf e}(\sigma)}{\d\sigma}=-\cQ^2_{\bf e}<0\, ,
\ee
where we have used \eqref{dualfields} and \eqref{GGmetric}.
Requiring that $\cT_{\bf e}$ does not become negative at $\sigma_* = \infty$ further restricts the range to $\eta \in (1/2,1]$. Moreover, for EFT strings we expect that a WGC-like bound is saturated asymptotically for $\sigma\rightarrow \infty$, so that an asymptotic relation of the form $M_{\rm P}  \cQ_{\bf e} \sim \gamma \cT_{\bf e}$ with a constant $\gamma$ is satisfied. It is easy to see that this only happens if  $\cQ_{\bf e} \sim \sigma^{-1}$.

 Therefore, the only reasonable option for EFT strings seems to be a physical charge $\cQ_{\bf e}$ that vanishes asymptotically like $\sigma^{-1}$, with a similar behaviour for the tension $\cT_{\bf e}$. As such, the proper distance diverges logarithmically as $\sigma \rightarrow \sigma_*= \infty$, and using \eqref{flowQ2} one obtains a K\"ahler potential which asymptotically takes the form $K \sim - n \log \sigma$, for $n \in \mathbb{R}^+$. This is indeed the case for all the string theory examples analysed in section \ref{s:examples}, where we have that asymptotically
\be
K \simeq - \log P(s)\, ,
\label{Klog}
\ee
 with $P(s)$ some homogeneous function of positive integral degree on the saxions. Hence, we can use the WGC and the existence of EFT strings as a physical motivation for the behaviour of the K\"ahler potential observed in string theory compactifications. Notice that then the dual saxions $\ell_i$ decrease as $\sigma^{-1}$ along the trajectory \eqref{sssfl}, just like \eqref{flowQ2} and the tension $\cT_{\bf e}$. Finally, using \eqref{Klog} one may argue that EFT string flows are asymptotically geodesic, see appendix \ref{ap:geodesic}. 

Let us remark that, in practice,  not any choice of charges $e^i$ corresponds to an EFT string, nor to a point at infinite distance. Indeed, as we will discuss in section \ref{s:instantons}, some choices of $e^i$ correspond to saxionic flows along which some non-perturbative effects become relevant and destroy the axionic shift symmetry. Interestingly, for such cases the string flow reaches the field space boundary for a finite value of the parameter $\sigma$ in \eqref{sssfl}, at least at the classical level.

To sum up, as we proceed along the string flow \eqref{imt} towards the core of an EFT string, we probe regions of the EFT field space along which the string becomes tensionless and weakly coupled, and the associated axionic symmetry becomes more and more exact. This picture may provide the wrong impression that at the string core we reach a regime in which strings are tensionless and the axionic symmetry exact. This is however not so, because \eqref{tsol} is not the actual flow realised by an EFT with a finite cut-off $\Lambda$. Instead, an approximation of \eqref{tsol} is realised, which tends towards \eqref{tsol} as we increase $\Lambda$. Taking this observation into account leads in fact to an interesting RG flow interpretation of the string solution, which we now turn to discuss.

\subsection{RG flow interpretation}
\label{sec:RGflow}

In an EFT with a given cut-off $\Lambda$ the string backreaction profile will not look like \eqref{solsplit}, but rather like a coarse-grained profile in which only 4d Fourier modes up to momentum $\Lambda$ can enter. From this kind of observation stems the EFT interpretation of the backreaction of extended objects, used in \cite{Michel:2014lva,Polchinski:2015bea} to interpret $p$-brane backreaction in string theory. Following the same philosophy, a 4d string backreaction can be interpreted in terms of a classical $\beta$-function for the 4d string couplings to the bulk fields, and in particular for its effective charge and tension. 

To apply this philosophy to our context, let us consider the linear energy density  ${\cal E}(r)$ of the EFT string solution contained in a disk of radius $r$, as computed in  section \ref{sec:linearen}. As follows from \eqref{linearE}, ${\cal E}(r)$ is a sum of two contributions, namely the energy contained in the string backreaction ${\cal E}_{\rm back}$ and the localised contribution ${\cal E}_{\rm string}$. While the sum ${\cal E}(r)$ is a physical, fixed quantity that only depends on the radius $r$, the contribution of each term depends on the energy cut-off $\Lambda$ of the EFT describing the axionic string. 

\begin{center}
	\begin{figure}[tb]
		\centering
		\includegraphics[width=8cm]{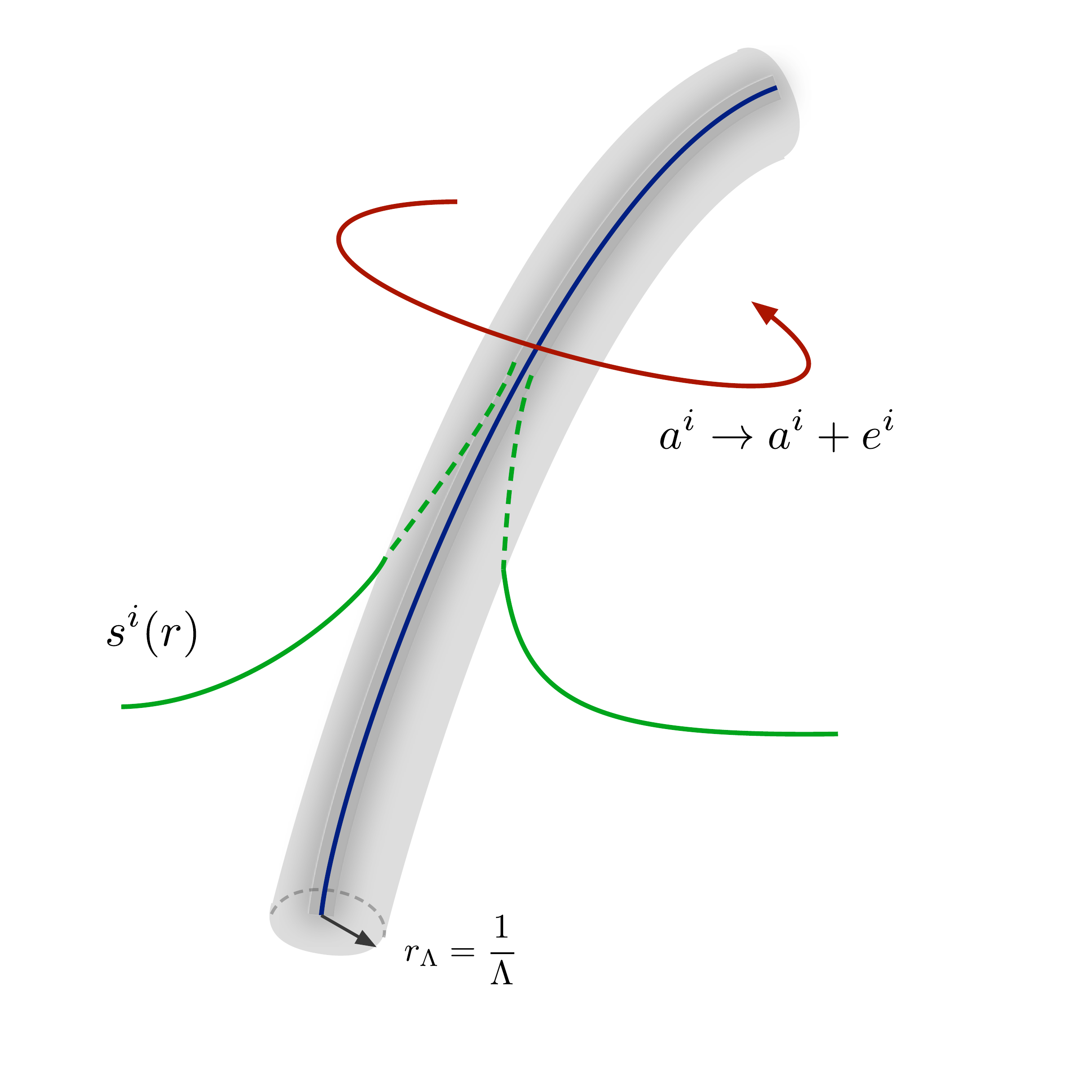}
		\caption{Axionic string backreaction \eqref{imt} and \eqref{axionmon}. In gray, we highlighted the region $r \leq r_{\Lambda}$ that cannot be resolved within the EFT.  \label{f:axionicstring}}   
	\end{figure}
\end{center}

From this observation one can derive interesting consequences. On the one hand, if we take the formal limit $\Lambda \rightarrow \infty$, then ${\cal E}_{\rm back}$  captures the full energy of the profile \eqref{solsplit}, and ${\cal E}_{\rm string}$ reaches its minimum. On the other hand, if we take $\Lambda \sim 1/r$ the EFT essentially sees a flat backreaction profile and no energy can be stored in the term ${\cal E}_{\rm back}$. Equivalently, for a given cut-off $\Lambda$ the choice of radius $r = 1/\Lambda$ implies that ${\cal E}_{\rm back}(r) = 0$, and the whole contribution to the linear energy is stored in ${\cal E}_{\rm string}$. Notice that this radius is the minimal distance from the string that can be resolved by the EFT and, by the previous observation, the point from where we can read the string couplings \eqref{sttringcoup} at the cut-off-scale $\Lambda$, directly from the string backreaction (see Fig.~\ref{f:axionicstring}).

In other words, the string backreaction \eqref{tsol} realises the RG flow of its couplings, with the cut-off-scale setting the distance $1/\Lambda$ at which the couplings are defined. The bulk profile along the radial direction $r$ can then be interpreted as an RG flow of the brane couplings as one changes $\Lambda \sim 1/r$. In particular, we may identify the effective string tension at the scale $\Lambda$ with 
\be
\label{Teff}
\cT_{\rm str}(\Lambda)=\cE(r_\Lambda)=M_{\rm P}^2\,e^i\ell_i(r_\Lambda)\, ,
\ee 
where $r_\Lambda \equiv 1/ \Lambda$ is the radius where the coarse-grained flow stops.\footnote{More precisely, one should use the backreacted metric to compute the distance from the string core. By using \eqref{Klog},  \eqref{stringwarp}  gives a warping $e^{2D}\sim P(s)$. Applied along the saxionic flow \eqref{imt}, this produces a logarithmic correction to the flat-space identification $r_\Lambda \sim 1/ \Lambda$. In the following we will ignore such a logarithmic correction.}
In this sense, approaching the string location along the profile \eqref{solsplit} can be related to increasing the EFT cut-off, with the corresponding changes in the effective string tension and charge. This variation can be easily linked to our above discussion if we express the flow parameter as $\sigma=\frac1{2\pi}\log (\Lambda r_0)$, with $\Lambda$ the cutoff at which the effective string tension is defined. Then \eqref{monoT} becomes
\be\label{monoT2}
2\pi  \Lambda \frac{\d\cT_{\bf e}(\Lambda)}{\d\Lambda} =-\cQ^2_{\bf e}<0\, .
\ee
It follows from our previous discussion that for EFT strings $\cT_{\rm str}(\Lambda)$ decreases as $\Lambda \rightarrow \infty$. In particular, for K\"ahler potentials of the form \eqref{Klog} this monotonically decreasing behaviour is such that $\cT_{\rm str}(\Lambda) \rightarrow 0$ as $\Lambda \rightarrow \infty$. This can either be seen by looking at the asymptotic behaviour of the dual saxions $\ell_i$ coupling to the string, or by performing a direct 4d field theory computation of the string RG flow along the lines of \cite{Buonanno:1998kx}, see appendix \ref{ap:BD}.

\vspace{-1.25cm}

In light of these considerations, some comments are in order:

\begin{itemize}
   
    \item[-]
    Near the EFT cut-off scale $\Lambda$ all the effects at a much lower scale can be neglected. In this way we may reinterpret our results of section \ref{sec:validity}, in which the different potentials that appear in sensible setups do not modify the string solution near the string core. In this sense, the relevant EFT field space $\MM$ for our analysis can either contain flat directions or fields whose mass is much smaller than $\Lambda$. In the following, a point in $\MM$ will denote either a vacuum or a constant-field configuration in which the Hubble and mass scales of the potential are negligible compared to $\Lambda$.

    \item[-]
    Even if the string solution of section \ref{s:fundamental} is obtained in a supersymmetric setup, our conclusions should also hold for EFTs in which supersymmetry is spontaneously broken. Indeed, SUSY-breaking effects will modify the string solution, but if supersymmetry is restored at some scale $\Lambda_{\rm SUSY}$ then such a solution will look like the above $\half$BPS string solution at distances shorter than $r_{\rm SUSY} = 1/\Lambda_{\rm SUSY}$. Notice that our reasoning only depends on the asymptotic behaviour of the string solution near the string core, and is therefore insensitive to SUSY-breaking effects or any mass deformation well below the EFT cut-off $\Lambda$. In particular, for the sake of the argument that leads to the behaviour  $\cQ_{\bf e} \sim \sigma^{-1}$, the saturation of the WGC-bound only needs to be imposed asymptotically. In terms of the RG flow interpretation, this means that the reasoning also applies to EFT strings that may not be seen as BPS in the IR, but become so in the UV.

\item[-] 
An analogous reasoning can be applied in cases where the axionic shift symmetry is broken by a perturbative superpotential. Indeed, as 
 already discussed in section \ref{sec:validity},  we may consider a superpotential of the form \eqref{supoEFT} with $W_{\rm flux}$ breaking the axionic shift symmetry. However, if the flow still makes sense as a field space direction, the string solution of section \ref{s:fundamental} should be a good approximation near the string core. Then,  since our reasoning in this section only depends on the shift symmetries of the K\"ahler potential  and the subsequent properties of the string charge $\cQ_{\bf e}$, having $W_{\rm flux} \neq 0$ should not affect the fact that the string should be located at infinite field distance, as well as all the consequences that follow. Pictorially, 4d EFT string solutions whose axionic symmetry is broken by $W_{\rm flux}$ correspond to solutions with a radial branch cut, in which a membrane is placed to render the string operator gauge invariant. Away from the membrane location the radial flow of the saxion should be less and less distorted as we approach the string core, and the discussion of this section should follow through.

\end{itemize}

\subsection{String flows as paths in field space}
\label{sec:fspaths}

One of the main applications of EFT string solutions and their RG flow interpretation is to extract physical information about similar paths in the EFT field space $\MM$. That is, we would like to map the EFT string solution \eqref{tsol} to a family of constant-field configurations that describe a trajectory in field space
\be
\gamma_{\bf e} \equiv\{ \bar{\bm{t}} = \bar{\bm a}_0 + \ii \bar{\bm{s}}(\sigma)\} \subset \MM\, , \qquad \bar{\bm{s}}(\sigma) = \bar{\bm{s}}_0 + \sigma {\bf e} \, , \qquad \bar{\bm a}_0, \bar{\bm s}_0 = \text{const.}
\label{gammapath}
\ee
now in the absence of any backreacting string and for a fixed EFT cut-off $\Lambda$. Here we use barred quantities to denote constant-field configurations parametrised by $\sigma$, as opposed to e.g. the local value of the saxion $\bm{s}(r)$ along the string solution \eqref{imt}. 

A first question is if the physics of a constant-field configuration $\bar{\bm{t}} \in \gamma_{\bf e} \subset \MM$ is captured by a local space-time patch of the backreacted EFT string solution \eqref{tsol}, around the point $z\in \mathbb{C}$ such that $\bm{t}(z) = \bar{\bm{t}}$. The difference between these two cases is that in the string solution the fields are not constant. Because of the logarithmic profile, the derivatives of the EFT string solution set a length scale of order $1/r \equiv 1/|z|$, after which we start seeing the field variations. Accordingly, we expect that a local  patch of the string solution captures the physics of a constant-field configuration provided that we consider energies above $1/r$. Then, because our EFT only describes energies up to $\Lambda$, we are restricted to the energy range $1/r < E < \Lambda$. Notice that this is consistent with the fact that the coarse-grained string solution stops at a distance  $r_\Lambda \sim 1/\Lambda$ from the string core, so that at shorter distances we have a constant profile. 

Therefore, if we want to understand the physics of constant-field configuration at energies near $\Lambda$, one may use the EFT string solutions to probe a region near $\bar{\bm{t}}  \in \MM$. Notice that the paths covered by a solution at finite $\Lambda$ are large but of finite distance, so we never reach infinite distance points in this way. However, we may still use such finite distance paths to see the asymptotic behaviour of the theory along large field distances in $\MM$. This will be the philosophy underlying the following sections of the paper. 

As emphasised above, of particular interest to us is to test the SDC along large distance paths $\gamma_{\bf e} \subset \MM$. This involves considering the mass scale $m_*$ of a tower of states that lies above $\Lambda$, and therefore that it has been integrated out in our EFT description, so it is quite hard to guess the moduli dependence of $m_*$ purely from EFT data. Nevertheless, for regions of field space $\MM$ probed by EFT string solutions, it is natural to assume that  any microscopic completion of our EFT contains an energy threshold $E_{\bf e} > \Lambda$, for any fundamental axionic string charge {\bf e} that exists in the theory. This can be argued by means of the Completeness Conjecture \cite{Polchinski:2003bq}, which would predict the existence of the corresponding string state, and from associating $E_{\bf e}$ with the mass of the lightest string oscillation modes. When the charge {\bf e} corresponds to an EFT string that is  $\half$BPS in the UV, and we are in a region of $\MM$ where the corresponding axionic shift symmetry is a good approximation, this energy threshold should be simply given by $E_{\bf e}^2(\bar{\bm{s}}) \equiv T_{\bf e}(\bar{\bm{s}}) = M_{\rm P}^2\,e^i\bar{\ell_i}$. That is, the {\it probe} string tension $T_{\bf e}$ that is computed from dimensional reduction in string compactifications to 4d, see  section \ref{s:examples} for examples.

Both quantities $T_{\bf e}$ and $\cT_{\bf e}$ have the same functional dependence on the dual saxions \eqref{thetension}, as dictated by supersymmetry and axionic shift symmetry. In fact, when we evaluate $\cT_{\bf e}$ along its string solution as in section \ref{sec:stringsol}, what we are doing is to estimate $T_{\bf e}$ along the corresponding saxionic path \eqref{gammapath} in field space. The connection of $T_{\bf e}$ with the RG flow interpretation of the string solution is however less obvious, because $\cT_{\bf e}(\Lambda)$ depends on the cut-off while $T_{\bf e}$ does not; $\cT_{\bf e}(\Lambda)$ appears as a coupling of an EFT operator, while $T_{\bf e}^{1/2}$ corresponds to the mass of a closed oscillating string of radius $T_{\bf e}^{-1/2}$, and so it is beyond the EFT resolution scale. 

To clarify their relation let us consider a closed string on a loop of radius $L > 1/\Lambda$, that we add `on top of' a constant field configuration $\bar{\bm{t}} \in \MM$.  This closed string  will backreact as in \eqref{tsol} with and UV cut-off at $\Lambda$, and now also with an IR cutoff at approximately $1/L$, coming from the fact that the string forms a  closed loop. Indeed, beyond a distance $L$ from the string its  backreaction will quickly die off, and we will just see an approximately constant field configuration such that $\Im{\bm{t}} \simeq \bar{\bm{s}}$, which we identify with a point in $\MM$. Microscopically, we associate a total energy of order $T_{\bf e}(\bar{\bm{s}})L$ to this string loop. Then, similarly to \eqref{linearE}, we can split the total energy of the system as 
\be
T_{\bf e}(\bar{\bm{s}})L = \cT_{\bf e}\left(\Lambda,\bar{\bm{s}}\right) L +  E_{\rm back}\left(\Lambda,L,\bar{\bm{s}}\right)\, ,
\label{splitEloop}
\ee
 where the lhs is independent of the EFT cut-off $\Lambda$. In the limiting case in which $\Lambda \sim 1/L$ there is no backreaction,  and we recover the microscopic result $\cT_{\bf e}(1/L,\bar{\bm{s}}) = T_{\bf e}(\bar{\bm{s}})$. If we now increase $\Lambda$ a backreaction will be developed, such that the fields $\bm{t}$ will start flowing towards the string core. More precisely we will have a coarse-grained version of the flow \eqref{solsplit}, with $r_0 = L$ and $\bm{s}_0 = \bar{\bm{s}}$. As described above, this flow will probe the physics of different constant-field configurations, until the region that corresponds to $\bar{\bm{s}}_\Lambda\equiv \bm{s}(r_\Lambda \equiv 1/\Lambda) = \bar{\bm{s}} + \frac{\bf{e}}{2\pi} \log(\Lambda L)$. By using the symmetries of the solution \eqref{tsol}, or directly from the similarities between \eqref{monoT} and \eqref{monoT2} it follows that\footnote{One can illustrate this symmetry explicitly by considering a single-field model with a K\"ahler potential of the form $K=-n \log s$. Then the EFT string flow yields the expression
\be\nonumber
 \mathcal{T}^{-1}(\Lambda,s) = \cT^{-1}(\Lambda_0,s_0) + \frac{2 }{n e M_{\rm P}^2}\left(s - s_0 + \frac{e}{2\pi}\log \frac{\Lambda}{\Lambda_0} \right)\, .
\ee} 
\be
\cT_{\bf e} (\Lambda e^{2\pi \sigma}, \bar{\bm{s}}) = \cT_{\bf e} (\Lambda, \bar{\bm{s}}+\bf{e} \sigma) \, .
\label{Tsymmetry}
\ee
From here we have that $ \cT_{\bf e}(\Lambda,\bar{\bm{s}}) = \cT_{\bf e}(1/L,\bar{\bm{s}}_\Lambda)=  T_{\bf e}(\bar{\bm{s}}_\Lambda)$.
That is, the renormalised tension $\cT_{\bf e}(\Lambda,\bar{\bm{s}})$ describes $T_{\bf e}$ in the vicinity of the string core. Equivalently, one can see $\cT_{\bf e}^{1/2}(\Lambda,\bar{\bm{s}})$ as the smallest amount of energy needed to create a string of size $T_{\bf e}^{-1/2}$ in this whole configuration.

To sum up, by changing the cut-off $\Lambda$ in the above setting generates a saxionic profile that interpolates between $\bar{\bm{s}}$ and $\bar{\bm{s}}_\Lambda\equiv\bm{s}(r_\Lambda)$. In this way, we probe the physics of constant-field configurations that follow the same path in $\MM$, with $\cT_{\bf e}(\Lambda,\bar{\bm{s}})$ measuring how  $T_{\bf{e}}$ changes along such paths. The larger $\Lambda L$ is, the larger the path in $\MM$, although some limitations will be in place. For instance, if $\cT_{\bf e}(\Lambda,\bar{\bm{s}}) \leq \Lambda^2$, we would be describing a region where our assumption \eqref{EFTregime} does not hold, and so we would not expect to capture the physics of a constant-field configurations in our 4d EFT. In fact, as we argue below, for $\bar{\bm{t}} \in \MM$ such that $T_{\bf e}(\bar{\bm{s}}) \leq \Lambda^2$ the semiclassical description of our EFT should break down.

It follows that those flows that are useful to probe the physics of EFT constant-field configurations must satisfy  $\Lambda^2 < \cT_{\bf e}$. In terms of EFT string solutions, this condition translates into a restriction of the parameters $(\bm{s}_0, r_0)$ that enter \eqref{imt} or, in the above closed string configuration, of the parameters $(\bar{\bm{s}}, L)$. By requiring that $\Lambda^2 < \cT_{\bf e}(1/r, \bar{\bm{s}})$ for any $r \in (r_\Lambda, L)$, one in particular constrains the allowed values for $\bar{\bm{s}}_\Lambda$, and therefore sets a bound on the choice of constant-field configuration $\bar{\bm{t}} \in \MM$ probed by EFT string flows.  This is consistent with the fact that an upper bound for $\Lambda$ translates, via \eqref{Tsymmetry}, into a maximal field range along saxionic directions generated by EFT strings. In fact, the correspondence \eqref{Tsymmetry} suggests that, just as the 4d EFT stops being valid at some scale above  $\Lambda$, the same should happen along large distances in $\MM$ generated by EFT string flows. This observation is at the origin of why along large field space trajectories corresponding to EFT strings one recovers the physics predicted by the SDC, a link that will be made more precise in section \ref{s:conjecture}.

\subsection{EFT breakdown}
\label{sec:breakdown}

Our results regarding the effective string tensions $\cT_{\bf e}$ indicate how the mass of the lightest oscillation modes of the different EFT strings $T^{1/2}_{\bf e}(\bar{\bm{s}})$ vary as we move in the EFT field space $\MM$ along the paths \eqref{gammapath}. By assumption, such modes lie above the EFT cut-off scale $\Lambda$, and so we must impose $T_{\bf e}(\bar{\bm{s}}) > \Lambda^2$. This is reminiscent of the first inequality in \eqref{EFTregime}, which follows from assuming that EFT strings are fundamental objects. On general grounds, we expect that fundamental objects lie above the EFT cut-off scale, since the EFT semiclassical description should break down whenever we are able to resolve one of them. 

What the results of this section show is that, as we move in field space $\MM$ for fixed cut-off $\Lambda$ along the path \eqref{gammapath}, the corresponding threshold $E_{\bf e} \equiv T^{1/2}_{\bf e}$ will decrease monotonically towards zero, and so at some point $\bm{s}_{\rm break}$ we will have that $T_{\bf e} (\bm{s}_{\rm break}) = \Lambda^2$. Beyond such a point, the EFT string should trigger the breakdown of the EFT semiclassical description. In this way any point of infinite distance that corresponds to an EFT string flow endpoint will present at least one natural candidate for a tower of states that satisfy the Swampland Distance Conjecture, as such string will become tensionless as we proceed along the path, and therefore its oscillation modes massless. One may even argue that the SDC will be satisfied whenever the said EFT string satisfies the WGC, deriving this way the exponential behaviour of the cut-off, see section \ref{ss:implications} for details. 

Let us discuss in some detail why a string threshold $T_{\bf e}^{1/2}$ below the cut-off implies the breakdown of the EFT 
and whether this is really due to an infinite tower of states as the SDC requires. First of all, we can argue for the existence of new degrees of freedom signaling the EFT breakdown when $T_{\bf e}< \Lambda^2$ for weakly coupled strings. For this, consider some classical fluctuating string solution, like the closed string loop in eq.\eqref{splitEloop}. As in there, this classical state has center-of-mass energy  $E= T_{\bf e} L$, where $L$ can be identified with either the oscillation time-scale or the  length of the string. On the one hand, the local low-energy EFT regime seems legitimate for  $L\gtrsim r_\Lambda\equiv 1/\Lambda$, so the classical estimate for the mass of the lightest states will be $E\sim T_{\bf e}/\Lambda$. On the other hand, the self-consistency of our semiclassical description of these string states requires the characteristic string length scale $L$ to be bigger than its Compton wavelength $1/E$. Hence, the semiclassical description of the lightest states with $L\sim r_\Lambda$ is only justified if $T_{\bf e} >\Lambda^2$. In other words, the semiclassical description of this continuous family of states breaks down  if  $T_{\bf e}< \Lambda^2$. Rather, they should be treated as quantum states to be included in the EFT spectrum. 
Taking into account that the string is weakly coupled, we expect that the continuum spectrum of fluctuations  gets discretised and yields an infinite tower upon quantisation, as it happens for critical strings. However, this will remain an assumption for us.

Even if an asymptotically tensionless EFT string indicates the eventual breakdown of the EFT, it does not specify the nature of the breakdown. As said, the oscillation modes of the string provide a natural candidate for a tower of states realising the SDC, but there  could be other towers of lighter states that force the EFT breakdown before the string becomes tensionless. Indeed, this effect has already been observed in \cite{Lee:2019wij} (see also \cite{Baume:2019sry}) where these lighter towers of states were interpreted as decompactification limits. In fact, on general grounds we would expect that a tower of modes of mass $m_*$ like KK modes becomes light along an EFT string flow at least as fast or even faster than the string tension. Otherwise we would be able to decouple both scales and engineer 4d string theories with an infinite number of oscillation modes in a space which is approximately Minkowski. In section \ref{s:conjecture} we will propose a precise relation between the behaviour of the scale $m_*$ in terms of the string tension which, to the best of our knowledge, is satisfied in all string theory compactifications, and which fits nicely with the Emergent String Conjecture of \cite{Lee:2019wij}.

We have seen that any EFT string backreaction drives the scalars along large field space distances. In other words, EFT string flows dynamically explore different perturbative vacuum sectors of the theory. In fact, as we will see in explicit examples, by taking different string charges one can explore all the asymptotic limits 
of a given perturbative regime via EFT string flows. Again, the vanishing asymptotic tension is guaranteed by a  K\"ahler potential of the form \eqref{Klog}, found near weakly-coupled infinite distance points in string theory compactifications. It is thus tempting to speculate that all points of this sort should correspond to EFT strings, an idea upon which we will elaborate in section \ref{s:conjecture}. Notice that this proposal fits nicely with the results of  \cite{Grimm:2018ohb,Grimm:2018cpv,Corvilain:2018lgw} which classify infinite distance limits in terms of monodromies of particle charges. In our $\cN=1$ setup this monodromy is still there, and it is physically realised by the discrete shift \eqref{tmon}. As discussed in \cite{Lanza:2020qmt}, in our more general setup it does not act on BPS particles as in the ${\cal N}=2$ setups of \cite{Grimm:2018ohb,Grimm:2018cpv,Corvilain:2018lgw}, but instead on membranes.

\begin{center}
	\begin{figure}[htb]
		\centering
		\includegraphics[width=8.5cm]{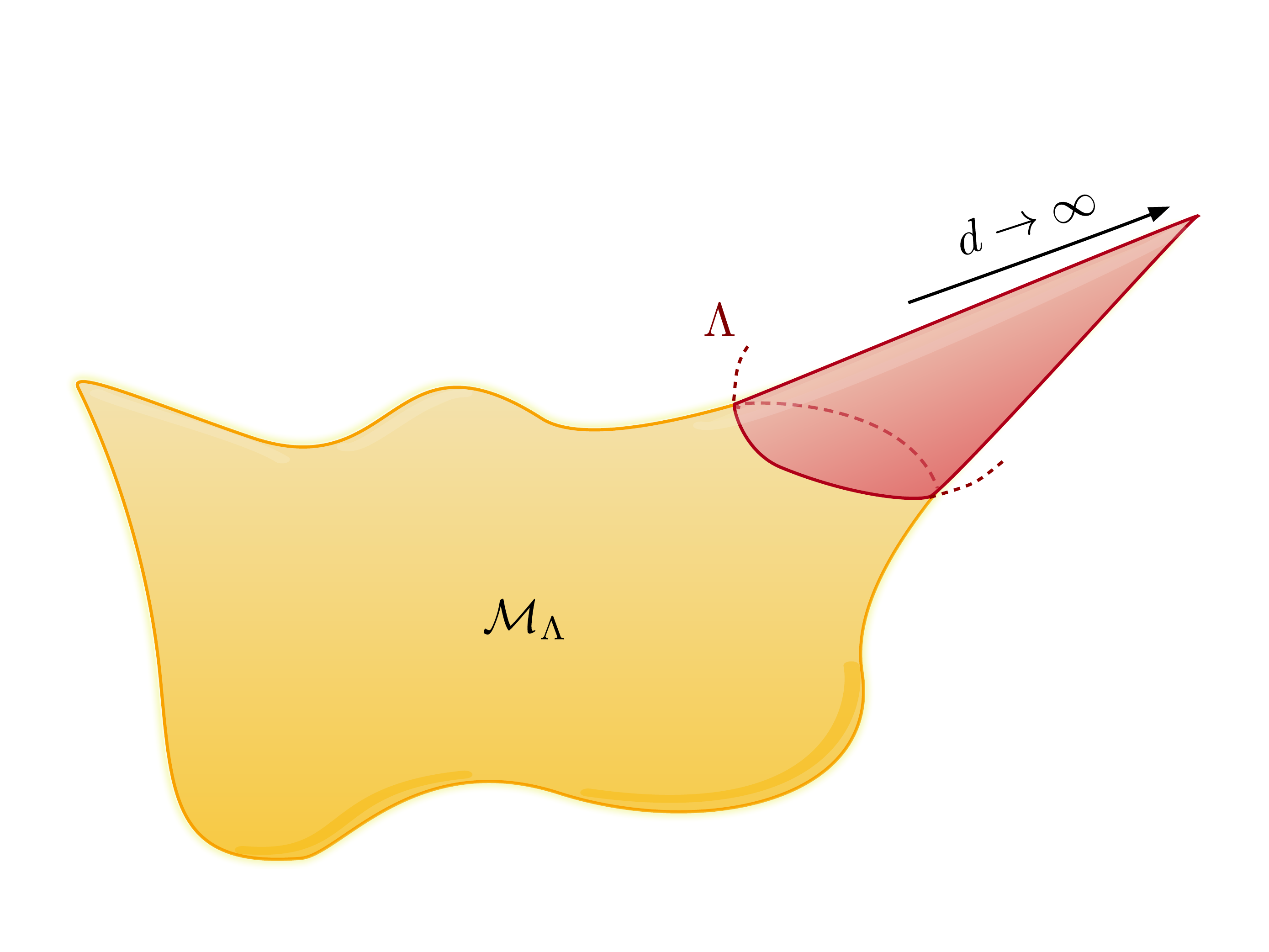}
		\caption{The moduli space $\mathcal{M}_\Lambda$ accessible within an EFT characterised by a cutoff $\Lambda$. \label{f:modspace}}   
	\end{figure}
\end{center}

Incidentally, this picture provides an interesting purely classical  realisation of the Emergence Proposal \cite{Grimm:2018ohb,Harlow:2015lma,Heidenreich:2017sim,Heidenreich:2018kpg}. For a fixed value of the EFT cut-off $\Lambda$, the EFT field space consistent with our  description, which we dub $\MM_\Lambda$, can only contain finite distances. Indeed, as explained above, when proceeding along the saxionic paths \eqref{gammapath} of infinite distance in $\MM$ generated by EFT strings, we will reach a set of points $\{\bm{t}_{\rm break}\}$ at which $T_{\bf e}({\bm{s}}_{\rm break}) = \Lambda^2$. At these points the EFT will break down, defining a boundary for $\MM_\Lambda$. Hence, at finite cut-off our EFT is associated with a finite scale-dependent field space $\MM_\Lambda$. As we lower $\Lambda$ and we integrate out 4d high energy modes, the paths $\gamma_{\bf e}$ will be allowed to reach smaller values for $e^i\ell_i$, which correspond to points at a larger field space distance and smaller string tensions. So $\MM_\Lambda$ will grow and, in the limit $\Lambda \rightarrow 0$, it will asymptote to the naive field space $\MM$, and points at infinite distance will emerge (see Fig.~\ref{f:modspace}). Around these points, the string couplings and the kinetic terms for the saxions coupling to the string will be connected, via \eqref{Tsymmetry}, with the string RG flow towards the UV, which dictates the asymptotic behaviour of the system.

%%%----------------------------

\vspace{-1.25cm}

\section{Strings, instantons and asymptotic limits}
\label{s:instantons}

The definition of EFT string implies a powerful statement about non-perturbative corrections in regions of the EFT field space $\MM$ with approximate axionic shift symmetries. In this section we analyse the interplay between instanton effects and string flows, specially in those cases where several axionic symmetries coexist. The underlying $\cN=1$ framework reveals a conical structure for the set of asymptotic limits of infinite distance, together with a discrete cone of EFT string charges generating them. This strengthens the correspondence between infinite distance asymptotic limits in $\MM$ and EFT string solutions, and motivates some of the  swampland criteria that will be proposed in the next section. The reader not interested in the general structure of EFT string charges may skip this material and jump to section \ref{s:conjecture}. Most of the definitions that emerge from the following analysis are summarised by table \ref{Tab:cones} and appendix \ref{app:flowdef}.

\subsection{Strings,  instantons and cones}
\label{sec:cones}

In a region of $\MM$ with approximate axionic symmetries in the field space metric, we may reach perturbative asymptotic regimes by taking limits in which linear combinations of saxions $s^i$ become very large. Indeed, notice that non-perturbative contributions coming from $\half$BPS instantons charged under the axionic symmetries are a sum of terms of the form
\be
 \cO( e^{2\pi \ii  m_i t^i })\, , \qquad m_i \in \bbZ\, .
 \label{orderinst}
\ee
Therefore, along paths of infinite distance like \eqref{sssfl} with $\sigma \rightarrow \infty$, several of these effects become negligible. Because these are typically the leading non-perturbative corrections, we expect the strength of other non-perturbative effects to become negligible as well. In particular, an approximate shift symmetry that corresponds to the discrete shift \eqref{tmon} should become exact if all non-perturbative effects such that
\be
\langle {\bf m}, {\bf e} \rangle  =  m_i e^i \neq 0 
\label{product}
\ee
asymptotically vanish in that limit. In other words, if $e^i$ correspond to the charges of an EFT string, all non-perturbative effects satisfying \eqref{product} should die off along the saxionic string flow \eqref{imt} towards $r \rightarrow 0$, as otherwise the axionic shift symmetry will be broken.

 In general, we can consider the instanton charges  $m^i$ as components of an element ${\bf m}$ of a lattice $M_{\mathbb{Z}}$ of rank $n=\#\text{(s)axions}$, dual to the lattice $N_{\mathbb{Z}}$ in which the string charges ${\bf e}$ take values. Similarly, the
chiral fields $t^i$ can be identified with a vector ${\bm t}\in N_{\mathbb{C}} \equiv  N_{\mathbb{Z}}\otimes \mathbb{C}$, subject to periodic identifications  ${\bm t}\simeq {\bm t}+{\bf n} $ with ${\bf n}\in N_{\mathbb{Z}}$. 
One then expects to be able to identify, for each perturbative regime associated with a chiral sector ${\bm t}$,   a set of chiral operators $\cO_{\bf m}\equiv e^{2\pi\ii \langle {\bf m},{\bm t}\rangle}$ which respect the axionic periodicities and   are exponentially suppressed as we proceed along the EFT string flows. Such operators are natural chiral observables on the asymptotic region and in concrete string theory models they typically enter the EFT as $\half$BPS instanton corrections.  In this sense ${\bf m}$ represents the corresponding `instanton charges'. 

From a purely EFT viewpoint, we can make the above ideas more precise by specifying a perturbative regime in the following way. For each region  with perturbative axionic symmetries in the field space metric, we assume that the breaking of the said symmetry is measured by a set of instanton charges $\cC_{\rm I}\subset M_{\mathbb{Z}}$, which  specifies the following set of asymptotic chiral observables:\footnote{The operators \eqref{chiralring} may transform under an additional duality group, which should be taken into account in order to construct duality invariant observables. We will come back to this point in section \ref{sec:duality}.}
  \be\label{chiralring}
  \cO_{\bf m}(x)\equiv e^{2\pi\ii \langle {\bf m},{\bm t}(x)\rangle}\quad~~~~ {\bf m}\in \cC_{\rm I}\, .
  \ee

The perturbative region is then identified by requiring that $\cO_{\bf m}(x)$ is exponentially suppressed for any ${\bf m}\in \cC_{\rm I}$. Since $ |\cO_{\bf m}|=e^{-2\pi \langle {\bf m},{\bm s}\rangle}$, we get the identification
  \be\label{asymptregion}
  \text{perturbative region:}\quad~~~ \langle {\bf m}, {\bm s}\rangle \gg 1\quad~~~ \forall {\bf m}\in \cC_{\rm I}\, ,
  \ee
  where ${\bm s}\equiv \Im{\bm t}\in N_{\mathbb{R}}$ collectively denotes the saxions $s^i$.   This means that the asymptotic region \eqref{asymptregion} can be identified as the {\em deep interior} of a saxionic cone $\Delta$, which is defined as follows:
  \be
\Delta \equiv \{{\bm s}\in N_{\mathbb{R}}\ | \ \langle {\bf m}, {\bm s} \rangle > 0\,,\ \forall {\bf m}\in \cC_{\rm I}\}\, .
\label{defDelta}
\ee
That is, we characterise $\Delta$ as the interior of the cone $\cC_{\rm I}^\vee$.\footnote{We recall that in general, given a vector space $V$ and its dual $V^*$, for any set $S\subset V$ one can define the (closed convex) cone $S^\vee \equiv\{ \eta\in V^* | \langle \eta, v \rangle\geq 0\,,\forall v\in S \}\subset V^*$.} It is easy to see that $\Delta$ is convex: if ${\bm s}$ belongs to $\Delta$, then  $\lambda {\bm s}$ also does for any $\lambda>0$, and if ${\bm s}_1,{\bm s}_2\in \Delta$ then ${\bm s}_1+{\bm s}_2\in \Delta$.

Another necessary property of this perturbative region is that any  asymptotic limit inside $\Delta$ such that $\langle {\bf m}, {\bm s}\rangle$ diverges to $+\infty$ for some choices of ${\bf m}\in \cC_{\rm I}$, while it remains constant for the remaining ones,   is an infinite distance limit. This condition clearly regards the  K\"ahler structure of the field space, and not just the holomorphic one, and is necessary to identify \eqref{asymptregion} with a proper physical perturbative region in which the axionic symmetries are perturbatively preserved.  
 
Reciprocally, in terms of $\Delta$ we can define $\cC_{\rm I}$ as the discretisation of the cone $\Delta^\vee$ dual to $\Delta$:  
\be\label{CIdef} 
\cC_{\rm I}=\Delta^\vee\cap M_{\mathbb{Z}}\, .
\ee
If ${\bf m}\in \cC_{\rm I}$ then, necessarily, $-{\bf m}\notin \cC_{\rm I}$. Hence, $\cC_{\rm I}$ has the structure of a discrete convex cone.  

We will see that this characterisation of a given perturbative regime is quite general and applies to all the string theory examples that  will be analysed in section \ref{s:examples}. In particular,  the elements of $\cC_{\rm I}$ are typically associated with $\half$BPS instantonic brane configurations appearing in the UV completion of the EFT. As a prototypical example, the reader can keep in mind the heterotic Calabi--Yau compactifications that will be considered in subsections \ref{sec:Hetexample} and \ref{sec:hetCY}. In the  large volume regime, the saxionic cone $\Delta$ contains the K\"ahler cone, whose corresponding set $\cC_{\rm I}$ can be identified with the Mori cone of effective curves, wrapped by world-sheet instantons. As one approaches the boundary of the saxionic cone one expects non-perturbative corrections to become relevant and then the perturbative description to break down, as it clearly happens in the heterotic K\"ahler cone example.

Notice, however, that the correspondence between points in $\cC_{\rm I}$ and   microscopic BPS instantons is in general   more subtle. For instance, there may be  walls of marginal stability, across which  certain instantons cease to be BPS. In such cases,  we will require that microscopic instantons in $\cC_{\rm I}$ should be at least asymptotically BPS,  along any asymptotic limit within $\Delta$. 
Furthermore, the existence of at least one microscopic BPS instanton (or multi-instanton) for  {\em each} point in $\cC_{\rm I}$ --  the ``BPS completeness" of the instanton sector -- is expected but in general not obvious. We will not find any explicit counterexample to this expectation, but a more systematic study of this important question would be worthwhile. Keeping these possible caveats in mind, we will refer to $\cC_{\rm I}$ as the set of BPS instanton charges relevant for $\Delta$.

The next step is to  compare $\Delta$ with the lattice of string charges $N_\bbZ$. In each perturbative regime associated with a saxionic cone $\Delta$, for any charge ${\bf e}\in N_\bbZ$ we may formally write down a supersymmetric  EFT contribution  of the form \eqref{stringS}, with tension $\cT_{\bf e}=M^2_{\rm P}|\langle{\bf e},{\bm \ell}\rangle| \equiv M^2_{\rm P}|  e^{ i}\ell_i|$, where ${\bm \ell}\in M_{\mathbb{R}}$ is the vector of dual saxions $\ell_i$, and which is a good approximation of the string tension in the perturbative regime \eqref{asymptregion}. In this description, two charges ${\bf e}$ and ${\bf e}'$ preserve the same $\half$ of supersymmetry if and only if $\langle {\bf e},{\bm \ell}\rangle$ and $\langle {\bf e}',{\bm \ell}\rangle$ have the same sign. In the following we will dub as $\half$BPS strings those that correspond to the choice $\langle {\bf e},{\bm \ell}\rangle >0$, for which $\cT_{\bf e}=M^2_{\rm P} \langle{\bf e},{\bm \ell}\rangle$, while the opposite choice will be dubbed as  anti-BPS strings. 

Now, in our 4d EFT description, a string charge {\bf e} is associated to a flow in the saxionic variables. If the string solution is $\half$BPS one expects that $\langle {\bf e},{\bm \ell}\rangle >0$ all along an EFT string flow, as this corresponds to a positive linear energy density \eqref{linearE} at different scales $\Lambda=1/r$ and with different flow parameters $({\bm s}_0, r_0)$. Since by varying the latter one can cover the saxionic cone $\Delta$ with each EFT string flow, one would expect that a set of EFT string charges that are mutually BPS have a positive tension for each charge and at each point of $\Delta$.

This observation motivates the definition of a distinguished set $\cC_{\rm S}$ of string charges ${\bf e}$ for which $\langle{\bf e},{\bm \ell}\rangle$ is positive for  any point of $\Delta$.  More precisely, one may express the saxionic cone in terms of dual variables as 
\be 
\label{cPdef0}
\cP \equiv \Big\{{\bm \ell} \in M_{\mathbb{R}} \ | \ \ell_i = - \half  \left. \frac{\del K}{\del s^i}\right|_{{\bm s} \in \Delta}\Big\}\, , 
\ee
which is nothing but the image of $\Delta$ under a Legendre transform, and is well-defined for a perturbative K\"ahler potential $K$ displaying the axionic shift symmetries associated with $\Delta$. Then $\cC_{\rm S}$ can be identified with the discretisation of the cone dual to $\cP$:
\be\label{CSP}
\cC_{\rm S}\equiv \cP^\vee\cap N_{\mathbb{Z}}\,.
\ee
The charges in $\cC_{\rm S}$  correspond to  potential BPS strings preserving the same $\half$ of supersymmetry all over $\Delta$ or, dually, $\cP$. On the other hand, the charges in $-\cC_{\rm S}$ would correspond to anti-BPS strings preserving the opposite $\half$ of supersymmetry. 

Besides being BPS, our definition of EFT string requires that the string backreaction remains within the perturbative region  \eqref{asymptregion}
as we approach the core of the string. In general, this is not guaranteed for any ${\bf e}\in \cC_{\rm S}$. To see this, let us first rewrite the saxionic flow \eqref{imt} as
\be\label{ssssflow}
\begin{aligned}
& {\bm s}={\bm s}_0-\frac1{2\pi}\,{\bf e}\log\lambda\equiv {\bm s}_0+\sigma\,{\bf e}  \quad~~\text{with \quad $\lambda\equiv \frac{r}{r_0}\equiv e^{-2\pi\sigma}$}\, .
\end{aligned}
\ee
Consider a flow that starts in the perturbative region \eqref{asymptregion}, that is $\langle {\bf m},{\bm s}_0\rangle\gg 1$ for any ${\bf m}\in\cC_{\rm I}$.  
The behaviour of chiral operators \eqref{chiralring} along the flow \eqref{ssssflow} is
\be
|\calo_{\bf m}|= e^{-2\pi \langle {\bf m}, {\bm s} \rangle} = e^{-2\pi \langle {\bf m}, {\bm s}_0 \rangle}\times \lambda^{ \langle {\bf m},{\bf e}\rangle } \, .
\label{inststr}
\ee  
If ${\bf e}\in\overline\Delta \equiv$ closure of $\Delta$, then $\langle {\bf m},{\bf e}\rangle \geq 0$ for any ${\bf m}\in\cC_{\rm I}$. If $\langle {\bf m},{\bf e}\rangle > 0$, then  all instanton effects in $\cC _{\rm I}$ asymptotically die off  as we take the limit $\lambda\rightarrow 0$, and if $\langle {\bf m},{\bf e}\rangle = 0$, they remain constant but still arbitrarily small for $\langle {\bf m}, {\bm s}_0 \rangle$ large enough. In both cases the flow  remains in the region  \eqref{asymptregion} for any $\lambda\in (0,1]$ and by assumption reaches the infinite distance boundary at the string core $\lambda= 0$. Such a charge ${\bf e}$  represents an EFT string, as defined in \eqref{CSEFT0}. 

If instead ${\bf e} \in \cC_{\rm S} - \cC^{\text{\tiny EFT}}_{\rm S}$ then there must be an instanton charge ${\bf m} \in \cC_{\rm I}$ such that $\langle{\bf m}, {\bf e}\rangle <0$. The associated non-perturbative effects grow along the string flow of charges ${\bf e}$, until they reach $\cO(1)$ contributions at a finite radial distance 
\be
\lambda_*=\frac{r_*}{r_0}=e^{-\frac{2\pi \langle{\bf m}, {\bm s}_0\rangle}{| \langle{\bf m}, {\bf e}\rangle|}}\, .
\label{lambdastr}
\ee
At this point we have that $\langle {\bf m}, {\bm s}(r_*) \rangle =0$, which means that ${\bm s}(r_*)$ necessarily belongs to the finite field distance boundary of $\overline\Delta$. Hence, for charges  ${\bf e} \in \cC_{\rm S} - \cC^{\text{\tiny EFT}}_{\rm S}$, the weakly-coupled EFT description breaks down along the string flow, due to some BPS instanton operator \eqref{chiralring} that becomes relevant and signals a non-perturbative regime. In this sense, we refer to strings corresponding to ${\bf e} \in \cC_{\rm S} - \cC^{\text{\tiny EFT}}_{\rm S}$ either as {\em strongly coupled} or non-EFT.

Therefore, we may identify the set of EFT strings associated with a given perturbative regime with the following lattice cone
 \be
 \cC^{\text{\tiny EFT}}_{\rm S}=\overline\Delta\cap N_{\mathbb{Z}}
 \label{CSEFT}\, .
 \ee
Notice that to arrive at this definition we have not imposed that $\cT_{\bf e} > 0$ along the flow, although as already mentioned in section \ref{sec:strinfdis} we expect this to be the case for any EFT string. In other words, we are lead to conclude that for a healthy EFT we must also have
\be
 \cC^{\text{\tiny EFT}}_{\rm S}\subset  \cC_{\rm S}\quad\Leftrightarrow\quad \overline\Delta\subset \cP^\vee\,.
\ee
As we will see, this condition is indeed satisfied for any string model considered in section \ref{s:examples}.

Similarly to the case of instantons, these EFT definitions could be challenged by their microscopic realisation. Indeed, in all string models that we will encounter the potential $\half$BPS string charges ${\bf e}\in \cC_{\rm S}$ admit a microscopic realisation in terms of wrapped branes. The BPSness of such brane configurations can in fact depend on additional UV information invisible to the EFT and experience the presence of walls of marginal stability, that could be crossed along a string flow. Needless to say, these general issues are crucial for understanding the possible BPS completeness of $\cC_{\rm S}$, and we will not attempt to exhaustively address them in  the present paper. However, all the examples that we will consider neatly suggest that these UV issues are in fact absent for the subsector of EFT string charges $\cC^{\text{\tiny EFT}}_{\rm S}$.

One can see that if $K$ takes the form \eqref{Klog} with $P$ homogeneous in the saxions, then $\cP$ is itself a cone. This is  what will happen in all the examples that we will consider. We will also see that $\cP$ can take different shapes, but by using \eqref{Klog} it is easy to see that  the  infinite distance limit obtained by homogeneously rescaling all  saxions always converge to the tip of $\cP$. This means that all BPS string tensions vanish in this limit. More generically, EFT string flows converge to  infinite distant points on the boundary $\cP$.  In particular, we have seen above that for a general K\"ahler potential of the form \eqref{Klog} the tension $\cT_{\bf e}=M^2_{\rm P}\langle {\bf e},{\bm \ell}\rangle$  vanishes asymptotically as
\be\label{univtensionflow}
\cT_{\bf e}\sim \frac{M^2_{\rm P}}{\sigma}\, ,\quad~~~ \text{for $\sigma\rightarrow\infty$}\, ,
\ee
along the flow \eqref{ssssflow}. Hence, it will end on the subset
\be\label{face}
\cF_{\bf e}=\overline\cP\cap H_{\bf e}\, ,\quad\quad \text{with}\quad H_{\bf e}\equiv\{{\bm\alpha}\in M_{\mathbb{R}}\ |\ \langle{\bm\alpha},{\bf e}\rangle=0 \}\,. 
\ee
In all the examples that we will consider, if ${\bf e}\in\cC_{\rm S}^{\text{\tiny EFT}}$ then $\cF_{\bf e}$ will be an infinite distance boundary face of $\cP$. $\cF_{\bf e}$ can be codimension-one, but can also have higher codimension, up to the maximal one, corresponding to $\cF_{\bf e}$ being just the tip of the cone $\cP$. The codimension of $\cF_{\bf e}$  counts the number  of linearly independent string charges ${\bf e}'$ whose  (probe) tensions $T_{{\bf e}'}=M^2_{\rm P}\langle {\bf e}',{\bm \ell}\rangle$ vanish asymptotically along the flow. We will say that the  string flow degeneracy is of order $p$ if it ends on a codimension-$p$ face $\cF_{\bf e}$.
We will also refer to order-one flows as {\em non-degenerate}, while flows of higher degeneracy will be dubbed {\em degenerate}.\footnote{If the K\"ahler potential takes the form \eqref{Klog}, we can also associate the degeneracy of the flow to the properties of the homogeneous function $P(s)$. Following \cite{Grimm:2018cpv}, one can divide the perturbative region into different growth sectors corresponding to a different ordering $s_1\gg s_2\gg \dots$ characterising what saxions grow faster. Within each growth sector, the function $P(s)$ will be approximated by a single monomial dominating asymptotically, which can be parametrised as  $P(s)\propto s_1^{n_1}s_2^{n_2-n_1}\dots s_i^{n_i-n_{i-1}}+\dots$. If we send for instance $s_i\rightarrow\infty$, $i=1,2,\dots j$, then  $n_j$ is the singularity type which is equivalent to sum over all the exponents of the saxions taken to the large field limit. If $P(s)$ is approximated by the same monomial for all the growth sectors, the perturbative regime only contains non-degenerate flows. Otherwise, it will contain degenerate flows. This latter case is associated to infinite distance limits in which the singularity type does not increase in the enhancement chain, so $n_i-n_{i-1}=0$ and some saxions are absent in some of the monomials. \label{f:singtype}}

Finally, we can also define the instantonic analogue of \eqref{CSEFT} as follows: \be\label{EFTinstcone0}
\cC^{\text{\tiny EFT}}_{\rm I}\equiv  \overline\cP\cap M_{\mathbb{Z}} \quad\subset\ \cC_{\rm I}\, ,
\ee
where the inclusion in $\cC_{\rm I}$ follows from \eqref{CIdef}.  
In appendix \ref{app:inst} it is shown how, analogously to what happens for EFT strings, the backreaction of  instantons of charges ${\bf m}\in\cC_{\rm I}^{\text{\tiny EFT}}$ generate acceptable EFT flow solutions, while the BPS instantons of charges  ${\bf m}\in\cC_{\rm I}-\cC_{\rm I}^{\text{\tiny EFT}}$ do not. 
We will dub these two classes as EFT and non-EFT instantons, respectively.  
Interestingly, only non-EFT BPS instantons  can generate the finite distance strong coupling effects along the flow of non-EFT strings discussed around \eqref{lambdastr}. This follows immediately from  \eqref{CSP} and $\cC^{\text{\tiny EFT}}_{\rm I}\subset \cP$, which imply that $\cC_{\rm S}\subset (\cC^{\text{\tiny EFT}}_{\rm I})^\vee$.  That is, if  ${\bf m}\in\cC_{\rm I}^{\text{\tiny EFT}}$, then  $\langle{\bf m},{\bf e}\rangle \geq 0$ for any ${\bf e}\in \cC_{\rm S}$.

We have summarised all the relevant definitions of the conical structures introduced above in Table~\ref{Tab:cones}.

\begin{table}[H]
	\centering
	\begin{tabular}{|c | c |}
		\hline 
		\rowcolor{gray!30} \xrowht{15pt} EFT data & Cone  
		\\ 
		\hline \xrowht{10pt}
		Saxionic cone &  $\Delta = \{{\bm s}\in N_{\mathbb{R}}\ | \ \langle {\bf m}, {\bm s} \rangle > 0\,,\ \forall {\bf m}\in \cC_{\rm I}\}$
		\\ 
		\xrowht{15pt}
	    EFT strings & $\cC^{\text{\tiny EFT}}_{\rm S}=\overline\Delta\cap N_{\mathbb{Z}}$
	    \\
		\xrowht{15pt}
		BPS instantons & $\cC_{\rm I}= \Delta^\vee\cap M_{\mathbb{Z}}$
		\\
	    \hline \xrowht{15pt}
	    Dual saxionic cone & $\cP = \Big\{{\bm \ell} \in M_{\mathbb{R}} \ | \ \ell_i = - \half  \left. \frac{\del K}{\del s^i}\right|_{{\bm s} \in \Delta}\Big\}$
	    \\
		\xrowht{15pt}
		EFT instantons & $\cC^{\text{\tiny EFT}}_{\rm I}=  \overline\cP\cap M_{\mathbb{Z}}$
		\\
		\xrowht{15pt}
		BPS strings & $\cC_{\rm S}= \cP^\vee\cap N_{\mathbb{Z}}$
		\\
		\hline 
	\end{tabular}
	\caption{EFT conical structures induced by strings and instantons.\label{Tab:cones}}
\end{table}

Notice that, in terms of the natural paring between string and instanton charges, the definition \eqref{CSEFT} of EFT string could be stated as 
\be\label{altCC}
\begin{aligned}
\cC^{\text{\tiny EFT}}_{\rm S}&=\{{\bf e}\in N_{\mathbb{Z}}|\langle {\bf m},{\bf e}\rangle \geq 0 \quad \forall {\bf m}\in\cC_{\rm I}\}\, .
\end{aligned}
\ee
Physically, given a string of charge ${\bf e}$ and instanton of charge ${\bf m}$, the pairing $\langle {\bf m},{\bf e}\rangle$ gives the magnetic charge of the instanton  under the two-form gauge field $\cB^{{\bf e}}_{2}\equiv e^i\cB_{2,i}$ electrically sourced by the string. Indeed, as discussed in more detail in appendix \ref{app:inst} -- see equation \eqref{instsource} -- this magnetic charge is measured by the flux 
\be
\frac1{2\pi}\int_{S^3}\cH_3^{\bf e}=\frac{1}{2\pi}e^i\int_{S^3}\cH_{3,i}=\langle {\bf m},{\bf e}\rangle\, ,
\ee
with $\cH_3^{\bf e} = \d \cB^{{\bf e}}_{2}$ and $S^3$ a three-sphere surrounding the instanton.  One may then reformulate \eqref{altCC} as follows:

\begin{center}
\emph{A BPS string belongs to $\cC^{\text{\tiny EFT}}_{\rm S}$ whenever all instantons in $\cC_{\rm I}$ carry non-negative \\ magnetic charge under the two-form field  $\cB^{{\bf e}}_{2}$ that couples to the string.} 
\end{center}

\noindent Dually,  we can also interpret the pairing $\langle {\bf m},{\bf e}\rangle$ as the axionic charge of the string corresponding to  the axion that couples `electrically' to the instanton:  if $a^{\bf m}\equiv m_ia^i$ denotes this axion, then $a^{\bf m}\rightarrow a^{\bf m} +\langle {\bf m},{\bf e}\rangle$ around the string. We can then  analogously say that  a BPS instanton  belongs to $\cC^{\text{\tiny EFT}}_{\rm I}$ if and only if all  strings in $\cC_{\rm S}$ carry non-negative axionic charge under the axion $a^{\bf m}$ that couples to the instanton.

\subsection{A simple example}
\label{sec:Hetexample}

Let us consider heterotic strings compactified in a Calabi--Yau three-fold $X$ and focus on the  large-volume, weak string coupling perturbative regime and the axionic symmetries that arise in this region of moduli space. The relevant  $\half$BPS instantons arise from world-sheet instantons wrapping holomorphic curves on $X$, and from NS5-branes wrapping $X$. Reversing their orientation, one obtains the corresponding set of anti-instantons. The cone of $\half$BPS instanton charges $\cC_{\rm I}$ is thus generated by effective curve classes, together with the NS5-brane charge. Similarly, mutually $\half$BPS 4d strings arise from NS5-branes wrapped on effective divisors, together with fundamental strings which are point-like in $X$. Their corresponding cohomology classes generate the discrete cone $\cC_{\rm S}$.\footnote{To be precise, one should take into account F1 charges induced by world-volume flux and curvature corrections  on the NS5-brane, which may translate into slightly different generators for the charge lattices. In the following we will ignore this effect, which plays no significant role in the discussion.} $\cC_{\rm I}$ and $\cC_{\rm S}$ are subsets of the dual lattices $M_{\mathbb{Z}}\simeq H_0(X,\mathbb{Z})\oplus H_4(X,\mathbb{Z})/\text{(torsion)}$ and  $N_{\mathbb{Z}}\simeq H^6(X,\mathbb{Z})\oplus H^2(X,\mathbb{Z})/\text{(torsion)}$, respectively.

The saxionic cone $\Delta$ corresponding to this perturbative regime is parametrised by the 4d dilaton and  the K\"ahler moduli. More precisely, the saxions are given by $s^i = (s^0 , s^a)$. The K\"ahler saxions $s^a$  arise from the decomposition of the (string frame) K\"ahler form
\be\label{Jexp}
J=  s^a[D_a]\, ,
\ee
in an integral basis $[D_a]$   Poincar\'e  dual to a set of divisors $D_a$. $J$ takes value in the K\"ahler cone $\calk(X)$, while the universal saxion $s^0$ is given by
\be\label{hetdil0}
s^0=\frac{1}{3!}\,e^{-2\phi}\kappa_{abc}s^as^bs^c=e^{-2\phi}V_X\, ,
\ee
where $\phi$ is the 10d dilaton, $\kappa_{abc}\equiv D_a\cdot D_b\cdot D_c $ are the triple intersection numbers of $X$ in this basis, and $V_X$ is the volume of $X$ in string units.
So, the relevant saxionic cone is
\be\label{hetDelta0}
\Delta=\mathbb{R}_{> 0}\oplus \calk(X)\,,
\ee
where $\mathbb{R}_{> 0}$ is parametrised by $s^0$.

The closure of the K\"ahler cone  is the cone  ${\rm Nef}^1 (X)$ generated by  nef divisors -- see appendix \ref{app:cones} for a summary of the relevant terminology. Applying  our general definition \eqref{CSEFT} to the present example we then see that $ \cC^{\text{\tiny EFT}}_{\rm S}$ is generated by the nef divisor classes and $H_0(X, \bbZ)_+\simeq \mathbb{Z}_{\geq 0}$.  
So, to sum up, we have the following sets of BPS charges
\begin{subequations}\label{hetlattices}
\begin{align}
\cC_{\rm S} & =\{\text{F1 $+$  NS5-branes on effective divisors}\}\simeq \mathbb{Z}_{\geq 0}\oplus \text{Eff}^1(X)_\mathbb{Z}\, , \label{hetlatticesA} \\
\cC_{\rm I}& =\{\text{NS5-branes on $X$ $+$   F1 on effective curves}\}\simeq \mathbb{Z}_{\geq 0}\oplus \text{Eff}_1(X)_\mathbb{Z}\, ,\label{hetlatticesB}\\
\cC^{\text{\tiny EFT}}_{\rm S} & =\{\text{F1 $+$  NS5-branes on nef divisors}\}\simeq \mathbb{Z}_{\geq 0}\oplus \text{Nef}^1(X)_\mathbb{Z}\, .\label{hetlatticesC}
\end{align}
\end{subequations}
The pairing $\langle {\bf m},{\bf e}\rangle$ between $\cC_{\rm I}$ and $\cC_{\rm S}$ just corresponds  to the geometrical intersection number. Notice that indeed, by definition of nef divisors, we can identify $\overline\Delta$ as the cone dual to the cone generated by $\cC_{\rm I}$. 
As will be discussed in the next section  an EFT string completeness hypothesis follows as a direct consequence of  Conjecture \ref{conj:DASC}. In this setup, such a hypothesis implies that every nef divisor admits an effective representative. This mathematical property has already been conjectured in \cite{Katz:2020ewz}, for similar underlying reasons. See subsection \ref{sec:sugrastrings} for the connection between our setup and the one in \cite{Katz:2020ewz}.

%In simple Calabi--Yau manifolds, like the ones considered in section \ref{s:examples}, the K\"ahler cone is rational polyhedral, which implies that it has a simple description in terms of the world-sheet instanton charges and the saxion vevs. This property is however known not to be true in general. Nevertheless, it was conjectured by Morrison in \cite{Coneconj} that if one restricts the K\"ahler cone to  a fundamental region under a duality group preserving the asymptotic structure, one recovers the rational polyhedral structure. This implies that any fundamental region of the K\"ahler cone, and therefore $\Delta$, is generated by a finite set of integral vectors ${\bf v}_I \in N_{\mathbb{Z}}$
%\be\label{RPcone}
%\Delta= \Big\{\sum_{I} s^I {\bf v}_I = s^0 {\bf v}_0 +  \sum_{a} s^a {\bf v}_a\ \Big|\ s^0, s^a \in\mathbb{R}_{>0}\Big\} \, .
%\ee

%In the following we will assume that Morrison's Cone conjecture holds true, which greatly simplifies the structure of $\Delta$. 

In our perturbative regime, the leading contribution to the K\"ahler potential for the above saxions reads
\be\label{hetK}
K=-\log s^0 -\log \left(\frac{1}{3!}\kappa_{abc}s^as^bs^c\right)+\ldots\, ,
\ee
which has indeed the form \eqref{Klog}. 
So the dual saxions $\ell_i=(\ell_0,\ell_a)$ are given by
\be\label{hetlin}
\ell_0=\frac1{2s^0}=\frac{e^{2\phi}}{2V_X}\,, \qquad \ell_a=\frac{1}{4V_X}\kappa_{abc}s^bs^c\, .
\ee
It is not so simple to describe  the structure of $\cP$ and $\cC^{\text{\tiny EFT}}_{\rm I}$ in general terms, see \eqref{cPdef0} and \eqref{EFTinstcone0}. So let us for now focus on a specific example, which will allow us to better illustrate some features of these class of models. We will come back to the general discussion in subsection \ref{sec:duality}.

Let us take the Calabi--Yau studied in \cite{Candelas:1994hw}.  This three-fold can be regarded as an elliptic fibration over $\mathbb{P}^2$, and is such that $b_2(X)=2$. The cone of effective divisors ${\rm Eff}^1(X)$ is generated by two divisors $B$ and  $L$: $B$ is the image of the $\mathbb{P}^2$ base under the global section of the elliptic fibration, while $L$ is the `vertical' divisor obtained by pulling back the base $\mathbb{P}^1\subset \mathbb{P}^2$ by using the fibration projection. The nef cone ${\rm Nef}^1(X)$ is instead generated by $L$ and the `horizontal' divisor $H\equiv B+3L$.  In the nef basis, the  intersection numbers are summarised by the formal polynomial
\be
\cI_X=9H^3+3H^2 L+H L^2\, ,
\ee   
where the coefficient of each monomial gives the value of the corresponding triple intersection.
The dual Mori cone $\overline{\text{Eff}}_1(X)$ is generated by the effective curves 
\be
h=L^2\, , \qquad l=B L=H L -3L^2\, .
\ee
The curve $h$ can be identified with elliptic fibre, while $l$ can be identified with the push-forward of a $\mathbb{P}^1$ in the base through the global section of the elliptic fibration.

The K\"ahler cone is simplicial and is generated by the Poincar\'e dual classes $[H]$ and $[L]$. We can then expand the (string frame) K\"ahler form as follows
\be
J=s^1 [H] + s^2 [L]\ \in\ H^2(X,\mathbb{R})\, ,
\ee 
where the K\"ahler saxions $s^1, s^2 \in \bbR_{>0}$ measure the volumes of the curves $h$ and $l$ respectively and, together with \eqref{hetdil0},  parametrise $\Delta$. The K\"ahler potential \eqref{hetK} takes the form
\be\label{hetKexample}
K = -\log s^0 -\log \Big[3  (s^1)^3+3 (s^1)^2s^2+ s^1(s^2)^2 \Big]+\ldots \, ,
\ee
and the dual saxions are 
\be\label{lhetex}
\begin{aligned}
\ell_0&=\frac{1}{2s^0}\, , \\
\ell_1&= \frac{9 (s^1)^2+6 s^1s^2+(s^2)^2}{6 (s^1)^3+6(s^1)^2s^2+2s^1(s^2)^2}\, ,\\
\ell_2&= \frac{3(s^1)^2+2s^1s^2}{6 (s^1)^3+6(s^1)^2s^2+2s^1(s^2)^2}\, .
\end{aligned}
\ee
\begin{center}
	\begin{figure}[htb]
		\centering
		\includegraphics[width=15cm]{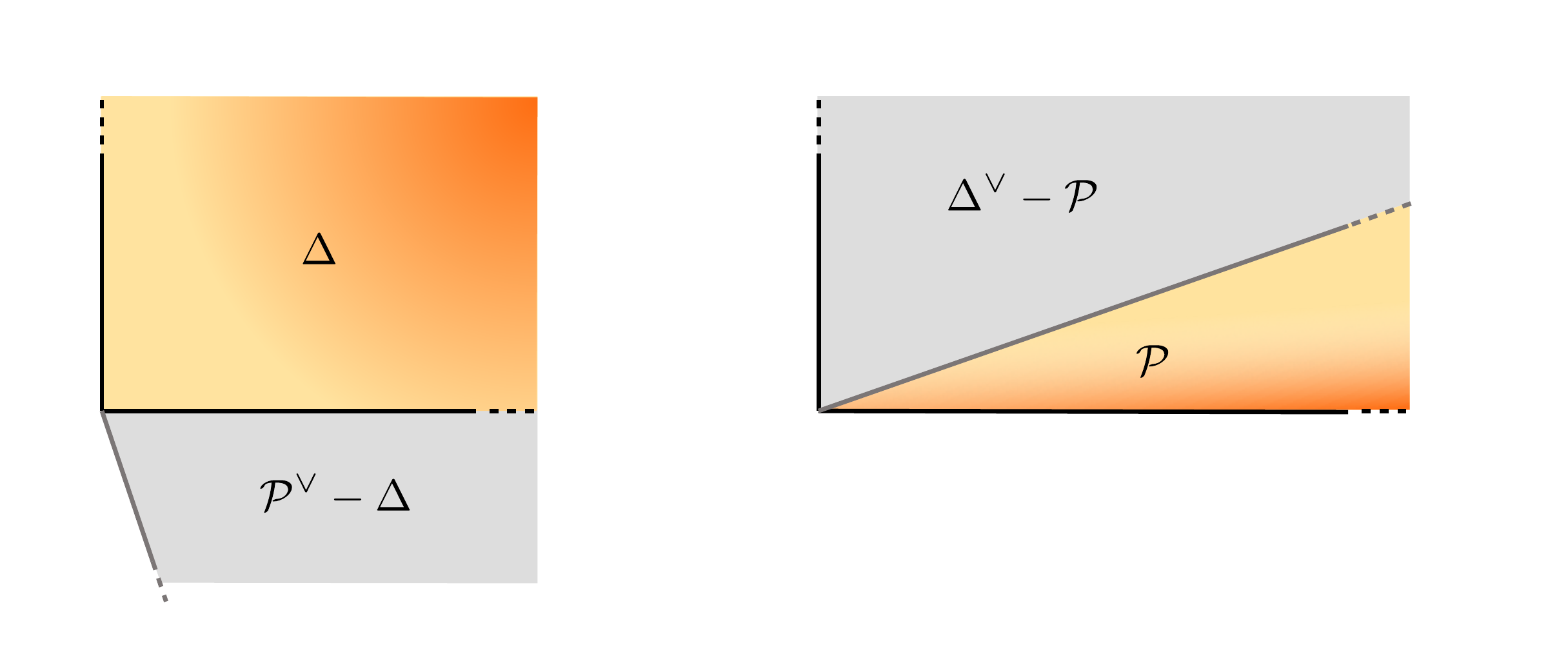}
		\caption{\small Saxionic cones and their duals $\Delta\subset \cP^\vee$ and  $\cP\subset \Delta^\vee$ of the two-K\"ahler-moduli heterotic model. Here we are suppressing the $\mathbb{R}_{> 0}$ piece of the moduli space that corresponds to $s^0$. The darker shaded regions correspond to the perturbative regime \eqref{asymptregion}. \label{f:excones0}}   
	\end{figure}
\end{center}

\begin{center}
	\begin{figure}[htb]
		\centering
		\includegraphics[width=14cm]{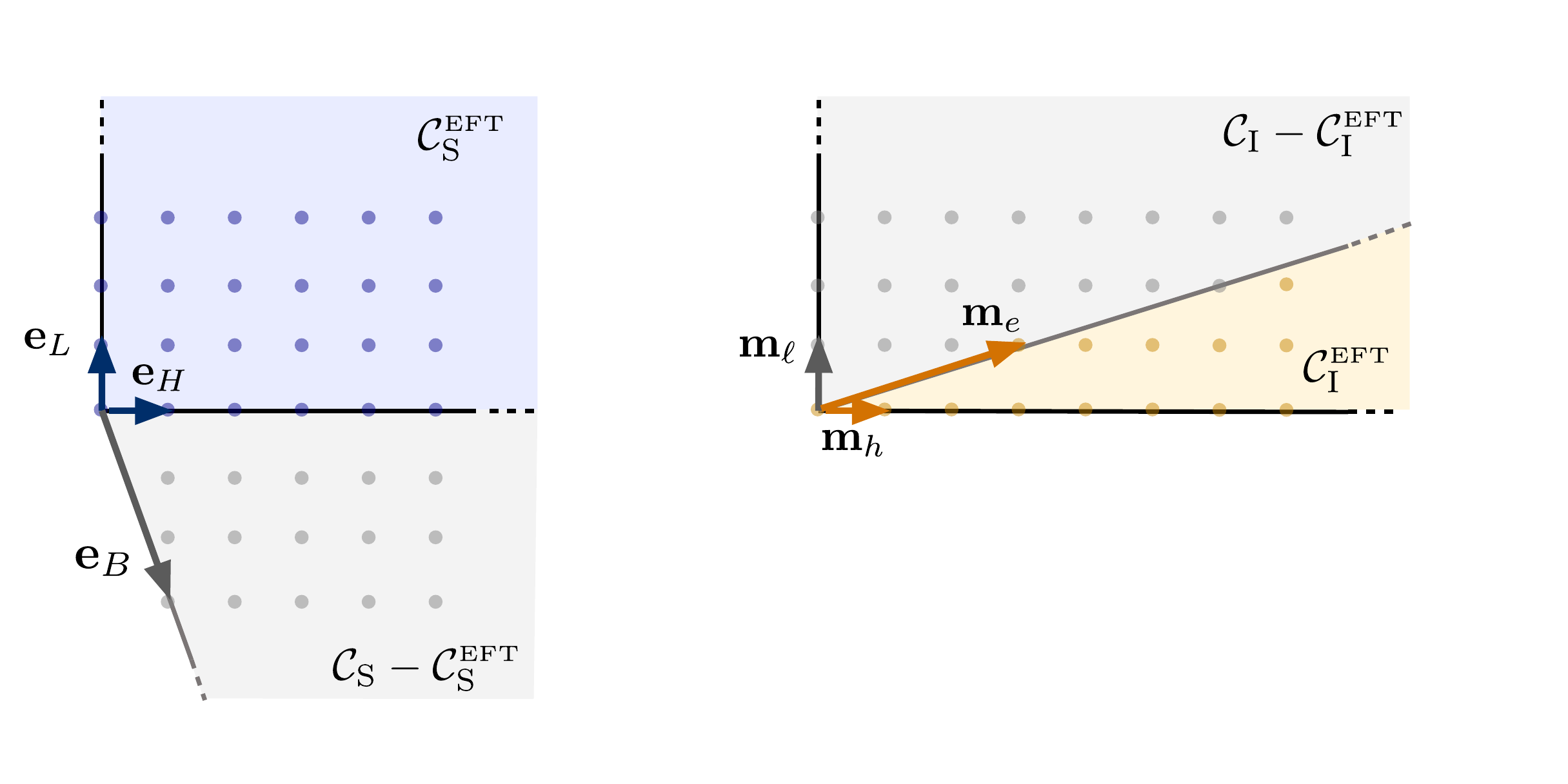}
		\caption{\small BPS and EFT string and instanton lattice cones $\cC^{\text{\tiny EFT}}_{\rm S}\subset \cC_{\rm S}$ and  $\cC^{\text{\tiny EFT}}_{\rm I}\subset \cC_{\rm I}$ of the two-moduli heterotic model. \label{f:exlattices0}}   
	\end{figure}
\end{center}

With these data, one can easily describe each of the continuous cones defined above:
\begin{subequations}
\label{contcones}
\bea
\Delta & = & \big\{(s^0, s^1, s^2)\in\mathbb{R}^3|s^0,s^1,s^2> 0\big\}\, , \\
\cP & = &\big\{(\ell_0, \ell_1, \ell_2)\in\mathbb{R}^3|\ell_0 > 0, \ell_1> 3\ell_2> 0\big\}\, , 
\eea
\end{subequations}
see figure \ref{f:excones0}, which also shows in darker colour the perturbative region \eqref{asymptregion} both in $\Delta$ and in $\cP$. The corresponding discrete cones of charges are:
\begin{subequations}
\label{disccones}
\bea
\cC_{\rm I} & =&  \big\{(m_0, m_1, m_2)\in\mathbb{Z}^3| m_0, m_1, m_2\geq 0 \big\} \simeq \langle\text{NS5}, h, l \rangle  \, ,  \\
\cC_{\rm S} & = &\big\{(e^0, e^1, e^2)\in\mathbb{Z}^3 | e^0, e^1 \geq 0, 3e^1+ e^2 \geq 0\big\}\simeq  \langle\text{F1}, B, L\rangle \, ,\\
\cC^{\text{\tiny EFT}}_{\rm S}  & =  & \big\{(e^0, e^1, e^2)\in\mathbb{Z}^3|e^0, e^1, e^2\geq 0 \big\}\simeq  \langle\text{F1}, H, L \rangle \, , \label{het_Cseft} \\
\cC^{\text{\tiny EFT}}_{\rm I}  & = &\big\{(m_0, m_1, m_2)\in\mathbb{Z}^3|m_0\geq 0, m_1\geq 3 m_2\geq 0\big\} \simeq \langle\text{NS5}, h,l+3h \rangle  \, ,
\eea
\end{subequations}
as illustrated in figure \ref{f:exlattices0}. Note that $\cC^{\text{\tiny EFT}}_S$ is dual to $\cC_{\rm I}$, as it happens in general, and that also $\cC^{\text{\tiny EFT}}_{\rm I}$ is  the dual of $\cC_{\rm S}$.  In particular,  $\cC_{\rm S}$   contains NS5-instantons wrapping effective divisors and, dually \cite{MR3019449}, $\cC^{\text{\tiny EFT}}_{\rm I}$ contains F1-instantons  wrapping `movable' curves -- see appendix \ref{app:cones}.  
The general differences between the  EFT and non-EFT strings and instantons have been discussed in general terms in subsection \ref{sec:cones} and in appendix \ref{app:inst},  respectively. We now illustrate them in this specific example, focusing on the string sector. 

Consider first the  EFT sector  $\cC^{\text{\tiny EFT}}_{\rm S}$, which is generated by the charges ${\bf e}_{\rm F1}=(1,0,0)$, ${\bf e}_{H}=(0,1,0)$ and ${\bf e}_{L}=(0,0,1)$. The corresponding strings are  {\em elementary}, in the sense that they cannot be decomposed as superposition of other BPS strings. The flows \eqref{ssssflow}  corresponding to ${\bf e}_{\rm F1}$ describes a zero coupling limit, in which the string coupling goes to zero, while the string frame internal metric remains constant. From the viewpoint of $\cP$,  these flows  end on the codimension-one face of $\cP$ defined by $\ell_0=0$, which is precisely  $\cF_{{\bf e}_{\rm F1}}$ as defined in \eqref{face}. On this face, precisely the strings with charges multiple of ${\bf e}_{\rm F1}$ become tensionless and the flow is non-degenerate. Also the string flows  associated with ${\bf e}_{L}$ are non-degenerate, as they end on the codimension-one face $\cF_{{\bf e}_{L}}\equiv \overline\cP\cap\{\ell_2= 0\}$, at which only the strings of charge proportional to ${\bf e}_{L}$ become tensionless. Instead, something different happens for the EFT flows generated by a string of charge ${\bf e}_{H}$. Indeed, in this case $s^1 \sim \sigma \rightarrow \infty$ and then from \eqref{lhetex} it is clear that both $\ell_1$ and $\ell_2$ vanish asymptotically along the flow.   Hence, both elementary strings become tensionless at the same rate:
\beq
\cT_{{\bf e}_{H}}\simeq \frac{3}{2\sigma}M_P^2\,, \quad \cT_{{\bf e}_{L}}\simeq\frac{1}{2\sigma}M_P^2 \,,
\eeq
 as we approach $\cF_{{\bf e}_{H}}$ along the flow.
In other words, these flows end on the codimension-2 face $\cF_{{\bf e}_{H}}$  and are degenerate of order-two.

\begin{center}
	\begin{figure}
		\centering
		\includegraphics[width=15cm]{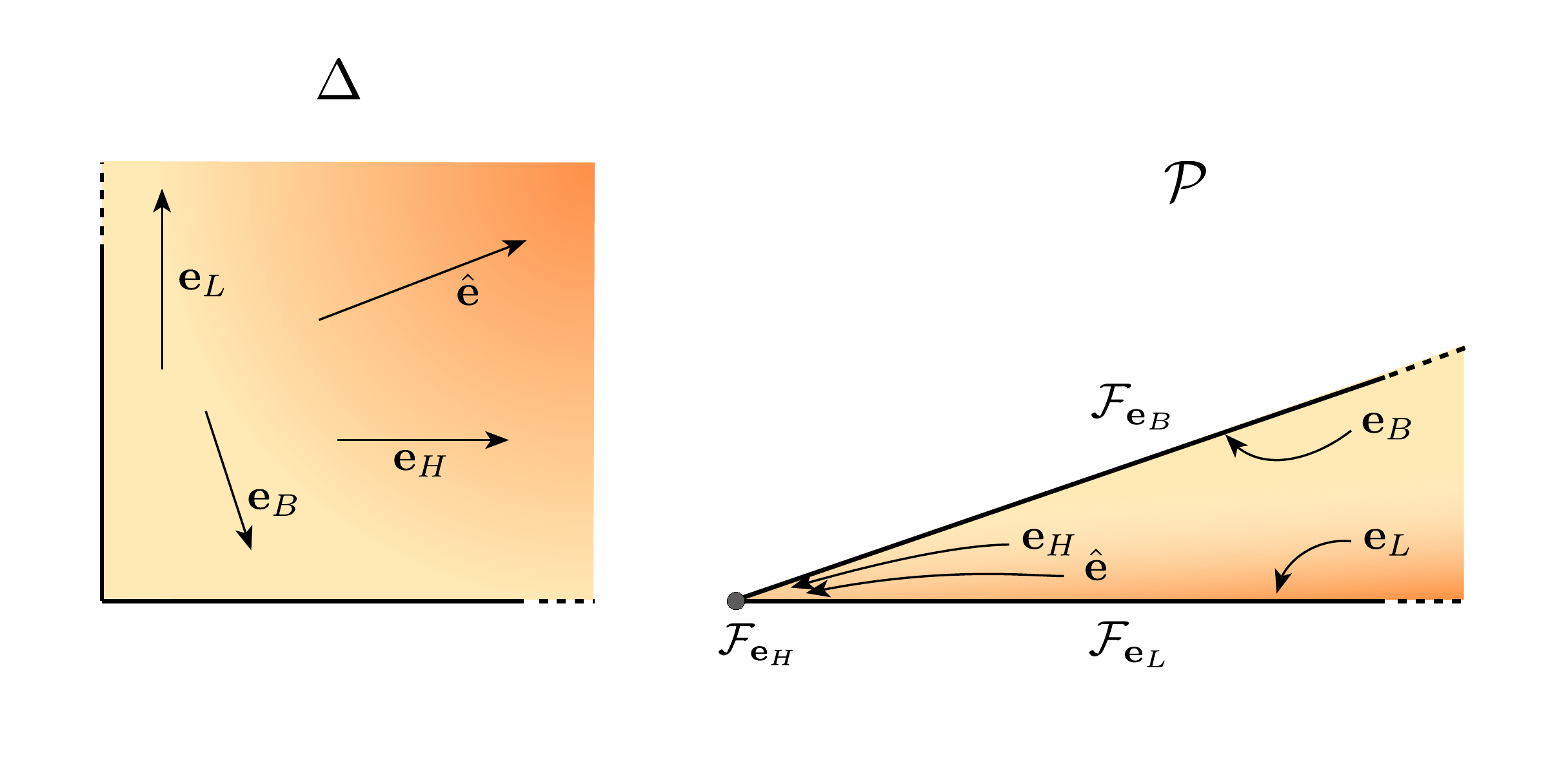}
		\caption{\small{Examples of saxionic string flows \eqref{ssssflow}.  On the left, string flows in the saxionic cone $\Delta$ (the $s^0$ direction is suppressed) and, on the right, their maps in the dual saxionic cone $\cP$ (the $\ell_0$ direction is suppressed). The charges ${\bf e}_H, {\bf e}_L, {\bf e}_B$ are elementary, while $\hat{\bf e}=(0,1,3)$ is not. The picture also shows that the ${\bf e}_B$  flow approaches the finite distance boundary $s^2=0$ of $\Delta$ and, correspondingly,  $\cF_{{\bf e}_B}$ lies in the strongly coupled light shaded region of $\overline\cP$.  
		\label{f:exflows0}}}   
	\end{figure}
\end{center}

Let us now consider the flow generated by the non-EFT charge  ${\bf e}_B=(0,1,-3)$.  The string flow \eqref{imt} reads 
\be
s^1=s_0^1-\frac1{2\pi}\log\frac{r}{r_0}\, ,\quad\quad s^2=s^2_0+\frac3{2\pi}\log\frac{r}{r_0}\, .
\label{Eflow}
\ee
We then see that $s^2(r)$ flows towards strong coupling as $r\rightarrow 0$ and reaches a strongly coupled regime at $r_*=r_0\exp{(-\frac{2\pi s^2_0}{3})}$, at which $s^2=0$ and $s^1=s_0^1+\frac13 s^2_0$. 
Geometrically, in this limit the base $\mathbb{P}^2$ collapses to zero size while the elliptic fibre volume remains finite.   
Even though $r_*/r_0$ can be made exponentially small by choosing large enough $s^2_0$, this behaviour is clearly opposite to the one of strings belonging to $\cC^{\text{\tiny EFT}}_{\rm S}$.  In the case of \eqref{Eflow} one must fine-tune the initial value $s_2^0$ to be large enough, the corresponding RG flow does not correspond to a marginally relevant operator, and it does not flow to weak coupling as we approach the string. Furthermore, differently from strings with charges in $\cC^{\text{\tiny EFT}}_{\rm S}$, the field-space distance travelled by the flow from $r_0$ to $r_*$ is classically finite, as one can explicitly check by using \eqref{flowQ2} and \eqref{d*}, see also figure~\ref{f:exflows0}.

Finally, one can compute the behaviour of the ${\bf e}_B$-string tension  around $r\simeq r_*$:
\be\label{Estringasy}
\begin{aligned}
\cT_{{\bf e}_B} =M^2_{\rm P}(\ell_1-3\ell_2) \simeq \frac{3 M^2_{\rm P}\left(s^2_0-\frac3{2\pi}\log\frac{r_0}{r}\right)^2}{27(s^1_0)^3+27(s_0^1)^2s^2_0 +9s_0^1(s^2_0)^2+(s^2_0)^3}\, ,
\end{aligned}
\ee
and conclude that it vanishes at $r=r_*$. In other words,  we see that also the non-EFT ${\bf e}_B$-flow reaches a face $\cF_{{\bf e}_B}$ of $\cP$ defined as in \eqref{face}. However, $\cF_{{\bf e}_B}$ is quite different from the faces corresponding to EFT strings, for two main reasons. First, $\cF_{{\bf e}_B}$ is at {\em finite} field space distance. Second,  according to the above arguments as we approach $\cF_{{\bf e}_B}$ we expect strong non-perturbative corrections to modify  \eqref{hetKexample} and break the axionic symmetry.  In fact, the BPS instanton of charges ${\bf m}_l=(0,0,1)$ becomes unsuppressed along the flow, since $\langle {{{\bf m}_l},{\bf e}_B}\rangle<0$. This matches with the fact that the ${\bf e}_B$-string is non-EFT and therefore does not satisfy \eqref{altCC}, showing a direct relation between non-EFT strings and non-EFT instantons in $\cC_{\rm I} -\cC^{\text{\tiny EFT}}_{\rm I} $. Let us also notice that such non-perturbative effects should in principle modify the flow \eqref{Eflow} significantly, and also obstruct the possibility to use the dual linear multiplet description in terms of two-form potentials $\cB_{2\,i}$. As a consequence, the ${\bf e}_B$-string flow cannot be straightforwardly described within our perturbative EFT description. In particular, near $r=r_*$ the formula \eqref{Estringasy} is no longer reliable and the string may actually remain tensionful. Figure \ref{f:exflows0} provides a pictorial illustrations of the above flows, and also of a non-elementary EFT one.

\subsection{Connection with the 5d viewpoint}
\label{sec:sugrastrings}

Even though the physical properties of strings in four- and five-dimensional theories are drastically different, the characterisation  \eqref{altCC} of the EFT string charges provides a direct link between our EFT strings and the {\em supergravity strings}  of five-dimensional $\cN=1$ theories given in \cite{Katz:2020ewz} -- see eq.(2.31) therein. 
The connection between the two settings can be made more concrete by compactifying the 5d M-theory models of \cite{Katz:2020ewz}  to four dimensions on a segment  \`a la Ho\v{r}ava-Witten \cite{Horava:1995qa,Horava:1996ma}. In this way one gets a strongly coupled $E_8\times E_8$ four-dimensional heterotic model, in which supersymmetry is broken by the $E_8$ sectors supported at the end-points of the Ho\v{r}ava-Witten segment. In this strong string coupling limit, the heterotic NS5 strings uplift to  M5-strings in M-theory.  In particular, the 4d EFT NS5 strings in   \eqref{hetlatticesC} uplift to the supergravity strings of  \cite{Katz:2020ewz}, which are indeed associated with nef divisors. Correspondingly, the heterotic F1-instantons in \eqref{hetlatticesB} uplift to BPS particles in the 5d M-theory models, which by definition  must have non-negative pairing with the supergravity strings, as in \eqref{altCC}. 

Notice that, on the other hand, the heterotic F1 strings in $\cC^{\text{\tiny EFT}}_{\rm S}$  do not uplift to M-theory strings. Rather, they uplift to M2-branes stretched along the Ho\v{r}ava-Witten segment, which then become membranes in 5d in the extreme decompactification limit. Correspondingly, the heterotic NS5-instantons do not uplift to 5d particles, but rather to M5-brane instantons.    The correspondence between the 4d cones ($\cC^{\text{\tiny EFT}}_{\rm S}$, $\cC_{\rm I}$ ) and 5d supergravity string and BPS particle charges is therefore not one-to-one, but their relation in terms of a non-negative pairing remains in both pictures.

In \cite{Katz:2020ewz} it is argued that supergravity strings are strings which cannot  survive a  rigid limit. This is clear for the above heterotic/M-theory strings associated  with nef effective divisors, which are then `movable' along the entire internal space -- see appendix \ref{app:cones}. Hence, these divisors are not  shrinkable and they do not survive a decoupling of gravity, that is, a decompactification of the internal space. Remarkably, we will see that this property actually holds for all the EFT strings of the string/M-theory models that we will discuss  in section \ref{s:examples}. For instance, also the F1 strings in \eqref{hetlatticesC} clearly satisfy this property, since in a decompactification limit they can be moved away from any local 4d physical sector surviving the decoupling of gravity.

%%%%%%%%%%%%%%%%%%%%%%%%%%%%%%%%%%%%%%%%%%%%%%%%%%%%%%%%%%%%%%%%%%%%%%%%%%%%%%%%%%%%%%%%%%%%%%%        

\subsection{Finiteness and conical structures}
\label{sec:duality}

The  heterotic model considered in subsection   \ref{sec:Hetexample} exhibits  some key features of the  broad class of models covered  by subsection \ref{sec:cones}. However, some of its particular features are not realised in general. Therefore we end up this section by discussing some general features of the conical structures relevant for string flows.

First of all, the saxionic cone of our heterotic example is not only  rational polyhedral (i.e.\ generated by a finite number of elements of $N_{\mathbb{Z}}$), but also simplicial, in the sense that its integral generators form a basis of $N_{\mathbb{R}}$. This property does not generically hold, since  $\Delta$ may be rational polyhedral but not simplicial, or it may even not be  rational polyhedral at all. For instance, let us consider the saxionic cone of a general heterotic compactification, which is given by \eqref{hetDelta0} and therefore it includes the K\"ahler cone of a general Calabi--Yau. On the one hand,  this K\"ahler cone can have a quite complicated structure \cite{wilson1992kahler} and in general does not satisfy these properties. On the other hand, the presence of an underlying rational polyhedral structure would appear natural from a physical viewpoint, as it would correspond to existence a finite number of appropriately quantised elementary EFT string charges generating all the other ones. In general, we dub the corresponding  EFT flows  as {\it elementary} flows. 

In fact, this intuition can still be  compatible with a non-rational polyhedral $\Delta$, if there exists  a `perturbative' duality group $\cG_{\mathbb{Z}}$   acting  linearly on $N_{\mathbb{Z}}$ and preserving  $\Delta$, with a rational polyhedral fundamental domain. This property is indeed expected to hold for the nef cone of Calabi--Yau three-folds, as  conjectured by Morrison in \cite{Coneconj}, see \cite{grassi1992automorphisms} for an explicit example. Morrison and Kawamata later extended this {\em cone conjecture}  to the  cone ${\rm Mov}^1(X)$ of movable divisors, see \cite{Morrison94beyondthe,Kawamata} for the precise statements and \cite{skauli2017cone} for a recent review.\footnote{Some implications of   the Morrison-Kawamata conjecture to the SDC have been recently discussed in \cite{Brodie:2021ain}, in the context of M-theory compactifications to five dimensions.} These conjectures admit a natural generalisation to a {\em saxionic cone conjecture} for our physical setting. Following \cite{Coneconj,Morrison94beyondthe} we define $\Delta_+$ as the convex hull of $\overline\Delta\cap N_{\mathbb{Q}}$, which is basically the cone obtained by eliminating from $\overline\Delta$ the irrational parts of its boundary. The saxionic cone conjecture can then be stated as follows: 

{\bf Saxionic Cone Conjecture}: for any  saxionic cone $\Delta$ associated with the asymptotic region of an EFT compatible with quantum gravity, there exists a  rational polyhedral cone $\Pi\subset \Delta_+$ such that the union of all its images  under the duality group $\cG_{\mathbb{Z}}$ covers $\Delta_+$.   
The polyhedral cone $\Pi$ can be chosen to be the closure of a fundamental saxionic domain.  

A non-trivial duality group $\cG_{\mathbb{Z}}$ implies that the  asymptotic sector of the moduli space should be identified with $\cM_{\rm pert}\equiv\cD/\hat\cG_{\mathbb{Z}}$, where $\cD\equiv \{{\bm t}\in N_{\mathbb{R}}+\ii\Delta\}$ is the tube domain parametrised by the chiral fields $t^i$, and $\hat\cG_{\mathbb{Z}}\equiv \cG_{\mathbb{Z}}\ltimes N_{\mathbb{Z}}$ is the complete perturbative duality group,    combining $\cG_{\mathbb{Z}}$  and the group of axionic integral shifts, which can be identified with $ N_{\mathbb{Z}}$. Correspondingly, the chiral observables should actually be given by $\cG_{\mathbb{Z}}$-invariant combinations of (possibly infinitely many) operators of the form \eqref{chiralring}, hence corresponding to multi-instanton effects.  Adapting the arguments of \cite{Coneconj,Morrison94beyondthe}, the validity of the saxionic cone conjecture would also allow one to partially compactify the infinite distance sector of $\cM_{\rm pert}$ by applying the general scheme of semi-toric local compactifications described in \cite{Looijenga}. This would be the starting point  to extend beyond the simplicial case the analysis made in  \cite{Lanza:2020qmt} for the structure of  the flux superpotential generated by EFT membranes \cite{Lanza:2019xxg} along the asymptotic limits associated with EFT strings.

 In the heterotic example of section \ref{sec:Hetexample}
the dual saxionic cone $\cP$ is simplicial, and then rational polyhedral, too. Again, this is not generically true, and actually $\cP$ can be non-polyhedral also in cases in which $\Delta$ is polyhedral. For instance, this happens in heterotic compactifications on Calabi--Yau spaces $X$ in which at least a codimension-one face of the K\"ahler cone $\cK(X)$ corresponds to a flopping contraction, according to the classification of \cite{wilson1992kahler}. In flopping contractions only curves collapse, while divisors do not. This implies that no NS5-strings become (classically) tensionless as we approach a flopping boundary wall of $\cK(X)$, and then there is no corresponding codimension-one face \eqref{face} in the boundary of $\cP$.

\section{4d strings and Swampland Criteria}
\label{s:conjecture}

In the previous sections we have discussed how, in a given perturbative regime, saxions are driven towards infinite field distance trajectories along flows generated by EFT strings. As a result, each EFT string can be associated with an asymptotic perturbative limit in the moduli space,  selected by the string charges. In this section, we will propose that the reverse is also true, namely that \emph{all} infinite field distance limits can be realised as an EFT string flow. Moreover, building on this conjecture we will propose a specific asymptotic behaviour of the EFT maximal possible cut-off $m_*$, associated with the appearance of an infinite tower of new light states, along such infinite field distance trajectories. The evidence for this second conjecture arises from a plethora of string theory examples, as will be shown in section \ref{s:examples}, and provides a very reduced set of possible values for the exponential rate of the SDC.

\subsection{The Distant Axionic String Conjecture}

Consider a 4d EFT compatible with a consistent theory of quantum gravity. We conjecture that the following  property is universally realised:

\vspace{1em}
\begin{conjecture}[{\bf Distant Axionic String Conjecture}]\
\label{conj:DASC}

\noindent
Every infinite field distance limit of a 4d EFT consistent with quantum gravity can be realised as an RG flow UV endpoint  of an EFT string.
\end{conjecture}
\vspace{1em}

This statement should be understood in terms of the triple correspondence between EFT string RG flows, backreacted solutions and paths in field space $\MM$ described in section \ref{sec:strinfdis}. By this correspondence, an EFT string of charge {\bf e} generates paths in $\MM$ of the form \eqref{gammapath} when the cut-off $\Lambda$ of the EFT is varied. The infinite distance point is approached as $\Lambda \to \infty$, which must be seen as a formal limit, since necessarily at some scale the 4d EFT description will stop being valid.
Clearly, EFT string RG flows define specific paths in field space. In this sense, the above conjecture claims that they are sufficient to reach any infinite distance singularity.

Given the discussion carried in the previous sections, this proposal is quite natural in a minimally supersymmetric setting.
Indeed, recall that an EFT string is a fundamental string magnetically charged under an axion satisfying \eqref{EFTregime} and exhibiting a perturbative continuous shift symmetry near its core. Moreover, all non-perturbative corrections breaking the shift symmetry should be suppressed near its core as defined in \eqref{CSEFT0}, which in turns implies a non-trivial positivity constraint on the allowed set of string charges, cf.\ eqs.~\eqref{CSEFT} and \eqref{altCC}. Therefore, an equivalent way to phrase Conjecture \ref{conj:DASC} is that all possible infinite distance regimes are associated with fundamental axionic symmetries. In such a case, by the Completeness Hypothesis\footnote{More precisely, we have assumed the string to be BPS in the UV, so we are actually assuming a BPS version of the Completeness Hypothesis, as in \cite{Katz:2020ewz}. It would be interesting to check if the BPS condition can be dropped.} we would expect that fundamental string operators with the corresponding magnetic charges appear in that regime of our EFT, and everything else follows from this.

The extension of the conjecture to non-supersymmetric setups is also quite natural if supersymmetry is spontaneously broken and restored at some scale $\Lambda_{\rm SUSY}$ below the EFT cut-off, following the observations at the end of section \ref{sec:RGflow}. If that is not the case, and if we nevertheless have an infinite-distance trajectory along which the quantum-corrected scalar potential is sufficiently flat, then Conjecture \ref{conj:DASC} becomes quite non-trivial. For instance, it would require the existence of an axion-like scalar that pairs up with the field describing the said trajectory, under which the string is magnetically charged. Notice that, in this sense, our conjecture is a sort of 4d reciprocal of Conjecture 4 of \cite{Ooguri:2006in}, for which every axion must come together with a radial mode (a saxion) to guarantee that the moduli space first homotopy group  vanishes.

The Swampland Distance Conjecture (SDC) \cite{Ooguri:2006in} predicts that along any infinite distance limit there appears an infinite tower of new  light states  whose lightest mass $m_*$ decreases exponentially as $e^{-\alpha\Delta\phi}$ with the geodesic field distance $\Delta\phi$, for some constant $\alpha$. 
From a purely EFT viewpoint, there is typically no way to identify neither such {\it tower scale} $m_*$ nor the corresponding constant $\alpha$ without knowing the UV completion of the EFT.
On the other hand, Conjecture 1 
implies a distinctive physical way of classifying the infinite distance limits as flows \eqref{ssssflow} generated by EFT strings. The corresponding tensions, whose renormalisation is also dictated by the flow,  then provide a natural cut-off scale which can be computed within the EFT. In certain cases, the strings themselves will provide the leading tower signaling the EFT breakdown, but in others, there will be additional towers of states getting light faster than the string. It is then natural to wonder what is the relation between $\cT$ and the leading tower scale $m_*$. The following proposal significantly restricts the range of possible answers to this question.

\vspace{1em}
\begin{conjecture}[{\bf Cut-off asymptotics}]\
\label{conj:cutoff}

\noindent
 Along an asymptotic limit specified by the RG flow of an EFT string, its tension $\cT$ goes to zero. The maximal EFT cut-off $m_*$ then scales like
\be\label{m*}
m_*^2\simeq M^2_{\rm P}A\left(\frac{\cT}{M^{2}_{\rm P}}\right)^{{\it w}}\quad~~~~~~ \text{for some positive integer ${\it w}= 1,2,\ldots $}
\ee 
with a coefficient $A$ depending on the non-flowing chiral fields. 
\end{conjecture}
\vspace{1em}

We will refer to the integer ${\it w}$ as the {\it scaling weight}  of the EFT string flow. In terms of the correspondence between RG flows and paths in field space $\MM$, Conjecture \ref{conj:cutoff} should be understood as follows. An EFT string of charge {\bf e} generates paths $\gamma_{\bf e} \subset \MM$ through the EFT RG flow. As discussed in section \ref{sec:fspaths}, the variation of the effective tension $\cT_{\bf e}$ with $\Lambda$ in \eqref{Teff} describes how the probe string tension $T_{\bf e}$ varies along $\gamma_{\bf e}$. The relation \eqref{m*} then provides the asymptotic variation of the tower scale $m_*$  along this same path in field space. While the conjecture could have been phrased in terms of the probe tension, and indeed in section \ref{s:examples} we will check the relation between the probe tension and $m_*$ in different string theory setups, let us recall that it is ${\cal T}_{\bf e}$ the quantity that more directly enters the EFT, and so it makes more sense to describe our swampland criteria in terms of it.

Notice that one can always compute, within the EFT, the relation between the flow parameter $\sigma$ and the field space distance travelled by the saxionic flow \eqref{ssssflow}. As discussed in appendix~\ref{ap:geodesic}, in our scheme EFT string flows can be argued to be asymptotically geodesic. Hence, knowing the scaling weight of the string flow would allow one to compute the exponential rate $\alpha$ appearing in the SDC.
Remarkably, in the string theory examples of section \ref{s:examples} we find that $A\simeq \mathcal{O}(1)$ for generic values of the non-flowing fields and only three possible values for the scaling weight, ${\it w}=1,2,3$.
Furthermore in all those examples we have observed the following intriguing convexity  of the scaling weight 
\be\label{covexindex}
{\it w}_{{\bf e}_1+{\bf e}_2}\leq {\it w}_{{\bf e}_1}+{\it w}_{{\bf e}_2} \quad~~~~~~\forall {\bf e}_1,{\bf e}_2\in \cC_{\rm S}^{\text{\tiny EFT}}\,,
\ee
which could provide an interesting organising principle of the allowed scaling degrees.

\subsection{Some implications}
\label{ss:implications}

The above proposals have interesting implications for the landscape of 4d EFTs consistent with quantum gravity.

\subsubsection*{Existence of EFT strings}

The first obvious consequence of our Conjecture  \ref{conj:DASC} is the existence of axionic strings in any EFT with a moduli space containing infinite distance points. As emphasised in subsection \ref{sec:validity}, our conjectures are based on the behaviour of EFT string tensions evaluated at the EFT cut-off $\Lambda$. As a result, they are insensitive to whether the asymptotic limits under consideration correspond to actual flat directions of the theory or to directions subject to some potential with scales below $\Lambda$. The only thing that matters is whether there is some geodesic in field space that can be taken arbitrarily large while staying below a finite energy scale. Since our conjecture characterises infinite field distances in terms of EFT strings, it thus implies the universal presence of such strings in EFTs consistent with quantum gravity describing large distances in field space.

Furthermore, we are not predicting a single EFT string but rather a complete spectrum of EFT strings associated to each asymptotic regime. This can be formulated in terms of the saxionic cone in \eqref{defDelta} and the cone of EFT string charges in \eqref{CSEFT} as follows:

{\bf EFT string completeness}:  for any EFT consistent with quantum gravity in a given asymptotic regime associated with a saxionic cone $\Delta$, any string charge in $\cC^{\text{\tiny EFT}}_{\rm S}$ is represented by an EFT string or by a superposition of them.  

Notice that this statement is conceptually different from the one that the standard Completeness Hypothesis \cite{Polchinski:2003bq} would suggest. Indeed, in our context the latter would imply the existence of a set of string operators in a particular EFT asymptotic regime. 
However, the string EFT description not only involves the existence of an operator with the appropriate charges, but also that the corresponding string flow is realised, along which the said operator is well-defined. As we have discussed in section \ref{sec:cones}, this last ingredient is quite restrictive, in particular when considering $\half$BPS strings.   What the EFT string completeness hypothesis implies is that for any charge in $\cC^{\text{\tiny EFT}}_{\rm S}$ there is both a string flow and a well-defined operator along it. For other charges like e.g.\ those in $\cC_{\rm S} - \cC^{\text{\tiny EFT}}_{\rm S}$ the existence of a flow would remain an open issue.

That these strings will be present for any 4d EFT depends on whether one can argue that every EFT contains weak coupling/infinite distance limits or not. In \cite{Ooguri:2006in}, it was indeed proposed that any moduli space should be non-compact, which seems a reasonable criterium for EFTs when evaluated at a high enough energy cut-off. This statement has no known counterexample for scale separated EFTs, i.e. that are truly four dimensional for an IR observer. Combining this with our conjecture, we would be predicting the universal presence of fundamental axionic strings in any 4d EFT consistent with quantum gravity!

\subsubsection*{Swampland Distance Conjecture}

A key feature of EFT strings is that the string tension goes to zero monotonically along the RG flow, so the string is getting lighter and lighter as we approach the infinite distance limit in moduli space. When a probe string tension $T_{\bf e}$ gets below  $\Lambda^2$, the EFT necessarily breaks down, as discussed in section \ref{sec:breakdown}.  As already pointed out in \cite{Lanza:2020qmt}, this observation can be used to provide a bottom-up derivation the SDC, which predicts the presence of an infinite tower of states becoming exponentially light in the proper field distance at every infinite distance limit in moduli space (see Conjecture 2 of \cite{Ooguri:2006in}). In particular, one can show that the exponential behaviour of the cut-off becomes just a consequence of having an EFT string satisfying  the WGC. 

To see this, consider the saxionic path \eqref{gammapath} in field space. The dependence of the probe string tension $T_{\bf e}$ with $\sigma$ can be obtained from the saxionic flow  $s^i(\sigma)=s_0^i+\sigma e^i$ generated by the exact backreaction of an EFT string of charge {\bf e}, as described in section \ref{sec:stringsol}. More precisely we have that  $T_{\bf e} (\sigma) = \cT_{\bf e} (\sigma) = M^2_{\rm P}\,e^{i}\ell_i (\sigma)$ satisfies \eqref{monoT}. By this correspondence, the field space distance is given by the integral of the string charge along the flow -- cf.~\eqref{d*}. Using \eqref{monoT}, this reads
\be
\mathrm{d}_\sigma =\frac{1}{M_{\rm P}}\int_0^{\sigma} \cQ_{\bf e}\d\sigma= \frac{1}{M_{\rm P}}\int^{\cT^0_{\bf e}}_{\cT_{\bf e}(\sigma)} \frac1{\cQ_{\bf e}}\d\cT_{\bf e}\leq  \frac1{\gamma} \log\frac{\cT^0_{\bf e}}{\cT_{\bf e}(\sigma)}\, ,
\ee
where we have imposed the WGC bound $M_{\rm P}\cQ_{\bf e}\geq \gamma\cT_{\bf e}$ with constant $\gamma$ in the last step.
Therefore, the maximum EFT cut-off consistent with the existence of the string goes as
\beq
\label{Lmax}
\Lambda_{\rm max}^2=\cT_{\bf e}(\sigma)< \cT_{\bf e}^0 \exp \left(-\gamma\  \mathrm{d}_\sigma \right)\, .
\eeq
This is one of the main implications of our proposal. We get that the WGC implies the exponential drop-off of the cut-off predicted by the SDC along paths generated by the EFT string flows. Using Conjecture \ref{conj:DASC} we can then argue for the universality of the result, as any infinite field distance can then be associated to a string flow. 

As follows from our experience in string theory examples and is captured by Conjecture \ref{conj:cutoff}, in most of the cases the string modes will not  correspond to the leading tower of states becoming light, but there can be additional towers at a scale $m_*<\sqrt\cT$. In those cases the string tension still provides an upper bound on the cut-off and, therefore, the extremality factor of the string represents a lower bound on the exponential rate of the SDC. More precisely, according to \eqref{m*}, the exponential behaviour  in \eqref{Lmax} implies an exponential behaviour for $m_*$:
\beq
m_*\leq m_*^0\exp(-\alpha\,{\rm d}_\sigma )\ , \quad \alpha=\frac{{\it w}\gamma}2\, ,
\eeq
where recall that ${\it w} \in \mathbb{Z}_{>0}$ and $\gamma$ is the extremality factor of the string. 

Notice that the exponential behaviour of the string tension could have also been derived by using the asymptotic logarithmic form of the K\"ahler potential in \eqref{Klog}, as the dependence of the string tension on the dual saxions is fixed by supersymmetry \eqref{Teff}. However, the power of the above result  is that we have alternatively derived the exponential behaviour of the tension without assuming any specific form of the K\"ahler potential, but rather by imposing the EFT string to satisfy the WGC. It is, therefore, a bottom-up model independent result, and no information about the asymptotic geometry of the internal space is required beyond the existence of an approximate axionic shift symmetry. What a K\"ahler potential of the form \eqref{Klog} does give us, is an argument for the EFT string flows to be asymptotically geodesic, see appendix \ref{ap:geodesic}. This allows us to relate the exponential rate $\alpha$ with the one of the tower of states of the SDC. Moreover, as shown in \cite{Lanza:2020qmt}, for a K\"ahler potential of the form \eqref{Klog} the charge-to-mass ratio of an extremal string can be written in terms of the exponents of the saxions in $P(s)$. More precisely $\gamma=\sqrt{\frac{2}{n}}$, with $n$  the so-called singularity type in \cite{Grimm:2018ohb,Grimm:2018cpv,Grimm:2019bey} (see footnote \ref{f:singtype}), which can be derived from the asymptotic scaling of $P(s)\sim \sigma^n$ for $\sigma\rightarrow\infty$ along the string flow \eqref{ssssflow}. Therefore we not only derive the SDC from the WGC for our EFT strings, but we also find a very constrained set of specific values for the SDC exponential rate.

The overall picture also fits well with refinements of the SDC, like the Emergent String Conjecture \cite{Lee:2019wij}. This conjecture distinguishes two classes of infinite distance limits:  decompactification limits or emergent critical string limits. Applied to our 4d setup, one would identify asymptotic limits with a scaling weight ${\it w} >1$ with decompactification limits, while those with scaling weight ${\it w}=1$ should contain the limits in which the EFT string is an emergent critical string. In fact, our scheme only demands that the 4d quantum field theory description must be broken above $\sqrt{\cT}$.  So even in the case ${\it w}=1$ the breakdown could either be triggered by an infinite number of EFT string oscillations or a tower of a different nature scaling at the same rate. Finally, notice that a determination of $w$ purely from the EFT description would allow us to estimate the EFT cut-off just using EFT data! We leave this exciting possibility for future research and continue with some observations regarding the value of $w$ observed in string compactifications in the following.

\subsubsection*{EFT membranes, the scaling weight and the species scale}

Four-dimensional EFTs with potentials below some cut-off scale $\Lambda$ are quite generic in string compactifications with fluxes, where the potential is multi-branched, and each branch is  not invariant under the discrete shift symmetry associated to the EFT string. In that case, the string is not a gauge invariant operator by itself, and a membrane needs to be attached to it to restore gauge invariance. By consistency, it is then expected that the string-membrane system remains as a localised operator along the string flow, just like the string is in the absence of a potential. In practice, this amounts to require that the membrane is a fundamental EFT object as long as the EFT string satisfies \eqref{EFTregime}, i.e.
\begin{equation}
\label{EFTmem}
    \Lambda^3 < {\cal T}_{\rm mem} < M_{\rm P}^2 \Lambda\, .
\end{equation}
Such membranes  were dubbed EFT membranes in \cite{Lanza:2019xxg,Lanza:2020qmt}, and their  relation with EFT strings was analysed in \cite[section 4]{Lanza:2020qmt}. There it was found that the tension of the lightest EFT membrane behaves asymptotically along the EFT string flow as
\begin{equation}
\label{asymem}
\frac{{\cal T}_{\rm mem}}{M_{\rm P}^3} \sim  \left(\frac{{\cal T}_{\rm str}}{M_{\rm P}^2}\right)^{n/2}\, ,    
\end{equation}
 up to some constant factor,  while the scaling of heavier membranes can differ by additional positive integral powers of ${\cal T}_{\rm str}/M_{\rm P}^2$. Here $n$ is the singularity type of the string flow introduced above.

This asymptotic behaviour has an interesting interpretation in terms of Conjecture \ref{conj:cutoff}. To see this, let us define the following EFT mass scale associated with the membrane:
 \be\label{mmem}
 E_{\rm mem}\equiv \frac{\cT_{\rm mem}}{M^2_{\rm P}}\,,
 \ee
which, as in \cite{Lanza:2019xxg,Lanza:2020qmt}, sets the energy scale of gravitational effects induced by the membrane as well as of the corresponding fluxes. Notice that the relation \eqref{asymem} can be written in the form \eqref{m*}, if we simply replace $(E_{\rm mem}, n)$ by $(m_*,w)$. We then see that Conjecture \ref{conj:cutoff} can be extended also to $E_{\rm mem}$ only if the total homogeneity degree of $P(s)$ is an {\em integer}. This is indeed what we have observed in all the string/M-theory models analysed in section \ref{s:examples}. In this sense, the extension of \eqref{m*} to $E_{\rm mem}$ may provide a physical interpretation of this observation and further support for the universal form \eqref{Klog} of the asymptotic K\"ahler potential.

 Now, because by definition along an EFT string flow we have that $\Lambda \leq m_*  \leq {\cal T}_{\rm str}^{1/2}$, the condition \eqref{EFTmem} requires that $E_{\rm mem}<m_*$. Conjecture \ref{conj:cutoff} and \eqref{asymem}  therefore imply that
\begin{equation}
\label{mmem-m*}
    \frac{E_{\rm mem}^2}{m_*^2} \sim \left(\frac{{\cal T}_{\rm str}}{M_{\rm P}^2}\right)^{n-w} < 1 \qquad \implies \qquad w \leq n\, .
\end{equation}
Additionally, requiring that $m_*^3 \leq {\cal T}_{\rm mem}$ along the string flow leads to  $n \leq 3{\it w}$. We therefore infer that the asymptotic behaviour of the K\"ahler potential restricts the scaling weight of the flow via the following inequality
\begin{equation}
    {\it w}\leq n \leq 3{\it w}\, .
    \label{qbound}
\end{equation}

As we will see in the next section, all the string examples that we analyse satisfy this inequality. In fact, 
in the examples analysed in section \ref{s:examples} we find very specific values for the scaling weight, namely
\beq
1 \leq {\it w} \leq 3\, ,
\label{rangeq}
\eeq
even if \eqref{qbound} could in principle allow for larger values. As already mentioned, the lower bound identifies $m_*$ with the mass scale of the EFT string modes, together with a tower of (typically KK) states that becomes massless equally fast. The upper bound is however a bit surprising, so let us try to rephrase it in more physical terms.

Let us assume that ${\it w}>1$ and that $m_*$ corresponds to the scale of a tower of states, as for instance the Emergent String Conjecture would predict. In terms of this tower one can estimate the species scale \cite{ArkaniHamed:2005yv,Distler:2005hi,Dimopoulos:2005ac,Dvali:2007hz,Dvali:2007wp} along the asymptotic limit as
\beq
\Lambda_{\rm sp} \sim \frac{M_{\rm P}}{\sqrt{N}}\, ,
\qquad  
\text{with}
\qquad
N\sim\frac{\Lambda_{\rm sp}}{m_*} \sim \frac{\Lambda_{\rm sp}}{{\cal T}^{\frac{{\it w}}2} M_{\rm P}^{1-{\it w}}}\, .
\eeq
We obtain that
\beq
\Lambda_{\rm sp}^2 \sim M_{\rm P}^2 \left(\frac{{\cal T}}{M_{\rm P}^2}\right)^{\frac{{\it w}}3}\, ,
\ee
so the range \eqref{rangeq} is equivalent to
\beq
m_*^2 \leq {\cal T} \leq \Lambda_{\rm sp}^2\, .
\label{qstring}
\eeq
In other words, if ${\it w} \leq 3$ we would be able to see the string as a quantum object below or around the species scale, so the semiclassical description of the string will break down at most at $\Lambda_{\rm sp}$.  From our experience with perturbative string theory, it is very suggestive that a quantum extended object appears before we reach the quantum gravity scale $\Lambda_{\rm sp}$, since it is indeed a string-like spectrum the one that cures the UV divergences associated to quantum field theory coupled to gravity. 
In addition, \eqref{qstring} suggests that the magnetic dual axion is no longer fundamental above $\Lambda_{\rm sp}$. Since we are associating infinite distance limits with axionic shift symmetries, this would imply that the non-compactness of the moduli space is only manifest below $\Lambda_{\rm sp}$. This nicely fits with the Emergence proposal \cite{Grimm:2018ohb,Heidenreich:2018kpg,Palti:2019pca} for which infinite field distance limits emerge in the IR from integrating out towers of light states up to the species scale.

\subsubsection*{Moduli space curvature at infinity}

In the original work  \cite{Ooguri:2006in}, several conjectures were proposed in addition to the SDC. 
In particular, Conjecture 3 of \cite{Ooguri:2006in} proposes that the scalar curvature near points at infinity is non-positive. This seems a natural consequence of the typical hyperbolic metric arising at large field distances, e.g. from a K\"ahler potential of the form $K=-\log s$. However, this conjecture is false in general, and a counterexample was first pointed out in \cite{trenner2010asymptotic}. In a 4d $\cN=1$ context, we will see that the failure of the conjecture is related to the presence of degenerate string flows. Nevertheless, even if this conjecture is false, we will draw a weaker condition along the same lines, namely that points at infinity have at least one negative holomorphic sectional curvature. This  statement follows from our Conjecture \ref{conj:DASC} and it is realised in all the string theory examples checked in this work.

To begin with, let us come back to our heterotic example in section \ref{sec:Hetexample}, which already presents a counterexample of the conjecture 3 in \cite{Ooguri:2006in}. We can compute the scalar curvature derived from the K\"ahler potential in \eqref{hetKexample}, obtaining
\beq
R=\frac{6 (81 s_1^6 + 162 s_1^5 s_2 + 135 s_1^4 s_2^2 - 45 s_1^2 s_2^4 - 
   18 s_1 s_2^5 - 2 s_2^6)}{s_2^3 (3 s_1 + s_2)^3}\ .
\eeq
One can check that this curvature is negative when $s_2\rightarrow \infty$ but positive when $s_1\rightarrow \infty$. As discussed in section \ref{sec:Hetexample}, there is a fundamental difference between both limits, as the former  corresponds to a non-degenerate string flow while the latter to a degenerate one. In other words, there is more than one EFT string becoming tensionless in the latter one. This suggests that it is precisely the existence of these degenerate flows what yields the positive curvature at certain regions near the infinite distance points. In fact, one can show that the curvature will always be negative along the non-degenerate flows and approach the value for the corresponding hyperbolic plane. This can be understood from a  more geometric perspective as follows. A perturbative regime in which all elementary string flows are non-degenerate corresponds to a factorisable K\"ahler potential to leading order. Hence, the field space factorises into different hyperbolic planes and the scalar curvature is therefore negative everywhere. Contrary, degenerate flows occur when the K\"ahler potential cannot be factorised in the sense that each elementary string flow selects a different leading term in the homogeneous function $P=e^{-K}$. In the language of \cite{Grimm:2018cpv}, the leading term of the K\"ahler potential is different in each growth sector, as explained in footnote \ref{f:singtype}.\footnote{This happens when the  singularity type does not increase in the enhancement chain. For example, in \eqref{hetKexample} the singular divisor located at $s^1= \infty$, i.e. $D_{s^1= \infty}$ has singularity type $n=2$ while the intersection $D_{s^1,s^2= \infty}$ has again $n=2$. Hence, $D_{s^2= \infty}$ is sort of behaving as a finite distance divisor with $n=0$. It makes sense then that the local geometry is not hyperbolic and even has positive curvature, which is something characteristic of finite distance divisors.}  

The above argument only shows that non-degenerate flows imply negative scalar curvature, but it would be interesting to show also the reverse, as examples suggest.
This would require a more detailed study of the interplay between positive scalar curvature and degenerate flows, which we leave for future work. 
Instead,
in the following, we will use 4d EFT string flows to prove a weaker condition than the conjecture in \cite{Ooguri:2006in}, namely that all holomorphic sectional curvatures are negative.

Indeed, let us consider an EFT string flow solution of the form \eqref{tsol}. On the one hand, the solution \eqref{tsol} maps a disc centered at the origin of $\mathbb{C}$ and with radius $r_0=|z_0|$ to a holomorphic disc $D$  in a perturbative region of the EFT field space $\MM$. The area of this disc computed with the pulled-back metric from $\MM$ corresponds to the linear energy density \eqref{Eback0}, which is nothing but the string tension at the cut-off scale $\Lambda_0 \sim 1/r_0$. The existence of the EFT string implies that such a tension is finite for some range of values for $r_0$, and so should be the corresponding disc area. 
On the other hand, the radial direction of the disc in $\mathbb{C}$, e.g. $|z| \in [0,r_0]$, is mapped via \eqref{solsplit} to an infinite distance trajectory on $\MM$, as discussed in section \ref{sec:strinfdis}. It follows that the holomorphic disc $D \subset \MM$ has negative curvature. In particular, for a physical string charge with asymptotic behaviour $\cQ_{\bf e} \sim \sigma^{-\eta}$ along the trajectory \eqref{sssfl}, one obtains a Gaussian curvature for the disc $D$ of the form 
\begin{equation}
 K_D  \sim - \eta \sigma^{2\eta-2} \, ,
\end{equation}
up to a constant positive numerical factor. Hence, for the range $\eta \in (0,1]$ leading to an infinite distance we obtain $K_D <0$. Remarkably, for the choice $\eta =1$ that corresponds to an asymptotically saturated WGC bound $M_{\rm P}\cQ_{\bf e} = \gamma\cT_{\bf e}$ as  realised by EFT strings, the curvature is asymptotically constant, while otherwise it decreases along the flow. If $\MM$ is a K\"ahler manifold, this implies that near the point at infinity corresponding to the EFT string RG flow UV endpoint there is at least one holomorphic sectional curvature which is negative. Finally, using Conjecture \ref{conj:DASC} one may conclude that all points of $\MM$ at infinite distance have this property.

Since the scalar curvature of a K\"ahler manifold is a sum of holomorphic bisectional  curvatures, the statement derived from Conjecture \ref{conj:DASC} is much weaker than Conjecture 3 of \cite{Ooguri:2006in}. Nevertheless, the former seems to be compatible with all the explicit constructions investigated up to date.

\subsubsection*{Global symmetries and generalisation to higher dimensions}

The identification of infinite distance limits with RG flows of EFT strings also makes manifest the relation between the SDC and the absence of global symmetries, which was first proposed in \cite{Grimm:2018ohb}. The EFT breakdown is required to avoid a 0-form continous axionic shift symmetry and $U(1)$ 1-form global symmetry from the dual $\cB_2$-field that would be restored otherwise in the limit, which would go against the swampland criterion of no global symmetries.  This also fits with the proposal in \cite{Gendler:2020dfp} for which there should be a gauge coupling of a $p$-form gauge field  vanishing at every infinite field distance limit. This allows one to merge the conjectures by having the same tower of states satisfying both the WGC the SDC and, therefore, the SDC exponential rate gets bounded by the WGC extremality factor \cite{Lee:2018spm,Gendler:2020dfp}. In our case, this gauge field is the $\cB_2$-field, and the string modes satisfy both the WGC and the SDC, so indeed the exponential rate is given by the extremality factor in \eqref{Lmax}.

It is also interesting to notice that, instead of using the Completeness Hypothesis, we can also argue for the existence of the magnetically charged string from requiring the absence of generalised global symmetries\footnote{Indeed, the absence of generalised global symmetries is equivalent to completeness of spectrum for compact connected gauge groups, see \cite{Heidenreich:appear}} along the path in $\mathcal{M}$, and not only at the infinite distance point. The axion is dual to a 2-form gauge field, and the string is required to break the associated 2-form global symmetry. From a topological perspective, this symmetry is associated to a non-trivial cobordism class\footnote{More concretely, it corresponds to $\Omega_1(\cM)$ where $\cM$ is the moduli space. The string flow provides the bordism that allows one to deform smoothly the circle in the moduli space associated to the axion to a point. We thank Miguel Montero and Jacob McNamara for this observation.} in the moduli space that gets trivialised once we add the string defect, in a similar spirit than in \cite{McNamara:2019rup,Dierigl:2020lai}. In the case at hand, this triviality of the cobordism group reduces to the more familiar notion of triviality of the first homotopy group, which was in fact conjectured in \cite{Ooguri:2006in}. Geometrically, it implies that the 1-cycle in the moduli space associated to the axion can be shrunk at infinite distance, as realised by the string flow.

It is also interesting to discuss possible generalisations of our proposal to higher dimensions. First,
it is easy to convince onself that the interpretation of the string backreaction as an RG flow outlined in \cite{Lanza:2020qmt} and section~\ref{sec:strinfdis} is not exclusive of four-dimensional theories. Indeed, in a generic $d\geq 4$-dimensional spacetime, a fundamental object of codimension 2, namely satisfying 
\be\label{EFTregimed}
 \Lambda^{d-2} <  \cT < M^{d-2} _{\rm P}\, ,
\ee 
will exhibit a backreaction similar to strings in 4d. A well-known case corresponds to D7-branes in the 10d EFT that is type IIB supergravity, see for instance  \cite{Denef:2008wq}. Therefore, a discussion analogous to the one carried in \cite{Lanza:2020qmt} and section~\ref{sec:strinfdis} should apply for any codimension 2 BPS object in a supersymmetric EFT, independently of the dimension. On the other hand, the connection between infinite distance limits and  codimension-two  BPS flows is not expected to be generically valid beyond our 4d  setting. For instance, in ten-dimensional IIA supergravity there are no codimension-two BPS objects whose backreaction flow could describe the infinite distance limit $e^\phi\rightarrow 0$. Hence, in general, one should at least drop one the two requirements: the BPSness or the codimension-2.

Keeping the codimension-2 of the object but dropping the BPS condition would still allow us to have an RG flow interpretation of the infinite distance limits, as only the codimension matters to obtain the logarithmic behaviour originated from the brane couplings being marginally relevant. However, we would loose computational control.  Given the relation with global symmetries outlined above,  it is in fact natural to expect the existence of codimension-2 objects whenever the singular loci are codimension-1 singularities of the moduli space, as these objects would be required to break the corresponding $(d-2)$-form global symmetry. The story gets more complicated when the moduli space is not complex, as happens in 10d IIA or in M-theory on a Calabi-Yau three-fold. In those cases, it might very well be that there are still some extended objects required to break an emergent $p$-form global symmetry at the infinite distance regimes, but that they have a larger codimension. If that is the case, the connection with RG flows might not be present in general. Hence, without further investigation, we cannot conclude anything about the fate of our conjectures in higher dimensions.

\section{String theory evidence}
\label{s:examples}

In this section, we check Conjectures \ref{conj:DASC} and \ref{conj:cutoff} for several classes of 4d $\cN=1$ string theory compactifications. In particular, we provide evidence for the universal presence of an EFT string at every infinite field distance limit as well as the correlation \eqref{m*} between the string tension and the tower scale $m_*$. As usual in string compactifications 4d string tensions are computed by dimensional reduction of different brane actions, and as such they describe the probe, scale-independent tension that we denote by $T$ in this paper. Nevertheless, in terms of saxion values $T$ has the same expression as the effective tension $\cT$. For simplicity and in order to facilitate the comparison of our results with expression like \eqref{thetension} and \eqref{m*}, we will denote both string tensions as $\cT$ in this section.

\subsection{Heterotic on Calabi-Yau three-folds}
\label{sec:hetCY}

We start by considering again the heterotic string Calabi--Yau compactifications  introduced in section \ref{sec:Hetexample}. 
In that section we have already discussed the saxionic sector describing the dilaton and K\"ahler moduli, as well as the corresponding string and instanton cones.
We now  discuss in some generality the behaviour of the relevant mass scales along the EFT string flows \eqref{ssssflow}, with charges taking values in \eqref{hetlatticesC}, and consider also the complex structure sector, which contributes to the K\"ahler potential \eqref{hetK} with the additional term
\be\label{hetCSK}
K_{\rm cs}=-\log\left(\ii\int_X\Omega\wedge \bar\Omega\right)\,.
\ee
For later convenience,  let us write the ansatz of the 10d string frame metric
\be\label{hetmetric}
\d s^2= e^{2A}\d s^2_4+l^2_{\rm s}\d s^2_X\, ,
\ee
where 
\be\label{hetU}
e^{2A} =\frac{l_s^2 M^2_{\rm P}\,e^{2\phi}}{4\pi  V_X}=\frac{l_s^2 M^2_{\rm P}}{4\pi s^0}\, ,
\ee
is fixed by requiring that $\d s^2_4$ is the 4d Einstein frame metric. 

\subsubsection*{F1 flows}

We first consider the flow generated by a string with $e_{\rm F1}=1$,  along which only $s^0$  changes as
\be\label{F1flow}
s^0(\sigma)=s^0_0+ \sigma\,,
\ee 
and then diverges as  $\sigma\rightarrow \infty$ (the non-elementary case $e_{\rm F1}>1$ is completely analogous).
Recalling that $s^0$ is defined as in \eqref{hetdil0}, and since all other chiral fields are kept fixed, one can see that this flow corresponds to sending $e^{\phi}\rightarrow 0$, while the string frame volume $V_X$ measured in string units remains constant, thus it is the well-known weak (string) coupling limit.
The corresponding EFT string tension, measured by the 4d Einstein frame metric, is just given by 
\beq
\cT_{\rm F1}=\frac{2\pi e^{2A}}{l_s^2}=\frac{M_{\rm P}^2}{2s^0}=M_{\rm P}^2 \ell_0\, ,
\eeq
which reproduces \eqref{Teff} upon using \eqref{hetU} and \eqref{lhetex} as expected. The string indeed becomes tensionless in the asymptotic limit $\sigma\rightarrow \infty$.

The KK scale measured by 4d Einstein frame metric vanishes as well. To see this, notice that  the 4d KK scale can be identified with $m_*=e^{A}/R_*$, where $R_*$ is the largest compactification length-scale measured in the string frame and $e^A$ is the Weyl-rescaling given in \eqref{hetU}. 
From here we obtain that
\be\label{hetKKF1}
m^2_*=\frac{e^{2A}}{R_*^2}=\frac{\pi M^2_{\rm P}}{\hat{R}_*^2 s^0}=\frac{2\pi}{\hat{R}_*^2}  \cT_{\rm F1}\, .
\ee
with $\hat{R}_* = 2\pi R_*/l_{\rm s}$. 
Since $\hat{R}_*$ does not change along the flow, we conclude that indeed the EFT string tension $\calt_{\rm F1}$ scales to zero with the same rate as the KK scale. This confirms Conjecture \ref{conj:cutoff} in \eqref{m*} with scaling weight ${\it w}=1$. Even though \eqref{hetKKF1} was derived in the large volume regime, we expect that it gives a reasonable estimate also for generic string size compactifications. In this case  $m_*$ is not necessarily the KK-scale but can  be identified with the string scale measured by the 4d Einstein frame metric. Hence  $m^2_*\sim \frac{e^{2A}}{l^2_{\rm s}}\sim\calt_{\rm F1}$. 

Furthermore, according to the general discussion around \eqref{asymem} the lightest membrane EFT mass scale $E_{\rm mem}\equiv\cT_{\rm mem}/M^2_{\rm P}$ will also scale as $m_*$, since the relevant contribution to the K\"ahler potential \eqref{hetK} is $-\log s^0$, which corresponds to the singularity type $n=1$. We summarise the scaling of the relevant quantities in Table~\ref{t:hetF1}.

\begin{table}[H]
	\centering
	\begin{tabular}{c |c| c|}
		\hhline{~|-|-|}
		& \cellcolor{gray!30}$\sigma^{-\frac12}$ & \cellcolor{gray!30}$\sigma^{-\frac16}$ \\ \hline
		\multicolumn{1}{|c|}{\cellcolor{gray!30}$\cT_{\rm F1}^{1/2}$}                                            & F1   &  \\ \hline
		\multicolumn{1}{|c|}{\cellcolor{gray!30}$m_*$}         &       $m_{\rm KK}$             &          \\ \hline
		\multicolumn{1}{|c|}{\cellcolor{gray!30}$E_{\rm mem}$}                                &      NS5 & \\
		\hline
		\multicolumn{1}{|c|}{\cellcolor{gray!30}$\mathcal{T}_{\rm mem}^{1/3}$}                                &       & NS5 \\\hline
	\end{tabular}
	\caption{Mass scalings along the flow generated by F1 strings.\label{t:hetF1}}
	
\end{table}

\subsubsection*{NS5 flows}

As discussed in section \ref{sec:Hetexample}, the set of EFT strings associated with large K\"ahler moduli limits correspond to NS5-branes wrapping effective nef divisors, see \eqref{hetlatticesC}. Suppose that $D\simeq e^a D_a$ is such a divisor. Then the corresponding string flow is
\be\label{hetDflow}
{\bm s}(\sigma)={\bm s}_0+{\bf e}\,\sigma\,.
\ee
Here ${\bm s}=s^a[D_a]\in \cK(X)$ and ${\bf e}=[D]=e^a[D_a] \in {\rm Nef}(X)_{\mathbb{Z}}$
and the corresponding EFT string tension, using \eqref{Jexp} and \eqref{hetlin}, is given by
\be
\label{Thet}
\cT_{\bf e}=\frac{\pi e^{2A}}{l_s^2}\int_{D}e^{-2\phi}J\wedge J=M_{\rm P}^2\, e^a\ell_a=\frac{3M^2_{\rm P}\,\kappa({\bf e},{\bm s},{\bm s})}{2\,\kappa({\bm s},{\bm s},{\bm s})}\, ,
\ee
where e.g.\ $\kappa({\bf e},{\bm s},{\bm s})\equiv \kappa_{abc}e^as^bs^c\equiv {\bf e}\cdot {\bm s}\cdot {\bm s}$. We clearly have $\cT_{\bf e}\sim 3M^2_{\rm P}/2\sigma$ for $\sigma\rightarrow\infty$, as expected.
Notice that, in order to keep the saxion  $s^0$ fixed along \eqref{hetDflow}, we need to take the strong coupling limit $e^{\phi}\rightarrow \infty$ as $\sigma\rightarrow\infty$. More precisely, since 
\be\label{hetdilaton}
e^{2\phi}=\frac{\kappa({\bm s},{\bm s},{\bm s})}{6 s^0}\, ,
\ee 
$e^{2\phi}$ must diverge as $\kappa({\bm s},{\bm s},{\bm s})$.
This means that there may appear new light degrees of freedom due to strong coupling effects, which become light faster than the KK scale associated to the Calabi--Yau threefold.  Let us see how this works for the $E_8\times E_8$ heterotic string, while we will consider the SO(32) model in section \ref{sec:typeI}. 
Its strong coupling limit is provided by the Ho\v{r}ava-Witten construction of M-theory compactified on $X\times S^1/\mathbb{Z}_2$ \cite{Horava:1995qa,Horava:1996ma}. The estimate of the new KK scale along the interval $S^1/\mathbb{Z}_2$ is  the same as for IIA/M-theory, and then coincides with the mass of a D0-brane.  Hence, taking into account that   the Einstein frame Weyl rescaling \eqref{hetU} remains constant along the flow, this gives a mass scale
\be
\label{mhet}
m^2_*=\frac{2\pi\, e^{2A-2\phi}}{l^2_{\rm s}}= \frac{3 M^2_{\rm P}}{\kappa({\bm s},{\bm s},{\bm s})}\, .
\ee
Notice that, since $s^0$ and the complex structure moduli are constant along the flow, $m_*$ scales as the  membrane EFT mass scale \eqref{mmem}: 
\beq
\label{mmem_het}
E_{\rm mem}^2\simeq  M^2_{\rm P} e^{K}= \frac{M^2_{\rm P}e^{K_{\rm cs}}}{s^0\,\kappa({\bm s},{\bm s},{\bm s})}= \frac{e^{K_{\rm cs}}}{s^0}\, m^2_*\, .
\eeq

By comparing \eqref{Thet} and \eqref{mhet}, we  see that the relation between $m^2_*$ and $\cT_{\bf e}$ will depend on the scaling of $\kappa({\bm s},{\bm s},{\bm s})$ and $\kappa({\bf e},{\bm s},{\bm s})$ in the asymptotic limit. Keeping in mind that the triple intersection of three nef divisors is always non-negative,  we can distinguish three cases, each one realising \eqref{m*} with a different scaling weight  ${\it w}$: 
\begin{enumerate}
    \item $\kappa({\bf e},{\bf e},{\bf e})> 0$. In this case \eqref{m*} is realised with scaling weight  ${\it w}=3$\, .
    
    \item $\kappa({\bf e},{\bf e},{\bf e}) = 0$, but  $\kappa({\bf e},{\bf e},{\bf e}') > 0$ for some  $ {\bf e}'\in {\rm Nef}(X)_{\mathbb{Z}}$. In this case ${\bf e}$ belongs to the boundary of  $\overline{\cK}(X)$ and  \eqref{m*} is realised with scaling weight  ${\it w}=2$.

    \item $\kappa({\bf e},{\bf e},{\bf e}') = 0$ for any   ${\bf e}'\in {\rm Nef}(X)_{\mathbb{Z}}$. Again ${\bf e}$ belongs to the boundary of  $\overline{\cK}(X)$ but now \eqref{m*} is realised with scaling weight  ${\it w}=1$.
 \end{enumerate} 

Before discussing in detail the above classification, we anticipate  that this matches the three main types of infinite distance singular limits that can be realised in a Calabi--Yau threefold \cite{Grimm:2018cpv,Corvilain:2018lgw} and are associated  with the different possible scalings of
$\kappa({\bm s},{\bm s},{\bm s})\sim\sigma^n$ and $\kappa({\bf e},{\bm s},{\bm s})\sim\sigma^{n-1}$  for $\sigma\rightarrow \infty$.  Indeed, from \eqref{Thet} and \eqref{mhet} it follows that the integer $n$ precisely coincides with the scaling weight ${\it w}$ appearing in \eqref{m*}. The integer $n$ was denoted the singularity type, and the three cases are known as Type IV, III and II, corresponding to $n=3,2,1$ respectively. This integer corresponds to the effective nilpotency order of the log-monodromy transformation in the mirror complex structure moduli space, and it was matched with the above properties of the intersection numbers in \cite{Corvilain:2018lgw}. Hence, the scaling weight of the flow is directly linked to the singularity type of the asymptotic limit. As expected, $n=3$ corresponds to a total decompactification limit, while $n=1$ is the limiting case in which the string modes become light at the same rate of a KK scale.
 
So far we have only considered the  KK scale \eqref{mhet} associated with the Ho\v{r}ava-Witten interval. In order to understand the possible role of the  Calabi-Yau KK mass scales, we need to better describe the geometries associated with the above classification. To this end, we can  borrow part of the results of \cite{Lee:2019wij}, which provides a complementary viewpoint on the above classification. 

Our three cases display then the following features:

{\em Case 1:} In this case  $\kappa({\bm s},{\bm s},{\bm s})\simeq \kappa({\bf e},{\bf e},{\bf e})\sigma^3$ and $\kappa({\bf e},{\bm s},{\bm s})\simeq \kappa({\bf e},{\bf e},{\bf e})\sigma^2$, and then the string-frame volume  diverges as $\sigma^3$. Hence $m^2_*\simeq  M^2_{\rm P}\sigma^{-3}$ and Conjecture \ref{conj:cutoff} is realised with scaling weight ${\it w}=3$. Note also that, since the Weyl rescaling \eqref{hetU} is constant, the Calabi--Yau KK mass scale  $m_{\rm KK}= e^{A}/(l_{\rm s} V_X^{1/6})$ asymptotically scales like $m^2_{\rm KK}\simeq \frac{M^2_{\rm P}}{s^0}\sigma^{-1}$, and then is much heavier than $m_*$.  Furthermore, from \eqref{Thet} it is clear that $\cT_{{\bf e}'}\simeq  M^2_{\rm P}\sigma^{-1}$ for {\em any} string charge ${\bf e}'\in {\rm Nef}(X)_{\mathbb{Z}}$. This means that, in  the terminology introduced around \eqref{face}, these flows are maximally degenerate in the K\"ahler sector. A concrete example of an EFT string generating such a flow is provided by the $H$-string in the two-moduli example of section \ref{sec:Hetexample}. 

{\em Case 2:} In this case, the self-intersection $C={\bf e}\cdot {\bf e}$  corresponds to a non-trivial Mori curve, since its intersection with any other nef divisor is non-negative. This case is in direct correspondence with the $J$-class A limit of \cite{Lee:2019wij}.
As proved in \cite{Oguiso93,wilson_1998} and emphasised in \cite{Lee:2019wij},  a Calabi--Yau three-fold realising 
this possibility corresponds to a $T^2$ fibration over a two-fold, where $C$ is a multiple of the $T^2$ fiber. An example is provided by the $L$-divisor  in the two moduli model of  section \ref{sec:Hetexample}. Along \eqref{hetDflow}  we asymptotically have $\kappa({\bm s},{\bm s},{\bm s})\simeq 3\kappa({\bf e},{\bf e},{\bm s}_0)\sigma^2$ and $\kappa({\bf e},{\bm s},{\bm s})\simeq 2\kappa({\bf e},{\bf e},{\bm s}_0)\sigma$, where $\kappa({\bf e},{\bf e},{\bm s}_0)=\int_C{\bm s}_0=\int_C{\bm s}>0$ measures the {\em constant} volume of $C$. Hence, from  \eqref{mhet} we get $m_*^2\simeq M^2_{\rm P}\sigma^{-2}$ and Conjecture \ref{conj:cutoff} is realised with scaling weight ${\it w}=2$. Since the  volume of $C$ is constant, the  scaling $\sim\sigma^2$ of the internal volume corresponds to an expansion of the four-dimensional base of the $T^2$ fibration. Hence, as in Case 1 we have  $m^2_{\rm KK}\simeq\frac{M^2_{\rm P}}{s^0}\sigma^{-1}$, which is asymptotically much heavier than $m^2_*$. Furthermore from \eqref{Thet} it follows that the order of degeneracy of the flow (defined below \eqref{face}) is given by the number of linearly independent charges ${\bf e}'$ (including ${\bf e}$) for which $\kappa({\bf e}',{\bf e},{\bf e})=0$.    

{\em Case 3:} This is related  to the $J$-class B limit of \cite{Lee:2019wij}. In this case ${\bf e}\cdot {\bf e}=0$ and then $\kappa({\bm s},{\bm s},{\bm s})\simeq 3\kappa({\bf e},{\bm s}_0,{\bm s}_0)\sigma$, while $\kappa({\bf e},{\bm s},{\bm s})\equiv  \kappa({\bf e},{\bm s}_0,{\bm s}_0)>0$ gives the constant volume of the nef divisor ${\bf  e}$. Then $m_*^2\simeq M^2_{\rm P}\sigma^{-1}$ and Conjecture \ref{conj:cutoff} is realised with scaling weight ${\it w}=1$. As discussed in \cite{Lee:2019wij}, this case corresponds to a K3 or $T^4$ fibration over a $\mathbb{P}^1$, where a multiple of the fibre can be identified with the divisor class ${\bf e}$. The scaling $\sigma$ of the internal volume corresponds  to the expansion of the base $\mathbb{P}^1$. Hence, we again get  $m^2_{\rm KK}\simeq \frac{M^2_{\rm P}}{s^0}\sigma^{-1}$ which, however, now scales like $m^2_*$ and could be taken as an alternative definition thereof. In  \cite{Lee:2019wij} it was proven  that if ${\bf e}$ and ${\bf e}'$ represent two non-proportional   nef divisors, then $C'={\bf e}\cdot {\bf e}'$ represents a non-trivial  curve. Applied to the present setting, this implies that for any EFT charge ${\bf e}'$ non-proportional to ${\bf e}$, we have that  $\kappa({\bf e}',{\bm s},{\bm s})\simeq 2\kappa({\bf e}',{\bf e},{\bm s}_0)\sigma$, where $\kappa({\bf e}',{\bf e},{\bm s}_0)\equiv \int_{C'} {\bm s}>0 $ is a constant measuring the volume of $C'$. Then from \eqref{Thet} we see that  the corresponding tension $\cT_{{\bf e}'}$ is not asymptotically vanishing  and the flow is non-degenerate.  

Note also that one can also check the index convexity \eqref{covexindex}. Since in this sector $1\leq {\it w}_{\bf e}\leq 3$, the only case that needs to be checked corresponds to both ${\bf e}_1$ and ${\bf e}_2$ of scaling weight 1 (that is, case 3). That is ${\bf e}_1\cdot {\bf e}_1={\bf e}_2\cdot {\bf e}_2=0$. It follows that ${\bf e}={\bf e}_1+{\bf e}_2$ has certainly vanishing triple self-intersection, $\kappa({\bf e},{\bf e},{\bf e})=0$, and then  ${\it w}_{\bf e}\leq 2$.

The above results are  summarised in the following table:

\begin{table}[H]
		\small\centering
		\makebox[\linewidth]{\begin{tabular}{c | c | c | c | c | c |}
			\hhline{~|-|-|-|-|-|}
			 & \cellcolor{gray!30}$\sigma^{-\frac32}$ & \cellcolor{gray!30}$\sigma^{-1}$ & \cellcolor{gray!30}$\sigma^{-\frac12}$ & \cellcolor{gray!30}$\sigma^{-\frac13}$ & \cellcolor{gray!30}$\sigma^{-\frac16}$ 
			 \\ \hline
			\multicolumn{1}{|c|}{\cellcolor{gray!30}$\cT_{\rm str}^{1/2}$}                    &  
			      &                       & NS5 & &
			\\ \hline
			\multicolumn{1}{|c|}{\cellcolor{gray!30}$m_*$}         &       M-th. KK  -- Case 1           &      M-th. KK  -- Case 2         &    M-th. KK  -- Case 3  & &         \\ \hline
			\multicolumn{1}{|c|}{\cellcolor{gray!30}CY $m_{\rm KK}$ }         &                        &                            &            CY  $m_{\rm KK}$    & &                                                            \\ \hline
			\multicolumn{1}{|c|}{\cellcolor{gray!30}$E_{\rm mem}$}                  &    NS5 -- Case 1      &             NS5 -- Case 2               &              NS5 -- Case 3           & &     \\ \hline
			\multicolumn{1}{|c|}{\cellcolor{gray!30}$\mathcal{T}^{1/3}_{\rm mem}$}        & &          &    NS5 -- Case 1      &             NS5 -- Case 2               &              NS5 -- Case 3               \\ \hline
		\end{tabular}}
		\caption{Mass scalings along the flows generated by NS5-strings in heterotic models.\label{tab:HetNS5}} 
	\end{table}

Notice that this table does not include the behaviour of any string tension beyond the one of the EFT string generating the flow. In principle there could  be non-EFT strings whose tension decreases faster than $\sigma^{-1}$. This happens for instance in the example of section \ref{sec:Hetexample}, where the  non-EFT string tension $\cT_{{\bf e}_B}$ decreases like $\sigma^{-3}$ along the flow generated by the EFT string charge ${\bf e}_{H}$. In this case we have that $\cT_{{\bf e}_B}^{1/2} \sim m_*$, and so the presence of the light non-EFT string does not change the result for $m_*$ obtained from our previous analysis. In the following we will assume that this is always the case, either because the non-EFT string tensions are asymptotically at or above $m_*^2$ or because they do not generate an infinite tower of oscillation modes. It would be interesting to check this expectation in other explicit examples.

\subsubsection*{Complex structure flows}

When some complex structure (CS) modulus is taken  to be large, we may approach an infinite distance locus where an approximate axionic shift symmetry arises. We can then choose a local parametrisation splitting the asymptotic complex structure moduli $\tau^\alpha\simeq \tau^\alpha+1$ into the axions  and saxionic partners, as in the general description proposed in the present paper. 
In particular, at the 4d EFT level, this perturbative regime admits a description in terms of dual $\cB_2$ potentials,  charged strings and corresponding BPS flows, as in section \ref{s:fundamental}.

The classification  of the possible infinite distance limits in the complex structure moduli space of a Calabi-Yau 3-fold can be borrowed from \cite{Grimm:2018ohb,Grimm:2018cpv}, leading to the three cases discussed above for the K\"ahler sector. Indeed, by mirror symmetry, any asymptotic region of the CS moduli space is expected to be equivalent to a corresponding asymptotic region of the K\"ahler structure moduli space of the mirror Calabi--Yau. Hence, the two mirror descriptions should share the same saxionic cone structure and spectrum of strings (henceforth dubbed CS strings), whose flows describe  the possible infinite-distance limits. In particular, it is clear that the CS EFT strings  are mirror to the NS5 EFT strings discussed above. By exploiting the SYZ description of mirror symmetry \cite{Strominger:1996it}, one can also  more concretely identify the 4d CS strings with  10d KK-monopoles wrapping internal divisors. 

The correspondence between CS strings and NS5 strings cannot be naively extended to the estimate of the relevant mass scales along the corresponding flows. In particular, the saxion $s^0$ defined in \eqref{hetdil0} does {\em not} contain the  flowing CS moduli and then not only the K\"ahler moduli but also the dilaton remains constant along the flow. Correspondingly, the strong coupling  mass scale  \eqref{mhet} does not become asymptotically light.  Absent some scale that plays the role of the M-theory scale in NS5-flows, this suggests that along CS flows the tower scale $m_*$ is mirror to the Calabi--Yau KK-mass scales appearing in table \ref{tab:HetNS5}. This identification would imply that Conjecture \ref{conj:cutoff} is always realised with scaling weight ${\it w}=1$.   This conclusion can be easily checked in simple toroidal orbifold models, see appendix~\ref{sec:CSflow_tor} for an explicit example. It would be interesting to perform a direct estimate of $m_*$ for general Calabi--Yau geometries, to confirm whether our naive expectation for ${\it w}$ is realised or not.

\subsection{Type I on Calabi-Yau three-folds}
\label{sec:typeI}

Type I models are related to heterotic models  by S-duality. Hence, their Calabi--Yau compactifications are very similar  to the heterotic ones, described in subsections \ref{sec:Hetexample} and \ref{sec:hetCY}. In particular, 
the complex structure moduli have  K\"ahler potential \eqref{hetCSK}, and the remaining moduli sector is still described by a K\"ahler potential of the form \eqref{hetK}. However, the saxions $s^a$ and $s^0$ are defined differently  from the heterotic ones \eqref{Jexp} and \eqref{hetdil0}. Rather, they are the image of \eqref{Jexp} and \eqref{hetdil0} under S-duality 
\be\label{typeIsaxions}
{\bm s}=s^a[D_a]\equiv e^{-\phi}J\quad,\quad s^0\equiv e^{-\phi} V_{X}= \frac1{3!}e^{2\phi}\kappa({\bm s},{\bm s},{\bm s})\,,
\ee
where $J$ denotes the string frame K\"ahler form of the compactification space $X$. Hence, the EFT  saxionic, string and instanton sectors are the same as  in the heterotic case while their microscopic interpretations are different and related by S-duality. In particular, the 10d dilaton and the internal string frame volume are given by the following saxionic functions
\be\label{typeIdilvol}
e^{2\phi}=\frac{6 s^0}{\kappa({\bm s},{\bm s},{\bm s})}\,,\quad V_X=\sqrt{\frac{6 (s^0)^3}{\kappa({\bm s},{\bm s},{\bm s})}}\,.
\ee
Note also that the Weyl-rescaling $e^{2A}$ appearing in the string-frame ansatz \eqref{hetmetric} now takes the form
\be
e^{2A}=\frac{l^2_{\rm s}M^2_{\rm P}\sqrt{6}}{4\pi\sqrt{s^0\kappa({\bm s},{\bm s},{\bm s})}}\,.
\ee

Consider for instance the flow generated by a D1-string, which is identical to the F1 flow \eqref{F1flow} of the heterotic model. In the type I case, from \eqref{typeIsaxions} it is clear that the asymptotic limit $s^0\rightarrow \infty$, with $s^a$ fixed, implies that   $e^\phi\rightarrow \infty$. Hence, the flow cannot be described microscopically within the perturbative type I model, but rather we need to use the S-dual description of subsection \ref{sec:hetCY}. 

Consider now D5-brane strings, which are S-dual to the NS5-strings considered in subsection \ref{sec:hetCY}. We saw that the heterotic NS5-string flows make the 10d dilaton diverge and in subsection \ref{sec:hetCY} we used the M-theory uplift of the $E_8\times E_8$ model to study them. The type I models now allow us to extend that discussion to the SO(32) model as well. Indeed, the D5-brane strings flows are as in \eqref{hetDflow}, while $s^0$ remains constant. Hence from \eqref{typeIdilvol} it follows that $e^{2\phi}$ vanishes as $1/\kappa({\bm s},{\bm s},{\bm s})$ along the flow and the 10d type I model remains weakly coupled. On the other hand, from \eqref{typeIdilvol} it also follows that $V_X\rightarrow 0$ as $1/\sqrt{\kappa({\bm s},{\bm s},{\bm s})}$, and the classical definition of the saxions becomes questionable.  

We can be more specific. The string-frame K\"ahler form can be written as a function of the saxions as follows
\be
J=\sqrt{\frac{6s^0}{\kappa({\bm s},{\bm s},{\bm s})}}\,{\bm s}\,.  
\ee
We can now adopt the same classification in terms of intersection numbers  as in subsection \ref{sec:hetCY}. Consider first Case 2, the internal space can be regarded as a $T^2$-fibration. In the type I setting, the classical volume of the $T^2$-fibre  vanishes asymptotically as $\sigma^{-1}$, while the base volume remains constant. Hence, we certainly cannot trust the classical description and we should further go to some other dual picture. For instance,  we can  imagine to perform a fibre-wise double T-duality and go to a D3-string of the kind  described in the following subsection \ref{sec:IIBF}. More directly, we could try to identify the relevant $m^2_*$ with the winding scale associated with the vanishing $T^2$-fiber:
\be\label{typeIwind}
m^2_*=\frac{e^{2A}\text{Vol}(T^2)}{\ell_{\rm s}^2}\sim \frac{M^2_{\rm P}}{\sqrt{s^0}\,\sigma^2}\quad~~~~~\text{for $\sigma\rightarrow \infty$}\,.
\ee
This is consistent with Conjecture \ref{conj:cutoff}, with scaling weight ${\it w}=2$ as in the $E_8\times E_8$ case, although the two microscopic interpretations of $m^2_*$ are completely different. 

In Case 1 the entire classical string frame volume goes to zero as $\sigma^{-\frac32}$. We can make an  estimate of $m_*^2$  by restricting to a toroidal  model which scales homogeneously and using the winding mass formula \eqref{typeIwind}. Since $\kappa({\bm s},{\bm s},{\bm s}) \sim \sigma^3$, we now get $m_*^2\sim M^2_{\rm P}\sigma^{-2}$ as in Case 2. This supports again Conjecture \ref{conj:cutoff} with scaling weight ${\it w}=2$. Note that this differs from the scaling weight ${\it w}=3$ found in the $E_8\times E_8$ model.    

Case 3 is slightly different. The internal space can be regarded as a $T^4$/K3-fibration over $\mathbb{P}^1$, in which the string frame $\mathbb{P}^1$ volume diverges as $\sqrt{\sigma}$, while the fibre's volume tends to zero as $1/\sigma$. We can then try to estimate $m_*^2$ by identifying it with the base KK-mass scale
\be
m^2_*=\frac{e^{2A}}{l^2_{\rm s}{\rm Vol}(\mathbb{P}^1)}\sim \frac{M^2_{\rm P}}{s^0 \sigma}\quad~~~~~\text{for $\sigma\rightarrow \infty$}\,.
\ee
This confirms Conjecture \ref{conj:cutoff} with scaling weight ${\it w}=1$. Alternatively, we may use the winding mass estimate \eqref{typeIwind} associated with a shrinking $T^4$-fiber, getting the same scaling weight.  

Finally, one may perform a naive estimate of the relevant scales under the CS-string flows, by applying a reasoning similar to the one in heterotic models. As in there, absent some other  scales that could play the role of $m_*$, one would obtain that Conjecture \ref{conj:cutoff} is realised with scaling weight ${\it w}=1$. See appendix~\ref{sec:CSflow_tor} for a check of this conclusion in an explicit model.

\subsection{Type IIB/F-theory compactifications}
\label{sec:IIBF}

We now consider IIB/F-theory compactifications to four-dimensions and its weak-coupling limit, see for instance \cite{Denef:2008wq} for a review. In these models the compactification space $X$ is K\"ahler and the axio-dilaton $\tau=C_0+\ii e^{-\phi}$ is non-constant and undergoes non-trivial monodromies around 7-branes. It is also useful to think of such compactification in terms of a dual M-theory compactification over a Calabi--Yau four-fold $Y$ that is elliptically fibred over the base space $X$. From the IIB viewpoint, it is convenient to adopt the 10d Einstein-frame description. We will then use an ansatz of the form \eqref{hetmetric} where, however, $\d s^2$ denotes the 10d Einstein frame metric, which will be used to measure 10d length scales in this subsection. Note that the Weyl-rescaling is now
\be\label{WeylFth}
e^{2A}=\frac{l^2_{\rm s}M^2_{\rm P}}{4\pi V_X}\,.
\ee
The four-dimensional $\cN=1$ EFT has been discussed in detail in \cite{Denef:2008wq,Grimm:2010ks} in the constant warping approximation.  Warping effects  \cite{GKP} may also be incorporated  as in \cite{Shiu:2008ry,Frey:2008xw,Martucci:2009sf,Martucci:2014ska,Martucci:2016pzt}, but we will not consider them in the following. 

\subsubsection*{K\"ahler saxions and EFT strings}

We will first focus  on the saxionic sector that parametrises the K\"ahler sector. As usual, the K\"ahler moduli $v_a$ appear in  the expansion of the (Einstein frame) K\"ahler form
\be\label{FTJexp}
J=v_a[D^a]\,,
\ee
where $[D^a]$ is the Poincar\'e dual of a basis of divisors $D^a\in H_4(X,\mathbb{Z})$. Assume first for simplicity that $H^3(Y)=0$. In this case, the K\"ahler sector is parametrised by the EFT saxions 
\be\label{FKsaxions}
{\bm s}\equiv \frac12 J\wedge J\in H^4(X,\mathbb{R})\quad\Rightarrow\quad s^a\equiv \int_{D^a}{\bm s}=\frac12\int_{D^a}J\wedge J=\frac12\kappa^{abc}\,v_bv_c\,.
\ee
In order to better determine the corresponding saxionic cone $\Delta$, we have to specify the relevant instanton charges. In the large volume regime these  are naturally associated with Euclidean D3-branes wrapping effective divisors.  This leads to the identification 
\be
\cC_{\rm I}={\rm Eff}^1(X)_{\mathbb{Z}}\ ,
\ee
see appendix \ref{app:cones} for the notation. Since $h^{2,0}(X)=0$ for a three-fold  $X$ which is the base of Calabi--Yau four-fold, the divisors generate $H_4(X,\mathbb{Z})$ and then we can identify the  lattice $M_{\mathbb{Z}}$ with the  torsion-free homology group $H_4(X,\mathbb{Z})_{\text{t.f.}}$. 

The K\"ahler potential for the chiral fields $t^a=a^a+\ii s^a$ associated with the saxions 
$s^a$ is obtained by inverting \eqref{FKsaxions} to express 
\be\label{KpotFt}
K_{\rm ks}=-2\log \int_X J\wedge J\wedge J=-2\log \kappa({\bm v},{\bm v},{\bm v})\,,
\ee
where $\kappa({\bm v},{\bm v},{\bm v})\equiv \kappa^{abc}v_av_bv_c$, as a function of $s^a=\Im t^a$. From \eqref{dualfields} one can then compute the dual saxions
\be\label{ellFth}
\ell_a=\frac{3v_a}{\kappa({\bm v},{\bm v},{\bm v})}=\frac{\int_{C_a}J}{2V_X}\,,
\ee
where $C_a\in H_2(X,\mathbb{Z})$ is a basis of two-cycles dual to $D^a$ (i.e.\ $C_a\cdot D^b=\delta^a_b$) and $V_X=\frac{1}{3!}\kappa({\bm v},{\bm v},{\bm v})$ is the (Einstein frame) volume of $X$.

From our general definition \eqref{defDelta}, we can therefore identify $\Delta$ with the interior of the dual of the cone of effective  divisors Eff$^1(X)$. This coincides with  the interior of the cone in $H^{2,2}(X,\mathbb{R})$ generated by the Poincar\'e dual of {\em movable} curves ${\rm Mov}_1(X)$  \cite{boucksom2013pseudo} -- see also appendix \ref{app:cones} and \cite{xiao:tel-01679333} for a review:
\be\label{FtheoryDelta}
\Delta\simeq {\rm Int}[{\rm Mov}_1(X)]\quad~~~~\text{(via Poincar\'e duality)}\,.
\ee
 By applying  \eqref{CSEFT}, we then see that the EFT string charges can be identified with the set of movable curves
\be\label{FtheoryEFTst}
\cC^{\text{\tiny EFT}}_{\rm S}={\rm Mov}_1(X)_{\mathbb{Z}}\,.
\ee
Physically, the corresponding EFT strings correspond to D3-branes wrapping movable curves inside $X$. Roughly speaking, a curve is movable if it is part of a family of curves that spans the entire $X$.   More general BPS strings correspond instead to D3-branes wrapping effective curves: $\cC_{\rm S}={\rm Eff}_1(X)_{\mathbb{Z}}$.\footnote{Since ${\rm Mov}_1(X)$ is a subset of $\overline{\rm Eff}_1(X)$ but not of ${\rm Eff}_1(X)$, a priori a movable curve  is {\em pseudo}-effective  but not necessarily    effective. The EFT string completeness proposed in section \ref{ss:implications}  suggests that  in F-theory models any movable curve admits an effective representative.} Here the relevant lattice is $N_\mathbb{Z}=H_2(X,\mathbb{Z})_{\rm t.f.}$, and the pairing with the dual lattice $M_\mathbb{Z}$ generated by the instanton charges coincides with the intersection number between four and two cycles. One can also check that a probe D3-brane wrapping an effective curve $C=e^a C_a$ gives an effective string of tension $\cT_{\bf e}=M^2_{\rm P}\langle {\bf e},{\bm\ell} \rangle$, in agreement the expected EFT value.  

 The correspondence between elements of the K\"ahler cone $\cK(X)$ and elements of the saxionic cone \eqref{FtheoryDelta} is defined through the map $J\mapsto {\bm s}\equiv \frac12 J\wedge J$ (combined with Poincar\'e duality). Note that this map  is injective   but {\em not} generically surjective \cite{fu2014relations}.  Hence, in general, the saxionic cone $\Delta$ parametrises a space which is larger than $\cK(X)$. In particular, there will be EFT string charges in \eqref{FtheoryEFTst} that generate flows that take us away from the K\"ahler cone. When this happens, we cross a boundary of $\cK(X)$ and at this point a curve $C\subset X$ collapses. However, by construction no divisor collapses at any point of the EFT string flow, and from the 4d perspective we remain in the perturbative region \eqref{asymptregion} where all non-perturbative effects that break the axionic shift symmetries $\int_D C_4^{\rm RR}$ are suppressed. Therefore, on physical grounds we expect that the actual saxionic space of the theory is an extension of $\cK(X)$ that accommodates all the asymptotic limits generated by \eqref{FtheoryEFTst}.
 
This expectation is supported by the so-called `small modifications' of the space $X$, which  have been observed in different instances. These are essentially transitions to another space $X'$ through the contraction of a curve $C\subset X$ 
and the blow-up of a curve $C'\subset X'$, with no divisor collapsing along them. In the upstairs  elliptically fibered Calabi-Yau four-fold $Y$ they correspond to flops of the geometry. From the 4d viewpoint they correspond to  strings whose tension vanishes classically but may receive  quantum corrections -- see e.g.~\cite{Witten:1996qb,Mayr:1996sh} for related discussions.
 
It is therefore natural to glue together the K\"ahler cones of the spaces related by these transitions, to get an extended K\"ahler cone $\cK(X)_{\rm ext}$. Based on our 4d physical intuition of EFT strings, we would then  expect that $\cK(X)_{\rm ext}$ can be alternatively parametrised by the above saxionic cone, that is,  that  the map
\be\label{KextB}
J\in \cK(X)_{\rm ext}\quad\mapsto\quad {\bm s}\equiv \frac12 J\wedge J\in \Delta\, ,
\ee
is a bijection. We will not try to prove this expectation in full generality, but rather to collect some evidence supporting it.

Part of the evidence comes from the fact that one can construct a huge class of F-theory models \eqref{KextB} where the base three-fold $X$ belongs to the class of  {\em Mori Dream spaces} \cite{hu2000mori}.\footnote{We thank Antonella Grassi for discussions on Mori Dream spaces and related aspects.} This class  includes toric  and Fano spaces \cite{birkar2010existence} as particular subclasses and, basically by definition, enjoy the following  decomposition of the cone of movable divisors into K\"ahler chambers connected by small modifications: 
\be\label{mov1dec}
{\rm Mov}^1(X)=\bigcup_{\text{$X'$ small mod. of $X$}}{\rm Nef}(X')\quad \equiv\  \overline{\cK(X)}_{\rm ext}\,.
\ee
So, if $D\in {\rm Mov}^1(X)$, through a sequence of small transitions one can reach a K\"ahler chamber in which $D$ is nef. By combining this property together with some results of \cite{lehmann2016convexity,xiao:tel-01679333},  one can conclude that \eqref{KextB} is indeed a bijection.

Further support is provided by the F-theory weak coupling limit, in which the IIB compactification can be described in terms of a double cover   Calabi--Yau three-fold $\hat X$.  It is known that the different K\"ahler chambers subdividing $\cK_{\rm ext}(\hat X)$ are related by flops and that  $\overline{\cK_{\rm ext}}(\hat X)$  coincides with the cone of movable divisors \cite{kawamata1988crepant}, hence $\overline{\cK_{\rm ext}}(\hat X)={\rm Mov}^1(\hat X)$.  By restricting this identity to classes which are invariant under the orientifold involution  one gets  $\overline{\cK_{\rm ext}}(\hat X)_+={\rm Mov}^1(\hat X)_+$, where $\overline{\cK_{\rm ext}}(\hat X)_+$ can indeed be identified with the extended K\"ahler cone of the IIB compactification and ${\rm Mov}^1(\hat X)_+$ with the cone of movable divisors over the corresponding F-theory base space $X$. In fact,  our working assumption $H^3(Y)=0$ implies that $H^2(\hat X)_-=0$ and than  we expect the restriction to orientifold-even classes to trivialise to an identity map.

\subsubsection*{Examples}
   
We now consider two simple illustrative examples and discuss, without trying to be exhaustive,  the behaviour of the relevant mass scales under the flows generated by EFT string charges \eqref{FtheoryEFTst}. A more involved example of this sort is considered in appendix \ref{sec:Fresolved}.

\bigskip

{\em Example 1: $X=\mathbb{P}^3$}

\noindent Consider first the simplest possible  compactification space: $X=\mathbb{P}^3$. In this case $M_{\mathbb{Z}}\simeq H_4(X,\mathbb{Z})\simeq \mathbb{Z}$ is generated by the hyperplane divisor  $L\simeq \mathbb{P}^2$, which also generates the BPS instanton charges 
\be
\cC_{\rm I}={\rm Eff}^1(X)_{\mathbb{Z}}={\rm Mov}^1(X)_{\mathbb{Z}}={\rm Nef}^1(X)_{\mathbb{Z}}=\mathbb{Z}_{\geq 0}L\,.
\ee
Dually, $N_{\mathbb{Z}}\simeq H_2(X,\mathbb{Z})\simeq \mathbb{Z}$ is generated by the two-sphere  $C\simeq \mathbb{P}^1=L\cdot L$. This generates also the possible string charges 
\be
\cC_{\rm S}=\cC^{\text{\tiny EFT}}_{\rm S}={\rm Eff}_1(X)_{\mathbb{Z}}={\rm Mov}_1(X)_{\mathbb{Z}}=\mathbb{Z}_{\geq 0}C\,.
\ee
The K\"ahler form can be written as $J=v[L]$ and since $L^3=1$ the saxion is $s=\frac12 v^2$. The K\"ahler potential \eqref{KpotFt} then reduces to $K=-3\log s$, up to irrelevant constants. The possible EFT string flows \eqref{ssssflow} reduce to $s(\sigma)=s_0+e\sigma$, with $e\in\mathbb{Z}_{\geq 0}=\cC^{\text{\tiny EFT}}_{\rm S}$. Under such a flow, asymptotically, the volume scales homogeneously like $V_X\sim \sigma^{\frac32}$. This implies that  the KK-scale $m_*^2=\frac{e^{2A}}{l_{\rm s}^2 }V_X^{-\frac13}=\frac{M^2_{\rm P}}{4\pi }V_X^{-\frac43}$ scales as  $M^2_{\rm P}\sigma^{-2}$. This is in agreement with Conjecture \ref{conj:cutoff} with scaling weight ${\it w}=2$. 

The scaling of the lightest membranes can be easily inferred from scaling of the K\"ahler potential. Indeed, as follows from the general relation \eqref{asymem}, the scale \eqref{mmem} set by the lightest membranes falls off as $E_{\rm mem} \sim M_{\rm P} \sigma^{-\frac32}$, so $n=3\neq w$. Notice also that  the membrane tension scales as $\mathcal{T}_{\rm mem} \sim M^3_{\rm P} \sigma^{-\frac32}$ along the flow, and is then  heavier than $m_*^3$. 

\bigskip

{\em Example 2: $X=\{\mathbb{P}^1$ fibration over $\mathbb{P}^2\}$}

\noindent A more interesting example is provided by choosing as internal space $X$ the toric $n$-twisted $\mathbb{P}^1$ fibration over $\mathbb{P}^2$ defined by the gauged linear sigma model \begin{center}
    \begin{tabular}{ |r | c | c | c | c | c | c | c }
    \hline
                    & $u_1$ &   $u_2$ &  $u_3$ & $u_4$  &  $u_5$  &   FI   \\ \hline
          U(1)$_1$ &    1      &       1     & 1&     $-n$    &     0     &   $v_1>0$\\
    U(1)$_2$ &    0      &       0     &      0     &     1     &     1   &      $v_2>0$\\
    \hline
    \end{tabular}
\end{center} 
where $n=1,2,\ldots$. A detailed discussion about this base can be found in \cite{Denef:2008wq}. If $\cD_I=\{u_I=0\}$ denote the toric divisors, let us pick the following (non-independent) divisors
\be
D^1= \cD_1\simeq \cD_2\simeq \cD_3\quad,\quad D^2=\cD_5\simeq E+n D^1\quad,\quad E=  \cD_4\simeq D^2-nD^1\,.
\ee
The relevant lattices are $M_{\mathbb{Z}}=H_4(X,\mathbb{Z})\simeq \mathbb{Z}^2$ and $N_{\mathbb{Z}}=H_2(X,\mathbb{Z})\simeq \mathbb{Z}^2$.
The cone of nef divisors Nef$^1(X)_{\mathbb{Z}}$ is generated by $D^1$ and $D^2$, while the cone of effective divisors Eff$^1(X)_{\mathbb{Z}}=\cC_{\rm I}\subset M_{\mathbb{Z}}$ is generated by $D^1$ and $E$. The intersection numbers are summarised by
\be
\cI=(D^1)^2D^2+nD^1(D^2)^2+n^2(D^2)^3\, ,
\ee
where the coefficient of each monomial gives the corresponding intersection number. We can expand the K\"ahler form in the nef basis, to describe the K\"ahler cone
\be 
\cK(X)=\left\{J=v_1 [D^1]+v_2 [D^2]\in H^2(X,\mathbb{R})| v_1,v_2>0\right\}\,,
\ee where $v_1,v_2$ are the two K\"ahler moduli that measure the volumes of the dual effective curves 
\be 
C_1\simeq D^1\cdot E\simeq D^1\cdot(D^2-nD^1)\quad,\quad C_2\simeq D^1\cdot D^1\,,
\ee
which satisfy $C_a\cdot D^b=\delta_a^b$. Note also that $E\cdot D^2=0$. We can identify $C_2$ with the $\mathbb{P}^1$ fibre and $C_1$ with the push-forward of $\mathbb{P}^1$ in the base through the section defined by $u_4=0$. 

The saxions are given by the following expansion
\be\label{Fsx}
\begin{aligned}
{\bm s}&= \frac12 J\wedge J=s^1 [C_1]+s^2[C_2]\in H^4(X,\mathbb{R})\\
&\quad \text{with}\quad s^1=v_1 v_2+\frac12 nv_2^2 \,,\quad  s^2=\frac12\left(v_1+nv_2\right)^2\,.  
\end{aligned}
\ee
The actual saxionic cone is obtained by imposing \eqref{defDelta} with $\cC_{\rm I}$ generated by the effective divisors $D^1$ and $E$:
\be
\Delta=\left\{{\bm s}\in H^4(X,\mathbb{R})\,\big|\langle D^1,{\bm s}\rangle =s^1> 0\,,\ \langle E,{\bm s}\rangle =s^2-ns^1> 0\right\}\, .
\ee
From our definition \eqref{CSEFT} it is easy to see  that the cone of EFT string charges:
\be 
\cC_{\rm S}^{\text{\tiny EFT}}=\{{\bf e}=e^1\,C_1+e^2\,C_2\in {\rm Mov}_1(X)_{\mathbb{Z}}|\, e^1\geq 0\,,\ e^2-n e^1\geq 0 \}
\ee
 is generated by the charges  ${\bf e}_1\equiv C_1+n C_2=D^1\cdot D^2$ and ${\bf e}_2\equiv C_2$. 

Note that in this case we have ${\rm Mov}^1(X)={\rm Nef}^1(X)$ and then  \eqref{KextB} with $\cK(X)_{\rm ext}\equiv \cK(X)$ should be one-to-one. This can be explicitly checked by inverting \eqref{Fsx} into
\be\label{vXtwP}
v_1=\sqrt{2(s^2-ns^1)}\,,\quad v_2=\frac1{n}\left(\sqrt{2s^2}-\sqrt{2(s^2-ns^1)}\right)\,,
\ee
which indeed cover the entire K\"ahler cone as we vary ${\bm s}\in\Delta$. Furthermore
\be\label{VXtwP}
V_X=\frac{\sqrt{2}}{3n}\left[(s^2)^{\frac32}-(s^2-ns^1)^{\frac32}\right]
\ee
and then, up to an irrelevant constant, the explicit form of the K\"ahler potential \eqref{KpotFt} is
\be
K_{\rm ks}=-2\log\left[(s^2)^{\frac32}-(s^2-ns^1)^{\frac32}\right] \, ,
\ee
and the dual saxions are 
\be
\ell_1=\frac{3n\sqrt{s^2-ns^1}}{2\Big[(s^2)^{\frac32}-(s^2-ns^1)^{\frac32}\Big]}\,,\qquad \ell_2 =\frac{3\left(\sqrt{s^2}-\sqrt{s^2-ns^1}\right)}{2\Big[(s^2)^{\frac32}-(s^2-ns^1)^{\frac32}\Big]}\,.
\ee
Note that $\ell_1$ and $\ell_2$ can (classically) take any possible value and then  we can identify the dual saxionic cone $\cP$ with $\cK(X)$: $\cP=\{{\bm \ell}=\ell_1[D]^1+\ell_2[D]^2|\, \ell_1,\ell_2>0\}$.  

We are ready to discuss the mass scalings under the EFT string flows \eqref{ssssflow}. We first choose a  charge ${\bf e}=e^1C_1+e^2C_2$ with $e^1>0$ (and $e^2\geq ne^1$). In other words, by decomposing ${\bf e}$ in terms of the generators ${\bf e}_1$ and ${\bf e}_2$, we are assuming that the contribution of ${\bf e}_1$ is non-vanishing  (in particular, the most elementary choice is ${\bf e}={\bf e}_1$.) We can then easily see that the internal volume scales as $V_X\sim \sigma^{\frac32}$ 
for $\sigma\rightarrow\infty$. Furthermore, $v_2\sim \sqrt{\sigma}$ and, if $e^2>ne^1$, also  $v_1\sim \sqrt{\sigma}$. In any case, the estimate of the KK mass scaling goes as in the $\mathbb{P}^3$ example discussed above. In particular, Conjecture \ref{conj:cutoff} is verified with scaling weight ${\it w}=2$. Note also that all these flows, including the elementary one with ${\bf e}={\bf e}_1$, are degenerate of order 2 -- see the definition below \eqref{face} -- since both $\ell_1$ and $\ell_2$ vanish in the asymptotic limit. Moreover, along this flow, the membrane scale \eqref{mmem} can be easily obtained from \eqref{asymem}, with $n=3$.

It remains to discuss the  possibility  $e^1=0$, for which it is sufficient to restrict to the elementary charge  ${\bf e}={\bf e}_2=C_2$. The corresponding flow \eqref{ssssflow} becomes $s^1=s^1_0$ and $s^2=s^2_0+ \sigma$. Asymptotically for $\sigma\rightarrow\infty$, we have 
$V_X\sim s^1_0\sqrt{\sigma}$, $v_1\sim \sqrt{\sigma}$ and $v_2\sim \frac{s^1_0}{\sqrt{\sigma}}$.
This implies that the KK-scale  $m_*^2=\frac{e^{2A}}{\ell^2_{\rm s}v_1}$ associated with the growing $C_1\simeq \mathbb{P}^1$ scales like $1/\sigma$ and then realises \eqref{conj:cutoff} with scaling weight ${\it w}=1$. Note also that $\ell_2\sim 1/\sigma$, while $\ell_1$ remains asymptotically finite. Hence the flow is non-degenerate. Furthermore, the membrane scale \eqref{mmem} falls off as $E_{\rm mem} \sim M_{\rm P} \sigma^{-\frac32}$ along such a flow, realising \eqref{asymem} with $n=3$. One may be skeptic about the reliability of these conclusions, since the volume of the fibral curve $C_2$  shrinks to zero size. However, this EFT flow is a particular realisation of a  $J$-class A limit of \cite{Lee:2019tst,Klaewer:2020lfg}, in which 
it is argued that there exists a dual weakly coupled heterotic description in which the D3-brane  wrapping $C_2$ is dualised to a F1-string. The D3-brane  wrapping $C_2$ precisely corresponds to our EFT string of charge ${\bf e}={\bf e}_2=C_2$ and then its flow can be mapped to an F1 flow of section \ref{sec:hetCY}. 

In fact, the above results should qualitatively hold for more general $J$-class A limits with heterotic dual, in which $X$ is a more general   $\mathbb{P}^1$-fibration.  The D3-brane wrapping the $\mathbb{P}^1$-fibre is an example of elementary  EFT string that, borrowing the results of \cite{Lee:2019tst,Klaewer:2020lfg}, satisfies  \ref{conj:cutoff} with scaling weight ${\it w}=1$ and is non-degenerate. 

\bigskip

\subsubsection*{Weak-coupling limit and odd moduli}

Of course, one can also consider  asymptotic limits along other moduli space directions. Here we consider the weak-coupling limit, in which an  F-theory compactification  is well described in terms of a compactification on a double cover Calabi--Yau three-fold $\hat X$ in presence of O7-planes. 

In the weak coupling limit $\Im\tau\rightarrow\infty$, where $\tau=C_0+\ii e^{-\phi}$ is the IIB axio-dilaton. In this limit the associated exponentially suppressed non-perturbative corrections are generated by D(-1)-instantons and the strings carrying a corresponding axionic charges are given by D7-branes wrapping the entire internal space. 

It is interesting to observe that, as emphasised for instance in \cite{Collinucci:2008pf}, in these backgrounds there exists a ``half'' D$(-1)$-brane, whose path-integral contribution $e^{\ii\pi\tau}$ to the 4d EFT is invariant only under an even integral shifts of $\tau$: $\tau\simeq \tau+2$ (while $\tau\neq \tau+1$).\footnote{We stress that this $2\mathbb{Z}$ periodicity of $\tau$ is a 4d effect, which should be interpreted as a spontaneous breaking  of the  10d $\mathbb{Z}$  periodicity, due to the presence of the O7-plane.} Correspondingly, any consistent EFT string must generate an {\em even} monodromy of $\tau$. In microscopic terms, these means that these EFT correspond to stacks of an {\em even} number $e_{\rm D7}$ of D7-branes  wrapping the internal space. This conclusion is  consistent with the fact that such D7-branes should carry an Sp($e_{\rm D7}$) gauge group, since  they have four mutually transversal directions with respect to the O7-planes.       

Along the flow generated by such EFT strings, the dilaton changes as $\Im\tau=\Im\tau_0+\sigma e_{\rm D7}$. If $e_{\rm D7}$ were the only string charges supported by the D7-brane, the saxions $s^a$ defined in  \eqref{KextB} would not run. Since these correspond to Einstein frame volumes, all string frame volumes ${\rm vol}_{\rm st}(D)=(\Im\tau)^{-1}\int_D{\bf s}$ would actually vanish as $\sigma\rightarrow 0$, leading to a breakdown of the 10d supergravity description. One should then go to some 10d dual frame, for instance by picking a toroidal configuration and performing six T-dualities. In this way one would get a IIB compactification with O5-planes, in which the EFT string is represented by a D1 brane, obtaining a setting which is qualitatively analogous to the type I case discussed above, hence confirming Conjecture \ref{conj:cutoff} with scaling weight ${\it w}=1$.   

However, the D7-string  can generically support some induced D3-brane charge due to internal curvature corrections and world-volume fluxes. Such charges would activate a flow of the $s^a$ saxions as well, so that the string frame volume of some divisors would remain finite. The internal world-volume fluxes could also carry additional string charges associated with  a  non-trivial odd cohomology group $H^2(\hat X)_-$ which has been so far assumed to be trivial (since we assumed $H^3(Y)=0$ in the dual M-theory Calabi-Yau geometry). In the case of non-trivial $H^2(\hat X)_-$ there appear additional `odd' moduli  $\beta^\alpha=c^\alpha-\tau b^\alpha$ obtained by expanding $C_2-\tau B_2$ in a basis $[\tilde D_\alpha]\in   H^2(X,\mathbb{Z})_-$, Poincar\'e dual to a set of odd divisors $\tilde D_\alpha$. The fluxed D7-strings can then induce around them monodromies of the axions $\Re\beta^\alpha$.  

Note that from the integral  periodicity of the IIB NS-NS $B_2$ potential it follows that $b^\alpha\simeq b^\alpha-n^\alpha$ with $n^\alpha\in\mathbb{Z}$. This induces the identification
\be
\beta^\alpha\simeq \beta^\alpha+n^\alpha\tau\,.
\ee 
This {\em perturbative} duality is an example of the duality group $\cG_{\mathbb{Z}}$ appearing in the saxionic cone conjecture of section \ref{sec:duality}. It also acts  on the other moduli $t^a$, since if $H^2(X,\mathbb{Z})_-$ is non-trivial their definition  is modified by $b^\alpha$-dependent terms \cite{Grimm:2004ua}. More precisely, we have
\be
t^a\simeq t^a-\tilde\kappa^a_{\alpha\beta}n^\alpha\beta^\beta-\frac12\tilde\kappa^a_{\alpha\beta}n^\alpha n^\beta\,\tau
\ee
where $\tilde\kappa^a_{\alpha\beta}$ is the intersection number between the even divisor $D^a$ and the odd divisors $\tilde D_\alpha,\tilde D_{\beta}$. This  duality implies that one may isolate a fundamental region for this saxionic sector, for instance  $0\leq \Im\beta^\alpha \leq \Im\tau$ corresponding to $-1 \leq b^\alpha \leq 0$. This conical fundamental region is indeed rational polyhedral, in agreement with the saxionic cone conjecture of section \ref{sec:duality}.

A detailed study of this extended moduli sector and of the corresponding EFT strings originating from fluxed D7-branes  would require a careful definition of the flux quantisation conditions and of the corresponding string and instanton lattices.  We leave this task, together with a more accurate test of the validity of our general framework for these classes of models, to the future. 

Finally, one may address CS string flows. Following the same strategy as in the heterotic and type I models, one may perform a naive estimate of the relevant compactification scales by looking at the NS5 string flows in a mirror manifold. If there is no relevant scale below the mirror KK scales, this would again imply that  Conjecture \ref{conj:cutoff} is realised with scaling weight ${\it w}=1$. If however there is some lighter tower of states, this conclusion could be modified. For instance, one could consider the would-be mirror duals of type IIA D0-branes in the orientifold limit, which would be non-BPS D3-branes wrapping special Lagrangian three-cycles. If these states generate a tower of states, they would likely behave like M-theory scale in table \ref{tab:IIANS5}, implying the same scaling weights $w=1,2,3$ as in there. Again, it would be interesting to perform a more detailed analysis of these CS flows to elucidate which of these two options, or even if a third one, is realised.

\subsection{Type IIA Calabi-Yau  orientifolds}
\label{ss:IIA}

Let us now consider Type IIA compactified on the projection of  Calabi--Yau threefold $X$ under the O6 orientifold involution $\iota:  X\rightarrow  X$. We now summarise the relevant ingredients to describe its EFT,
focusing on the closed string moduli sector. 

The Calabi--Yau moduli are encoded in the (string frame) K\"ahler form $J$ and the (3,0)-form $\Omega$, whose  normalisation is  fixed by
\be\label{IIAnormOmega}
\frac{\ii}{8}\Omega\wedge \bar\Omega=\frac1{3!}J\wedge J\wedge J\, ,
\ee
and an overall constant phase has been fixed by the requirement that the O6-planes are calibrated by $\Re\Omega$.  $J$ and $\Omega$ must satisfy the projection condition $\iota^*J=-J$ and $\iota^*\Omega=\bar\Omega$, respectively.

There are two sets of chiral fields parametrising closed string moduli. As in the heterotic case, a first set of chiral fields is provided by the $b_2(X)_-$ complexified (string frame) K\"ahler moduli $t^a \equiv a^a+\ii s^a$ parametrising $B_2+\ii J$
\be
B_2+\ii J\equiv  t^a[D_a^+]\equiv  (a^a+\ii s^a)[D_a^+]\, ,
\label{bfieldIIA}
\ee
where now $[D_a^+]$ provide a basis for the odd cohomology classes $H^2( X; \mathbb{R})_-$.\footnote{ $[S]$ refers to the  Poincar\'e dual of a $p$-cycle $S$, which in our conventions is defined by $\int_S\omega=\int_X\omega\wedge [S] $, for any closed $p$-form $\omega$. Since the O6-plane involution inverts the orientation, Poincar\'e duality connects even/odd cycles with odd/even cohomology classes.} The remaining moduli, which include the dilaton, complex structure moduli and axionic partners, are encoded  in
a second set of $b_3(X)_+$ chiral fields  $\hat t^\alpha\equiv \hat a^\alpha+\ii \hat s^\alpha$, identified through the expansion
\be
C_3+\ii e^{-\phi}\Re\Omega\equiv \hat t^\alpha\,[\Sigma_\alpha^-]\equiv  (\hat a^\alpha+\ii \hat s^\alpha)\,[\Sigma_\alpha^-]\, ,
\label{expcpxIIA}
\ee
where $[\Sigma_\alpha^-]$ forms a basis for the even three-form classes $H^3(X; \mathbb{R})_+$.  Notice that the saxions $\hat  s^{\alpha}$ parametrise the dilaton and the complex-structure moduli of $X$ only through a non-trivial field redefinition, which furthermore depends also on the K\"ahler structure moduli because of the normalisation \eqref{IIAnormOmega}.

From the above description it follows that we can split the saxionic cone into 
\be\label{DeltaIIA}
\Delta=\Delta_{\rm K}\times \hat\Delta\, ,
\ee where $\Delta_{\rm K}$ is generated by the K\"ahler saxions $ s^a$,  while the remaining saxions $\hat s^\alpha$ take values in $\hat\Delta$. 
We will also use the index-free notation
\be\label{IIAsaxions}
\begin{aligned}
{\bm s}&\equiv [J]=s^a[D_a^+]\in H^2(X,\mathbb{R})_-\,,\\ \hat{\bm s}&\equiv e^{-\phi}[\Re\Omega]=\hat s^\alpha[\Sigma_\alpha^-]\in H^3(X,\mathbb{R})_+\,.
\end{aligned}
\ee

The string frame volume of $X$   depends only on the K\"ahler saxions $s^a$:
\be
V_X(s)= \frac1{3!}\,\int_X J\wedge J\wedge J=\frac1{3!}\,\kappa({\bm s},{\bm s},{\bm s})\,,
\ee
with $\kappa({\bm s},{\bm s},{\bm s})\equiv \kappa_{abc} s^as^bs^c$ as in the heterotic models. On the other hand, by  following \cite{Hitchin:2000jd} one can express  the odd class $[e^{-\phi}\Im\Omega]\in H^3(X,\mathbb{R})_-$  as a function of the  $\hat s^\alpha$ saxions only, and define an associated  Hitchin function
\be
\cH(\hat s)\equiv \frac\ii8 \int_X e^{-2\phi}\Omega\wedge\bar\Omega\quad~~~\rightarrow\quad~~~ \delta \cH=\frac12\int_X \delta \hat{\bm s}\wedge [e^{-\phi}\Im\Omega]\,.
\ee
Notice that $\cH(\hat s)$ is  homogeneous of degree two  and that the 10d dilaton can be written as
\beq\label{IIAdilaton}
e^{2\phi}=\frac{V_X(s)}{\mathcal{H}(\hat s)}\, .
\eeq

In fact, this description of the moduli is accurate  in a perturbative regime  in which the backreaction of fluxes and localised sources can be neglected and the warping  can be approximated to be constant. In such a case, the 10d string frame metric is as in \eqref{hetmetric} with
\be\label{IIAcomp}
 e^{2A}=\frac{l_{\rm s}^2M^2_{\rm P}}{2\pi\calh(\hat s)}\, ,
\ee
and the relevant terms contributing to the EFT K\"ahler potential are  \cite{Grimm:2004ua} 
\be\label{IIAK}
K=-\log V_X(s)-2\log\cH(\hat s)\, .
\ee
We also note also that  in this perturbative regime the entire moduli space of the closed string sector corresponds to  axionic symmetries, in the sense that all chiral fields can be split into axion + saxion. Hence, all of them can be dualised to linear multiples, as in \eqref{dualfields}.  In particular, the dual saxions are 
\be
\begin{aligned}
\ell_a&=-\frac{1}{2}\frac{\del K}{\del s^a}=\frac{1}{4 V_X}\int_{D_a^+}J\wedge J\equiv  \frac{3\,\kappa([D_a^+],{\bm s},{\bm s})}{2\,\kappa({\bm s},{\bm s},{\bm s})}\,,\\
\hat \ell_\alpha&=-\frac12\frac{\del K}{\del \hat s^\alpha}=-\frac{1}{2\cH}\int_{\Sigma_{\alpha}^-}e^{-\phi}\Im\Omega\,.
\end{aligned}
\ee
We stress that, by using the results of \cite{Hitchin:2000jd},  $\hat \ell_\alpha$ are functions of $\hat s^\alpha$ only.

The choice of basis $\{[D_a^+]\}_a$,  $\{[\Sigma_\alpha^+]\}_\alpha$ and their interpretation in terms Poincar\'e dual classes of cycles can be fixed by demanding that the axions $a^a$, $\hat{a}^\alpha$ have periodicity one. Similarly to the case of the axio-dilaton in type IIB compactifications with O7-planes, the periodicity of type IIA  axions is fixed by the set of instanton charges in the compactification, which depends on the precise action of the orientifold involution on the relevant homology classes. Consider a basis ${D}_i \in H_4(X; \mathbb{Z})$ of $b_2(X)$ divisors and associated basis ${C}^j\in H_2( X,\mathbb{Z})$ of two-cycles, such that ${D}_i \cdot {C}^j=\delta^j_i$. By taking linear combinations of the cycles $C^j$  we can then construct a set   of odd two-cycles $C^a_-$ that generate the orientifold-odd lattice $H_2( X,\mathbb{Z})_-$. We expect that there exists a set of dual even four-cycles $D_a^+ $, which satisfy $D_a^+ \cdot C^b_- = 2\delta^b_a$ and can be identified with the Poincar\'e dual of the classes $[D_a^+]$ in \eqref{bfieldIIA}. Note that in general $D_a^+ $ may only generate a sublattice of $H_4( X;\mathbb{Z})_+$. 
For instance, suppose that $\{C^a_-\}_a$ can be chosen to be  a subset of  $\{C^j\}_j$. In this case
the dual four-cycles would be of the form $D_a^+ \equiv D_a + \iota(D_a)$. The worldsheet instantons wrapping $C^a_-$ would correspond to crosscaps in the quotient geometry, as in \cite{Acharya:2002ag}, and so Dirac quantisation would imply that a  NS5-branes can only wrap four-cycles of the form $D^+ = D + \iota(D)$.

Similarly, one can always consider a symplectic basis of three-cycles $\Sigma_\mu, \tilde\Sigma^\nu \in H_3(X;\mathbb{Z})$, such that $ \Sigma_\mu \cdot \tilde\Sigma^\nu =\delta^\nu_\mu$. Again, we can identify a set of even three-cycles $\tilde\Sigma^\alpha_+$ generating 
$H_3(X;\mathbb{Z})_+$ and we expect that there exists a dual set of odd-three cycles $\Sigma_\alpha^-$, such that  $\Sigma_\alpha^- \cdot \tilde\Sigma^\beta_+ =2\delta^\beta_\alpha$. Their Poincar\'e duals can then be used in the expansion \eqref{expcpxIIA}.
Suppose for instance  that  $\{\tilde\Sigma^\alpha_+\}_\beta \subset \{\tilde\Sigma^\mu\}_\mu$. Then the minimal 4d instantons charges of the compactification will be given by Euclidean D2-branes wrapping each of the $\tilde\Sigma^\alpha_+$, whose worldvolume gauge group may be either $O(1)$ or $USp(2)$. Whenever a D2-brane wrapping $\tilde\Sigma^\alpha_+$ yields a gauge group $O(1)$, Dirac quantisation implies that the dual 4d strings are made up from D4-branes wrapping the odd three-cycles $\Sigma_\alpha^- \equiv \Sigma_\alpha - \iota(\Sigma_\alpha)$. Alternatively, if the gauge group is $USp(2)$ then we may choose $\Sigma_\alpha^- \equiv \Sigma_\alpha$ whenever $\iota(\Sigma_\alpha) = - \Sigma_\alpha$, but then it is more convenient to redefine $\tilde\Sigma^\alpha_+ \to 2 \tilde\Sigma^\alpha_+$ to account for the double instanton charge. In both cases we end up with an even intersection number  $\Sigma_\alpha^- \cdot \tilde\Sigma^\beta_+ =2\delta^\beta_\alpha$ in the covering space.

We can now discuss each sector from the viewpoint proposed in the present paper.

\subsubsection*{K\"ahler moduli and NS5 string flows}

The properties of this sector are very similar to  those discussed in the heterotic case, and so we will be short. One minor technical difference is due to the orientifold projection. So, we can identify $\Delta_{\rm K}$ with the orientifold-odd projection $\cK(X)_-$ of the Calabi--Yau K\"ahler cone $\cK(X)$. The corresponding BPS instantons are provided by orientifold-odd holomorphic curves, while EFT strings correspond to NS5-branes wrapping even nef divisors (which are assumed to be also effective, as in \cite{Katz:2020ewz}):
\be
\cC^{\rm K}_{\rm S}={\rm Nef}(X)^+_{\mathbb{Z}}.
\ee 
Furthermore, if $D\simeq e^aD_a^+$ is a generic effective divisor, by using \eqref{IIAcomp} one can easily check that the corresponding BPS string tension $\cT_{\bf e}=\frac{\pi e^{2A}}{2l^2_{\rm s}}\int_{D}e^{-2\phi}J\wedge J$ obtained by dimensional reduction  matches the EFT formula $\cT_{\bf e}=M^2_{\rm P}e^a\ell_a$.

Note also the analogy between \eqref{IIAdilaton} and \eqref{hetdilaton}. Hence, the dilaton  diverges along any flow generated by these EFT strings  as in the heterotic case, and the entire discussion on scaling of the relevant UV masses presented for that case can be applied {\em verbatim} also in this case. So we refer the reader to the previous subsection for details and just  emphasise that Conjecture \ref{conj:cutoff} is realised, with possible scaling weights ${\it w}=1,2,3$. 

Let us however point out that the membrane spectrum of the heterotic and IIA models is quite different. In particular, in IIA the lightest membranes correspond to D2-branes. However, this fact does not affect the universality of the asymptotic behaviour of the relevant  mass scales. We can then summarise the relevant mass scalings with the following table, which is basically the same as  table~\ref{tab:HetNS5}:

 \begin{table}[H]
        \small
		\centering
		\makebox[\linewidth]{\begin{tabular}{c|c | c | c|c|c|}
			\hhline{~|-|-|-|-|-|}
			 & \cellcolor{gray!30}$\sigma^{-\frac32}$ & \cellcolor{gray!30}$\sigma^{-1}$ & \cellcolor{gray!30}$\sigma^{-\frac12}$ 
			 & \cellcolor{gray!30}$\sigma^{-\frac13}$ 
			 & \cellcolor{gray!30}$\sigma^{-\frac16}$ \\ \hline
			\multicolumn{1}{|c|}{\cellcolor{gray!30}$\cT_{\rm st}^{1/2}$}                    &  
			      &                       & NS5 & &
			\\ \hline
			\multicolumn{1}{|c|}{\cellcolor{gray!30}$m_*$}         &       M-th. KK  -- Case 1           &      M-th. KK  -- Case 2         &    M-th. KK  -- Case 3    & &            \\ \hline
			\multicolumn{1}{|c|}{\cellcolor{gray!30}CY $m_{\rm KK}$ }         &                        &                            &            CY  $m_{\rm KK}$             & &                                                 \\ \hline
			\multicolumn{1}{|c|}{\cellcolor{gray!30}$E_{\rm mem}$}                  &    D2 -- Case 1      &             D2 -- Case 2               &              D2 -- Case 3       &   &           \\ \hline
			\multicolumn{1}{|c|}{\cellcolor{gray!30}$\cT_{\rm mem}^{1/3}$}   &       &                                   &  D2  -- Case 1    &  D2  -- Case 2 &  D2  -- Case 3                                                   \\ \hline
		\end{tabular}}
		\caption{Mass scalings along the flows generated by NS5-strings in IIA orientifold models.\label{tab:IIANS5}} 
	\end{table}

\subsubsection*{Complex structure sector}

The description of the saxionic cone factor $\hat\Delta$ appearing in \eqref{DeltaIIA} is more subtle. Naively, this should identify the possible values of  $\hat{\bm s}=e^{-\Phi}[\Re\Omega]\in  H^3(X,\mathbb{R})_+$. From this definition, it is clear that if $\hat{\bm s}\in \hat\Delta$, then also $\lambda\hat{\bm s}\in \hat\Delta$  for any $\lambda>0$. Hence, at least classically, $\hat\Delta$ is indeed a cone. Furthermore, by extending the results of \cite{Hitchin:2000jd} to our orientifolded setting, the moduli spaces described by $\hat{\bf s}$ can be locally identified with an open neighbourhood of $H^3(X,\mathbb{R})_+$. 
On the one hand, by combining this observation with the conical structure, we can imagine to subdivide $\hat\Delta$ into the union of open subcones of $H^3(X,\mathbb{R})_+$. 
On the other hand, a global definition of $\hat\Delta$ as in \eqref{defDelta} is not obvious. Part of the complication  is due to the fact that, at large volume, the BPS instanton sector $\hat\cC_{\rm I}$  corresponds to Euclidean D2-branes wrapping internal special Lagrangian cycles calibrated by $e^{-\phi}\Re\Omega$. Differently from previous examples, such instantonic branes can violate the BPS condition at some point of the accessible  saxionic domain, by crossing a stability wall.  As in \cite{GarciaEtxebarria:2008pi}, one may adapt the definition of \cite[Appendix B]{deBoer:2008fk} to our setting and distinguish two kinds of stability walls: {\it i)} threshold  stability walls, where the three-cycle  splits into mutually BPS three-cycles and {\it ii)} marginal stability walls, where such constituents are not mutually BPS. Applying the general scheme of section \ref{sec:cones}, we define the set of instanton charges $\hat\cC_{\rm I}$ in terms of the three-cycles that at most cross walls of threshold stability along the asymptotic limits contained in $\hat\Delta$.
The non-trivial part of this definition is that it should be compatible with the initial choice of $\hat\Delta$ through \eqref{defDelta}. In the following we will assume that such compatible pair $(\hat\cC_{\rm I}, \hat\Delta)$ can be found, as one can check in some simple examples.

Regarding 4d strings, BPS charges ${\bf \hat{e}}\in \cC_{\rm S}$ correspond to D4-branes wrapping SLag three-cycles calibrated by $-e^{-\phi}\Im\Omega$, with vanishing worldvolume flux. Their tension is given by
\begin{equation}
    \cT_{\bf \hat{e}} = -\frac{\pi e^{2A}}{l^2_{\rm s}}\int_{\Sigma} e^{-\phi} \Im \Omega   = M^2_{\rm P}\, \hat{e}^\alpha \hat{\ell}_\alpha \, .
        \label{TD4BPS}
\end{equation}
Now, just like for BPS instantons, the SLag condition could be satisfied only at some sublocus of the saxionic domain. However, a BPS string solution like that of section \ref{s:fundamental} requires that ${\bf \hat{e}}$ remains BPS along its own flow, which in particular constrains the class of three-cycles that can host an EFT string. In particular, crossing a marginal stability wall along a string flow would signal an inconsistency, as the system would leave the BPS locus. If the string is non-EFT, non-perturbative effects unaccounted for in the solution of section \ref{s:fundamental} could in principle fix this apparent inconsistency, but for the case of EFT strings this seems unlikely to happen. We therefore conclude that a necessary condition for EFT strings made up of D4-branes wrapping SLag three-cycles is that they can only cross threshold stability walls along their flow.   This conclusion provides yet another constraint on the choice of $\hat\Delta$, through the definition \eqref{CSEFT}, or equivalently \eqref{altCC}.

Unfortunately, our current general understanding of this saxionic sector, and of the corresponding EFT string sector,  is quite limited. Hence we will illustrate the above observations and discuss the EFT string flows in a simple concrete toroidal model. There we will indeed see that indeed {\em any} EFT string charge populating  $\cC_{\rm S}^{\text{\tiny EFT}}$ admits a SLag representative at any point in $\hat\Delta$. We believe that this is not a lucky accident, but rather a general property of the EFT string sector.        
  
\bigskip    

{\em Toroidal orbifolds}

\smallskip

\noindent Let us consider the case where $X = (T^2 \times T^2\times T^2)/\Gamma$, with $\Gamma \in SU(3)$ some orbifold action, and an orientifold involution of the form $\iota:(z_1,z_2,z_3)\rightarrow (\bar z_1,\bar z_2,\bar z_3)$. For simplicity, we will focus on the case where $\Gamma = \mathbb{Z}_2 \times \mathbb{Z}_2'$, with the choice of discrete torsion of \cite{Cvetic:2001nr}, and a choice of complex structure of the form $\tau_i = \ii \frac{R_{2i}}{R_{2i-1}}$. Nevertheless, our discussion can be easily generalised to other cases, see e.g. \cite{Blumenhagen:2005mu,Blumenhagen:2006ci,Marchesano:2007de} for reviews. 

With these choices, the holomorphic $(3,0)$-form reads
\be
\begin{aligned}
\Omega&=R_1R_3R_5\, \d z_1\wedge\d z_2\wedge \d z_3\\
&=(R_1\d y_1+\ii R_2\d y_2)\wedge (R_3\d y_3+\ii R_4\d y_4)\wedge(R_5\d y_5+\ii R_6\d y_6)\, .
\end{aligned}
\ee
A basis for even three-form classes $[ \Sigma_\alpha^-]$ is given by 
\be
\begin{aligned}
[\Sigma_0^-] =4\d y_1\wedge \d y_3\wedge \d y_5\, , &\quad [\Sigma_1^-]=-4\d y_1\wedge \d y_4\wedge \d y_6\, ,\\
[\Sigma_2^-] =-4\d y_2\wedge \d y_3\wedge \d y_6\, , &\quad [\Sigma_3^-] =-4\d y_2\wedge \d y_4\wedge \d y_5\, ,
\end{aligned}
\label{even3form}
\ee
while one for odd three-form classes $[\tilde{\Sigma}_+^\beta]$ is
\be
\begin{aligned}
[\tilde\Sigma^0_+]=-2\d y_2\wedge \d y_4\wedge \d y_6\, , & \quad [\tilde\Sigma^1_+] =2\d y_2\wedge \d y_3\wedge \d y_5\, ,\\
[\tilde\Sigma^2_+] =2\d y_1\wedge \d y_4\wedge \d y_5\, , &\quad [\tilde\Sigma^3_+] =2\d y_1\wedge \d y_3\wedge \d y_6\, .
\end{aligned}
\label{odd3form}
\ee
Notice that their Poincar\'e duals satisfy $ \Sigma_\alpha^-  \cdot \tilde\Sigma^\beta_+ = 2\delta_\alpha^\beta$. 
The relative factor of 2 in between these two basis takes into account that D2-branes wrapping the even three-cycles $\tilde{\Sigma}_+^\beta$ dual to \eqref{odd3form} yield $O(1)$ instantons. By using  this basis in the expansion \eqref{expcpxIIA}, we can identify    $\hat{a}^\alpha = \frac{1}{8} \int_{T^6}  C_3 \wedge [\tilde{\Sigma}^\alpha_+]$ and
\be
\begin{aligned}
\hat s^0=\frac{1}{4} e^{-\phi}R_1R_3R_5\,, & \quad \hat s^1=\frac{1}{4} e^{-\phi}R_1R_4R_6\, ,\\
 \hat s^2=\frac{1}{4}e^{-\phi}R_2R_3R_6\,, & \quad \hat s^3=\frac{1}{4}e^{-\phi}R_2R_4R_5\, .
\end{aligned}
\ee
The saxions $\hat{s}^\alpha$ measure the internal volume of $O(1)$ instantons wrapping calibrated three-cycles, while $2\hat{s}^\alpha$ measure the volume of $U(1)$ instantons. Similarly, $2\hat{a}^\alpha$ would be the axion with unit periodicity in the unorientifolded theory, while $\Re \hat{t}^\alpha =  \hat{a}^\alpha$ is the actual unit periodicity axion upon orientifolding. In terms of these fields we have that
\be
\cH(\hat{s}) = \frac{\ii}{32} \int_{T^6}e^{-2\phi}  \Omega \wedge \bar{\Omega}  = 8 \sqrt{\hat{s}^0\hat{s}^1\hat{s}^2\hat{s}^3}\, ,
\ee
and so the K\"ahler potential reads
\be
K = - \sum^3_{\alpha=0}\log \Im \hat t^\alpha + \dots
\ee
where we have omitted the piece for the K\"ahler moduli, which will play no role in the following. The dual saxions are then given by 
\be
\hat{\ell}_\alpha = \frac{1}{2\hat{s}^\alpha}\, .
\ee

From here it follows a very simple structure for the saxionic cones of this sector
\begin{subequations}
\label{IIAcontcones}
\bea
\hat\Delta & = & \big\{(\hat{s}^0, \hat{s}^1, \hat{s}^2, \hat{s}^3)\in\mathbb{R}^4| \hat{s}^0,\hat{s}^1,\hat{s}^2,\hat{s}^3 > 0\big\}\, , \\
\hat\cP & = &\big\{(\hat{\ell}_0, \hat{\ell}_1, \hat{\ell}_2, \hat{\ell}_3)\in\mathbb{R}^4|\hat{\ell}_0, \hat{\ell}_1, \hat{\ell}_2, \hat{\ell}_3 > 0\big\}\, ,
\eea
\end{subequations}
and it is instructive to see how the cone structure for 4d instantons and strings discussed in section \ref{sec:cones} arises in this context. As discussed above both objects arise by wrapping D2- and D4-branes, respectively, on calibrated three-cycles.\footnote{One could also consider D4-branes and D6-branes wrapping coisotropic five-cycles, similarly to the construction of \cite{Font:2006na}. Since a priori these objects do not result into new 4d charges, we will not consider them in the following.} As usual in these kinds of models one may obtain a good idea of the spectrum of BPS objects by considering factorisable three-cycles of the form
\be
\Pi = 2 (n^1,m^1) (n^2, m^2) (n^3,m^3)\, , \qquad n^i, m^i \in \mathbb{Z}
\label{f3cycle}
\ee
where we follow the notation of \cite{Cvetic:2001nr}.\footnote{More precisely, a three-cycle $(n^1,m^1) (n^2, m^2) (n^3,m^3)$ corresponds, in the covering space $T^6$, to the class 
\be
\Pi = \left( n^1 a_1 + m^1 a_2 \right) \times \left( n^2 a_3 + m^2 a_4 \right) \times \left( n^3 a_5 + m^3 a_6 \right)\, ,
\ee 
where $a_i$ is the one-cycle class of $T^6$ along the coordinate $y_i$.} If this three-cycle is not invariant under the orientifold projection one needs to add the orientifold image, which lies at
\be
\iota(\Pi) = 2 (n^1,-m^1) (n^2, -m^2) (n^3,-m^3)\, .
\ee

Consider a Euclidean D2-brane wrapping the even three-cycle $\tilde\Sigma=\Pi+\iota(\Pi)$. 
Both $\Pi$ and $\iota(\Pi)$ will be simultaneously BPS if $\int_{\Pi} e^{-\phi}\Re \Omega > 0$ and $\int_{\Pi} e^{-\phi}\Im \Omega = 0$. The second condition is equivalent to
\be
\tilde\varepsilon \equiv  m^1m^2m^3 \hat{\ell}_0 - m^1n^2n^3 \hat{\ell}_1 - n^1m^2n^3 \hat{\ell}_2 - n^1n^2m^3 \hat{\ell}_3 = 0 \, . 
\label{BPSD2}
\ee
In the generic case 
\be 
I_\Pi \equiv  \Pi \cdot \iota(\Pi) = 8 n^1n^2n^3m^1m^2m^3 
\ee 
does not vanish and then \eqref{BPSD2} describes a marginal stability wall. This can for instance be seen by adapting to the present context the Fayet-Iliopoulos analysis of \cite{Cremades:2002te}, and geometrically it boils down to applying the angle theorem \cite{Lawlor1989,Douglas:1999vm} to the present setup. On the one hand, in the region of moduli space in which $I_\Pi \tilde\varepsilon < 0$ an open string tachyonic mode will develop between the D2-branes, and they will recombine into a single smooth object representing $\tilde\Sigma$. On the other hand for $I_\Pi \tilde\varepsilon > 0$ the system will be non-BPS and $\tilde\Sigma$ will not have a BPS representative. Instantons of this sort will not belong to the instanton cone $\hat\cC_{\rm I}$ defined above, as for any $\Pi$ with  $I_{\Pi}\neq 0$ there will be some asymptotic limit crossing the marginal stability wall. The way to construct factorisable cycles that belong to $\hat\cC_{\rm I}$ is to set $I_\Pi =0$, while choosing  wrapping numbers $n^i, m^i$ such that \eqref{BPSD2} has non-trivial solutions. For instance one may fix $m^1=0$, and then $n^1, n^2, n^3, m^2 > 0$, $m^3<0$. The resulting system has a wall of threshold stability at $|m^2n^3| \hat{\ell}_2 = |n^2m^3| \hat{\ell}_3$, and at both sides of the wall $\Pi$ and $\iota(\Pi)$ recombine into a single BPS three-cycle. Finally, one can construct the simplest elements of $\hat\cC_{\rm I}$ by choosing wrapping numbers such that all the coefficients in \eqref{BPSD2} vanish, while still imposing that $\int_{\Pi} e^{-\phi}\Re \Omega > 0$. It is easy to see that, if the most general D2-brane instanton charge is of the form $
[\tilde\Sigma]=\hat{m}_\alpha [\tilde{\Sigma}_+^\alpha]$,
 then the  cone of instanton charges $\hat\cC_{\rm I}$ reads
\be
\hat{\cC}_{\rm I} = \left\{ \hat{\bf m}=(\hat{m}_0,  \hat{m}_1, \hat{m}_2, \hat{m}_3) \in \mathbb{Z}_{\geq 0}^4  \right\}\, .
\label{CIT6}
\ee
Again, these are not the most general charges for a $\half$BPS instanton, but those charges that do not belong to  \eqref{CIT6} will not be BPS everywhere on the perturbative asymptotic region associated with saxionic cone $\hat{\Delta}$. 

A similar story applies to 4d strings made up from D4-branes wrapping an odd three-cycle of the form $\Sigma=\Pi-\iota(\Pi)$. Now  the simultaneous BPS  conditions for $\Pi$ and $-\iota(\Pi)$ will amount to $\int_{\Pi} e^{-\phi}\Im \Omega < 0$ and $\int_{\Pi} e^{-\phi}\Re \Omega = 0$, the latter condition being equivalent to
\be
\varepsilon \equiv   n^1n^2n^3 \hat{s}^0 - n^1m^2m^3 \hat{s}^1 - m^1n^2m^3 \hat{s}^2 - m^1m^2n^3 \hat{s}^3 = 0 \, .
\label{BPSD4}
\ee
This again indicates the locus of a marginal stability wall for the generic case $I_\Pi \neq 0$.
The total string charge of this system is given by
\be
[\Sigma]= \hat e^\alpha[\Sigma_\alpha^-]  = m^1m^2m^3 [\Sigma_0^-] - m^1n^2n^3 [\Sigma_1^-] - n^1m^2n^3 [\Sigma_2^-] - n^1n^2m^3 [\Sigma_3^-]\, ,
\label{IIAchargestring}
\ee
so that along the flow \eqref{ssssflow} generated by these charges the stability parameter $\varepsilon$ varies as
\be
\varepsilon(r) = \varepsilon_0 +  \half I_\Pi \, \sigma(r)\, .
\ee
Therefore, along the flow of a D4-brane with charges \eqref{IIAchargestring}, the quantity $I_\Pi \varepsilon$ will increase its value until the wall of marginal stability is crossed and the system becomes non-BPS. Clearly, such a pathological behaviour is not admissible for an EFT string, and one should look for a more specific set of charges. As for instantons, this is achieved by considering factorisable three-cycles such that $I_\Pi =0$, and so the D4-branes cross walls of threshold stability or no stability wall at all. If we characterise the 4d string charges of this sector as $\hat{e}^\alpha [\Sigma_\alpha^-]$, this procedure selects the following cone
 \be
 \cC^{\text{\tiny EFT}}_{\rm S} = \left\{ \hat{\bf e}=(\hat{e}^0,  \hat{e}^1, \hat{e}^2, \hat{e}^3 )\in \mathbb{Z}_{\geq 0}^4  \right\}\, ,
\label{CST6}
\ee
 which coincides with the definition \eqref{CSEFT}. In fact, due to the simplicity of this setup, we have that $\overline\Delta = \cP^\vee$ and so $\cC^{\text{\tiny EFT}}_{\rm S} = \cC_{\rm S}$.

Let us consider a few simple string flows in this setup, in order to check Conjecture \ref{conj:cutoff}. For instance one may consider one of the generators of the cone \eqref{CST6}, like for instance $\hat{\bf {e}} = (1,0,0,0)$. This corresponds to a D4-brane wrapping the three-cycle $\Sigma=\Pi-\iota(\Pi)$, with  $\Pi = 2(0,1)(0,1)(0,1)$. Together they generate the elementary flow $\hat{s}^0=\hat{s}^0_0 +  \sigma $, while $\hat s^1,\hat s^2,\hat s^3 $ and the K\"ahler moduli remain fixed. It is clear that $\cT_{\hat{\bf e}} =  M^2_{\rm P} \hat{\ell}_0 \, \sim\,  M^2_{\rm P} \sigma^{-1}$, as in \eqref{univtensionflow}. Furthermore, $e^{2\phi}, e^{2A} \sim \sigma^{-\frac12}$ and  the radii evolve as  $R_1, R_3, R_5 \sim \sigma^{\frac14}$ and  $R_2, R_4, R_6 \sim \sigma^{-\frac14}$.  
 Since we are drawn to a weak coupling limit, the lightest tower of modes correspond to KK states along the larger radii $R_1, R_3, R_5$ or winding modes along the smaller radii $R_2, R_4, R_6$. In both cases we obtain $m_*^2 \sim M_{\rm P}^2 \sigma^{-1}$,
realising Conjecture \ref{conj:cutoff} with ${\it w}=1$. The result is summarised in table \ref{tab:IIAflow}. The same result applies to other elementary flows. 

\begin{table}[ht]
\centering
\begin{tabular}{c| c| c| c| c| }
\hhline{~|-|-|-|-|}
    & \cellcolor{gray!30}$1/\sigma^{1/2}$ & \cellcolor{gray!30}$1/\sigma^{1/6}$ & \cellcolor{gray!30}$\sigma^{1/6}$ & \cellcolor{gray!30}$\sigma^{1/2}$ \\ \hline
    
\multicolumn{1}{|c|}{\cellcolor{gray!30}$\cT_{\rm str}^{1/2}$}   & D4                                       &                                          &                                        &                                        \\ \hline
\multicolumn{1}{|c|}{\cellcolor{gray!30}$m_*$}         & $m_{\rm KK}$                                 &                                          &                                        &                                        \\ \hline
\multicolumn{1}{|c|}{\cellcolor{gray!30}$E_{\rm mem}$} & D$p$                                       &                                          &                                        & NS5                                    \\ \hline
\multicolumn{1}{|c|}{\cellcolor{gray!30}$\cT_{\rm mem}^{1/3}$}   &                                          & D$p$                                       & NS5                                    &                                        \\ \hline
\end{tabular}
\caption{Elementary flow $ \hat s^i\rightarrow \infty$\label{tab:IIAflow}}
\end{table}

Non-elementary flows may be constructed by considering bound states of the above elementary charges, or also by wrapping D4-branes on $\Sigma=\Pi-\iota(\Pi)$ with $\Pi$ factorisable. For instance we can take $\Pi = 2(0,1)(n^2,m^2)(-n^3,m^3)$
with $n^2, n^3, m^2, m^3 >0$. Together with its orientifold image, this system corresponds to the string charges $\hat{\bf {e}} = (m^2m^3,n^2n^3,0,0)$, and as discussed above it only crosses wall of threshold stability along $\hat{\Delta}$. It generates the flow  
\be
\hat{s}^0   = \ \hat{s}_0^0 +  m^2m^3 \sigma\, ,  \qquad  \hat{s}^1   = \ \hat{s}^1_0 +  n_b^2n_b^3 \sigma\, ,
\label{flowa}
\ee
while $\hat s^2,\hat s^3$ and the K\"ahler moduli remain fixed. Again one can check that $\cT_{\hat{\bf e}}$ scales as in \eqref{univtensionflow}. Furthermore $e^{2\phi} \sim e^{2A} \sim \sigma^{-1}$, the first two radii scale as $R_1 \sim \sigma^{\frac12}$, $R_2 \sim  \sigma^{-\frac12}$ while  the other four radii remain constant.
 The lightest tower of the compactification again corresponds to KK states or winding modes, now along the radii $R_1$ and $R_2$ respectively. In both cases  $m_*^2 \sim M_{\rm P}^2 \sigma^{-2}$ realising Conjecture \ref{conj:cutoff} with ${\it w}=2$. Similar results are obtained for non-elementary flows of this sort, and even bound states of them. Some possibilities are collected in table \ref{tab:IIAnone}.

Notice that in both of these examples one may consider BPS D2-brane instantons on  cycles $\tilde\Sigma=\tilde\Pi+\iota(\tilde\Pi)$ with $\tilde\Pi$ of the form \eqref{f3cycle}, and such that $\langle {\bf m},{\bf e}\rangle\equiv\frac12 \Sigma \cdot \tilde\Sigma$ is negative. However, if that is the case, the marginal stability parameter $I_{\tilde\Pi} \tilde\varepsilon$ will grow along the corresponding string flow, and the instanton will become non-BPS. Following the general philosophy of section \ref{sec:cones} we neglect their effect, which is encoded in the fact that they do not belong to $\hat{\cC}_{\rm I}$. It would be interesting to confirm this expectation by taking into account their precise effect on the EFT, possibly along the lines of \cite{GarciaEtxebarria:2008pi}.

Finally, we may consider 4d membranes made up of D$p$-branes wrapping even cycles, whose volume stays constant along each string flow. Their tension $\cT_{\rm mem}$ then  evolves as $ e^{3A-\phi} l_{\rm s}^{-3}$ and one can easily check that it scales in agreement with  \eqref{asymem}. For the case of an elementary flow, this implies $E_{\rm mem}\sim m_*\sim \sigma^{-1/2}$, i.e. $n=w=1$, as shown in table \ref{tab:IIAflow}.

\begin{table}[H]
	\centering
	\begin{tabular}{| c |c |c |c| }
		\hhline{|-|-|-|-|}
		\cellcolor{gray!30} String charge & \cellcolor{gray!30}${\it w}=1$ & \cellcolor{gray!30}${\it w}=2$
		& \cellcolor{gray!30}${\it w}=3$ 
		\\ 
	    \hhline{|-|-|-|-|}
		\cellcolor{gray!30} $(e^1,0,0,0,0,0,0)$  & M-th. KK &  &  
		\\ 
		\hline
		\cellcolor{gray!30} $ (e^1,e^2,0,0,0,0,0)$  &  & M-th. KK &  
		\\ 
		\hline
		\cellcolor{gray!30} $ (e^1,e^2,e^3,0,0,0,0)$  &  &  &  M-th. KK
		\\ 
		\hline
		\cellcolor{gray!30} $ (0,0,0,\hat{e}^0,0,0,0)$  & $m_{\rm KK}$, $m_{\rm w}$ &  &  
		\\
		\hline
		\cellcolor{gray!30} $ (0,0,0,\hat{e}^0,\hat{e}^1,0,0)$  &  & \multirow{3}*{$m_{\rm KK}$, $m_{\rm w}$}  &  
		\\
		\hhline{-|~|~|~|} 
		\cellcolor{gray!30} $ (0,0,0,\hat{e}^0,\hat{e}^1,\hat{e}^2,0)$   &  &  &  
		\\
		\hhline{-|~|~|~|} 
		\cellcolor{gray!30} $ (0,0,0,\hat{e}^0,\hat{e}^1,\hat{e}^2,\hat{e}^3)$   &  &  &  
		\\
		\hline
		\cellcolor{gray!30} $ (e^1,0,0,\hat{e}^0,0,0,0)$  &  & $m_{\rm KK}$ & 
	    \\
	    \hline
		\cellcolor{gray!30} $ (e^1,e^2,0,\hat{e}^0,0,0,0)$  &  & $m_{\rm KK}$, M-th. KK & 
	    \\
		\hline
		\cellcolor{gray!30} $ (e^1,e^2,e^3,\hat{e}^0,\hat{e}^1,\hat{e}^2,\hat{e}^3)$  &  &  & $m_{\rm KK}$, M-th. KK 
		\\
		\hline
	\end{tabular}
	\caption{Behavior and identification of $m_*$ in comparison with the string tension for some choices of string charges. We have collected $({\bf e}, \hat{\bf e})$ with the components $e^a,\,\hat{e}^\alpha  \in \mathbb{Z}_{>0}$. Strings with charge ${\bf e}$ drive a flow within the K\"ahler moduli sector; strings with charges $\hat{\bf e}$ trigger a flow involving the dilaton and the complex structure moduli. The scalings are invariant for permutations of NS5 string charges $e^a$ or D4 string charges ${\hat e}^\alpha$. It can be checked that other mixed linear combinations of ${\bf e}$ and $\hat{\bf e}$ do not deliver scaling weights ${\it w}>3$.}
	\label{tab:IIAnone}
\end{table}

\subsection{M-theory on \texorpdfstring{\hbox{$G_2$}}{G2} spaces}
\label{ss:Mth}

We finally consider an M-theory compactification on a smooth $G_2$-holonomy space $X$ -- see for instance \cite{Beasley:2002db,Acharya:2004qe} for a summary on the relevant quantities defining the corresponding EFT. The 11d space-time metric reads 
\be
\d s^2=e^{2A}\d s^2_4+l^2_{\rm M}\d s^2_X\, ,
\ee
where $\d s^2_X$ is the dimensionless $G_2$-holonomy metric, $l_{\rm M}$ is the M-theory Planck length appearing in the 11d Einstein-Hilbert  term $\frac{2\pi}{l^9_{\rm M}}\int R$. The Weyl rescaling necessary to a get 4d Einstein frame metric is
\be\label{MUM}
e^{2A}=\frac{l^2_{\rm M}M^2_{\rm P}}{4\pi V_X}\, ,
\ee
where $V_X$ is the internal volume in $l_{\rm M}$-units.
The information contained in the $G_2$ metric is  completely encoded in the corresponding associative 3-form $\Phi$, normalised so that
\be
\frac17\int_X\Phi\wedge *\Phi=V_X\,.
\ee
 
Here $*\Phi$ can be identified with the co-associative four-form and can be written as a function of $\Phi$. The $G_2$ holonomy condition is equivalent to imposing $\d\Phi=0$ and $\d*\Phi=0$. In a smooth large volume regime, all chiral coordinates are of the (s)axionic type: the chiral fields $t^i\equiv a^i+\ii s^i$ are identified by expanding the complex combination $[A_3+\ii \Phi]\in H^3(X,\mathbb{C})$, where $A_3$ is the (flat) M-theory potential, into an integral cohomology basis  
\be\label{chiralM}
{\bm t}= {\bm a}+\ii {\bm s}\equiv [A_3+\ii \Phi]=t^i[\Sigma_i]\in H^3(X,\mathbb{C}) \,,
\ee 
where $[\Sigma_i]$ are  Poincar\'e dual to a basis of four-cycles $\Sigma_i\in H_4(X;\mathbb{Z})_{\text{t.f.}}$, with the subscript ``t.f." selecting the torsion free part. Up to irrelevant constants, the K\"ahler potential is given by 
\be\label{KG2}
K=-3\log \int_X \Phi\wedge*\Phi\, .
\ee
Here  both $\Phi$ and $*\Phi$ must be considered as functions of the saxions $s^i$. The dual saxions $\ell_i$ defined in \eqref{dualfields} are given by the expansion 
\be\label{Mell}
 {\bm\ell}=\ell_i[\tilde\Sigma^i]=\frac{1}{2 V_X}[*\Phi]\in H^4(X,\mathbb{R})\, ,
\ee
where $\tilde\Sigma^i$ is a basis of three-cycles dual to the four-cycles  $\Sigma_i$,  such that $\Sigma_i\cdot \tilde\Sigma^j=\delta_i^j$.

\subsubsection*{Strings, instantons and cone structure}

The effective four-dimensional BPS instantons and strings correspond to M2-branes wrapping  associative (i.e.\ $\Phi$-calibrated) three-cycles $C$ and M5-branes wrapping coassociative (i.e.\ $*\Phi$-calibrated) four-cycles $S$, see for instance \cite{Halverson:2015vta} for a recent discussion. In order to make explicit the  link with our general discussion of section \ref{s:instantons}, we must make the identifications $M_{\mathbb{Z}}=H_3(X,\mathbb{Z})_\text{t.f.}$ and $N_{\mathbb{Z}}=H_4(X,\mathbb{Z})_\text{t.f.}$. The BPS instanton and string charges take then values in
\be
{\bf m}\equiv [C]=m^i[\Sigma_i]\in H_3(X,\mathbb{Z})_\text{t.f.}\, , \qquad {\bf e}\equiv [S]=e_i[\tilde\Sigma^i]\in H_4(X,\mathbb{Z})_\text{t.f.}\, .
\ee
By direct dimensional reduction one can  check that the  action of a Euclidean M2-brane wrapping $C$ takes the form  $S_{\rm M2}=-2\pi\ii\langle {\bf m}, {\bm t}\rangle$ and then generates corrections of the form \eqref{chiralring}, as in \cite{Moore:1999gb}. Similarly,  wrapping  an M5-brane on $S$ one gets a string with tension $\cT_{\bf e}=M^2_{\rm P}\langle {\bf e},{\bm \ell}\rangle$, in agreement with the expected 4d EFT formula.  
 
However, in order to identify $\cC_{\rm I}\subset H_3(X,\mathbb{Z})_{\rm t.f.}$ and $\cC_{\rm S}\subset H_4(X,\mathbb{Z})_{\rm t.f.}$ one may encounter the M-theory counterpart of the walls of marginal stability discussed for the IIA models. In order to address this issue one first needs to better specify the saxionic space.  Unfortunately, differently from the classical moduli of Calabi--Yau K\"ahler structures, not so much is known about the global structure of the moduli space of classical $G_2$-holonomy structures.  It is known that  this moduli space can be locally identified with an open subset of $H^3(X,\mathbb{R})$  \cite{joyce1996a},  parametrised by the saxions ${\bm s}\equiv [\Phi]$. This local subset can always be extended to a conical one, since an overall constant positive rescaling of the metric preserves the $G_2$-holonomy and rescales $\Phi$ (although as we approach the `tip' of the cone the classical approximation clearly breaks down). Regarding the global structure, it is known that ${\bm s}\equiv [\Phi]$ must satisfy a list of positivity conditions, which have been enumerated in \cite{Halverson:2015vta} -- see eqs.(3.5)-(3.9) therein. In the same paper, motivated by the analogy with the K\"ahler case, it was suggested that these conditions could be actually sufficient to completely characterise the $G_2$-holonomy space. One of these conditions is 
\be\label{Mscone}
\langle {\bf m},{\bm s}\rangle =\int_{C}\Phi> 0\quad\text{with\quad ${\bf m}\equiv [C]$}\,,
\ee
for any $C$ admitting an associative representative. From our perspective, this condition is naturally related to our definition of saxionic cone \eqref{defDelta}.  However, as already point out in \cite{Halverson:2015vta}, the existence of an associative representative can in fact be $\Phi$-dependent. This issue is just the M-theory counterpart of the walls of  mariginal stability encountered in the IIA models. According to our general prescription, we rather assume that we can pick a saxionic cone $\Delta$ defined by \eqref{Mscone} for any three-cycle $C$ asymptotically admitting an associative representative along any infinite distance limit inside $\Delta$. The maximal set of BPS instanton charges  $\cC_{\rm I}$ is then defined as in \eqref{CIdef}. 

It would be clearly important to test and possibly make more precise this  definition of saxionic cone. Unfortunately,   to our knowledge, the present understanding of the moduli space of $G_2$-holonomy spaces is still too poor to address these issues.  For instance,  even the convexity of the domain of the possible values of  saxions defined by ${\bm s}\in[\Phi]$ is not obvious at all. Furthermore,  the deep interior of $\Delta$ may include regimes in which the classical geometrical description of the compactification space breaks down, still preserving the approximate axionic symmetry. $\Delta$ may in fact cover different $G_2$-manifolds, related by phase transitions which do not affect the approximate shift symmetry.
We have already encountered similar effects in the IIB/F-theory context in section \ref{sec:IIBF}.
\footnote{Take for instance the non-compact $G_2$ spaces constructed in \cite{bryant1989construction,Gibbons:1989er} and studied in \cite{Atiyah:2001qf,Acharya:2004qe}, which provide local descriptions of resolved  codimension-seven singularities. The simplest examples include  resolutions of cones which can be described as $\mathbb{R}^3$ fibrations over a coassociative base $S=S^4$ or $S=\mathbb{P}^2$. The  singular conical geometries are recovered by taking ${\rm Vol}(S)\rightarrow 0$. In this limit no non-trivial three-cycle shrinks to zero-size (the associative three-cycles are rather given by the non-compact $\mathbb{R}^3$ fibres). Hence, we are still  deep inside the saxionic cone $\Delta$ and the axionic symmetries are perturbatively preserved.}

In the present setting, EFT strings should correspond to  M5-branes wrapping  coassociative 4-cycles $S$ which have non-negative intersection number with all three-cycles corresponding to BPS instanton charges:   $C\cdot S=\langle {\bf m},{\bf e}\rangle \geq 0$, for any ${\bf m} =[C] \in \cC_{\rm I}$. Notice that \eqref{KG2} is again of the form \eqref{Klog}, with $P(s)$ homogeneous of degree $7$. Hence, as in all our previous examples, the string tension of such  EFT strings tends to zero along the corresponding  RG flows, and it is interesting to compare it with other asymptotically vanishing energy scales. Indeed, we expect the appearance of light KK modes since the volume of the associative three-cycles $C$ with strictly positive  $C\cdot S>0$ become infinite along the string flow of charge ${\bf e}=[S]$. We now illustrate these effects by focusing on Joyce's models -- see also \cite{Xu:2020nlh}  for related results in TGS models.

\subsubsection*{Joyce's model}

We now analyse  these effects in  
Joyce's  compact models \cite{joyce1996a,joyce1996b,joyce2000compact}, which are obtained as the resolutions $X$ of toroidal orbifolds $T^7/\Gamma$. More precisely, $\Gamma$ should preserve the associative three-form 
\be\label{orbPhi}
\begin{aligned}
\Phi&=\eta_{123}+\eta_{145}+\eta_{167}+\eta_{246}-\eta_{257}-\eta_{347}-\eta_{356}\,.
\end{aligned}
\ee
Here $\eta_{abc}\equiv \eta_a\wedge  \eta_b\wedge \eta_c$ where $\eta_a=R_a\d y_a$ (no sum over repeated indices) is  the vielbein of  $T^7$ parametrised by  periodic coordinates  $y_a\simeq y_a+1$, $a=1,\ldots,7$. The radii $R_a$ must be restricted in order to be compatible with the $
\Gamma$ projection and parametrise the `untwisted' moduli sector. The corresponding saxions measure the volume of associative three-cycles $C$ obtained as projection of $ \Gamma$-invariant $T^3$'s inside $T^7$, which contribute to $\cC_{\rm I}$. Analogously, the projection of $\Gamma$-invariant $T^4$'s inside $T^7$ contribute $\cC_{\rm S}$. The moduli space and the spectrum of BPS instantons and strings receives  contributions also from the `twisted sector' coming from the resolution of the orbifold singularities. In order to better illustrate what can happen, let us focus on the concrete model in which  $\Gamma=\mathbb{Z}_2\times\mathbb{Z}_2\times\mathbb{Z}_2$ whose generators $\langle \alpha,\beta,\gamma\rangle$ act on the toroidal coordinates  in the following way
\be\label{Joyceproj}
\begin{aligned}
\alpha&:  (y_1,\ldots,y_7)\mapsto (y_1, y_2, y_3, -y_4, -y_5, -y_6, -y_7)\,,\\
\beta&:  (y_1,\ldots,y_7)\mapsto(y_1, -y_2, -y_3, y_4, y_5,
\frac12-y_6, -y_7)\,,\\
\gamma&:  (y_1,\ldots,y_7)\mapsto(-y_1, y_2, -y_3, y_4,\frac12-y_5, y_6,\frac12-y_7)\ .\\
\end{aligned}
\ee
A detailed discussion of its properties can be found in \cite{joyce2000compact}, which we now summarise. First of all, it is immediate to check that the 7 three-forms  appearing in \eqref{orbPhi} are separately invariant under \eqref{Joyceproj}. In fact, they generate $H^3(T^7/\Gamma,\mathbb{Z})$. Correspondingly, we can identify seven associative 3-cycles $C^a\simeq T^3$, $a=1,\ldots,7$. For instance, $C^1$ is parametrised by $(y_1,y_2,y_3)$ (while the other coordinates are constant), $C^2$ is parametrised by $(y_1,y_4,y_5)$, and so forth.
Analogously, one can construct seven coassociative 4-cycles $S_a$, calibrated by 7  four-forms appearing in $*\Phi$. These associative 3- and 4-cycles survive after the singularities have been repaired. Correspondingly, one can introduce seven `untwisted' saxions 
\be
\label{Mth_s}
s^a=\int_{C^a}\Phi\,.
\ee

The resolution of the singularities provides 36 additional 3-cycles $\tilde C^\alpha$ (and 36 4-cycles). The singular locus of $T^7/\Gamma$ is the disjoint union of 12 $T^3$ - take for instance the $T^3$ defined by $y^4=y^5=y^6=y^7=0$. Around each of these $T^3$ orbifolds singularities $T^7/\Gamma$ looks  like $\mathbb{T}^3\times \mathbb{C}^2/\mathbb{Z}_2$. One can then locally smooth-out  each singularity by blowing-up a two-sphere $l\simeq S^2$ at the origin of $\mathbb{C}^2/\mathbb{Z}_2$, getting a four-dimensional space $\cS$ which has  the same topology of the hyperk\"ahler Eguchi-Hanson space. 
Around each resolved singularity $X$ looks like $T^3\times \cS$ and by taking all the possible combinations of the form $S^1_{(A)}\times l$, with $S^1_{(A)}\subset T^3$ for $A=1,2,3$, one generates $3\times 12=36$ additional 3-cycles $\tilde C^\alpha$, and correspondingly,  36 additional saxionic coordinates:
\be
\label{Mth_sbu}
\tilde s^\alpha=\int_{\tilde C^\alpha}\Phi\, . 
\ee
Analogously, one can construct 36 additional  4-cycles $\tilde S_\alpha$  of the form $T^2\times l$, with $T^2\subset T^3$.

We will not try to fully explore the general structure of the saxionic cone, but rather focus on the EFT string flows \eqref{ssssflow} associated with purely `untwisted' charges ${\bf e}=(e^1,\ldots, e^7)$. 
The twisted saxions $\tilde s^\alpha$ do not change along these flows and then are asymptotically subleading, as in the resolved $T^6/(\mathbb{Z}_2)^3$ F-theory model discussed in appendix \ref{sec:Fresolved}. Note that, since $\Phi$ is a calibration, $|\tilde s^\alpha|$ represents a lower bound for the volume of   $\tilde C^\alpha$, which is saturated if the $\tilde C^\alpha$ is calibrated. In this regime, one may then use the approximate EFT worked out in \cite{Lukas:2003dn,Barrett:2004tc}, but do not actually need it. Indeed, in order to estimate the relevant light energy scales   it is sufficient to focus on the leading contribution to the K\"ahler potential, which is just given by
\be
\label{Mth_K}
K=-\log(s^1\ldots s^7)+\ldots\,.
\ee
so the dual saxions read
\begin{equation}
\label{Mths_ell}
   \ell_a = \frac{1}{2 s^a}\,.
\end{equation}
This observation also holds for more general Joyce's models, not just the one discussed above. 

As usual, the tension of an EFT-string  scales according to the universal behaviour \eqref{univtensionflow} along its own flow. Consider for instance a string originating an M5-brane wrapping a four-cycle $\tilde S = e^a \tilde{S}_a $, where $\tilde{S}_a$ is a basis of four-cycles  dual to $C^a$. The string tension is given by
\begin{equation}
\label{Mths_Ts}
   \mathcal{T}_{\bf e} = \frac{2\pi}{l_M^2} e^{2A} \int_{\tilde{C}} * \Phi = M_{\rm P}^2 e_a \ell^a\,.
\end{equation}
where we have used \eqref{MUM} and \eqref{Mell}.
 
In order  to estimate  the behavior of the other relevant energy scales is straightforward and proceeds as in the previous examples.
The natural candidates for $m_*$ are the lightest among KK, whose masses are set by 
\begin{equation}
\label{Mths_KKw}
m^2_{\rm KK} \simeq \frac{M_{\rm P}^2}{R^2_* V_X}\,, 
\end{equation}
where $R_*$ is the radius that grows faster along the string flow.  However, the EFT breaking may be caused also by the light M2 branes wrapped on internal two-cycles. In order to estimate the corresponding mass scale, let us preliminary notice that, from \eqref{Joyceproj}, the orbifold should not allow for any non-trivial bulk two-cycles. Thus, M2 particles may only originate from the blown-up two-spheres, which combine with the torus $S^1$'s to form the 3-cycles $\tilde C^\alpha$. As already noticed, $|\tilde s^\alpha|$ provides a lower bound for the volume of the corresponding 3-cycle $\tilde C^\alpha$.  It follows that, along the flows generated by untwisted charges,  for any  blown-up two-cycle $l$ we can write a lower bound Vol$(l)\geq \frac{c}{R_*}$ for some constant $c$. Since the mass of the corresponding wrapped M2-brane is given by $m_{\rm M2}=2\pi e^A {\rm Vol}(l)/l_{\rm M}$,  this bound implies that    
\begin{equation}
\label{Mths_mM2}
m^2_{\text{M2}} \gtrsim  \frac{ c^2  M_{\rm P}^2}{R^2_* V_X}\,. %\qquad m^2_{\rm w} \simeq \frac{M_{\rm P}^2R^2_*}{V_X}\,.
\end{equation}
Comparing it to \eqref{Mths_KKw} we conclude that $m_{\rm M2}/m_{\rm KK}\gtrsim c$ and  we can then identify $m_*$ with the KK-mass $m_{\rm KK}$.

The explicit relation between the radii $R_a$ and the saxions $s^a$ is given in Appendix~\ref{sec:Mthtoroidal}. Notice that the saxions can be written as products of three radii, since they parametrise the volume of three-cycles. This implies that the radii are homogeneous functions of degree $1/3$. Using that the volume $V_X=\frac17e^{-K/3}$ is an homogenous function of degree $7/3$ due to \eqref{Mth_K}, we conclude that $R^2_* V_X$ is an homogeneous function of degree three. This implies that it scales asymptotically as
 \beq
 R^2_* V_X\sim \sigma^w\ ,\quad \text{with }w\leq 3\,,
 \eeq
 along the string flow.  As shown in appendix \ref{sec:Mthtoroidal}, $w$ is not only bounded by $3$ but is actually integral: $w=1,2,3$. Combined with \eqref{Mths_KKw}, this determines the scaling weight of the flow as
 \beq
 m_*^2=m_{\rm KK}^2\simeq  M^2_{\rm P}\left(\frac{\mathcal{T}_{\bf e}}{M^2_{\rm P}}\right)^w\,,
 \eeq
 realising Conjecture \ref{conj:cutoff}.  
 The specific result for $w$ depends on the choice of ${\bf e}$ and  is summarised by table   \ref{t:Mth} for the different string flows. It is easy to see that an elementary flow will have $w=1$ as only one saxion is sent to infinity and therefore $R^2_* V_X\sim \sigma$ (see table \ref{t:mth2}). Analogously, the non-elementary string flow in which all saxions are sent to infinity will have a scaling weight equal to the homogeneity degree of $R^2 V_X$, i.e. $w=3$. 
 The derivation of the scaling weight for other types of flows can be found in Appendix \ref{sec:Mthtoroidal}.

\begin{table}[H]
	\centering
	\begin{tabular}{|c |c |c| c| }
		\hline
		\cellcolor{gray!30} String charge & \cellcolor{gray!30}${\it w}=1$ & \cellcolor{gray!30}${\it w}=2$
		& \cellcolor{gray!30}${\it w}=3$ 
		\\ \hline
		\cellcolor{gray!30} ${\bf e}_{a}$  & $m_{\rm KK}$ &  &  
		\\ \hline
		\cellcolor{gray!30} ${\bf e}_{ab}$  &  &  \multirow{3}*{$m_{\rm KK}$}  &  
		\\ 
		\hhline{-|~|~|~|}
		\cellcolor{gray!30} ${\bf e}_{abc} \notin \mathcal{C}_{3}$  &  & &
		\\ 
		\hhline{-|~|~|~|}
		\cellcolor{gray!30} ${\bf e}_{abcd} \in \mathcal{C}_{4}$  &  &  &  
		\\ 
		\hline
		\cellcolor{gray!30} ${\bf e}_{abc} \in \mathcal{C}_{3}$  & \multirow{5}*{ } &  &  \multirow{5}*{$m_{\rm KK}$}
		\\
		\hhline{-|~|~|~|} 
		\cellcolor{gray!30} ${\bf e}_{abcd} \notin \mathcal{C}_{4}$  &  &  &  
		\\
		\hhline{-|~|~|~|} 
		\cellcolor{gray!30} ${\bf e}_{abcde} $  &  &  &  
		\\ 
		\hhline{-|~|~|~|} 
		\cellcolor{gray!30} ${\bf e}_{abcdef} $  &  &  &  
		\\
		\hhline{-|~|~|~|} 
		\cellcolor{gray!30} ${\bf e}_{abcdefg}$  &  &  &  
		\\ 
		\hline
	\end{tabular}
	\caption{Behavior of $m_*$ in comparison with the string tension.  Here we have introduced ${\bf e}_{a_1 a_2 \ldots a_m} = c^1 {\bf e}_{a_1} + c^2 {\bf e}_{a_2} + \ldots + c^m {\bf e}_{a_m}$, where $c^a \in \mathbb{Z}_{>0}$ and ${\bf e}_{a} $ denotes a basis of elementary string charges: $({\bf e}_{a})^b = \delta_{a}^b$. We have further introduced the sets $\mathcal{C}_{3} = \{ {\bf e}_{123}, {\bf e}_{145}, {\bf e}_{167} , {\bf e}_{246}, {\bf e}_{257},{\bf e}_{347},{\bf e}_{356}\}$ and $\mathcal{C}_{4} = \{ {\bf e}_{4567}, {\bf e}_{2367}, {\bf e}_{2345} , {\bf e}_{1367}, {\bf e}_{1346},{\bf e}_{1256},{\bf e}_{1247}\}$.}
	\label{t:Mth}
\end{table}

\begin{table}[H]
	\centering
	\begin{tabular}{c | c | c |}
		\hhline{~|-|-|}
		& \cellcolor{gray!30}$\sigma^{-\frac12}$ & \cellcolor{gray!30}$\sigma^{-\frac16}$  \\ \hline
		\multicolumn{1}{|c|}{\cellcolor{gray!30}$\cT_{\rm str}^{1/2}$}                                             & M5  &   \\ \hline
		\multicolumn{1}{|c|}{\cellcolor{gray!30}$m_*$}         &       $m_{\rm KK}$       &                \\ \hline
		\multicolumn{1}{|c|}{\cellcolor{gray!30}$E_{\rm mem}$}                                &      M2, M5 & \\ \hline
		\multicolumn{1}{|c|}{\cellcolor{gray!30}$\mathcal{T}^{1/3}_{\rm mem}$}                                &   &    M2, M5  \\ \hline
	\end{tabular}
	\caption{Mass scalings along the flow generated by an elementary string obtained from a wrapping M5-brane.\label{t:mth2}}
	
\end{table}
 
Finally, we will also have membranes coming from either M2 branes spanning three external spacetime directions or M5 branes wrapped on  three-cycles $\tilde C$ such that $\int_{\tilde C}\Phi$ remains constant along the string flow. The membrane scale is given by
\beq
E_{\rm mem}\sim e^{\frac12K}M_{\rm P}\sim \cT_{\bf e}^{n/2}
\eeq
where $n$ is an integer that can take values up to the homogeneity degree of $e^K$, i.e. $n\leq 7$. 
It is interesting to notice that in these M-theory examples, even if $n$ can be up to seven, we still find that the scaling weight $w$ is not bigger than three!

\section{Conclusions}
\label{s:conclusions}

In this work we have studied the physics emerging at large field distances of 4d $\cN=1$ EFTs by characterising the asymptotic limits in terms of fundamental axionic BPS strings, dubbed EFT strings. Whenever such an EFT string is present in the theory, it induces a classical backreaction profile on the saxionic partner to the axion under which the string is magnetically charged, that drives the saxion to infinite distance at the string core. Hence, such a backreaction profile can be mapped to a particular infinite distance boundary of the field space selected by the string charges. As in \cite{Goldberger:2001tn,Michel:2014lva} we can also interpret this backreaction profile  as an RG flow of the brane couplings, which tells us how the string tension behaves as we change the cut-off of the theory. 
We arrive then to a further correspondence between string RG flows and infinite distance limits in field space such that going to the UV corresponds to probing large field distances. 
Whether the converse is true, meaning whether every infinite distance limit can be mapped to a string RG flow, is rather non-trivial and will depend on whether there is an axionic shift symmetry emerging at every infinite distance limit. Based on string theory evidence, we propose that this is indeed the case; 
a proposal that we encode in the following conjecture:
\begin{center}
\emph{Distant Axionic String Conjecture} (DASC): Every infinite field distance limit of a 4d EFT consistent with quantum gravity can be realised as an RG flow UV endpoint  of an EFT string.
\end{center}
The RG flow triggered by an EFT string implies that the string tension goes to zero monotonically at the UV/large field distance limit. Hence, our conjecture implies the universal presence of a string becoming tensionless at every infinite distance limit. When the string tension $\cT$ becomes below the EFT cut-off-squared, the semiclassical description of the string breaks down and the EFT gets sensitive to the string excitation modes. Hence, ${\cT}^{1/2}$ acts as the maximum value that the EFT cut-off could take, which occurs at a smaller energy scale the larger the distance we travel in field space. Therefore, this implies an upper bound on the field distance that can be described by an EFT with a fixed energy cut-off. If the string satisfies the WGC, we have also shown, following \cite{Lanza:2020qmt}, that the tension behaves exponentially on the proper field distance. This is the key ingredient behind the derivation of the SDC from our proposal. In fact, this allows us to give a lower bound for the exponential rate of the SDC tower in 4d $\cN=1$ EFTs in terms of the string extremality factor. The same exponential behaviour arises if we use the typical asymptotic log-structure \eqref{Klog} of the K\"ahler potential arising in string compactifications instead of the WGC, which suggests that this log-structure might indeed be a universal quantum gravity feature at the asymptotic limits. 

Our identification between string RG flows and infinite field distance limits also allow us to provide a peculiar realisation of the Emergence proposal \cite{Grimm:2018ohb,Heidenreich:2018kpg,Palti:2019pca,vanBeest:2021lhn} as well as understanding the presence of the string (and consequently, the SDC) as a quantum gravity obstruction to restore a global symmetry at infinite distance in moduli space, as also proposed in \cite{Grimm:2018ohb,Gendler:2020dfp}. In addition to the continuous axionic shift symmetry which becomes exact at infinity, there would also be a $U(1)$ 2-form global symmetry emerging as the gauge coupling of the 2-form gauge field goes to zero. Furthermore, we also show that the Conjecture 3 proposed in \cite{Ooguri:2006in} about having negative scalar curvature is false in general, although a weaker version survives; namely, the existence of the string flow implies that  points at infinity have at least one negative holomorphic sectional curvature.

More generally, there could be other infinite tower of states becoming light at a faster rate at the asymptotic limit, which would therefore signal the EFT breakdown before it
gets sensitive to the string modes. Interestingly, based again on the string theory evidence, we propose that the leading tower of states always satisfies the following relation to the string tension:
\beq
\label{conj2}
m_*^2\simeq M^2_{\rm P}A\left(\frac{\cT}{M^{2}_{\rm P}}\right)^{{\it w}}\quad~~~~~~ \text{for some positive integer ${\it w}= 1,2,\ldots $}
\eeq
for some constant $A$. 
In other words, we propose that each EFT string is characterised by an integral scaling weight ${\it w}$, which dictates the behaviour of $m_*$ in terms of the string tension $\cT$ along the corresponding asymptotic flow.    

We provide evidence for our two conjectures by analysing large classes of examples in 4d $\cN=1$ string theory compactifications. This includes compactifications of heterotic on Calabi-Yau three-folds, Type IIA/B on orientifolds, F-theory on elliptically fibred Calabi-Yau four-folds and M-theory on certain G2 manifolds. Surprisingly, we only find three possible values for the scaling weight  ${\it w}$ in all these classes of examples, namely ${\it w}=1,2,3$. It would be interesting to understand the physical meaning of such constrained set of possibilities which, as we have pointed out, imply that $\cT^{1/2}$ is at or below the species scale. Given the link between the SDC and the existence of dualities, a tantalising possibility is that the value of ${\it w}$ is characterising the possible dualities that can get manifest at the asymptotic limits, which would give us information about the nature of the dual tower. 

It is also important to notice that not all BPS charges correspond to EFT strings. For this to happen, non-perturbative corrections breaking the shift symmetry must get suppressed along the RG flow, which imposes certain positivity constraints on the EFT string charges. In particular, a BPS string is an EFT string  whenever all BPS instantons relevant for the asymptotic regime carry non-negative magnetic charge under the 2-form gauge field that couples to the string. This leads us to characterise the asymptotic region of an EFT in terms of a saxionic cone $\Delta$ hinting a possible  semi-toric structure of the kind discussed in \cite{Coneconj}.  Our physical framework then  naturally suggests  a saxionic generalisation of the Morrison-Kawamata cone conjectures \cite{Coneconj,Morrison94beyondthe,Kawamata}. In physical terms, this saxionic cone conjecture basically states that the full set of EFT string charges $\cC_{\rm S}^{\text{\tiny EFT}}$ (and hence $\Delta$) is generated by a finite number of elementary duality-inequivalent charges.  Furthermore, our definition of EFT strings shares many similarities with the definition of 5d supergravity strings in \cite{Katz:2020ewz}. In particular, all our examples clearly indicate  that also our EFT strings only exist in supergravity theories and do not survive any rigid/decompactification limit.  

The DASC implies that any  charge in $\cC^{\text{\tiny EFT}}_{\rm S}$ should be physically realised  by an EFT string, or by a superposition of them. In specific models this physically motivated {\em EFT completeness}  translates into  some non-trivial  mathematical conjectures. For instance, in large volume heterotic compactifications  we identified  a sector of   $\cC_{\rm S}^{\text{\tiny EFT}}$ with NS5-branes wrapping  nef divisors. The EFT completeness would then imply that on Calabi-Yau three-folds all nef divisors should admit an effective representative, as already conjectured by similar arguments in \cite{Katz:2020ewz}. As another example,  in large volume F-theory models on  elliptically fibred Calabi--Yau four-folds a sector of  $\cC_{\rm S}^{\text{\tiny EFT}}$ is identified with D3-branes wrapping movable curves of the base $X$ of the elliptic fibration. Hence,  the EFT completeness similarly implies that in such bases $X$ any movable curve  should admit an effective representative. It would be interesting to better investigate these mathematical `byproduct'  conjectures, and extend them to other geometrical structures representing microscopically the possible EFT strings. Our perspective on the F-theory models has also highlighted  a potential relevance of Mori Dream spaces \cite{hu2000mori} in this context,  which deserves a more  thorough study.

It would also be interesting to understand better the physics associated to non-EFT strings. In those cases, the string flow brings us to a finite boundary of the saxionic cone, where we cannot trust anymore the EFT description as some instanton corrections become unsupressed. In certain cases, these strings even become tensionless  at these finite distance boundaries, at least at the classical level. Hence, they are similar to the tensionless strings that characterise the  conformal field theories engineered at finite field distance points in higher dimensions.  Furthermore, it would be important to better investigate the role of possible walls of BPS stability. As we have illustrated in some concrete examples, they may appear in the microscopic realisation of the 4d BPS instantons and strings and can crucially affect the identifications of the corresponding cones of charges.

Our results are valid even in the presence of a scalar potential, as long as the potential energy remains below the EFT cut-off along the infinite distance limit. Analogously, they also apply to setups with SUSY spontaneously broken below the EFT cut-off. Hence, it is justified  to apply our bounds on the field range directly to phenomenological models, including large field inflation, since we also give a precise concrete lower bound for the exponential rate of the tower. The generalisation to higher dimensions is, however, unclear. In principle, the same story holds in higher dimensions if we can replace the EFT string by some BPS codimension-2 object. In general, however, BPS and codimension-2 are conditions that are not always compatible, and one of the two must be dropped.
As a first goal, it would be interesting to formulate our conjecture in terms of (possibly non-BPS) codimension-2 objects whenever the moduli space contains codimension-1 singularities, as the presence of these objects can be argued from the absence of global symmetries.  It would also be interesting to investigate the possible connection between  the RG-flow interpretation of our EFT strings and the holographic perspective on the infinite distance limits proposed in \cite{Grimm:2020cda,Grimm:2021ikg}. 

Our work emphasises the relevance of extended objects to understand the quantum gravity principles underlying the SDC, even if these extended objects do not give rise to the leading tower of states becoming light. This motivates to revisit the physics at asymptotic limits in higher dimensions to check whether some relation along the lines of \eqref{conj2} holds for arbitrary dimension. Our work also heavily uses the map between large field distances when moving in moduli space and large field distances induced by the presence of low codimension objects in an EFT. This latter perspective might hopefully shed some light into the search of a bottom-up rationale for the SDC, regardless of string theory.

\vspace{-.1cm}
 
\section*{Acknowledgements}

\vspace{-.1cm}

We would like to thank Andreas Braun, I\~naki Garc\'ia-Etxebarria, Antonella Grassi, Thomas Grimm, Jim Halverson, Jacob McNamara, Ernesto C.~Mistretta, Miguel Montero, \'Angel Uranga,  Cumrun Vafa, Timo Weigand and Max Wiesner for useful comments and discussions. SL is supported by the Dutch Research Council (NWO) via a Start-Up grant; during the initial stage of the project, SL was also supported by a fellowship of Angelo Della Riccia Foundation, Florence and a fellowship of Aldo Gini Foundation, Padova. FM is supported through the grants SEV-2016-0597 and PGC2018-095976-B-C21 from MCIU/AEI/FEDER, UE. LM is supported in part by the MIUR-PRIN contract 2017CC72MK\_003. IV is supported by Grant 602883 from the Simons Foundation.

%%%%%%%%%%%%%%%%%%%%%%%%%%%%%%%%%%%%%%%%%%%%%%%%%%%%%%%%%%%%%%%%%%%%%%%%%%%%%

\appendix

%%%%%%%%%%%%%%%%%%%%%%%%%%%%%%%%%%%
%%%%%%%%%%%%%%%%%%%%%%%%%%%%%%%%%%%

\section{Terminology}
\label{app:glossary}

In this appendix we collect the basic definitions made and used throughout the main text. In \ref{app:flowdef} we gather the definitions that characterise the string flows introduced in section~\ref{s:fundamental}, and whose casuistics was analysed in section \ref{sec:cones}. In \ref{app:cones} we summarise the terminology regarding cones in algebraic K\"ahler geometry used in sections  \ref{s:instantons} and \ref{s:examples}.

\subsection{Glossary of strings and their flows}
\label{app:flowdef}

\noindent\textbf{BPS string solution}: holomorphic map $t^i:D \to \MM$ from a disc $D \in \mathbb{C}$ to the EFT field space $\MM$ that  saturates the bound \eqref{EBPSbound}, assuming a vanishing or negligible superpotential for the chiral fields $t^i =a^i + \ii s^i$. The string charges $e^i$ are encoded in the monodromy \eqref{tmon} around the string core.

\vspace{1em}
\noindent\textbf{Fundamental string}: string solution with a singular core that cannot be resolved with a 4d quantum field theory approach, so the string corresponds to a fundamental localised object in the theory. Its tension satisfies \eqref{EFTregime} for any admissible choice of EFT cut-off $\Lambda$. 

\vspace{1em}
\noindent\textbf{BPS axionic string}: BPS string with a flow that, in the limit \eqref{infs}, displays a holomorphic profile of the form \eqref{tsol}, and a metric warp factor \eqref{stringwarp} where the K\"ahler potential has a continuous shift symmetry $a^i\to a^i+ e^i \times$const.% . 

\vspace{1em}
\noindent\textbf{EFT string}: BPS fundamental axionic string whose continuous shift symmetry is exact at the perturbative level, and all non-perturbative corrections are suppressed for any value in $t^i(D)$. In a given asymptotic EFT regime with a saxionic cone \eqref{defDelta} the cone of EFT string charges $\cC^{\text{\tiny EFT}}_{\rm S}$ can be characterised by \eqref{CSEFT}, or equivalently by \eqref{altCC}.

\vspace{1em}
\noindent\textbf{String RG flow}: In an EFT with a cut-off $\Lambda$, the actual EFT string profile is a coarse-grained approximation of  \eqref{tsol}, involving Fourier modes up to $\Lambda$. Within a disc of radius $r_\Lambda \sim 1/\Lambda$ around the string core the profile is approximately constant, and all of its energy is described by the couplings of the localised string operator \eqref{stringS}. Changing $\Lambda$ induces a flow of such couplings, which can be measured by evaluating the solution \eqref{imt} at $r_\Lambda$. In this sense, the saxionic profile \eqref{imt} towards $r \to 0$ is understood as an EFT string RG flow towards the UV.

\vspace{1em}
\noindent\textbf{Effective string tension $\cT(\Lambda)$}: Coupling that appears in the EFT piece of the action \eqref{stringS} describing a string. This coupling is computed by evaluating \eqref{thetension} on the EFT string solution \eqref{imt} at the distance $r_\Lambda \sim 1/\Lambda$, where $\Lambda$ is the EFT cut-off scale. As such, it varies along the string RG flow. The asymptotic behaviour of $\cT_{\bf e}$ along formal limit $\Lambda \to \infty$ can be used to understand the behaviour of the probe string tension $T_{\bf e}$ along the saxionic direction \eqref{sssfl} selected by its charges. $\cT$ also depends on the flow parameters $({\bm s}_0, r_0)$ in \eqref{imt} which may be related to a point ${\bm s}$ in saxionic field space via some IR regularisation mechanism. 

\vspace{1em}
\noindent\textbf{Probe string tension $T({\bm s})$}: Tension that is associated to 4d string but it is computed in a microscopic completion of the theory, e.g. by performing dimensional reduction in string compactifications. For EFT strings in axionic regimes it also takes the form \eqref{thetension}, but it only depends on the expectation values for the saxions and it is independent of the EFT cut-off scale $\Lambda$. $T$ signals a breakdown of the semiclassical description of the 4d EFT whenever $T \leq \Lambda^2$.

\vspace{1em}
\noindent\textbf{Elementary string flow}: A BPS string flow is dubbed \emph{elementary} if it is generated by a BPS string charge that cannot be decomposed into a positive linear combination of other BPS string charges. Otherwise we call it \emph{non-elementary}.

\vspace{1em}
\noindent\textbf{Flow degeneracy order}: An EFT string flow describes a one-dimensional path in the dual saxionic cone $\cP$ defined in \eqref{cPdef0}, and it necessarily ends in one of its boundary faces $\mathcal{F}_{\bf e}$, see \eqref{face}. We say that the string flow degeneracy is of order $p$ if it ends on a codimension-$p$ face $\mathcal{F}_{\bf e}$. A string flow is called \emph{non-degenerate} if $p=1$, otherwise the flow is called \emph{degenerate}.

\vspace{1em}
\noindent\textbf{Singularity type}:
Given an K\"ahler potential of the form \eqref{Klog} and a saxionic EFT string flow \eqref{sssfl}, the corresponding singularity type $n$ is identified  by the asymptotic scaling $P(s)\sim \sigma^n$ along the flow  -- see also footnote \ref{f:singtype}.

%%%%%%%%%%%%%%%%%%%%%%%%%%%%%%%%%%%
%%%%%%%%%%%%%%%%%%%%%%%%%%%%%%%%%%%

\subsection{Algebraic Geometry}
\label{app:cones}

Here we summarise our notations for some relevant cones in algebraic K\"ahler geometry that we use throughout this paper. We mostly follow the notation of \cite{lehmann2016convexity,xiao:tel-01679333}. The following cones are obtained as positive real spans of various types of curves and divisors, and possibly closures thereof. They are subsets of $\mathbb{R}$-spans  $N^1(X)_{\mathbb{R}}$ and $N_1(X)_{\mathbb{R}}$ of the N\'eron-Severoni groups $N^1(X)_{\mathbb{Z}}$ (of divisors modulo numerical equivalence) and $N_1(X)_{\mathbb{Z}}$ (of curves modulo numerical equivalence) -- see for instance \cite{LazarsfeldI}. The cones we  consider are:
\begin{subequations}
\begin{align}
\label{Not_Eff}
\text{Eff}^1(X) &\equiv\{\text{cone generated by effective divisors}\}\,,
\\
\label{Not_Eff1b}
\text{Eff}_1(X) &\equiv\{\text{cone generated by effective curves}\}\,,
\\
\label{Not_Nef1}
\text{Nef}^1(X) &\equiv\{\text{cone generated by  nef divisors}\}\,,
\\
\label{Not_Kcone}
 \text{Amp}^1(X)&\equiv\{\text{cone generated by ample divisors}\}\,,
\\
\label{Not_Mov1}
\text{Mov}^1(X) &\equiv\{\text{cone  generated by movable divisors}\}\,,
\\
\label{Not_Mov1b}
\text{Mov}_1(X)&\equiv\{\text{cone generated by movable curves}\}\,.
\end{align}
\end{subequations}
We recall that an effective divisor is a positive linear combination  holomorphic  subvarieties of complex codimension one. Similarly, and   effective curve is  a positive linear combination of holomorphic subvarieties of complex dimension one.
Furthermore the closure of $\text{Eff}^1(X)$, $\overline{\text{Eff}}^1(X)$, is known as cone of pseudo-effective $\mathbb{R}$-divisors, while $\overline{\text{Eff}}_1(X)$ is also know as Mori cone of pseudo-effective $\mathbb{R}$-curves, often denoted $\overline{\text{NE}}(X)$. The intersection defines a natural pairing between $N^1(X)_{\mathbb{R}}$ and $N_1(X)_{\mathbb{R}}$. By using this pairing, we can define $\text{Nef}^1(X)$ as the dual of the Mori cone or, equivalently,  of $\text{Eff}_1(X)$:
\be
\text{Nef}^1(X)=\text{Eff}_1(X)^\vee\,.
\ee
Notice that, by definition, $\text{Nef}^1(X)$ is closed.
By Kleiman criterion -- see e.g. \cite{LazarsfeldI} -- one can then identify  $\text{Amp}^1(X)$ (which is open) with the interior of $\text{Nef}^1(X)$:
\be
\text{Amp}^1(X)={\rm Int}[\text{Nef}^1(X)]\,.
\ee
and, viceversa, $\text{Nef}^1(X)$ can be identified with the closure of the cone of ample divisors: $\text{Nef}^1(X)=\overline{\text{Amp}}^1(X)$. 
Notice also that  $\text{Amp}^1(X)$ is Poincar\'e dual to the K\"ahler cone: 
\be
\cK(X)\equiv \{\text{K\"ahler cone}\} \simeq  \text{Amp}^1(X) \quad~~~~~~~ \text{(via Poincar\'e duality)}\,.
\ee
A detailed discussion on the cones of movable divisors and curves can be found in \cite{xiao:tel-01679333}. For our purposes, it is sufficient to loosely define  movables divisors and curves as those that can be moved across the entire space $X$. Notice the sequence of inclusions 
\be
\text{Amp}^1(X)\ \subset\  \text{Nef}^1(X)\ \subset\ \text{Mov}^1(X)\ \subset\   \overline{\text{Eff}}^1(X)\,,
\ee
while in general nef and movable $\mathbb{R}$-divisors could be non-effective (but only pseudo-effective). Analogously, we have $\text{Mov}_1(X)\ \subset\   \overline{\text{Eff}}_1(X)$. Furthermore, by \cite{boucksom2013pseudo} we can identify $\text{Mov}_1(X)$ with the cone dual to the cone of effective divisors:
\be
\text{Mov}_1(X)=\text{Eff}^1(X)^\vee\,.
\ee

Finally, all the above cones are defined over the real numbers. The corresponding integral cones are then obtained by intersection with $N^1(X)_{\mathbb{Z}}$ and  $N_1(X)_{\mathbb{Z}}$, for example
\be
{\text{Eff}}^1(X)_{\mathbb{Z}}\equiv {\text{Eff}}^1(X)\cap N^1(X)_{\mathbb{Z}}\,,\quad \text{Mov}_1(X)_{\mathbb{Z}}\equiv \text{Mov}_1(X)\cap N_1(X)_{\mathbb{Z}}\,, \quad \ldots 
\ee
In most of our examples $N^1(X)_{\mathbb{Z}}=H^2(X,\mathbb{Z})_{\text{t.f.}}$ and $N_1(X)_{\mathbb{Z}}=H_2(X,\mathbb{Z})_{\text{t.f.}}$, where the subscript t.f.\ selects the torsion free part of the group.

%%%%%%%%%%%%%%%%%%%%%%%%%%%%%%%%%%%
%%%%%%%%%%%%%%%%%%%%%%%%%%%%%%%%%%%

%%%%%%%%%%%%%%%%%%%%%%%%%%%%%%%%%%%
%%%%%%%%%%%%%%%%%%%%%%%%%%%%%%%%%%%

\section{Gravitational contributions to the energy density}
\label{app:graveff}

Equation \eqref{linearE} has been obtained by using the energy-momentum tensor of the localised string and of the backreacted (s)axionic fields. In this appendix we discuss the inclusion of the gravitational contribution to $\cE(r)$, showing that it does not affect \eqref{linearE}. We will follow and generalise \cite{Dabholkar:1990yf}, which focused on a simple dilatonic model equivalent to the single axionic case with K\"ahler potential $K=-\log\Im t$.

As in \cite{Dabholkar:1990yf}, we start from the total stress-energy pseudo-tensor $\Theta_{mn}$, which includes also the contribution of the gravitational stress-energy pseudo-tensor, computed with respect to the Minkowski vacuum. This is defined as follows (see for instance \cite{Weinberg:1972kfs}). One writes the metric as $g_{mn}=\eta_{mn}+h_{mn}$, where $\eta_{mn}$ is the Minwkoski metric, and then splits the Einstein tensor into $G_{mn}=G^{(1)}_{mn}+\Delta G_{mn}$, where $G^{(1)}_{mn}=R^{(1)}_{mn}-\frac12 \eta_{mn}\eta^{pq}R^{(1)}_{pq}$ is  the contribution  linear in $h_{mn}$,  with
\be 
 R^{(1)}_{mn}=\frac12 \left(\del_m\del_p h^p{}_n+\del_n\del_p h^p{}_m-\del^p\del_p h_{mn}-\del_m\del_n h^p{}_p\right)\ .
\ee
Then, the Einstein equation $G_{mn}=M^{-2}_{\rm P}T_{mn}$ can be written in the form $G^{(1)}_{mn}=M^{-2}_{\rm P}\Theta_{mn}$, where $\Theta_{mn}=T_{mn}-M^2_{\rm P}\Delta G_{mn}$ is the total stress-energy pseudo-tensor. Notice that, on-shell, one can use the Einstein equation to identify  $\Theta_{mn}$ with $M^2_{\rm P}G^{(1)}_{mn}$. 
For our metric ansatz \eqref{metric}, we then get the following non-vanishing contributions to $\Theta_{mn}$  
\be
\Theta_{\mu\nu}=2\eta_{\mu\nu} M^2_{\rm P} \del_z\del_{\bar z} e^{2D}\equiv \frac12 \eta_{\mu\nu}M^2_{\rm P}\frac{\del^2 e^{2D}}{\del y^\alpha \del y^\alpha}\ ,
\ee
where we have split $x^m=(x^\mu,y^\alpha)$, with $x^\mu=(t,x)$ and $y^1+\ii y^2=z$. This form of $\Theta_{mn}$ is valid for any solution of the Einstein equations of the form \eqref{metric}, and we assume  it to be valid for an appropriate  localised regularisation of our string solution \eqref{tsol}. The overall normalisation  adopted in \eqref{stringwarp} ensures that the metric reduces to the canonical Minkowski one at $r_0$. We can then obtain the total linear  energy density by integrating $\Theta_{tt}$ over the disk of radius $r_0$.  The result can be written as a boundary term
\be\label{Etotder}
\begin{aligned}
\cE_{\rm tot}&=\int_{D(r_0)} \d^2y\, \Theta_{tt}=-\frac12 M^2_{\rm P} \int_{D(r_0)}  \d^2y \frac{\del^2 e^{2D}}{\del y^\alpha \del y^\alpha}\\
&=-\frac12 M^2_{\rm P} \oint_{r=r_0} \d\theta \frac{\del e^{2D}}{\del\log r}=-\pi M^2_{\rm P}\frac{\del e^{2D}}{\del\log r}\Big|_{r=r_0}\ ,
\end{aligned}
\ee
which is then independent of the localised regularisation of the solution. 
In the last step of \eqref{Etotder},  we have assumed that the K\"ahler potential is invariant under axionic shifts and the string flow solution is axially symmetric, as in section \ref{sec:dualform}.
From \eqref{stringwarp} and \eqref{imt} 
we then obtain 
\be
\frac{\del e^{2D}}{\del\log r}\Big|_{r=r_0}=- \frac{\del K(s(r))}{\del\log r}\big|_{r=r_0}=- \left[\frac{\del K}{\del s^i}\frac{\del s^i}{\del \log r}\right] \Big|_{r=r_0}=\frac{1}{\pi}e^i\ell_i(r_0)\ .
\ee
Hence, we arrive at 
\be\label{grEtot}
\begin{aligned}
\mathcal{E}_{\rm tot}(r_0)&=M^2_{\rm P}\, e^i\ell_i(r_0)\ ,
\end{aligned}
\ee
which is the same as \eqref{linearE} at $r=r_0$. Since $r_0$ is arbitrary, we conclude that \eqref{linearE} is valid also if we take into account gravitational effects, as anticipated.  In other words,  the gravitational field does not contribute to the effective tension of the string. This conclusion is confirmed by the perturbative field theory calculation of appendix \ref{ap:BD}.

Equation \eqref{grEtot} generalises what  was obtained in \cite{Dabholkar:1990yf} for the simple dilatonic model. As in \cite{Dabholkar:1990yf}, this result may be interpreted as a  `non-renormalisation' of the string tension, in the following sense: \eqref{grEtot}  means that the string backreaction inside the disk of radius $r_0$, for fixed boundary conditions $\ell_i(r_0)=\ell^0_i$, does not `renormalise' the `bare' string tension $M^2_{\rm P}e^i\ell^0_i$ which one would get from a probe string on the unperturbed vacuum satisfying the same boundary conditions. On the other hand, in the EFT viewpoint adopted in  the present paper and \cite{Lanza:2020qmt}, we may identify $r_0$ with a floating cut-off length scale $r_\Lambda$ and obtain a scale dependent effective string tension.

%%%%%%%%%%%%%%%%%%%%%%%%%%%%%%%%%%%
%%%%%%%%%%%%%%%%%%%%%%%%%%%%%%%%%%%

\section{Geodesic paths in field space and string flows}
\label{ap:geodesic}

In Section~\ref{s:fundamental} we computed the scalar flows  induced by fundamental strings. The paths that the chiral fields draw in field space are given by \eqref{tsol}. In this section we address the question of whether those paths are geodesic in field space.

As a preliminary remark, it is worth noticing  that the solution \eqref{tsol} is independent of the field space metric. Thus, generically, the trajectories are {\em not} geodesic. In order to see this explicitly, let us focus on the radial saxionic trajectories described by  \eqref{imt} and keep the axions $a^i$ fixed. The geodesic equations that can be obtain from the action \eqref{effaction}, assuming that the K\"ahler potential depends solely on the saxions $s^i$, read
\be
\label{geo_gen}
\frac{{\rm D}^2 s^i}{\d\lambda^2}\equiv\frac{{\rm d}^2 s^i}{\d\lambda^2}+\Gamma^i{}_{jk}\frac{{\rm d} s^j}{\d\lambda}\frac{{\rm d} s^k}{\d\lambda} = 0\,,
\ee
where $\lambda$ is an affine parameter and $\Gamma^i{}_{jk}$ are the standard Christoffel symbols associated with the saxionic metric $\cG_{ij}$ defined \eqref{GGmetric}. Identifying $\lambda$ with the proper field distance \eqref{d*} and employing the explicit saxionic flow \eqref{imt}, we obtain that the geodesic equation \eqref{geo_gen} is generically violated in following way
\be
\label{geo_fi}
\frac{{\rm D}^2 s^i}{\d\lambda^2}= f^i\quad~~~~~\text{with}\quad f^i\equiv \frac{M^2_{\rm P}}{2\cQ_{\bf e}^2}\left(\frac{\cG^{ij}}{M^2_{\rm P}}-\frac{e^ie^j}{\cQ_{\bf e}^2}\right)\del_j\cQ_{\bf e}^2\,.
\ee
Hence the deviation of the radial flow from a geodesic is encoded in the `force' $f^i$, which satisfies the orthogonality condition $\cG_{ij}e^i f^j=0$. Its norm is given by
\be
\label{flowforce}
\|{\bm f}\|^2\equiv \cG_{ij}f^if^j=\frac{M^2_{\rm P}}{4\cQ_{\bf e}^4}\left(\frac{\cG^{ij}}{M^2_{\rm P}}-\frac{e^ie^j}{\cQ_{\bf e}^2}\right)\del_i\cQ_{\bf e}^2\del_j\cQ_{\bf e}^2\,.
\ee
Thus, the radial flow is geodesic if and only if
\be
\label{geo_f0}
\|{\bm f}\|=0\,.
\ee

For moduli spaces described by a single chiral field, only one saxion can flow. The physical charge of the string inducing such a flow is $\mathcal{Q}^2_{\bf e} = M_{\rm P}^2 \mathcal{G}_{ss} e^2$. Thus, \eqref{geo_f0} is trivially satified. For higher dimensional moduli spaces, separable K\"ahler potentials $K=\sum_i K^{(i)}(s^i)$ still deliver geodesic flows if they are elementary, namely only one string charge among $e^i$ is non-vanishing, while more general flows may not be geodesic. 

On the other hand we would now like  to argue that if the K\"ahler potential on the saxionic  cone $\Delta$ takes the form \eqref{Klog} then, under some additional reasonable assumption, the EFT flows  are {\em asymptotically geodesic}. Let us first consider an EFT charge ${\bf e}\in \cC_{\rm S}^{\text{\tiny EFT}}$ which sits within $\Delta$ (and not on its boundary). By writing the flow in the form \eqref{ssssflow} and using the homogeneity of  $\cG_{ij}({\bm s})\equiv\frac12 \del_i\del_jK({\bm s})$ (with $\del_i\equiv \frac{\del}{\del s^i}$) and  $\cQ^2_{\bf e}({\bm s})\equiv M^2_{\rm P}\cG_{ij}({\bm s}) e^i e^j$, in the asymptotic limit $\sigma\rightarrow\infty$ we have 
\begin{subequations}
\begin{align}
\cQ^2_{\bf e}({\bm s})&=\cQ^2_{\bf e}({\bm s}_0+\sigma{\bf e})=\frac1{\sigma^2}\left[\cQ^2_{\bf e}({\bf e})+M^2_{\rm P}\cO(\sigma^{-1})\right]\label{Qexp}\,,\\
(\del_i\cQ^2_{\bf e})({\bm s})&=\frac12 M^2_{\rm P}e^je^k (\del_i\del_j\del_k K)({\bm s}_0+\sigma{\bf e})=\frac1{2\sigma^3} M^2_{\rm P}\left[e^je^k (\del_i\del_j\del_k K)({\bf e})+\cO(\sigma^{-1})\right]\nonumber\\
&=-\frac2{\sigma^3}M^2_{\rm P}\left[e^j \cG_{ij}({\bf e})+M^2_{\rm P}\cO(\sigma^{-1})\right]\,.
\end{align}
\end{subequations}
Notice that we can consider  $\cG_{ij}({\bf e})$  non-degenerate since  ${\bf e}\in\Delta$ and we are assuming that $K=-\log P$ is well defined over the entire $\Delta$. Hence,  also  $\cQ^2_{\bf e}({\bf e})$ is well defined. Furthermore,  $\cQ^2_{\bf e}({\bf e})=\frac{n}2$, where $n>0$ is the singularity type which in this case it is simply the homogeneity degree of $P$.  We can then similarly obtain
\be
\begin{aligned}
(e^i\del_i\cQ^2_{\bf e})({\bm s})&=-\frac{2}{\sigma^3}\left[\cQ^2_{\bf e}({\bf e})+\cO(\sigma^{-1})\right]\,,\\
\left(\cG^{ij}\del_i\cQ^2_{\bf e}\del_j\cQ^2_{\bf e}\right)({\bm s})&=\frac{4 M^2_{\rm P}}{\sigma^4}\left[\cQ^2_{\bf e}({\bf e})+\cO(\sigma^{-1})\right]\,,
\end{aligned}
\ee
Together with \eqref{Qexp}, these imply that 
\be
\label{Geodf2}
\lim_{\sigma\rightarrow\infty}\|{\bm f}\|^2({\bm s}_0+\sigma{\bf e})=0 \,.
\ee

This argument does not immediately work if ${\bf e}$ sits  on the boundary of $\Delta$, since $\cG_{ij}({\bf e})$ may not be well defined.   However, in this case it is reasonable to assume that one can go to some saxionic base in which part of the saxions do not participate at all in the flow. Correspondingly, one can restrict $\Delta$ to a lower-dimensional cone $\hat\Delta$, which contains the corresponding restriction $\hat{\bf e}$ of the original charge.  The above reasoning can then be repeated, reaching the same conclusion. In particular, if one can restrict to a one-dimensional $\hat\Delta$, then the radial flow is exactly geodesic.

Let us  check the validity of this argument in some concrete example. Consider first the separable K\"ahler potential
\be
K = -\sum_i n_i\log s^i\,.
\ee
It is not difficult to verify  that  elementary flows are exactly geodesic and  any non-elementary flow is asymptotically geodesic.

As another example, consider the K\"ahler potential \eqref{hetKexample} that was examined in Section~\ref{sec:Hetexample}. For the choice of charges ${\bf e}=(e^0,e^1,e^2) = (1,0,0)$, it can be easily checked that the flow is exactly geodesic. Instead, choosing ${\bf e}= (0,1,0)$ we obtain 
\be
\|{\bm f}\|^2 = M_{\rm P}^4 \frac{9 (s^1)^2 (s^2)^3 [3 (s^1)^2 + 3 s^1 s^2 + (s^2)^2]^2}{[9 (s^1)^3 + 9 (s^1)^2 s^2 + 3 s^1 (s^2)^2+ (s^2)^3]^3}\,,
\ee
Hence, the associated string flow is not generically geodesic. However, asymptotically we have $s^1 \sim \sigma \to \infty$ while $s^0,s^2$ are fixed, and then
\be
\|{\bm f}\|^2 \sim \frac{M_{\rm P}^4}{\sigma^3} \to 0\,.
\ee
Thus, such an elementary flow can be regarded as asymptotically geodesic. Analogously, for an elementary charge ${\bf e}=(0,0,1)$ we get
\be
\|{\bm f}\|^2 = M_{\rm P}^4 \frac{9 (s^1)^4 [3 (s^1)^2 + 3 s^1 s^2 + (s^2)^2]^2}{s^2 (3 s^1 + s^2)[9 (s^1)^3 + 9 (s^1)^2 s^2 + 3 s^1 (s^2)^2+ (s^2)^3]^3}\,.
\ee
Asymptotically along the corresponding flow we have $s^2 \sim \sigma \to \infty$, while $s^0,s^1$ are fixed, so that
\be
\|{\bm f}\|^2 \sim \frac{M_{\rm P}^4}{\sigma^4} \to 0\,.
\ee
Therefore, also this elementary flow is asymptotically geodesic. 

Finally, one can explicitly check that \eqref{Geodf2} holds for generic non-elementary charges $(e^0,e^1,e^2) \in \mathcal{C}_{\rm S}^{\text{\tiny{EFT}}}$, with $e^i \in \mathbb{Z}_{>0}$.

%%%%%%%%%%%%%%%%%%%%%%%%%%%%%%%%%%%
%%%%%%%%%%%%%%%%%%%%%%%%%%%%%%%%%%%

\section{String RG flows in Field Theory}
\label{ap:BD}

On field theoretical grounds, fundamental objects are localised operators that couple to the bulk fields via their tensions and charges. As such, their couplings to the bulk fields are generically subjected to renormalisation and, subsequently, to an RG flow. At the same time, the inclusion of extended objects within effective descriptions modifies the background geometry, reflecting on nontrivial solutions for the bulk fields. As an example, the BPS solution \eqref{tsol} may be considered as the effect of the backreaction upon the dual saxions $\ell_i$ induced by a single fundamental string.

The link between these two seemingly unrelated descriptions were threaded in \cite{Michel:2014lva}, where it was shown that the RG flow induced by extended objects coincides with the effect of their backreaction. In the following we will delve into this identification restricting to the case of supersymmetric strings introduced in section~\ref{s:fundamental}. As we will shortly see, the tension of BPS strings obeys a \emph{non-renormalisation condition}, generalising \cite{Dabholkar:1990yf,Buonanno:1998kx}. From this viewpoint, we will show how the BPS solution \eqref{tsol} may be regarded as the RG flow of the dual saxions.

In $\mathcal{N}=1$ locally supersymmetric theories, the coupling of a BPS-fundamental string \cite{Lanza:2019xxg,Lanza:2019nfa} to the bulk supergravity is easily written in terms of linear multiplets. The components of each linear multiplet $L_i$ comprise a real scalar $\ell_i$, dubbed \emph{dual saxion} in section~\ref{s:fundamental}, a real gauge two-form $\mathcal{B}_{2\,i}$ and matter Weyl fermions $\psi^\alpha_i$. The full action describing the coupling of a fundamental string to a bulk supergravity whose matter content is solely encoded within linear multiplets is
\be
\label{FT_action_tot}
S = S_{\rm bulk} + S_{\rm string}\,,
\ee
with the bulk part
\be
\label{FT_dualaction}
S_{\rm bulk} = M^2_{\rm P}\int \frac{1}{2}R*1 -\frac12\int \cG^{ij}(\ell)\left( M^2_{\rm P}\,\d\ell_i\wedge *\d\ell_j+\frac{1}{M^2_{\rm P}}\cH_{3\, i}\wedge * \cH_{3\, j} \right)\,,
\ee
and the localised string contribution
\be
\label{FT_stringS}
S_{\rm string}=-\int_\cS \sqrt{-h}\, \mathcal{T}_{\rm str} +e^i \int_\cS \cB_{2\, i}\, , \qquad {\rm with} \quad \mathcal{T}_{\rm str} = M^2_{\rm P}\,\int_\cS  |e^i\ell_i|\,.
\ee

From a field theoretical perspective, the string tension appearing in \eqref{FT_stringS} plays the role of a coupling parameter and, as such, is subjected to renormalisation. In order to study its renormalisation, we employ the background field method and expand the bulk fields $g_{\mu\nu}$, $\ell_i$ and $\cB_{2\,i}$ around some \emph{fixed} background values $g^0_{\mu\nu}$, $\ell^0_i$ and $\cB_{2\,i}^0$:
\be
\label{FT_bgexp}
g_{\mu\nu}(u,v)= g^0_{\mu\nu} + h_{\mu\nu}(u,v), \qquad \ell_{i}(u,v) = \ell_{ i}^0+ \hat\ell_{i}(u,v)\,, \qquad \cB_{2\, i}(u,v) = \cB_{2\,i}^0+ \hat\cB_{2\, i}(u,v)\,.
\ee
Here we have denoted with $h_{\mu\nu}$, $\hat\ell_i$ and $\hat\cB_{2\,i}$ the variations of the fields around their background values and, employing the $SO(1,1)\times SO(2)$ symmetries of the system, these are assumed to solely depends on the coordinates ${\bf u} = (u,v)$, which parametrise the directions transverse to the string.\footnote{The coordinates $(u,v)$ are the real counterpart of the complex coordinates $(z,\bar z)$ introduced in \eqref{metric}.} Close enough to the string, while still at distances $r> \Lambda^{-1}$, its backreaction on the field metric is negligible and the background metric may be just considered Minkowski: $g_{\mu\nu}^0 = \eta_{\mu\nu}$. We will further assume the string to be static, stretching along the $(t,x)$-spacetime directions at fixed $(u,v)$-position and that the only nontrivial component of $\cB_{2\,i}$ are $\cB_{01\, i} \equiv \cB_i$, with the background expansion $\cB_{i}(u,v) = \cB_{i}^0+ \hat\cB_{2 i}(u,v)$.

\begin{center}
	\begin{figure}[htb]
		\centering
		\includegraphics{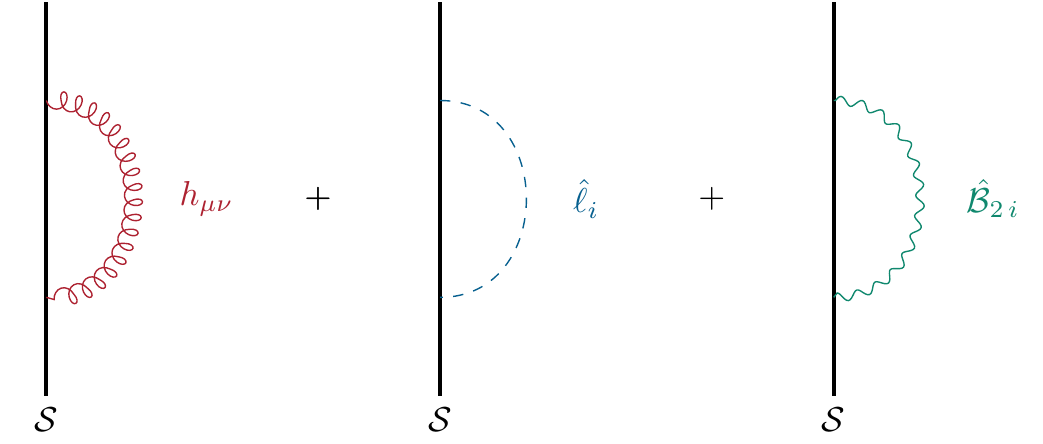}
		\caption{\small{The self interactions of a string due to the graviton $g_{\mu\nu}$, dual saxion $\ell$ and gauge two-form $\cB_{2}$.}
			\label{Fig:FT_selfstrings}}
	\end{figure}
\end{center}

Let us introduce the {\em bare} string action
\begin{equation}
\label{FT_stringSb}
S_{\rm string}^{(0)}  = - \int_{\mathcal{S}} {\rm d}t\,{\rm d}x\, \cT_{\rm bare}  + e^i \int_\mathcal{S} \cB_{2\,i}^0\,,
\end{equation}
which is \eqref{FT_stringS} evaluated on the background and with the bare tension
\be
\label{FT_Tbare}
\cT_{\rm bare} = M_{\rm P}\,  e^i\, \ell^0_i\,.
\ee
Owing to the coupling of the string to the background perturbations $h_{\mu\nu}$, $\hat \ell_i$ and $\hat \cB_{2\,i}$, interactions of the string as in Fig~\ref{Fig:FT_selfstrings} are possible. These self-interactions are expected to renormalise the string tension. In order to explicitly compute the renormalisation induced by the Feynman diagrams in Fig.~\ref{Fig:FT_selfstrings}, we proceed as in \cite{Buonanno:1998kx}. 

We regard the strings as \emph{external} sources for the bulk fields. Namely, we rewrite the contribution \eqref{FT_stringS} of the string to the bulk action by expanding it linearly in field perturbations
\begin{equation}
\label{FT_SstrJ}
S_{\rm string}  = S_{\rm string}^{(0)}    - \int \d^4 x\, \left(h^{\mu\nu} J_{\mu\nu}  + \hat\ell_i J_{\ell}^{i}  + \hat{\mathcal{B}}_{i} J_{\bf e}^i \right)+\ldots
\end{equation}
where we introduced the `currents'
\begin{subequations}
	\label{RFCs_J}
	\begin{align}
	J_{\mu\nu}  &= - \frac{\delta}{\delta h^{\mu\nu}} S_{{\rm string}} = \frac{|e^j \ell_j^0|}2 \begin{pmatrix}
	1 & 0 & 0 & 0 \\0 & -1 & 0 & 0 \\ 0& 0& 0 & 0 \\ 0 & 0 & 0 & 0
	\end{pmatrix} \delta \left(\mathcal{S}\right) \,,
	\\
	J_{\ell}^i  &= -\frac{\delta}{\delta \hat\ell_i }  S_{{\rm string}}= M_{\rm P}^2 e^{i} \delta \left(\mathcal{S}\right) \,,
	\\
	J_{\bf e}^i  &= \frac{\delta}{\delta \hat\cB_i }  S_{{\rm string}}= e^{i}\delta \left(\mathcal{S}\right) \, ,
	\end{align}
\end{subequations}
which play the role of classical sources for the graviton, dual saxions and two-forms. 

It is convenient to momentarily collect the field variations as $\psi^{\cA} \equiv (h^{\mu\nu}, \hat\ell_i, \hat\cB_i^{\mu\nu})$ and the sources as $J_{\cA} \equiv (J^{\mu\nu},J_{\ell}^i,J_{\bf e}^i)$. Then, the bulk action \eqref{FT_dualaction} is recast as
\begin{equation}
S =  S_{\rm string}^{(0)}   + \int \d t\, \d x\, \d^2 {\bf u}   \int \d^2 {\bf u}' \left(  \frac12 \psi^{\cA} ({\bf u}) Q_{\cA,\cB}({\bf u};{\bf u}') \psi^{\cB} ({\bf u}')   - J_{\cA}({\bf u})\psi^{\cA} ({\bf u})  \right)\,.
\end{equation}
Integrating out the $\psi^{\cA}$, we arrive at
\begin{equation}
\label{FT_SJJ}
S = S_{\rm string}^{(0)}  + \frac{\ii}2 \int \d t\, \d x\, \d^2 {\bf u}  \int \d^2 {\bf u}' J_{\cA} ({\bf u}) \Delta^{\cA,\cB}({\bf u};{\bf u}') J_{\cB} ({\bf u}')  \,,
\end{equation}
with $\Delta^{\cA,\cB} = \ii (Q_{\cA,\cB})^{-1}$ the field propagators in two-dimensional Euclidean space. Explicitly, the propagators for the graviton $h_{\mu\nu}$, dual saxions $\ell_i$ and two-forms $\cB_i $ propagators are respectively given by
\begin{subequations}
	\label{FT_Prop}
	\begin{align}
	\Delta^{\mu\nu, \rho\sigma}  ({\bf u};{\bf u}')  &= ( \eta^{\mu\rho}\eta^{\nu\sigma}+\eta^{\mu\sigma}\eta^{\nu\rho} - \eta^{\mu\nu}\eta^{\rho\sigma}) \Delta  ({\bf u};{\bf u}')\,,
	\\
	\Delta_{i j}^{\ell}  ({\bf u};{\bf u}') &= \mathcal{G}_{ij}(\ell^0) \Delta  ({\bf u};{\bf u}')\,,
	\\
	\Delta_{i j}^{\cB}  ({\bf u};{\bf u}') &= -M_{\rm P}^4 \mathcal{G}_{ij}(\ell^0) \Delta  ({\bf u};{\bf u}')\,,
	\end{align}
\end{subequations}
where $\Delta  ({\bf u};{\bf u}')$ is the  two-dimensional scalar propagator
\begin{equation}
	\label{FT_Prop0}
	\Delta  ({\bf u};{\bf u}') = \frac{\ii}{4\pi} \log |{\bf u}-{\bf u}'| \equiv \frac{\ii}{4\pi} \log r\,.
\end{equation}
The second term that appears in \eqref{FT_SJJ} can be rewritten as
\begin{equation}
\label{FT_SJJb}
\frac{\ii}2 \int_{\mathcal{S}} \d t\, \d x\,   J_{\cA} ({\bf u}) \Delta^{\cA,\cB}({\bf u};{\bf u}) J_{\cB} ({\bf u})  \,,
\end{equation}
and is responsible for the renormalisation of the string tension. However, as is clear from the behaviour of the two-dimensional Euclidean propagator \eqref{FT_Prop0}, the contribution \eqref{FT_SJJb} entails a propagator evaluated at zero distance. Thus, \eqref{FT_SJJb} is divergent both in the infrared and in the ultraviolet and has to be appropriately regularised. The simplest regularisation is achieved by converting the propagator \eqref{FT_Prop0} to momentum space and using a sharp momentum cutoff as follows:
\begin{equation}
\Delta^{\rm reg}  ({\bf u};{\bf u}) = \int_{\mu \leq |p| \leq \Lambda} \frac{{\rm d}^2 p}{(2 \pi)^2} \tilde \Delta (p) = - \frac{{\rm i}}{M_{\rm P}^2}\int_{\mu \leq |p| \leq \Lambda} \frac{{\rm d}^2 p}{(2 \pi)^2 p^2} = - \frac{{\rm i}}{2 \pi M_{\rm P}^2} \log \frac{\Lambda}{\mu}\,.
\end{equation}
where $\mu$ and $\Lambda$ denote, respectively, an IR energy scale and the EFT cutoff. Therefore, \eqref{FT_SJJb} would contribute to the string tension with
\begin{equation}
\label{FT_SJJbb}
\frac{\ii}2 \int_{\mathcal{S}} \d t\, \d x\,   J_{\cA} ({\bf u}) \Delta^{\cA,\cB}({\bf u};{\bf u}) J_{\cB} ({\bf u})  = (C_{\rm g}+C_{\ell}+C_{\cB}) \log \frac{\Lambda}{\mu}\, ,
\end{equation}
where we have singled out the contribution from gravity $C_{\rm g}$, the dual saxions $C_\ell$, and the gauge two-forms $C_{\cB}$. Therefore, the renormalised string tension is the bare tension to which one needs to add appropriate counterterms to cancel the divergences originating from \eqref{FT_SJJbb};
\be
\label{FT_Tren}
\cT_{\rm ren}=\cT_{\rm bare} +  (C_{\rm g}+C_{\ell}+C_{\cB}) \log \frac{\Lambda}{\mu}\, .
\ee

However, the explicit computations of the contributions in \eqref{FT_SJJbb} reveals that
\be
\label{FT_TrenC}
C_{\ell}=  - C_{\cB} = \frac{1}{4\pi} M_{\rm P}^2 \cG_{ij}(\ell_0) e^i e^j = \cQ^2_{\bf e} (\ell_0) \,, \qquad C_{\rm g}=0\,.
\ee
As a result, the string tension is \emph{de facto not renormalised}; namely, given a background specified by the value $\ell_i^0$, the renormalised string tension coincides with its bare value, computed upon the same background:
\begin{equation}
\label{FT_Tnonren}
	\cT_{\rm ren} (\ell^0)=\cT_{\rm bare} (\ell^0)\,.
\end{equation}

The (non-)renormalisation of the tension can be alternatively understood as the cumulative effect of the backreaction due to bulk fields, which \emph{dresses} the bare tension with additional contributions. Loosely speaking, the tension of a string evaluated at a distance $r_* > \delta$ receives contributions from the energy stored within the annulus $\delta < r < r_*$ compared to the string tension evaluated at the cutoff length $\delta$.\footnote{Compare this case to the renormalisation of the electron mass. In presence of an electric field, the electron mass evaluated at the length scale $\delta_2$ gets renormalised, with respect to the mass at $\delta_1<\delta_2$, as $m(\delta_2) - m (\delta_1) = \cE_{\rm el}$, where $\cE_{\rm el}$ is the energy of the electric field stored within $\delta_1$ and $\delta_2$ \cite{Dirac:1938nz}. However, this argument is not general and holds when the energy comes solely from gauge fields -- see \cite{Buonanno:1998kx,Cannella:2008nr} for a more detailed discussion.} However, extended sources such as strings do contribute to the stress-energy tensor also with localised contributions. As clarified in \cite{Buonanno:1998kx,Cannella:2008nr}, these also ought to be taken into account in the renormalisation of the bare tension.  Therefore, the renormalised tension is generically written as
\be
\label{FT_Trenb}
\cT_{\rm ren}=\cT_{\rm bare}+ \cE_{\rm bulk}(\delta) + \cE_{\rm loc}(\delta)\,.
\ee
The bulk contribution $\cE_{\rm bulk}(\delta)$ takes into account the energy of the bulk fields stored between $\delta$ and $r_* > \delta$. Singling out the different contributions, we write
\be
\label{FT_bulkb}
\cE_{\rm bulk}(\delta)=\cE^{\rm g}_{\rm bulk}(\delta)+\cE_{\rm bulk}^{\ell}(\delta)+\cE^{\cB}_{\rm bulk}(\delta)\, .
\ee
From \eqref{Eback} we recall that
\be
\label{FT_Ebulk}
\cE_{\rm bulk}^{\ell}(\delta)+\cE^{\cB}_{\rm bulk}(\delta)=M^2_{\rm P}\int_\delta^{r_*}\cJ_\MM=e^iM^2_{\rm P}[\ell_i (r_*)-\ell_i(\delta)]
\ee
and, employing the BPS-condition \eqref{H3ell}, we further recognise that
\be
\label{FT_bulkc}
\cE_{\rm bulk}^{\ell}(\delta)=\cE^{\cB}_{\rm bulk}(\delta)=\frac12e^i M^2_{\rm P}[\ell_i(r_*)-\ell_i(\delta)]\,.
\ee
The localised contributions to the renormalisation come from gravity and the dual saxion:
\be
\label{FT_loc}
\cE_{\rm loc} (\delta)= \cE_{\rm loc}^{\rm g}(\delta) + \cE_{\rm loc}^{\ell}(\delta)\,.
\ee
As noticed in \cite{Buonanno:1998kx}, these are related to the bulk contributions appearing in \eqref{FT_Ebulk} as follows:
\be
\label{FT_locb}
\cE_{\rm loc}^{\rm g}(\delta)= - \cE_{\rm bulk}^{\rm g}(\delta){\tiny }\,, \qquad \cE_{\rm loc}^{\ell}(\delta) = -2 \cE_{\rm bulk}^{\ell}(\delta) = - eM^2_{\rm P}[\ell(r_*)-\ell(\delta)]\,.
\ee
Clearly, inserting \eqref{FT_bulkb} and \eqref{FT_loc} in \eqref{FT_Trenb} the non-renormalisation for the tension is recovered.

The non-renormalisation of the string tension has the following implication. If we wish to evaluate the string tension at a reference radius $r$, then its value is simply given by the bare tension \eqref{FT_Tbare}, as it appears in the action \eqref{stringS}, evaluated at $r$:
\be
\cT(\ell(r)) = M_{\rm P}^{2} e^i \ell_i(r)\,,
\ee
without further corrections. In the language introduced above, $\ell_i(r)$ have to take the background values $\ell^0_i$ at the distance $r$ from the string. From the solitonic solution \eqref{imt} we immediately get the `flow' of the string tension
\be
\label{FT_ellflowb}
\cT(\ell(r)) = \frac{M_{\rm P}^{2} e^i \ell_i(r_0)}{1-\frac{e^i \ell_i(r_0)}{\pi}\log\frac{r}{r_0}}\,,
\ee
where $r_0$ is a fixed distance from the string.

\section{EFT instantons}
\label{app:inst}

For each instanton charge ${\bf m}\in\cC_{\rm I}$ appearing in the possible non-perturbatively generated chiral operators \eqref{chiralring}, one can  write down BPS (singular) solutions, generalising the D(-1) solution discussed in \cite{Gibbons:1995vg}. These can be described using the Wick-rotated Euclidean version of the bulk terms in \eqref{dualaction}
\be\label{dualaction2}
\frac12\int \cG^{ij}\left( M^2_{\rm P}\,\d\ell_i\wedge *\d\ell_j+\frac{1}{M^2_{\rm P}}\cH_{3\, i}\wedge * \cH_{3\, j} \right)\, .
\ee
In Lorentzian signature,  the dualisation of the 2-forms $\cB_{2\,i}$ to the dual axions $a^i$ is obtained by relaxing the closure condition $\d\cH_{3\, i}=0$ and   introducing the Lagrange multiplier $\int \d a^i\wedge \cH_{3\, i}$. In the  Wick-rotated action we then have to add instead $-\ii\int \d a^i\wedge \cH_{3\, i}$.%, with $a^i=\Re t^i$. 

Suppose we want to add to this action a source term  $2\pi\ii m_i t^i(x_0)$, corresponding to a point-like instanton located at the point $x^\mu_0$. The relevant terms in the action are then given by 
\be\label{bulk+inst}
\frac12\int \cG^{ij}\left( M^2_{\rm P}\,\d\ell_i\wedge *\d\ell_j+\frac{1}{M^2_{\rm P}}\cH_{3\, i}\wedge * \cH_{3\, j} \right)-\ii\int \d a^i\wedge \cH_{3\, i}-2\pi\ii\, m_i t^i(x_0)\, ,
\ee
so by integrating out $a^i$ one gets the modified Bianchi identity  
\be\label{instsource}
\d\cH_{3\, i}=2\pi m_i\delta_4(x_0)\, ,
\ee
which implies that $\int_{S^3}\cH_{3\, i}=2\pi m_i$ on any three-sphere $S^3$ surrounding $x_0$.
By keeping the boundary term, the action \eqref{bulk+inst} reduces to
\be
\frac12\int \cG^{ij}\left( M^2_{\rm P}\,\d\ell_i\wedge *\d\ell_j+\frac{1}{M^2_{\rm P}}\cH_{3\, i}\wedge * \cH_{3\, j} \right)+2\pi\, m_i s^i(x_0)-2\pi\ii m_i a^i_{\infty}\, ,
\ee
where $a^i_{\infty}$ are the expectation values of the axions at infinity. By some simple manipulations, using \eqref{instsource}  and recalling that $\d s^i=-\cG^{ij}\d\ell_i$ and $s^i=\frac12\frac{\del F}{\del \ell_i}$, with $F=K+2\ell_is^i$ (following the conventions of ~\cite{Lanza:2019xxg}), we can write the relevant terms  in the effective action in BPS form
\be\label{Euclaction}
S_{\rm E}=-\frac1{2M^2_{\rm P}}\int \cG^{ij}\left( \cH_{3\, i} +M^2_{\rm P}\,*\d\ell_i\right)\wedge *\left( \cH_{3\, j} +M^2_{\rm P}\,*\d\ell_j\right)-2\pi\ii m_i t^i_{\infty}\, .
\ee
It is then clear that configurations satisfying  the BPS condition
\be\label{instBPScond}
*\d\ell_i=-\frac1{M^2_{\rm P}}\cH_{3\, i}\, ,
\ee
extremise the action and do not provide any contribution to the energy-momentum tensor. Hence, \eqref{instBPScond} on a Euclidean flat metric $g_{\mu\nu}=\delta_{\mu\nu}$ gives a complete solution of the gravitational equations of motion. 
Combining \eqref{instBPScond} and \eqref{instsource} one gets
\be
\d*\d\ell_i= -\frac{2\pi }{M^2_{\rm P}}\,m_i \delta_4(x_0)\, ,
\ee
which given the boundary conditions has unique solution:
\be\label{instflow}
\ell_i(x)=\ell_i^{\infty}+\frac{m_i}{2\pi M^2_{\rm P}\, |x-x_0|^2}\, ,
\ee 
with $|x|^2=\delta_{\mu\nu}x^\mu x^\nu$. The corresponding profile of $\cH_{3\, i}$ is  fixed by \eqref{instBPScond}. Furthermore, on these BPS  solutions \eqref{Euclaction}  reduces to $S_{\rm E}|_{\rm BPS}=-2\pi\ii \langle {\bf m},{\bf t}_\infty\rangle$.  Hence  these instantons will generate non-perturbative effects proportional to $e^{-S_{\rm E}|_{\rm BPS}}=e^{2\pi\ii \langle {\bf m},{\bf t}\rangle}|_{\infty}$, as expected.   

Remember that in our picture the instanton charges  ${\bf m}\in \cC_I$ generate  the dual cone $\Delta^\vee$ and that the positivity of the EFT string tensions over the entire saxionic cone $\Delta$ is equivalent to $\overline\cP\subset \Delta^\vee$, where $\cP$ is defined in \eqref{cPdef0}. We will assume that $\cP$ is a cone, as it is automatically the case if we consider K\"ahler potentials of the form \eqref{Klog}. Furthermore, we will assume that this cone is convex.\footnote{In string models we expect $\cP$  to be  conical at least in some approximate sense, `around the tip' of $\cP$. Instead the convexity of $\cP$, while easily realised in explicit models, is not automatic and one can in fact find explicit counterexamples thereof.}  Now, if $\overline\cP$ is a {\em strict} subcone of $\Delta^\vee$, then we can distinguish two possible classes of flows: either ${\bf m}\in\overline\cP$ or ${\bf m}\in \Delta^\vee - \overline\cP$. In the first case   the instanton backreaction \eqref{instflow} never exits  $\cP$ (if ${\bm \ell}_\infty\in\cP$).  Note also that, even though at a distance $|x-x_0|\sim M^{-1}_{\rm P}$  the saxions $\ell^i(x)$ become of order one and enter a strongly coupled region, this happens at distances well below the EFT cutoff length scale  $\Lambda^{-1}\gg M^{-1}_{\rm P}$. Hence, the solution remains weakly coupled and trustable in the region in which the EFT makes sense. This motivates us to indentify the instanton lattice subcone
\be\label{EFTinstcone}
\cC^{\text{\tiny EFT}}_{\rm I}\equiv\{{\bf m}\in \cC_{\rm I}\cap\overline\cP\} \quad\subset\ \cC_{\rm I}\, .
\ee
In the explicit heterotic  example discussed in section \ref{sec:Hetexample}, $\cC^{\text{\tiny EFT}}_{\rm I}$  is generated by an NS5-brane wrapping $X$ and world-sheet instantons wrapping the nef divisor intersections $h$ and $e$.

Let us now look at the case where  ${\bf m}\in \cC_{\rm I} - \cC^{\text{\tiny EFT}}_{\rm I}$. Then the solution \eqref{instflow} exits the domain $\cP$ as we approach $x_0$ from infinity. In particular, the smaller we choose the asymptotic values  $\ell_i^{\infty}$, the larger is the radial distance $|\Delta x|_*=|x-x_0|_*$ at which the solution exits the perturbative domain $\cP$. The value of $|\Delta x|_*$ is directly related to the tension of a possible string whose charge ${\bf e}$ belongs to $\cP^\vee - \overline\Delta$, or equivalently ${\bf e}\in \cC_{\rm S} - \cC^{\text{\tiny EFT}}_{\rm S}$. Indeed, for such a charge we would have $\langle {\bf e}, {\bm \ell}^{\infty}\rangle >0$ while  $\langle {\bf m},  {\bf e}\rangle <0$ since ${\bf m}\notin \cC^{\text{\tiny EFT}}_{\rm I}$.  
This means that at the distance
\be
|\Delta x|^2_*=-\frac{\langle {\bf e}, {\bf m}\rangle}{2\pi \langle {\bf e}, {\bm \ell}^{\infty}\rangle }\equiv \frac{|\langle {\bf e}, {\bf m}\rangle|}{2\pi\, \cT^\infty_{{\bf e}}}\, ,\quad~~~\text{for ${\bf e}\notin\cC^{\text{\tiny EFT}}_{\rm S}$}\, ,
\ee 
the probe tension of the string of charge ${\bf e}$ vanishes, and the dual saxions $\ell_i$ attain the boundary of their domain $\cP$.  
Notice that such a string ${\bf e}\notin\cC^{\text{\tiny EFT}}_{\rm S}$ would precisely correspond to a non-EFT, whose RG flow is expected to receive strong non-perturbative corrections generated by BPS instantons of charges ${\bf m}\notin  \cC^{\text{\tiny EFT}}_{\rm I}$. 

To sum up, instanton solutions corresponding to  ${\bf m}\in \cC_{\rm I} - \cC^{\text{\tiny EFT}}_{\rm I}$ break down along their flow. 
Given the analogy with   string solutions corresponding to ${\bf e}\in \cC_{\rm S} - \cC^{\text{\tiny EFT}}_{\rm S}$, we see a direct correlation between the existence of non-EFT strings and non-EFT instantons. The latter are the ones relevant along the flows of the formers, and viceversa.  This may provide an EFT criterion to distinguish between the available non-perturbative corrections, possibly refining the conjecture of \cite{Palti:2020qlc}.

\section{EFT string flows in toroidal orbifolds}
\label{app:toroidal}

In this appendix we check conjectures \ref{conj:DASC} and \ref{conj:cutoff} for compactifications related to toroidal orbifolds, in which we have better control over the spectra of KK towers. In \ref{sec:CSflow_tor} we consider EFT string flows on the complex structure sector of heterotic and Type I compactifications, in \ref{sec:Fresolved} we analyse the F-theory model discussed in \cite{Denef:2005mm} and in \ref{sec:Mthtoroidal} we investigate in detail the M-theory Joyce's model introduced in section~\ref{ss:Mth}. For toroidal orbifold EFT string flows in the context of type IIA see subsection \ref{ss:IIA}.

\subsection{Heterotic/type I complex structure flows in toroidal orbifolds}
\label{sec:CSflow_tor}

Let us examine the string flows on the complex structure sector in heterotic and Type I four-dimensional compactifications. For definiteness, we first focus on the heterotic models, considering the simple case of compactification over the toroidal orbifold $X=T^6/(\mathbb{Z}_2\times \mathbb{Z}_2')$. We denote the internal complex coordinates $z^a=y^{2a-1}+\tau^{a} y^{2a}$ ($a=1,2,3$), with $y^m\simeq y^m+1$ and the complex structures $\tau^a$, which are identified under the orbifold action as follows:
\begin{equation}
(z^1,z^2,z^3)\simeq (-z^1,-z^2,z^3)\simeq (z^1,-z^2,-z^3)\,.
\end{equation}
Below we shall focus on the `untwisted' sector of moduli space only. However, it is worth noticing that the orbifold allows for the presence of fixed points, which can be resolved resulting in the introduction of a set of twisted blow-up moduli. However, the blow-up moduli may be considered to lie in the bulk of the moduli space and thus are not expected to significantly affect the asymptotic limits of the untwisted sector.

The internal {\em string frame} metric which appears in \eqref{hetmetric} is
\begin{equation}
\label{het_tor_met}
{\rm d} s^2_X=\sum^3_{a=1} \frac{2 \,s^a}{\Im \tau^{a}}\d z^a\d\bar z^{\bar a}=\sum^3_{a=1} \frac{2 \,s^a}{\Im \tau^{a}}\left[(\d y^{2a-1}+\Re\tau^a\, \d y^{2a})^2+(\Im\tau^a)^2 (\d y^{2a})^2\right]
\end{equation}
where we have introduced the K\"ahler moduli $s^a$. The six radii $(R_1,\ldots,R_6)$ (measured in string units) can be easily expressed in terms of the K\"ahler and complex structure moduli $(s^a, {\rm Im} \tau^a)$ as
\begin{equation}
\label{hetT6R}
R_{2 a-1}=\frac{ 1}{2\pi}\sqrt{\frac{ 2 s^a}{\Im\tau^a}}\quad,\quad R_{2a}=\frac{ |\tau^a|}{2\pi} \sqrt{\frac{ 2 s^a}{\Im\tau^a}}
\end{equation}
The holomorphic three-form and the K\"ahler form are 
\begin{equation}
\label{het_Omega_J}
\Omega=\d z^1\wedge \d z^2\wedge \d z^3\,,\qquad J = s^a \omega_a
\end{equation}
where $\omega_a$ integrally quantised untwisted 2-forms are given by
\begin{equation}
\omega_a= \d y^{2a-1}\wedge\d y^{2a}=\frac{\ii}{2 \Im\tau^a}\, \d z^a\wedge \d \bar z^a
\end{equation}
In comparison with the more general models examined in Section~\ref{sec:hetCY}, here the only non-vanishing intersection number is $\kappa_{123} = 1$.

Let us now pass to examine the four-dimensional $\mathcal{N}=1$ chiral spectrum. A first set of chiral coordinates is given by the complex structure moduli $\tau^a$, with the identification $\tau^a\simeq \tau^a+1$. The K\"ahler moduli $s^a$ introduced in \eqref{het_Omega_J} combine with the $B_2$-axions delivering another set of chiral fields $t^a=a^a+\ii s^a$. 
Finally, the ten-dimensional dilaton $\phi$ enters the definition of an additional 4D chiral field, the universal modulus $t^0 =a^0+\ii s^0$, with
\begin{equation}
\label{T6hats}
s^0 = e^{-2\phi} V =e^{-2\phi} s^1s^2s^3 \,.
\end{equation}
The K\"ahler potential is given by the sum of  \eqref{hetK} and \eqref{hetCSK} and takes the form
\begin{equation}
\begin{aligned}
	K &= -\log s^0 -\sum^3_{a=1}\log s^a-\sum^3_{a=1}\log \Im  \tau^a\,.
\end{aligned}
\end{equation}

The dual saxions $\ell_i=(\ell_0,\ell_a,\lambda_a)$, corresponding to the saxions $s^i=(s^0,s^a,\Im\tau^a)$ respectively, can be straightforwardly computed from the definition \eqref{dualfields}:
\begin{equation}
\ell_0=\frac{1}{2 s^0}\quad,\quad \ell_a=\frac{1}{2 s^a}\quad,\quad \lambda_a=\frac{1}{2 \Im\tau^a}
\end{equation}
Let us then distinguish the corresponding string charges as ${\bf e}=(\hat e, \vec e,\vec\epsilon)=(\hat e, e_a,\epsilon_a)$.

The flow of the saxion $s^0$ induced by an F1 string was already analyzed in Section~\ref{sec:hetCY}. Instead, the flow which drives the saxions $s^a$ towards large values is induced by NS5 strings and can be easily inferred from the general discussion of Section~\ref{sec:hetCY}. Indeed, being $\kappa_{123}$ the only non-vanishing intersection number, the toroidal orbifold case corresponds to the Case 3 of Table~\ref{tab:HetNS5}: the string tension scales in the same manner as both the M-theory KK scale and the tension of the lightest membranes.

However, the toroidal case offers the opportunity to examine more explicitly the flow of the complex structure moduli $\tau^a$. The strings inducing such flows are identified with 10D KK monopoles wrapping some internal divisors and have charges ${\bf e}=(0,\vec 0,\vec\epsilon)$. The corresponding flow \eqref{ssssflow} becomes
\begin{equation}
{\rm Im}\, \tau^a={\rm Im}\, \tau^a_{0} + \epsilon^a \sigma
\end{equation}
and drives ${\rm Im}\, \tau^a$ to large values when $\sigma \to \infty$. Thus, along these flows, the string tensions scale as
\begin{equation}
\mathcal{T}_{\rm str}=M^2_{\rm P}\epsilon^a\lambda_a \sim \frac{M^2_{\rm P}}{\sigma} \to 0\,.
\end{equation}

First, let us consider an elementary flow, say generated by $\vec\epsilon=(1,0,0)$. (More general flows with $\vec\epsilon$ of the form  $(\epsilon^1,0,0)$, $(0,\epsilon^2,0)$ and $(0,0,\epsilon^3)$ lead to identical conclusions.) In this case only  $\tau^1$ flows, driving ${\rm Im}\,\tau^1 \to \infty$. Recalling \eqref{hetT6R} we see that $R_1\sim (\Im\tau)^{-\frac12}$ and $R_2\sim (\Im\tau)^{\frac12}$. Hence, the tower scale $m_*$ can be identified with  either  the lightest  KK mode along $R_2$ or  the lightest winding mode  along $R_1$.  The resulting behaviour is qualitatively the same. For instance, by using the KK scale we get:
\begin{equation}\label{hetKKtorus}
m_{*}^2 = m_{\rm KK}^2= \frac{(2\pi)^2 e^{2A}}{R_{2}^2\, l^2_{\rm s}} \sim \frac{M^2_{\rm P}}{\sigma }
\end{equation}
where we have employed \eqref{hetU}. 
The lightest membranes correspond to NS5-branes wrapping the $y^1$-direction and a two-cycle along $(y^3,\ldots, y^6)$. The corresponding scale \eqref{mmem} falls off as
\begin{equation}
\label{mmem_het_tor_el}
E^2_{\rm mem}\simeq  M^2_{\rm P} e^{K} \sim \frac{M^2_{\rm P}}{\sigma}\,.
\end{equation}

Thus, the tension of an elementary string scales in the same way as both $m^2_*$ and the membrane scale $E_{\rm mem}^2$. Then, \eqref{m*} and \eqref{mmem-m*} are realised with ${\it w}=n=1$. These results are collected in Table~\ref{tab:het_tor_el}.

\begin{table}[H]
	\centering
	\begin{tabular}{c | c | c|}
		\hhline{~|-|-|}
		& \cellcolor{gray!30}$\sigma^{-\frac12}$ & \cellcolor{gray!30}$\sigma^{-\frac16}$  \\ \hline
		\multicolumn{1}{|c|}{\cellcolor{gray!30}$\mathcal{T}_{\rm F1}^{1/2}$}                                            & KK monopole    &   \\ \hline
		\multicolumn{1}{|c|}{\cellcolor{gray!30}$m_*$}         &       $m_{\rm KK}$               &        \\ \hline
		\multicolumn{1}{|c|}{\cellcolor{gray!30}$E_{\rm mem}$}                                &      NS5 & \\ \hline
		\multicolumn{1}{|c|}{\cellcolor{gray!30}$\mathcal{T}^{1/3}_{\rm mem}$}                                &     & NS5 \\ \hline
	\end{tabular}
	\caption{Asymptotic mass scalings along   CS flow $\Im\tau^1\sim\sigma\rightarrow \infty$.\label{tab:het_tor_el}}
\end{table}

It is also instructive to consider more general CS flows, along which more  saxions $\Im\tau^a$  are sent to large values. It is easy to realise that the asymptotic behaviour of $m_*$ is  always as in \eqref{hetKKtorus}, and then the scaling weight is ${\it w}=1$ for all these flows. Instead, the scaling of the lightest relevant membrane scales changes. Consider for instance a flows generated by $\epsilon^a>0$, $\forall a$. The relevant scalings are summarised in the following table, and imply that \eqref{mmem-m*} is realised with $n=3$.

\begin{table}[H]
	\centering
	\begin{tabular}{c | c | c |}
		\hhline{~|-|-|}
		& \cellcolor{gray!30}$\sigma^{-\frac32}$ & \cellcolor{gray!30}$\sigma^{-\frac12}$\\ \hline
		\multicolumn{1}{|c|}{\cellcolor{gray!30}$\mathcal{T}_{\rm F1}^{1/2}$}                                            &  &  KK monopole \\ \hline
		\multicolumn{1}{|c|}{\cellcolor{gray!30}$m_*$}       &  &       $m_{\rm KK}$                       \\ \hline
		\multicolumn{1}{|c|}{\cellcolor{gray!30}$E_{\rm mem}$}                                &      NS5 & \\ \hline
		\multicolumn{1}{|c|}{\cellcolor{gray!30}$\mathcal{T}^{1/3}_{\rm mem}$}                                &     & NS5 \\\hline
	\end{tabular}
	\caption{Asymptotic mass scalings along   CS flow $\Im\tau^a\sim\sigma\rightarrow \infty$, $\forall a$.}
\end{table}

One can similarly work out the case of flows associated with only two non-vanishing charges $\epsilon^a$, e.g. $\vec\epsilon=(1,1,0)$. They all realise \eqref{mmem-m*} with $n=2$.

These findings also hold for the complex structure sector of the Type I compactification over the toroidal orbifold $X=T^6/(\mathbb{Z}_2\times \mathbb{Z}_2')$. In fact, as outlined in Section~\ref{sec:typeI}, the complex structure moduli of Type I EFTs are described by the same K\"ahler potential as those of heterotic EFTs. Therefore, the relations among the mass scales for complex structure flows is the same between the two models.

\subsection{F-theory K\"ahler flows on the resolved base  \texorpdfstring{\hbox{$T^6/(\mathbb{Z}_2)^3$}}{T6/Z2to3}}
\label{sec:Fresolved}

\noindent Here we discuss the F-theory model discussed in \cite{Denef:2005mm}. Its  base space $X$   can be identified with one of the possible resolutions of $T^6/(\mathbb{Z}_2)^3$, which is the orientifold projection of a $T^6/(\mathbb{Z}_2\times \mathbb{Z}_2)$ orbifold and can be identified with $\mathbb{P}^1\times \mathbb{P}^1\times \mathbb{P}^1$. The elliptic fiber degenerates to a $I^*_0$ Kodaira singularity along $12$ planes, each representing one $O7$ plus 4 $D7$ branes. These intersect along  $4\times 4\times 3=48$  lines at which the Weierstrass model become non-minimal, which in turn intersect at $64$ points, representing  O3-planes. The singular lines can be resolved in different ways by blowing up 48 exceptional divisors $E_{1\alpha,2\beta}, E_{2\beta,3\gamma}, E_{3\gamma,1\alpha}$, with $\alpha,\beta,\gamma=1,\ldots,4$. The different possible resolutions of the base are related by flips.
As described in detail in \cite{Denef:2005mm}, these flips of $X$ can be regarded as orientifold projections of  flops of the Calabi--Yau double cover $\hat X$, connecting the different possible resolutions of the   orbifold $T^6/(\mathbb{Z}_2\times \mathbb{Z}_2)$. 

As in  \cite{Denef:2005mm}, one can choose as basis $D^a$, $a=1,\ldots, 51$, of $H_4(X,\mathbb{Z})$ the effective divisors  $(R_1,R_2,R_3,E_{1\alpha,2\beta}, E_{2\beta,3\gamma}, E_{3\gamma,1\alpha})$, where $R_1,R_2,R_3$ correspond to 3 $\mathbb{P}^1\times \mathbb{P}^1$ movable divisors of the base. Note that the divisors $R_1,R_2,R_3$ are `non-elementary'  effective divisors. For instance, we can decompose $R_1$ in the following linearly equivalent way 
\be\label{lindec}
R_1\simeq D_{1\alpha} +\sum_{\beta}E_{1\alpha,2\beta}+\sum_{\gamma}E_{3\gamma,1\alpha}\quad~~~~ \forall\alpha=1\,\dots,4\,,
\ee
where $D_{1\alpha}$ are 4  rigid effective divisors 
 wrapped by  4 stacks of  7-branes, each supporting a pure SO$(8)$ SYM sector. 
By cyclically permuting the indices of \eqref{lindec} one gets similar relations for $R_2$ and $R_3$, in terms of effective divisors $D_{2\beta}$  and $D_{3\gamma}$. One can then take  $D_{i\alpha}$ and $E_{i\alpha,j\beta}$ as generators of the (non-simplicial) cone Eff$^1(X)$. 
If in \eqref{FKsaxions} we use the above basis of divisors $D^a$ to identify the set of saxions  $s^a=\{s^i,s^{i\alpha,j\beta} \}$,  the saxionic cone is defined by
\be
\begin{aligned}
\Delta=\left\{\begin{array}{l}s^{1\alpha,2\beta}>0 \quad\text{(for any $\alpha,\beta$)}\quad\text{and $(1,2,3)$ cyclic permutations},\\
 s^1-\sum_{\beta}s^{1\alpha,2\beta}-\sum_{\gamma}s^{3\gamma,1\alpha}>0\quad\text{(for any $\alpha$)}\quad \text{and $(1,2,3)$ cyclic permutations}\,.
 \end{array}
 \right.
\end{aligned}
\ee 

In order to identify the EFT strings, we pick the following dual basis of curves $C_a=(C_i,C_{i\alpha,j\beta})$:
\be
C_1=R_2 R_3\,,\quad  C_{1\alpha,2\beta}=-E_{1\alpha,2\beta} R_3\quad\quad\text{and $(1,2,3)$ cyclic permutations}
\ee
such that $C_a\cdot D^b=\delta_a^b$. Note that $C_{i\alpha,j\beta}$ are not effective. Then, a BPS string corresponding to the charges  \be
{\bf e}=e^a\, C_a= e^i\,C_i+e^{1\alpha,2\beta}C_{1\alpha,2\beta}+e^{2\beta,3\gamma}C_{2\beta}+e^{3\gamma,1\alpha}C_{3\gamma,1\alpha}\in {\rm Eff}_1(X)_{\mathbb{Z}}
\ee
is an EFT string if ${\bf e}$ belongs to
\be\label{DenefDelta}
\begin{aligned}
\cC_{\rm S}^{\text{\tiny EFT}}=\left\{\begin{array}{l}e^{1\alpha,2\beta}\geq 0 \quad\text{(for any $\alpha,\beta$)}\quad\text{and $(1,2,3)$ cyclic permutations},\\
 e^1-\sum_{\beta}e^{1\alpha,2\beta}-\sum_{\gamma}e^{3\gamma,1\alpha}\geq 0\quad\text{(for any $\alpha$)}\quad \text{and $(1,2,3)$ cyclic permutations}\,.
 \end{array}
 \right.
\end{aligned}
\ee
Note that these comments, and in particular $\cC_{\rm S}^{\text{\tiny EFT}}$,  are independent of the chosen resolution. 

The simplest possible choice of EFT string is given by $e^{i\alpha,j\beta}=0 $, so that  $e^i\geq 0$. According to \eqref{FtheoryEFTst}, these EFT strings correspond to D3-branes wrapping the movable curve ${\bf e}=e^i C_i$. In this case the corresponding flow $s^{i}=s^i_0+e^i\sigma$, $s^{i\alpha,j\beta}\equiv s^{i\alpha,j\beta}_0$ describes a rescaling of the movabile divisors $R_i$, while the volume of the exceptional divisors $E_{i\alpha,j\beta}$ remains constant. The K\"ahler form can be expanded in the following way 
\eqref{FTJexp}:
\be
J=\sum_iv_i[R_i]+\sum_{\alpha,\beta}v_{1\alpha,2\beta}[E_{1\alpha,2\beta}]+\sum_{\beta,\gamma}v_{2\beta,3\gamma}[E_{2\beta,3\gamma}]+\sum_{\gamma,\alpha}v_{3\gamma,1\alpha}[E_{3\gamma,1\alpha}]\,,
\ee
but  intersection numbers and volumes of the exceptional divisors depend on the  K\"ahler cone chamber of $\cK(X)_{\rm ext}$.
For simplicity, we restrict to  the  chamber  corresponding to the `symmetric' resolution \cite{Denef:2005mm} of all singularities, in which
the intersection numbers are given by
\be
\begin{aligned}
\cI=&\,R_1R_2R_3+\Big(-\sum_{\alpha\beta}E^2_{1\alpha,2\beta}R_3-\ldots\Big)+\Big(2\sum_{\alpha\beta}E^3_{1\alpha,2\beta}+\ldots\Big)\\
&+\Big[-\frac12\sum_{\alpha\beta\gamma}E_{1\alpha,2\beta}(E^2_{2\beta,3\gamma}+E^2_{3\gamma,1\alpha})-\ldots\Big]+\frac12\sum_{\alpha\beta\gamma}E_{1\alpha,2\beta}E_{2\beta,3\gamma}E_{3\gamma,1\alpha}
\end{aligned}
\ee
where $\ldots$ are $(1,2,3)$ cyclically permuted terms and the half-integral values is due to the presence of 64 O3-planes. The K\"ahler cone is defined by the following conditions \cite{Denef:2005mm}: $v_i+\sum_\alpha v_{i\alpha,j\beta}> 0$, $v_{1\alpha,2\beta}- v_{2\beta,3\gamma}- v_{3\gamma,1\alpha}>0$ and $(1,2,3)$ cyclic permutations, from which we can also extract the generators of ${\rm Eff}_1(X)$.
Note that the these conditions imply that $v_{i\alpha,j\beta}<0$, which is compatible with the fact that the curves $C_{i\alpha,j\beta}$ are {\em anti}-effective.
Given the structure of the symmetric resolution, it is consistent to further restrict to the subcone of $\cK(X)$ in which  $v_{i\alpha,j\beta}\equiv v_{i,j}$ and $s^{i\alpha,j\beta}\equiv s^{i,j}$ for any $\alpha,\beta=1,\ldots,4$. The relation between the saxions and the K\"ahler moduli then reduces to 
\be
\begin{aligned}
\label{Ex3_sv}
s^1&=v_2v_3-8 v^2_{2,3} 
\,,\\
s^{2,3}&=- v_1 v_{2,3} +v_{2,3}^2-v_{1,2}^2-v_{3,1}^2  -2 v_{2,3} (v_{1,2}+v_{3,1}) + 2 v_{1,2} v_{3,1} 
\end{aligned}
\ee
and $(1,2,3)$ cyclic permutations. 

We first consider a flow generated by a set of positive charges $e^1,e^2,e^3>0$, say ${\bf e}=[C_1]+[C_2]+[C_3]$. In this case all the saxions associated to the movable divisors $R_i$ flow as $s^{i}=s^i_0+\sigma$ with $\forall\, i$, while the remaining saxions are kept fixed to constant values, $s^{i,j}\equiv s^{i,j}_0$. Asymptotically as $\sigma \to \infty$, it can be seen that along the flow the volumes $v_i$ of the curves $C_i$ scale as $v_i \sim \sqrt{\sigma}$, while the volumes of the curves $C_{i\alpha,j\beta}$ fall off as $|v_{i,j}| \sim \frac{1}{\sqrt{\sigma}}$. Thus, the internal volume scales as $V_X\sim \sigma^{\frac32}$. Recalling \eqref{WeylFth}, we now can estimate the KK-scale as $m_*^2 \sim \frac{e^{2A}}{l_{\rm s}^2 R^2_{*}} \sim \frac{M^2_{\rm P}}{V_X R^2_{*}}$, where $R_*$ denotes the Einstein frame radius which grows faster along the flow, namely $R_*^2 \sim v_i \sim \sqrt{\sigma}$. Thus, $m_*^2 \sim M^2_{\rm P}\, \sigma^{-2}$ and from \eqref{ellFth} one can explicitly check that the tension $\mathcal{T}_{\bf e}=M^2_{\rm P}e^i\ell_i$ of the string generating the flow scales as $ \mathcal{T}_{\bf e}\sim M_{\rm P}^2\, \sigma^{-1}$, as expected. Therefore, Conjecture \ref{conj:cutoff} is realised with scaling weight ${\it w}=2$.

Consider now the flow corresponding to one vanishing $e^i=0$, say ${\bf e}=[C_1]+[C_2]$. Only the saxions $s^1$ and $s^2$ flow as $s^i = s^i_0 + \sigma$ for $i=1,2$, while the remaining ones are fixed $s^3 \equiv s^3_0$, $s^{i,j}\equiv s^{i,j}_0$. Only the volume of the curve $C_3$ is driven to large value, as $v_3 \sim \sigma$, while the volume $v_{1,2}$ falls off as $v_{1,2} \sim \sigma^{-1}$; the volumes of all the other curves are not affected by the flow and we can assume them to be fixed at some finite values. Since the full internal volume scales as $V_X \sim \sigma$ and the radius with maximal growth scales as $R_*^2 \sim \sigma$, we can estimate $m_*^2 \sim M_{\rm P}^2\sigma^{-2}$ and thus, as above,  Conjecture \ref{conj:cutoff} holds with scaling weight ${\it w}=2$.

Finally, consider an elementary charge, say ${\bf e}=[C_1]$. In this case, solely the saxion $s^1$ flows as $s^1 = s^1_0 + \sigma$, with all the other saxions are fixed at constant values. Asymptotically along the flow, the volumes of the curves $C_2$ and $C_3$ grow as $v_2 \sim v_3 \sim \sqrt{\sigma}$, while the volumes of $C_1$, $C_{1,2}$, $C_{3,1}$ shrink as $v_1 \sim v_{1,2} \sim v_{3,1} \sim \frac{1}{\sqrt{\sigma}}$. This implies that the volume scales as $V_X \sim \sqrt{\sigma}$ and the maximal radius as $R^2_* \sim \sqrt{\sigma}$. Hence $m_*^2 \sim M_{\rm P}^2\,\sigma^{-1}$ and  we see that Conjecture \ref{conj:cutoff} is now satisfied with scaling weight ${\it w}=1$.

In the case of some non-vanishing charges $e^{i\alpha,j\beta}>0$,     the volumes of the corresponding exceptional divisors $E_{i\alpha,j\beta}$  grow along the flow. However, from \eqref{DenefDelta} it follows that also $e^i,e^j>0$ and then the volume of the moving divisors $R_i,R_j$ should also increase along the flow. Hence, we expect the above conclusions to be qualitatively valid in these cases as well, but it would be interesting to explicitly test our expectations in all these cases as well as in other models along the same lines.

Let us now investigate the mass scalings along the flows that involve the complex structure moduli. The model exhibits only three complex structure moduli which are packaged within three chiral fields $\tau^i$, $i=1,2,3$. The identification $\tau^i \simeq \tau^i +1$ singles out their real parts ${\hat a}^i \equiv {\rm Re}\,\tau^i$ as the axions, while the imaginary parts ${\hat s}^i \equiv {\rm Im}\,\tau^i$ are their saxionic partners. These determines the size of the internal (Einstein frame) radii as
\begin{equation}
\label{IIB_CS_Ri}
    R^{2i-1}\simeq \frac{1}{\sqrt{\hat{s}^i}}\,, \qquad R^{2i}\simeq \sqrt{\hat{s}^i}\,,
\end{equation}
up to irrelevant constant, for fixed K\"ahler moduli and in the limit of negligible blown-up modes.

The complex structure moduli space is then described by the K\"ahler potential
\begin{equation}
\label{IIB_CS_K}
    K = - \log \hat{s}^1 \hat{s}^2 \hat{s}^3\,.
\end{equation}
Therefore, the dual saxions \eqref{dualfields} are 
\begin{equation}
\label{IIB_CS_l}
    {\hat \ell}_i = \frac{1}{2 \hat{s}^i}\,.
\end{equation}
The strings which drive the complex structure flows display the tension
\begin{equation}
\label{IIB_CS_T}
    \mathcal{T}_{\hat{\bf e}} = M_{\rm P}^2 \hat{e}^i{\hat \ell}_i\,.
\end{equation}
Here we have collected the charges as $\hat{\bf e} = (\hat{e}^1,\hat{e}^2,\hat{e}^3)$.

Consider a string with elementary charge $\hat{\bf e}=(1,0,0)$ (the other elementary flow cases $\hat{\bf e}=(0,1,0)$ and $\hat{\bf e}=(0,0,1)$ can be treated analogously). The string induces a flow that drives $\hat{s}^1$ to the boundary of the moduli space as $\hat{s}^1 \sim \sigma \to \infty$, while keeping the remaining saxions fixed $\hat{s}^2 = \hat{s}^2_0$, $\hat{s}^3 = \hat{s}^3_0$. The tower scale is set by the lightest KK modes. Such a scale can be estimated by noticing that, along the flow, the radius, among \eqref{IIB_CS_Ri}, that diverge more rapidly is $R^2$ as $(R^2)^2\equiv R_*^2 \sim \hat{s}^1 \sim \sigma$. Thus, employing \eqref{WeylFth}, we get $m_*^2 \sim \frac{M_{\rm P}^2}{R_*^2} \sim M_{\rm P} \sigma^{-1}$. Alternatively, the tower scale is equivalently set by the winding modes, owing to a shrinking internal radius $(R_1)^2 \equiv R_*^2 \sim \frac{1}{\hat{s}^1} \sim \sigma^{-1}$. Thus we recognise that Conjecture~\ref{conj:cutoff} here holds with the scaling weight ${\it w}=1$.

Flows induced by non-elementary string charges $\hat{\bf e}=(\hat{e}^1,\hat{e}^2,\hat{e}^3)$, with $\hat{e}^i>0$, also realise Conjecture~\ref{conj:cutoff} with the scaling weight ${\it w}=1$. As above, the tower of states can be identified with either the KK or winding modes. In fact, along any of these flow it is possible to single out an internal radius among \eqref{IIB_CS_Ri} which grows as $R_{*}^2\sim \sigma$ determining the lightest KK modes, and another which shrinks as $R_{*}^2\sim \sigma^{-1}$, which instead determines the lightest winding modes. In both cases the tower scale falls off as $m_*^2 \sim M_{\rm P}^2 \sigma^{-1}$. Being this the same behaviour as the string tension, Conjecture~\ref{conj:cutoff} is verified with ${\it w}=1$.

\subsection{M-theory flows on the toroidal orbifold  \texorpdfstring{\hbox{$T^7/(\mathbb{Z}_2)^3$}}{T7/Z2to3}}
\label{sec:Mthtoroidal}

\noindent In this section we delve into the details of the analysis of string flows in M-theory toroidal orbifold models. We consider Joyce's models \cite{joyce1996a,joyce1996b,joyce2000compact}, which are  introduced in section~\ref{ss:Mth}. Specifically, the compactification manifold with $G_2$-holonomy is a toroidal orbifold $T^7/\Gamma$, and we here choose the orbifold $\Gamma$ to be $\Gamma=\mathbb{Z}_2\times\mathbb{Z}_2\times\mathbb{Z}_2$. As  in section~\ref{ss:Mth}, we parametrise the toroidal coordinates with $y_a$, $a=1,\ldots,7$, with the identification $y_a\simeq y_a+1$, and we denote the radii of the torii with $R_a$. The generators of the orbifold $\mathbb{Z}_2\times\mathbb{Z}_2\times\mathbb{Z}_2$ act on the coordinates $y_a$ as in \eqref{Joyceproj}. 

The model is characterised by a set of seven saxions $s^a$, defined by 
\be
\label{Mth_sa}
s^a=\int_{C^a}\Phi\,.
\ee
Here $\Phi$ is the associative three-form \eqref{orbPhi} that is invariant under the orbifold action \eqref{Joyceproj} and $C^a$ is a basis of three-cycles of $H_3(T^7/\Gamma,\mathbb{Z})$. For example, the cycle $C^1$ can be chosen to span the directions $(y_1,y_2,y_3)$ that are preserved under \eqref{Joyceproj}, $C^2$ spans $(y_1,y_4,y_5)$, and similarly with the remaining. Explicitly, plugging \eqref{orbPhi} in \eqref{Mth_sa}, we find the following relation among the saxions $s^a$ and the radii $R_a$:
\begin{equation}
\label{Mth_sR}
\begin{aligned}
&s^1=R_1 R_2 R_3\,,\quad   & s^2 = R_1 R_4 R_5\,,\quad   &  s^3 =  R_1 R_6 R_7 \quad   & s^4 = R_2 R_4 R_6\,,\\
&s^5=R_2 R_5 R_7\,,\quad   & s^6 = R_3 R_4 R_7\,,\quad   &  s^7 =  R_3 R_5 R_6\,,
\end{aligned}
\end{equation}
which can be inverted as
\begin{equation}
\label{Mth_Rs}
\begin{aligned}
&R_1=\frac{(s^1s^2s^3)^\frac13}{(s^4s^5s^6 s^7)^\frac16}\,,\quad   R_2=\frac{(s^1s^4s^5)^\frac13}{(s^2s^3s^6 s^7)^\frac16}\,,\quad   R_3=\frac{(s^1s^6s^7)^\frac13}{(s^2s^3s^4 s^5)^\frac16}\,,\\
&R_4=\frac{(s^2s^4s^6)^\frac13}{(s^1s^3s^5 s^7)^\frac16}\,,\quad   R_5=\frac{(s^2s^5s^7)^\frac13}{(s^1s^3s^4 s^6)^\frac16}\,,\quad   R_6=\frac{(s^3s^4s^7)^\frac13}{(s^1s^2s^5 s^6)^\frac16}  \,,\\
&R_7=\frac{(s^3s^5s^6)^\frac13}{(s^1s^2s^4 s^7)^\frac16}\,.
\end{aligned}
\end{equation}
Thus, the internal volume is expressed in terms of the saxions as follows:
\begin{equation}
\label{Mth_V}
V_X = R_1 R_2 \ldots R_7 = (s^1 s^2 \ldots s^7)^{\frac13}\,.
\end{equation}

The K\"ahler potential acquires the simple form
\begin{equation}
    K = - \log (s^1 s^2 \ldots s^7)\,,
\end{equation}
and the dual saxions \eqref{dualfields} are
\begin{equation}
\label{Mth_ell}
   \ell_a = \frac{1}{2 s^a}\,.
\end{equation}

In 4D, strings originate from M5 branes which wrap cycles in $H_4(T^7/\Gamma,\mathbb{Z})$. By choosing a basis of four-cycles $\tilde{C}_a$ dual to $C^a$, the tension of a string that stem from an M5 brane wrapped over the cycle $\tilde C = e^a \tilde{C}_a $ can be computed as follows
\begin{equation}
\label{Mth_Ts}
   \mathcal{T}_{\bf e} = \frac{2\pi}{l_M^4} e^{2A} \int_{\tilde{C}} * \Phi = M_{\rm P}^2 e_a \ell^a\,.
\end{equation}
In the last step we have employed \eqref{MUM} and then re-expressed the tension in terms of the dual saxions \eqref{Mth_ell}. We will also collect ${\bf e}=(e_a)$.

In the following, in order to investigate the validity of Conjecture~\ref{conj:cutoff} and compute the scaling weight ${\it w}$ for various string flows, we need to specify the scale $m_*$ at which the 4D EFT breaks down and that has to be related to the string tension \eqref{Mth_Ts}. The natural candidates are the lightest among KK, whose masses are set by
\begin{equation}
\label{Mth_KKw}
m^2_{\rm KK} \simeq \frac{M_{\rm P}^2}{R^2_* V_X}\,,
\end{equation}
up to irrelevant constants. Here $R_*$ generically denotes any of the radii $R_a$. The EFT breaking scale $m_*$ is then set by any of the scales \eqref{Mth_KKw} which displays the fastest fall off along the given string flow. 

For completeness, in the flows examined below we will also relate the mass scale \eqref{mmem} set by the lightest membranes. In 4D, membranes can originate from either M2 branes spanning three external spacetime directions or M5 branes wrapped on linear combinations of three-cycles $C^a$. In either case, the scale \eqref{mmem} is
\begin{equation}
\label{Mth_mmem}
   E_{\rm mem} \sim e^{\frac K2} M_{\rm P}=  \frac{M_{\rm P}}{\sqrt{s^1 s^2 \ldots s^7}}\,.
\end{equation}

Let us now examine some relevant string flows and how Conjecture~\ref{conj:cutoff} is realised.

\begin{description}
\item[${\bf e} = (1,0,0,0,0,0,0)$] Elementary flows are generated by strings with a single non-vanishing, positive charge $e^a$. Such a string can be obtained from an M5 brane wrapped over any of the four-cycles $\tilde{C}_a$. As an example, we may consider the flow generated by the string with charge $e^1 >0$, while $e^a = 0$ $\forall a \neq 1$. The flow generated by this string drives only the saxion $s^1$ to infinite distance, that is $s^1 \sim e^1 \sigma \to \infty$. The string generating the flow becomes tensionless with a fall off $\mathcal{T}_{\rm str} \sim M_{\rm P}^2 \sigma^{-1}$. As can be easily seen from \eqref{Mth_Rs}, along such a flow  $R_1, R_2, R_3 \sim \sigma^{\frac13}$, while $R_4, R_5, R_6, R_7 \sim \sigma^{-\frac16}$. Thus, the lightest modes among those in \eqref{Mth_KKw} are the KK modes associated to $R_* = R_1, R_2, R_3$. Their masses scale as
\begin{equation}
m_*^2 = m_{\rm KK}^2 \simeq \frac{M_{\rm P}^2}{\sigma}\,.  
\end{equation}
Therefore, Conjecture~\ref{conj:cutoff} is realised with the scaling weight ${\it w}=1$. It can be easily checked that the scale \eqref{Mth_mmem} set by lightest membranes displays an analogous scaling along this elementary flow. These findings are summarised in the following table:

\begin{table}[H]
	\centering
	\begin{tabular}{c | c | c |}
		\hhline{~|-|-|}
		& \cellcolor{gray!30}$\sigma^{-\frac12}$ & \cellcolor{gray!30}$\sigma^{-\frac16}$  \\ \hline
		\multicolumn{1}{|c|}{\cellcolor{gray!30}$\cT_{\rm str}^{1/2}$}                                             & M5  &   \\ \hline
		\multicolumn{1}{|c|}{\cellcolor{gray!30}$m_*$}         &       $m_{\rm KK}$       &                \\ \hline
		\multicolumn{1}{|c|}{\cellcolor{gray!30}$E_{\rm mem}$}                                &      M2, M5 & \\ \hline
		\multicolumn{1}{|c|}{\cellcolor{gray!30}$\mathcal{T}^{1/3}_{\rm mem}$}                                &   &    M2, M5  \\ \hline
	\end{tabular}
	\caption{Mass scalings along the flow generated by an elementary string obtained from an M5 wrapped over the four-cycle $\tilde{C}_a$.\label{t:Mth_el}}
\end{table}

\item[${\bf e} = (1,1,0,0,0,0,0)$] A non-elementary string with only non-null charges $e^1 = e^2 =1$ induces the flow $s^1 \sim s^2 \sim \sigma \to \infty$, while the other saxions remain fixed. The tower scale $m_*$ is set by lightest KK modes, associated to $R_* = R_1$ in \eqref{Mth_KKw}. These scale as
\begin{equation}
m_*^2 = m_{\rm KK}^2 \simeq \frac{M_{\rm P}^2}{\sigma^2}\,,
\end{equation}
realising Conjecture~\ref{conj:cutoff} with scaling weight ${\it w}=2$. 

\begin{table}[H]
		\centering
		\makebox[\linewidth]{\begin{tabular}{c | c | c | c |}
			\hhline{~|-|-|-|}
			  & \cellcolor{gray!30}$\sigma^{-1}$ & \cellcolor{gray!30}$\sigma^{-\frac12}$ & \cellcolor{gray!30}$\sigma^{-\frac13}$ 
			 \\ \hline
			\multicolumn{1}{|c|}{\cellcolor{gray!30}$\cT_{\rm str}^{1/2}$}                    
			      &         &  M5 &
			\\ \hline
			\multicolumn{1}{|c|}{\cellcolor{gray!30}$m_*$}        &    $m_{\rm KK}$     & &
			\\ \hline
			\multicolumn{1}{|c|}{\cellcolor{gray!30}$E_{\rm mem}$}        &    M2, M5     & &
			\\ \hline
			\multicolumn{1}{|c|}{\cellcolor{gray!30}$\mathcal{T}^{1/3}_{\rm mem}$}        &   & &  M2, M5     
			\\ \hline
		\end{tabular}}
		\caption{Mass scalings along the flows generated by the string with charges $(1,1,0,0,0,0,0)$.\label{t:Mth_nonel_1}}
\end{table}

\item[${\bf e} = (1,1,1,0,0,0,0)$] Consider a non-elementary string with only non-null charges $e^1 = e^2 = e^3 = 1$. This string induces a flow that drives $s^1 \sim s^2 \sim s^3 \to \infty$. The tower scale $m_*$ is here identified with the KK modes associated to $R_* = R_1$ and their masses fall off as
\begin{equation}
m_*^2 = m_{\rm KK}^2 \simeq \frac{M_{\rm P}^2}{\sigma^{3}}\,.  
\end{equation}
Thus, Conjecture~\ref{conj:cutoff} is realised with scaling weight ${\it w}=3$. The relevant scales are summarised in the following table:

\begin{table}[H]
		\centering
		\makebox[\linewidth]{\begin{tabular}{c | c | c| }
			\hhline{~|-|-|}
			  & \cellcolor{gray!30}$\sigma^{-\frac32}$ & \cellcolor{gray!30}$\sigma^{-\frac12}$ 
			 \\ \hline
			\multicolumn{1}{|c|}{\cellcolor{gray!30}$\cT_{\rm str}^{1/2}$}                    
			      &         &  M5
			\\ \hline
			\multicolumn{1}{|c|}{\cellcolor{gray!30}$m_*$}        &    $m_{\rm KK}$     & 
			\\ \hline
			\multicolumn{1}{|c|}{\cellcolor{gray!30}$E_{\rm mem}$}        &    M2, M5     & 
			\\ \hline
			\multicolumn{1}{|c|}{\cellcolor{gray!30}$\mathcal{T}^{1/3}_{\rm mem}$}        &        &  M2, M5
			\\ \hline
		\end{tabular}}
		\caption{Mass scalings along the flows generated by the string with charges $(1,1,1,0,0,0,0)$.\label{t:Mth_nonel_2}} 
\end{table}

\item[${\bf e} = (1,1,0,1,0,0,0)$] Consider non-elementary strings with only non-null charges $e^1=e^2=e^4=1$. Since $s^1 \sim s^2 \sim s^4 \sim \sigma \to \infty$, the tower scale $m_*$ can be either fixed by the KK scale (for $R_* = R_1, R_2, R_4$). These scale as
\begin{equation}
m_*^2 = m_{\rm KK}^2 .  
\end{equation}
Conjecture~\ref{conj:cutoff} is realised with scaling weight ${\it w}=2$. The behavior of the relevant scales along such a flow is summarised in the table below.

\begin{table}[H]
		\centering
		\makebox[\linewidth]{\begin{tabular}{c | c | c | c|}
			\hhline{~|-|-|-|}
			  & \cellcolor{gray!30}$\sigma^{-\frac32}$ & \cellcolor{gray!30}$\sigma^{-1}$ & \cellcolor{gray!30}$\sigma^{-\frac12}$ 
			 \\ \hline
			\multicolumn{1}{|c|}{\cellcolor{gray!30}$\cT_{\rm str}^{1/2}$}                    
			    &  &         &  M5
			\\ \hline
			\multicolumn{1}{|c|}{\cellcolor{gray!30}$m_*$}      &  &    $m_{\rm KK}$      & 
			\\ \hline
			\multicolumn{1}{|c|}{\cellcolor{gray!30}$E_{\rm mem}$}        &    M2, M5     &  &
			\\ \hline
			\multicolumn{1}{|c|}{\cellcolor{gray!30}$\mathcal{T}^{1/3}_{\rm mem}$}        &   &  & M2, M5  
			\\ \hline
		\end{tabular}}
		\caption{Mass scalings along the flows generated by the string with charges $(1,1,0,1,0,0,0)$.\label{t:Mth_nonel_3}} 
\end{table}
\item[${\bf e} = (1,1,1,1,1,1,1)$] As last example, consider non-elementary strings with all non-null charges. For example, choose $e^a=1$ $\forall a$, resulting in a flow that drives all the saxions to infinite distance $s^a \sim \sigma \to \infty$. The tower scale $m_*$ is fixed by the KK scale (for $R_*$ given by any of the radii in \eqref{Mth_Rs}), which falls off as
\begin{equation}
m_*^2 = m_{\rm KK}^2 \simeq \frac{M_{\rm P}^2}{\sigma^{3}}\,.  
\end{equation}
Thus, as in the latter two cases above, Conjecture~\ref{conj:cutoff} is realised with scaling weight ${\it w}=3$. However, in this case, the membrane cutoff scale $\mathcal{T}^{1/3}_{\rm mem}$ falls off faster than the string scale $\mathcal{T}^{1/2}_{\rm str}$:
\begin{equation}
\mathcal{T}^{1/3}_{\rm mem} \sim \frac{M_{\rm P}}{\sigma^{\frac76}} \sim M_{\rm P} \left(\frac{\mathcal{T}_{\rm str}^{1/2}}{M_{\rm P}}\right)^{\frac73}\,.
\end{equation}
Indeed, from the Type IIA perspective, this flow corresponds to a large string coupling limit $g_{\rm s} \gg 1$ and one needs to employ the full M-theory description. The relevant scales for such a flow are summarised in the Table below.

\begin{table}[H]
		\centering
		\makebox[\linewidth]{\begin{tabular}{c | c | c | c | c |}
			\hhline{~|-|-|-|-|}
			  & \cellcolor{gray!30}$\sigma^{-\frac72}$ & \cellcolor{gray!30}$\sigma^{-\frac32}$ & \cellcolor{gray!30}$\sigma^{-\frac76}$ & \cellcolor{gray!30}$\sigma^{-\frac12}$ 
			 \\ \hline
			\multicolumn{1}{|c|}{\cellcolor{gray!30}$\cT_{\rm str}^{1/2}$}                    
			    &  &     &    &  M5
			\\ \hline
			\multicolumn{1}{|c|}{\cellcolor{gray!30}$m_*$}      &  &    $m_{\rm KK}$  &   & 
			\\ \hline
			\multicolumn{1}{|c|}{\cellcolor{gray!30}$E_{\rm mem}$} &  M2     &        &  &
			\\ \hline
			\multicolumn{1}{|c|}{\cellcolor{gray!30}$\mathcal{T}^{1/3}_{\rm mem}$}        &   &  & M2 & 
			\\ \hline
		\end{tabular}}
		\caption{Mass scalings along the flows generated by the string with charges $(1,1,1,1,1,1,1)$.\label{t:Mth_nonel_4}} 
\end{table}
\end{description}

The analysis above can be similarly carried for other non-elementary charges, leading to the classification of flows as in Table~\ref{t:Mth}.

%\bibliographystyle{jhep}
%\bibliography{references}

\providecommand{\href}[2]{#2}\begingroup\raggedright\endgroup

\end{document}